# The Abstract Structure of Quantum Algorithms

William J. Zeng

Oriel College

University of Oxford

A thesis submitted for the degree of

*Doctor of Philosophy*

Trinity 2015

To my family:
Bernard, Christopher, Grace, and Teresa.

*One should be light like a bird, and not like a feather.*
- Paul Valery

# Acknowledgements

Research is humbling work, and I mean this in the proudest way. I am deeply grateful for the time I've been given to pursue it. In no way would this have been possible without the dedication, encouragement, and support of many.

I begin with thanks to Michel Devoret and Rob Schoelkopf at Yale, and Andreas Wallraff at ETH Zurich, all of whom, during my undergraduate years, made quantum computation a reality for me. Prof. Devoret has my particular thanks, as, with considerable patience and insight, he witnessed my (at times bouncing) trajectory over the field since I first wrestled with the abelian hidden subgroup algorithm in his office in the summer of 2008. This is where my fascination with the structure of quantum algorithms began.

My mature study of the subject is with the tools and language taught to me in huge part, by Samson Abramsky, Bob Coecke, Chris Heunen, Aleks Kissinger, and Jamie Vicary. The support and self-directed agency that my supervisors Bob and Jamie encouraged in my studies were crucial. Bob has an independent sense, which, while rare enough as it is, is all the rarer with his match of openness and generosity; and I don't just mean at Club 13, Beijing. Jamie's eye for detail and deliberate approach were especially invaluable in guiding my initial transition from student-with-ideas to researcher. Our first work directly together (Section 4.2 in this thesis) gave me the initial confidence that has culminated in this thesis. Special thanks also goes to Stefano Gogioso. Even though we only have been working together for a year, he has been an important catalyst, and the backlog of our ideas is ever growing. How long would my working papers (esp. on Mermin non-locality) have sat alone in my office desk if he and Andre Ranchin had not continued to ask over them. Indeed this sense of shared ideas is one of the aspects I am most grateful for.

The approach to the framework presented here - for me an initially broad leap of mathematical background - would not have been in any way possible without my cohort in the Quantum Group at Oxford, especially Miriam Backens, Katriel Cohn-Gordon, Brendan Fong, Amar Hadzihasanovic, Dan Marsden, Shane Mansfield, Nadish de Silva, and Vladimir Zamdzhiev, who were learning alongside me. Other researchers in our group, including Niel de Beaudrap, Matty Hoban, Clare Horsman,

Kohei Kishida, and Ray Lal have made useful sounding boards for the work developed here, and I thank them for this.

There are many others young researchers that I have met at conferences and workshops during the course of my thesis study, who have helped me situate and clarify the ideas presented here. A, by no means exhaustive list, includes Ning Bao, Alex Kubica, Shaun Maguire, Raghu Mahajan, Sandra Rankovic, Jonathan Skowera, Nathaniel Thomas, and Michael Walter, and I am happy to count them as my friends. I would also like to especially thank Stephen Jordan and Ronald de Wolf for their suggestions and interest in the quantum algorithms work presented in Section 4.2.

My work in the final chapter of this thesis, on structural connections between quantum computation and a model in natural language processing, required quick acclimatization in a new field. For this I thank Stephen Clark, Dimitri Kartsaklis, Tamara Polajnar, Mehrnoosh Sadrzadeh and everyone else in the DisCoCat project for their openness and inclusiveness.

My academic life has benefited from the support of many other dear friends throughout my time at Oxford. My particular gratitude is to Merritt Moore, Ankur Desai, David Furlong, Andrew Lanham, Victor Pontis, Bary Pradelski, Max Roser, Spencer Salovaara, Lucas Zwirner, my entire OUBC family, my Rhodie family, Wire & String attendees and many others who I have had the privilege to invite to dinner. May I leave you un-puzzled, MM.

Finally, I am also glad to thank the students that I have had throughout the years as they have taught me so much.

> *And every science, when we understand it not as an instrument of power and domination but as an adventure in knowledge pursued by our species across the ages, is nothing but this harmony, more or less vast, more or less rich from one epoch to another, which unfurls over the course of generations and centuries, by the delicate counterpoint of all the themes appearing in turn, as if summoned from the void.*
>
> \- Alexandre Grothendieck

Thank you all, as I start to play my small part.

# Abstract

Quantum information brings together theories of physics and computer science. This synthesis challenges the basic intuitions of both fields. In this thesis, we show that adopting a unified and general language for process theories advances foundations and practical applications of quantum information.

Our first set of results analyze quantum algorithms with a process theoretic structure. We contribute new constructions of the Fourier transform and Pontryagin duality in dagger symmetric monoidal categories. We then use this setting to study generalized unitary oracles and give a new quantum blackbox algorithm for the identification of group homomorphisms, solving the GROUPHOMID problem. In the remaining section, we construct a novel model of quantum blackbox algorithms in non-deterministic classical computation.

Our second set of results concerns quantum foundations. We complete work begun by Coecke et al. [32, 34], definitively connecting the Mermin non-locality of a process theory with a simple algebraic condition on that theory's phase groups. This result allows us to offer new experimental tests for Mermin non-locality and new protocols for quantum secret sharing.

In our final chapter, we exploit the shared process theoretic structure of quantum information and distributional compositional linguistics. We propose a quantum algorithm adapted from [109] to classify sentences by meaning. The clarity of the process theoretic setting allows us to recover a speedup that is lost in the naive application of the algorithm.

The main mathematical tools used in this thesis are group theory (esp. Fourier theory on finite groups), monoidal category theory, and categorical algebra.

# Contents

# List of Figures

# Nomenclature

| | |
|---|---|
| SMC | symmetric monoidal category, Def. 2.2.1. |
| †-SMC | dagger symmetric monoidal category, Sec. 3.1. |
| QPT | quantum-like process theory, Def. 3.2.6. |
| †-FA | dagger Frobenius algebra, Def. 3.3.4. |
| †-CFA | dagger commutative Frobenius algebra, Sec. 3.3.2. |
| †-SCFA | dagger special Frobenius algebra or classical structure, Def 3.3.5. |
| CPM | completely positive maps, Sec. 3.7. |
| †-qSFA | dagger quasi-special Frobenius algebra, Def. 4.1.3. |
| †-qSCFA | dagger quasi-special commutative Frobenius algebra. |
| GROUPHOMID | group homomorphism identification problem, Def. 4.2.9. |
| QCRel | the QPT model in sets and relations, Def 4.3.21. |

# Chapter 1

# Overview

Despite almost two decades of research, we still seek new and useful quantum algorithms. This is of interest where the meaning of useful ranges from "able to generate experimental evidence against the extended Church-Turing thesis" to "commercially viable". Better languages, frameworks, and techniques for analyzing the structure of quantum algorithms will aid in these attempts. One such programme initiated by Abramsky, Coecke, et. al de-emphasizes the role of Hilbert spaces and linear maps and instead focuses on topological flows of information within quantum-like systems [6, 31, 37]. This approach captures all the familiar structure of quantum computation - from teleportation to quantum secret-sharing - and locates the particular setting of Hilbert spaces as an instance of more general **process theories** [33, 35, 36].

This thesis develops a process theoretic approach to the structure of quantum algorithms and protocols, furthering the "search for structure" in both the foundations and applications of quantum information. We briefly clarify the relationship of this approach to these two domains.

### Foundations

We consider quantum computation (quantum theory on finite dimensional Hilbert spaces) as an instance of a larger class of process theories. These theories start from a process based axiomatic structure that emphasizes the information processing that occurs in quantum systems. This work thus contributes to existing literature that presents an information theoretic foundation for quantum theory, such as Hardy [54], Chiribella et al. [26], and work in generalized probabilistic theories by Barrett [17]. Process theoretic generalization focuses on the information processing features of quantum mechanics that generate properties like the quantum Fourier transform (Section 4.1), unitary oracles (Section 4.2), and Mermin non-locality (Chapter 5). Further, it allows us to consider analogs of complex quantum protocols in alternative physical theories that inhabit new mathematical settings (Section 4.3).

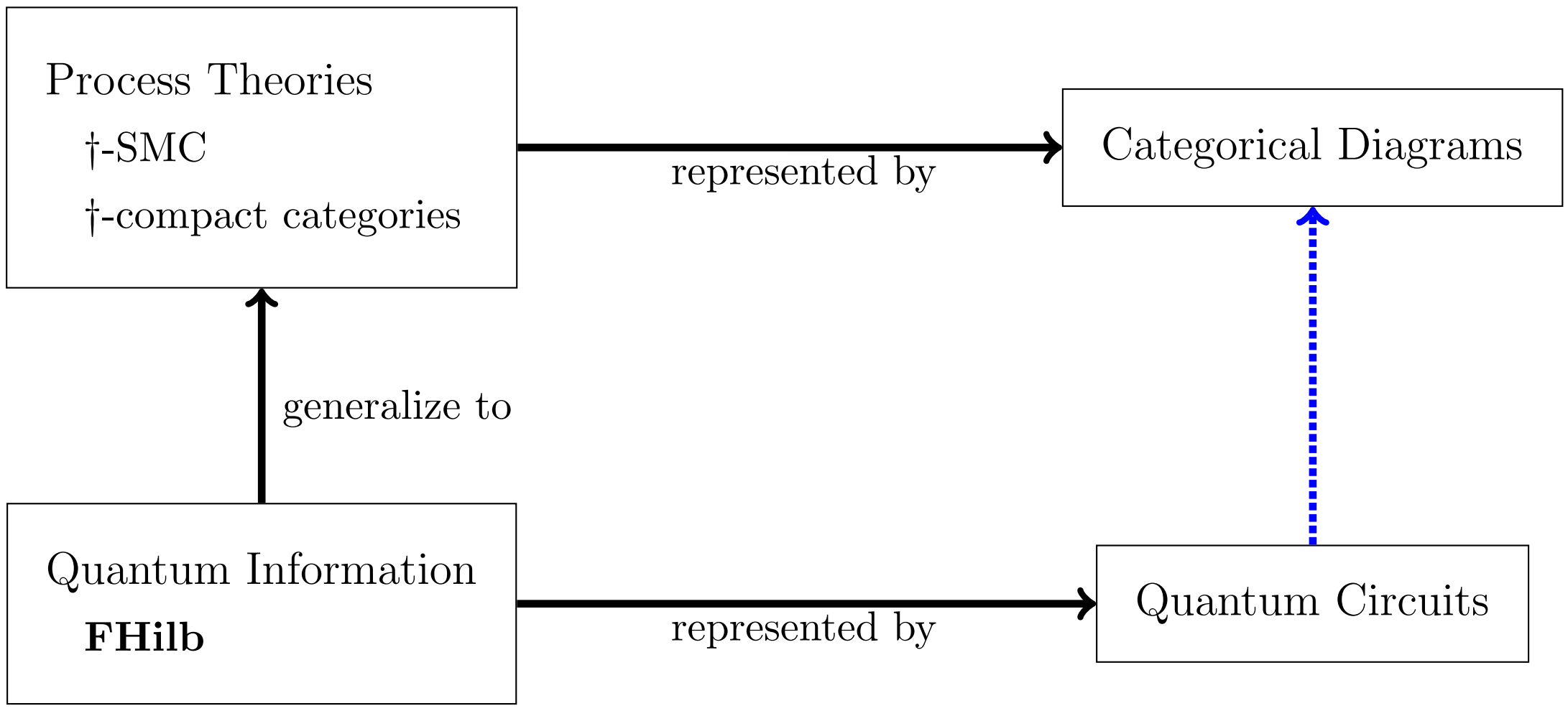

Figure 1.1: Schematic of the approach taken in this thesis. Viewing quantum information as a process theories provides a more powerful and diagrammatic approach to presenting quantum protocols.

### Applications

While there already exist many intriguing applications for quantum information - as quantum computers or quantum communications networks - our knowledge of these applications is often based on isolated techniques. As it has become necessary to consider more advanced applications, the language we have used to describe quantum mechanics has similarly advanced. Differential equations led to matrix mechanics and then to quantum circuits, which have formed the basis for most known quantum algorithms as well as the basis for the current generation of quantum programming languages, e.g. LIQUi$|\rangle$ [106] and Quipper [48]. Still, these languages are, by a structural standard, more akin to quantum assembly language than high-level programming languages.

By generalizing from quantum information to process theories (Figure 1.1), we are able to lift quantum circuit diagrams into a more powerful notation for specifying and verifying algorithms (Chapters 4 and 6). This approach handles algorithm analysis with high-level abstractions that are appropriate for advanced applications. As we will show, it allows the function of large classes of blackbox algorithms to be verified with a few diagrammatic arguments. As such, these techniques have an important role to play as we seek to best exploit the structure of quantum theory.

## Outline

The first two chapters introduce background material. Chapter 2 covers the basic categorical diagrams for process theories and connects them to examples, including quantum circuits. Chapter 3 adds known structures to define a **quantum-like process theory** (QPT). Diagrams for QPTs have additional power on top of quantum circuits that allows direct reasoning about observables, phases, complementarity and measurements. Examples for quantum teleportation and the controlled-not gate are provided.

Chapter 4 contains the first of the main results of this thesis as it applies QPTs to the study of quantum algorithms in three ways. In Section 4.1, we develop a connection between the Fourier transform and strongly complementary observables in a QPT. This connection is of interest in quantum information where the Fourier transform plays a key role in many algorithms. The role of strongly complementary observables pinpoints exactly where this particular transformation appears in the structure of quantum theory. The connection is also of independent mathematical interest, as we give new categorical proofs of Pontryagin duality, the convolution theorem, and the Fourier inversion theorem in arbitrary dagger symmetric monoidal categories. This construction allows us to investigate other mathematical settings where Fourier-like transforms occur, such as in the category of sets and relations.

In Section 4.2, we use the QPT framework and its connection to the Fourier transform to analyze quantum blackbox algorithms. We begin with a purely abstract construction and proof of unitarity for oracles in arbitrary QPTs. This construction directly connects these unitary oracles with complementary observables. Then, as warmup, we overview Vicary's process theoretic verifications and generalizations of the Deutsch-Jozsa, single-shot Grover's, and hidden subgroup algorithms from [104]. This allows us to introduce a new blackbox quantum algorithm for the group homomorphism identification problem (GROUPHOMID) which, in many cases, offers a large speedup over the optimal classical algorithm. These algorithms are specified and verified using the tools of process theories.

Section 4.3 leverages the QPT presentation of these algorithms to construct a toy model of the Deutsch-Jozsa algorithm, the single-shot Grover's algorithm, and the GROUPHOMID algorithm in the category of sets and relations: a QPT we call QCRel. In a particular sense, these provide non-deterministic classical models for the structure of these quantum algorithms. This section also includes some new mathematical results on self-conjugate comonoid homomorphisms in sets and relations.

Chapter 5 investigates the connection between Mermin non-locality and strongly complementary observables that was first introduced by Kissinger et al. [32]. We extend this work to cover all arbitrary QPTs by developing the connection between locality and phase groups from Coecke et al. [34]. This leads us to an easy to check algebraic condition on the phase groups of an QPT that acts as an indicator for Mermin non-locality or locality. In particular, this allows us to show that QCRel (the QPT in sets and relations) is Mermin local despite much of its quantum-like structure. In terms of quantum theory, these results show how to construct a large class of Mermin non-locality experiments for any number of parties, that have access to any number of measurements, on systems of arbitrary dimension. This is a powerful generalization over the restricted cases that are currently known and we use it further suggest a large class of new quantum secret sharing protocols.

Chapter 6 uses recent work in a process theoretic framework from computational linguistics, called distributional compositional (DisCo) linguistics [28], to investigate quantum algorithms for computational linguistic tasks. The shared QPT structure of quantum theory and DisCo linguistics makes quantum algorithms particularly apt for this domain, and allows to us improve on the naive application of such algorithms. As an example, we use classical preprocessing to adapt a quantum algorithm for clustering into a sentence classification algorithm with an improved speedup.

We then conclude in Chapter 7 with a brief outlook on future work on blackbox algorithms, the computational complexity of process theories in general, and applications to quantum programming languages.

## Related Papers

Material presented in this thesis has developed out of several publications and working papers over the last few years. Much of the material in this thesis has appeared in print in these original sources and specific citations are included in the abstracts of each chapter. Some of this work was completed jointly with other researchers and these sections are also highlighted throughout the next. Selections from these joint papers are only those that this author contributed significantly towards. We provide a list of these papers in chronological order for convenience:

1. Zeng, William. Diagramming Quantum Algorithms: The Fourier transform. In *Proceedings of The XXIX International Colloquium on Group-Theoretical Methods in Physics : Posters*, Tianjin, China, August 2012.

2. Zeng, William and Vicary, Jamie. Abstract structure of unitary oracles for quantum algorithms. *Electronic Proceedings in Theoretical Computer Science* 172, 2014, pp. 270-284.

3. Zeng, William. Models of Quantum Algorithms in Sets and Relations. In preparation.

4. Gogioso, Stefano and Zeng, William. Mermin Non-locality in Abstract Process Theories. *Proceedings of the 12th Intl. Workshop on Quantum Physics and Logic*, Oxford, July 2015.

5. Gogioso, Stefano and Zeng, William. Fourier transforms from strongly complementary observables. In preparation.

6. Zeng, William and Coecke, Bob. Quantum algorithms for compositional natural language processing. In preparation.

# Chapter 2

# Categories and Diagrams

**Chapter Abstract**

This chapter introduces basic background material. We introduce the relevant categorical definitions and show how symmetric monoidal categories can be interpreted as process theories, using quantum circuits as a motivating example. We then review several other examples of process theories from the literature.

## 2.1 Monoidal categories

Monoidal categories (sometimes called tensor categories) provide an abstract structure for processes that are equipped with both sequential and parallel composition. One might be tempted to think of sequential composition as time-like and parallel composition as space-like, but we should be careful to examine this notion in cases like, for example, quantum theory (See Example 3.2.8). We call a process $f : A \to B$ a **morphism** from some input $A$ to output $B$. The $A$ and $B$ are called **objects** and we sometimes say that $f$ is a morphism **between** them. For morphisms $f : A \to B$ and $g : B \to C$, their **composite** is a morphism $g \circ f : A \to C$. This data can be structured into a category, which then embodies the notion of sequential composition.[1] We will sometimes leave the objects of morphisms implicit when they can be inferred from context.

**Definition 2.1.1.** A **category** $\mathbf{C}$ is a set of objects $\mathrm{Ob}(\mathbf{C})$ and a set of morphisms $\mathrm{Arr}(\mathbf{C})$ between them, such that for all $A, C, B, D \in \mathrm{Ob}(\mathbf{C})$ and all $f : A \to B$, $g : B \to C$, and $h : C \to D$ in $\mathrm{Arr}(\mathbf{C})$:

[1] As a foundational comment, categories in the broadest mathematical literature do not in general require their objects and morphism to be sets. Categories where they do, as defined and used in this thesis, are *small* categories. These details relate to the role categories can play in mathematical foundations [76].

- for every pair of morphisms $f, g$, their composite $g \circ f : A \to C$ is also in $\mathrm{Arr}(\mathbf{C})$;
- composition is associative:

$$h \circ (g \circ f) = (h \circ g) \circ f \tag{2.1}$$

- for every object $A$ there is an $\mathrm{id}_A : A \to A$ in $\mathrm{Arr}(\mathbf{C})$ called the **identity morphism** such that for all $f$:

$$\mathrm{id}_B \circ f = f = f \circ \mathrm{id}_A. \tag{2.2}$$

A category can thus be thought of as encoding processes where sequences can be associatively composed and where we always have access to a “do-nothing“ process, which is the identity morphism. Note that the objects of a category are somewhat superfluous, as they are in one-to-one correspondence with the identity morphisms. Due to this correspondence, we refer interchangeably to an object and its identity morphism. It is because categories are focused on morphisms that we see them as encoding a process theory.

Morphisms and their compositions can be represented in string diagrams:

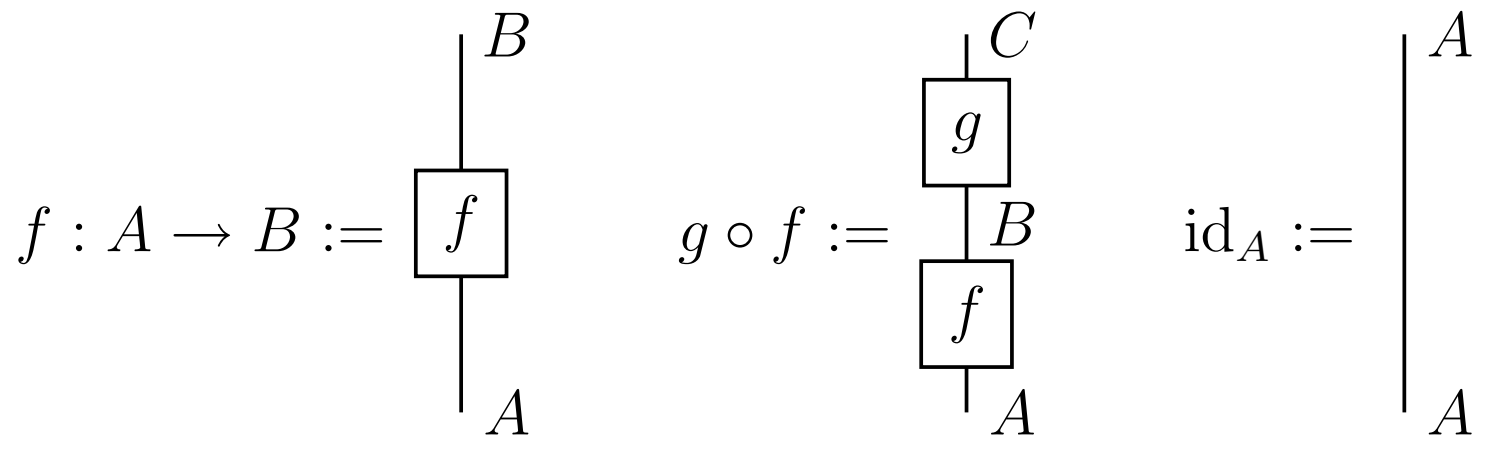

(2.3)

Here, vertical connectivity (read from bottom to top) represents the flow of morphism composition.

**Definition 2.1.2.** A (strict)[2] **monoidal category** is a category $\mathbf{C}$ equipped with a **categorical tensor** $(- \otimes -) : \mathbf{C} \times \mathbf{C} \to \mathbf{C}$ and a **unit object** $I \in \mathrm{Ob}(\mathbf{C})$ that obey:

$$(A \otimes B) \otimes C = A \otimes (B \otimes C), \tag{2.4}$$

$$I \otimes A = A = A \otimes I. \tag{2.5}$$

[2]Throughout this thesis we take monoidal categories to be strict, i.e. those associators and unitors are identities. In fact, every monoidal category is monoidally equivalent to a strict monoidal one [63].

Tensor composition is represented by side-by-side placement, and the unit object is the "empty" diagram:

$f \otimes g :=$ 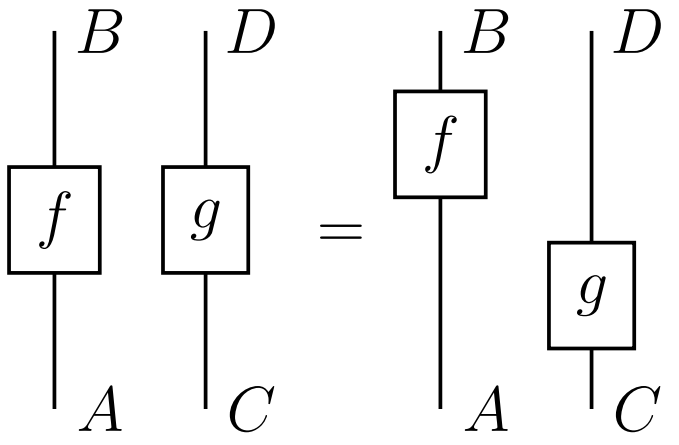 $\text{id}_I :=$ (2.6)

To interpret this, we consider the objects $A, B$ as **systems**, so that morphisms $f : A \to B$ are **processes** from one system to another. Thus $A \otimes C$ is thought of as a composite system with composite morphism $f \otimes g$ that acts independently on its parts. The identity object is interpreted as the empty system, which matches its diagram. We can also define states of systems.

**Definition 2.1.3.** A **state** of $A \in \text{Ob}(\mathbf{C})$ is a morphism $|\psi\rangle : I \to A$ drawn as

$|\psi\rangle :=$ 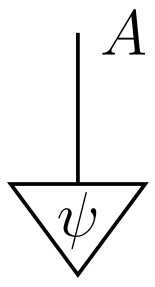 (2.7)

The state morphism can be thought of as a *preparation* process to create that state: it is a process that starts with nothing and has output of type system. The tensor product of two states, $|\psi\rangle \otimes |\phi\rangle$, corresponds to the usual notion of product state from quantum theory (Example 3.2.3).

**Definition 2.1.4.** In a monoidal category, **effects** on an object $A$ are morphisms $E : A \to I$.

Similarly to the way states act as preparations, an effect can be thought of as a *test* process. It takes as input some system and outputs the outcome of that test on the system. Thus, the preparation of a system in a certain state $|\psi\rangle$, which is then tested against effect $E$, has representation:

$E \circ |\psi\rangle =$ 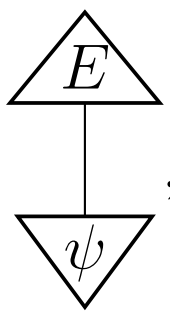 , (2.8)

which is intended to look suggestively like an inner product. Said another way, effects turn systems into outcomes of tests on that system. Coecke and Paquette [36] provide an instructive example of this interpretation using a monoidal category model of cooking. After introducing additional structure, we will return to the details of effects in quantum-like theories in Section 3.1.

## 2.2 Symmetric monoidal categories

**Definition 2.2.1.** A monoidal category is **symmetric** (is an SMC) when it has isomorphisms $\sigma_{A,B} : A \otimes B \to B \otimes A$ that satisfy the following graphical equations:

$\sigma_{A,B} :=$ 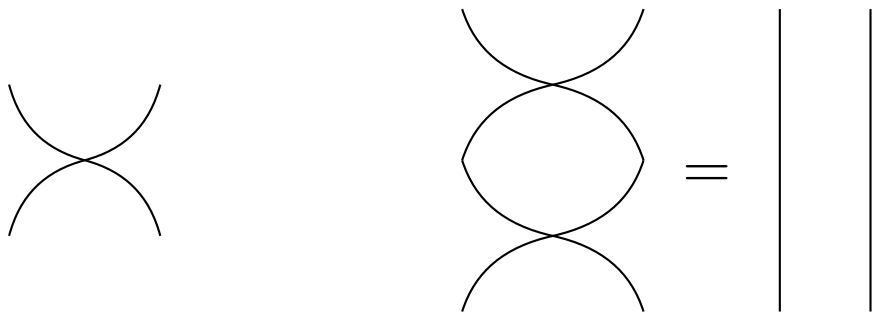 (2.9)

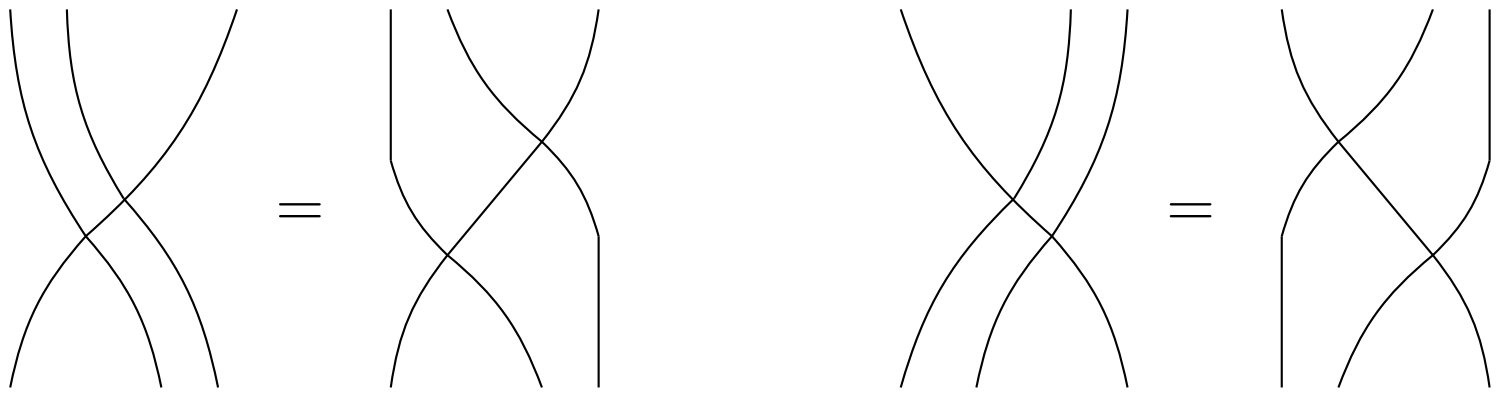 (2.10)

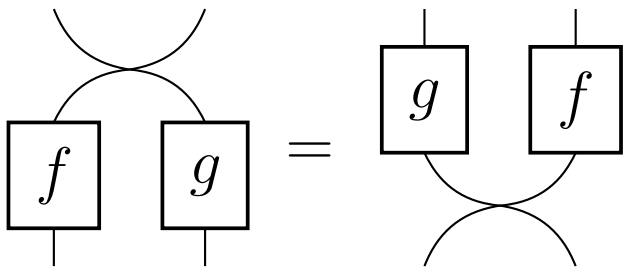 (2.11)

where Equation 2.11 for all $f, g$ is the naturality of $\sigma$.

**Examples 2.2.2.** The following are some explicit examples of SMC's:

- **Hilb** and **FHilb**, the category where objects are (finite dimensional) Hilbert spaces; morphisms are linear maps; the categorical tensor is the tensor product; the unit object $I = \mathbb{C}$.
- **Vect** and **FVect**, the category where objects are (finite dimensional) vector spaces; morphisms are linear maps; the categorical tensor is the tensor product; the unit object $I = \mathbb{C}$.
- **Rel** and **FRel**, the category where objects are (finite) sets; morphisms are relations; the categorical tensor is the cartesian product; the unit object is the singleton set, i.e. $I = \{\bullet\}$.
- **Set** and **FSet**, the category where objects are (finite) sets; morphisms are functions; the categorical tensor is the cartesian product; the unit object is the singleton set, i.e. $I = \{\bullet\}$.
- Given a finite group $G$, its representations form an SMC **Rep(G)**, where objects are finite dimensional representations of $G$; morphisms are intertwiners for the group action;[3] the categorical tensor is the tensor product of representations; the unit object is the trivial action of $G$ on the 1-dimensional vector space.

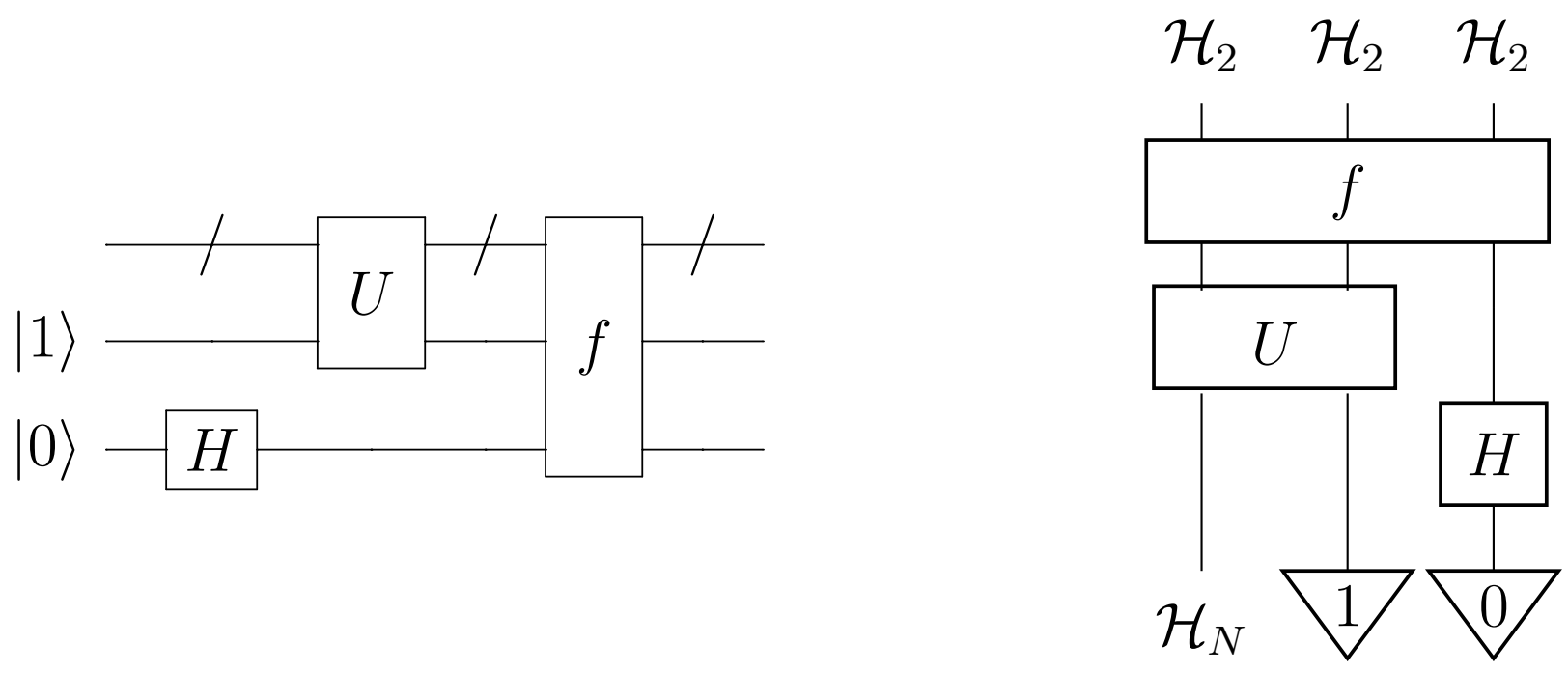

Figure 2.1: On the left we have a quantum circuit (read left to right) and its corresponding categorical diagram in **FHilb** on the right (read bottom to top). In both depictions, the boxes are linear maps and the wires are Hilbert spaces. In quantum circuits these are implicitly qubits, with a slash used to denote products of qubits. In the categorical diagram we explicitly write spaces or leave them generic.

Some of these examples have customary process theoretic interpretations, such as **Rel** as a setting for nondeterministic classical processes and **FHilb** for quantum computation. We'll elaborate on these interpretations in Section 2.3.

It is important to note that the graphical notation we have introduced to describe morphisms in SMCs is not merely a notational convenience. We can reduce all the structural rules for SMCs down to simple diagrammatic equivalence, as the following theorem of Joyal and Street shows.

**Theorem 2.2.3.** *[62, Thm 2.3] A well-formed equation between morphisms in a symmetric monoidal category follows from the axioms if and only if it holds in the graphical language up to isomorphism of diagrams.*

This emphasizes the power of the diagrammatic presentation. Rather than needing to check the many different rewrite rules that form the diagrammatic axioms, we need only check that the diagrams are isomorphic. This will become especially important as we introduce more structure in Chapter 3.

### 2.2.1 Symmetric monoidal categories & quantum circuits

This thesis applies the structure and graphical calculi for SMCs to the study of protocols and algorithms for and inspired by quantum theory. For this purpose, it can be useful to think of the SMC graphical calculus as mathematical scaffolding that underlies quantum circuit diagrams. See Figure 2.1 for a concrete example. Note that state preparation is included as a part of the categorical diagram while it is external labeling in the quantum circuit. In **FHilb** we have $I = \mathbb{C}$, and so the states, by Definition 2.1.3, of some Hilbert space $\mathcal{H}$ are exactly maps $|\psi\rangle : \mathbb{C} \to \mathcal{H}$. These are recognized as the usual quantum state vectors $|\psi\rangle \in \mathcal{H}$. This allows states to be manipulated graphically as well, to advantages discussed later. In uncovering the SMC scaffolding of quantum circuits, we can improve it, introducing techniques to reason graphically about more advanced structures that better capture salient features of these processes. We cover these new features in Chapter 3.

[3]Linear maps that commute with the action of $G$.

### 2.2.2 Other categorical definitions

We make use of several other standard categorical concepts, whose definitions are reproduced here.

**Definition 2.2.4.** From any category $\mathbf{C}$ we can construct a dual category $\mathbf{C^{op}}$ where $\mathrm{Ob}(\mathbf{C^{op}}) = \mathrm{Ob}(\mathbf{C})$ and for every morphism $f : A \to B$ in $\mathrm{Arr}(\mathbf{C})$ there is a morphism $g : B \to A$ in $\mathrm{Arr}(\mathbf{C^{op}})$ and vice versa.

Loosely speaking, we can think of $\mathbf{C^{op}}$ as a version of $\mathbf{C}$ where we have turned all the morphisms around so that inputs become outputs. In general this will be a different kind of category than $\mathbf{C}$.

**Definition 2.2.5.** Given a category $\mathbf{C}$ and $A, B \in \mathrm{Ob}(\mathbf{C})$, the **homset** $\mathrm{Hom}(A, B)$ is the set of morphisms in the category of type $f : A \to B$.

We can think of $\mathrm{Hom}(A, B)$ as all the processes that input system $A$ and output $B$. Sometimes, to specify the category, we write $\mathbf{C}(A, B)$ for the homset $\mathrm{Hom}(A, B)$ in $\mathbf{C}$.

As we often consider the relationship between categories, we recall the notion of a structure preserving map between categories. In our context they can be thought of as homomorphisms between process theories.

**Definition 2.2.6.** Given two categories $\mathbf{C}$ and $\mathbf{D}$, a **functor** $F : \mathbf{C} \to \mathbf{D}$ consists of

1. A mapping on objects $F : \mathrm{Ob}(\mathbf{C}) \to \mathrm{Ob}(\mathbf{D}) :: A \mapsto F(A)$
2. A mapping on arrows $F : \mathbf{C}(A, B) \to \mathbf{D}(F(A), F(B)) :: f \mapsto F(f)$ that preserves composition and identities, i.e.

(a) For $f : A \to B$ and $g : B \to C$ then $F(g \circ f) = F(g) \circ F(f)$

(b) $F(1_A) = 1_{F(A)}$

Monoidal and symmetric monoidal functors also preserve their respective additional structures. For further details, see Coecke and Paquette's introductory "Categories for the practising physicist" [36].

## 2.3 Process theories

We have seen that strict symmetric monoidal categories capture the structure of diagrammatic languages like quantum circuit diagrams.[4] For this reason we introduce the following definition:

**Definition 2.3.1.** A **process theory** is a strict symmetric monoidal category where objects are interpreted as systems, morphisms are interpreted as processes, morphism composition is interpreted as sequential process composition, and the monoidal product is interpreted as parallel process composition.

Process theories and associated structures on them form the mathematical setting of this thesis.

Monoidal and symmetric monoidal categories were introduced as "categories with multiplication" by MacLane in [77] and later formalized as string diagrams by [62]. There are other examples, besides quantum information, where diagrammatic concepts in computer science and physics have been formalized using SMCs, and these can all be considered as examples of the process theory considered here. In general, this approach is most useful when processes have both a natural diagrammatic (*geometric*) presentation, but also have an *algebraic* interpretation.

In physics, Penrose's tensor notation [86] is, in modern language, precisely the diagrammatic representation of the symmetric monoidal category **FVect** of finite dimensional vector spaces and linear maps. Feynman diagrams are representations of morphisms in the symmetric monoidal category of positive energy representations of the Poincaré group [15]. The symmetric (and braided) monoidal category framework is an important foundation for $n$-dimensional topological quantum field theories.

[4] While this thesis chooses to use strict SMC's as process theories, it may be reasonable in other settings to consider non-strict monoidal categories. These kinds of process theories are certainly still well-formed, but do not have as clean a diagrammatic representation. We would, for example, need to keep track of how many copies of "white space" (the monoidal unit) are introduced around a diagram.

Atiyah [10] introduces the idea that one can think of a TQFT as a symmetric monoidal functor between the category of n-cobordisms and finite dimensional vector spaces: $T : \mathbf{nCob} \to \mathbf{FVect}$. In fact, Abrams [2] and Kock [70] show two-dimensional TQFT's are equivalent to commutative Frobenius algebras, which we introduce later in Definition 3.3.4. Reshetikhin, Turaev, Baez, and Lauda discuss how this perspective also plays an important role in the study of generalized knot invariants [92] and quantum groups [15].

We also find examples in computer science and control theory. Petri nets, which present naturally in diagrams, have been connected to linear logic through their connection with monoidal categories by several authors [3, 78, 95]. In particular Marti and Meseguer show how each Petri net, by closure under sequential and parallel composition, makes a suitable kind of symmetric monoidal category [78, 81]. Baez and Erbele [13] and Bonchi et al. [21] show that signal-flow diagrams from control theory can be seen as morphisms in the category $\mathbf{FinRel_k}$ whose objects are finite dimensional vector spaces of the field $k$, whose arrows are linear relations, and whose monoidal product is the direct sum. In fact the internal monoids, comonoids, and bialgebras that are described in Chapter 3 also have a natural interpretation in this setting as investigated by Baez, Erbele [13] and Bonchi et al. [22]. Bonchi et al. have further used these structures to axiomatize generalized linear algebra using diagrams [21].

In recent work, Baez and Fong have formalized passive linear networks (electrical circuits consisting of inductors, capacitors, and resistors) using symmetric monoidal categories [14]. Here the relationship between a category of circuits and the Lagrangian representing them is presented as a dagger functor between dagger compact categories. These dagger compact categories are symmetric monoidal categories with additional structure covered Chapter 3. SMC based string diagrammatic theories also appear in interactive theorem proving [51], parallel programming [82], programming language semantics [79], and natural language processing [38]. This final connection is elaborated on and leveraged in Chapter 6.

The planar diagrams presented in this chapter can be understood as part of a larger $n$-categorical hierarchy. Joyal and Street's [62] original work describe monoidal categories as coherent under planar isotopy, and the coherence of higher classes of these graphical languages can be regarded as geometric isotopies in higher dimensions, e.g. coherence for SMCs is up to a 4-dimensional isotopy. Selinger's monoidal category survey provide many examples [99]. A general reference for the connections between computation, topology, and physics that emerge from symmetric monoidal categories is the Rosetta Stone by Baez and Stay [12].

# Chapter 3

# Structures in Process Theories

**Chapter Abstract**

In this chapter we introduce the remaining background on categorical structures for process theories. From this background, we define a quantum-like process theory that comes equipped with a quantum-like interpretation through a generalized Born rule. Many of the properties developed here, while still general, are extensions of ideas from quantum theory, e.g. complementarity. Process theories that possess these properties can be considered quantum-like, and have diagrammatic representations that can be used as more powerful extensions of quantum circuits.

## 3.1 The Dagger

This section introduces concepts that expand categorical diagrams beyond quantum circuits. We see that abstract linear algebra can be introduced by adding a so-called dagger functor. In the case of **FHilb** this corresponds to the familiar notion of adjoint (complex-transpose). The addition of the dagger also allows us to take a perspective on quantum-like symmetric monoidal categories (not just **FHilb**).

**Definition 3.1.1.** A **dagger functor** on a category $\mathbf{C}$ is an involutive contravariant functor $\dagger : \mathbf{C} \to \mathbf{C}$ that is the identity on objects. A **dagger category** is a category equipped with a dagger functor.

Spelling out this definition, the dagger functor has the following properties for $f, g : \mathrm{Arr}(\mathbf{C})$ with suitable types and $A \in \mathrm{Ob}(\mathbf{C})$:

$$\left(f^{\dagger}\right)^{\dagger} = f \tag{3.1}$$

$$(g \circ f)^{\dagger} = f^{\dagger} \circ g^{\dagger} \tag{3.2}$$

$$\mathrm{id}_A^\dagger = \mathrm{id}_A \tag{3.3}$$

Thus for any $f : A \to B$ in a dagger category, its dagger or **adjoint** $f^\dagger : B \to A$ also exists in that category. We use the shorthand †-SMC for a dagger symmetric monoidal category. In †-SMCs the dagger and the monoidal product are required to cooperate, i.e. $(f \otimes g)^\dagger = f^\dagger \otimes g^\dagger$.[1]

The quantum setting of **FHilb** is a dagger category whose canonical dagger is the adjoint, and generalizing it in the manner of Definition 3.1.1 allows us to generalize many familiar terms, following Abramsky and Coecke [5]:

**Definition 3.1.2.** A morphism $f : A \to B$ in a dagger category is:

- **self-adjoint** when $f^\dagger = f$;
- a **projector** when self-adjoint and $f \circ f = f$
- an **isometry** when $f^\dagger \circ f = \mathrm{id}_A$;
- **unitary** when both $f^\dagger \circ f = \mathrm{id}_A$ and $f \circ f^\dagger = \mathrm{id}_B$;
- **positive** when $f = g^\dagger \circ g$ for some morphism $g : H \to K$.

These concepts both have meaning as mathematical objects in linear algebra and as information theoretic concepts in a process theory. For example, a projector is some process where multiple sequential applications have the same effect as a single application in the broadest possible sense.

Further, the dagger can be intuitively extended to an operation on diagrams: it flips the picture upside-down around the horizontal axes. As a visual aid, morphisms can now be drawn with broken symmetry as follows:

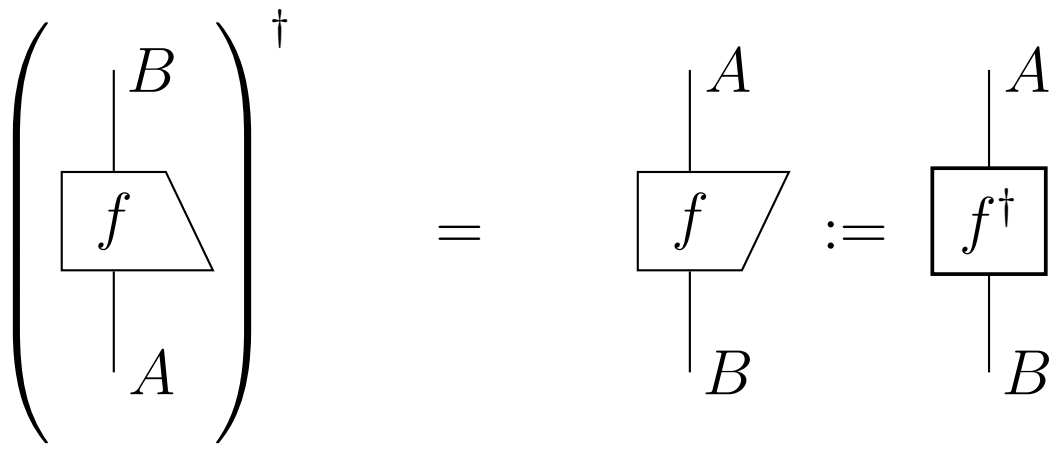

(3.4)

Thus the unitarity condition becomes:

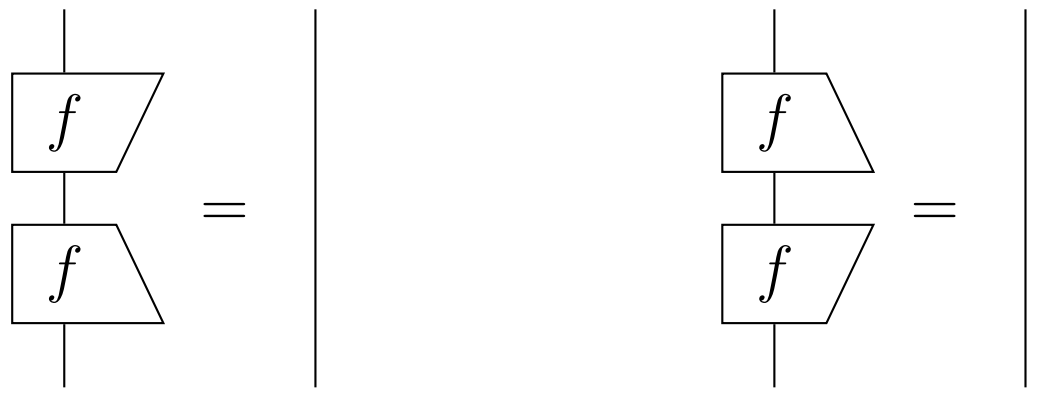

(3.5)

[1]In general monoidal dagger categories the structural morphisms of unitors and associators are required to be unitary as Definition 3.1.2.

The next two definitions, that of scalars and effects, further our understanding of †-SMCs with a generalized version of the Born rule that provides a basic notion of measurement.

**Definition 3.1.3.** A **scalar** in a SMC is a morphism $a : I \to I$. As these are maps from the empty diagram to the empty diagram they are unsurprisingly represented as:

(3.6)

This categorical view of scalars was made explicit in [5]. To gain perspective on why these properly represent scalars, we consider two facts. Firstly, Kelly [66, Prop 6.1] showed that these scalars form a commutative monoid under categorical composition, as is graphically expressed by:

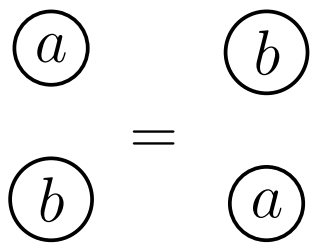

(3.7)

Secondly, in **FHilb** they correspond exactly to the complex numbers, as linear maps $\mathbb{C} \to \mathbb{C}$.

Recall from Definition 2.1.4, that, in a †-SMC, effects on an object $A$ are morphisms $\langle\psi| : A \to I$. The dagger gives a method for assigning an effect to every state and vice versa: just as some preparation $p : I \to A$ prepares system $A$ in that state, the effect $p^\dagger : A \to I$ eliminates the system $A$ by the process $p^\dagger$. This duality presents generalized **inner products** as compositions of states and effects that generalize the usual Dirac notation [5]:

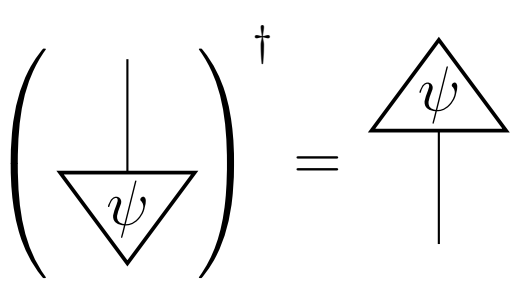

$\langle\phi| \circ |\psi\rangle = \langle\phi|\psi\rangle =$

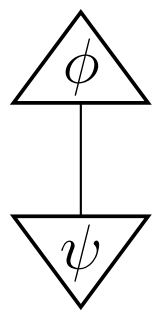

(3.8)

## 3.2 The generalized Born rule

Measurement in these process theories comes from a statement connecting probabilities and inner products. For a state $X : I \to A$ and an effect $Y : A \to I$ in a †-SMC, the **amplitude** of outcome $Y$ given preparation $X$ is:

$$a = Y \circ X : I \to I. \tag{3.9}$$

We could then define the operational probability $\text{Prob}(X|Y) = |a|^2$. While this clearly makes sense in **FHilb**, where amplitudes are complex numbers, one might ask under what general conditions the scalars $I \to I$ will square to our usual notion of probability. Vicary provides an answer in the following theorem:

**Theorem 3.2.1.** *[103, Thm 4.2] In a monoidal dagger-category with simple tensor unit, which has all finite dagger-limits and for which the self-adjoint scalars are Dedekind-complete, the scalars have an involution-preserving embedding into the complex numbers.*

For any category satisfying these conditions, this embedding can be used, along with appropriate normalization, to extract real-valued probabilities. Still, even if the scalars do not embed in the complex numbers one can generally handle the needed structure of complex conjugation to obtain a generalized Born rule. This is done using monoidal categories with duals.[2]

Duals in a process theory help us define the quantum-like properties of entanglement and conjugation.

**Definition 3.2.2.** In a †-SMC, a system $A$ has a[3] **dual** system $A^*$ if there exist processes

$$e_A : I \to A^* \otimes A \quad \text{and} \quad d_A : A \otimes A^* \to I, \tag{3.10}$$

such that the following equations hold:

$$(d_A \otimes \text{id}_A) \circ (\text{id}_A \otimes e_A) = \text{id}_A \qquad (\text{id}_{A^*} \otimes d_A) \circ (e_A \otimes \text{id}_{A^*}) = \text{id}_{A^*} \tag{3.11}$$

When duals are introduced, we can denote them by placing arrows on the objects:

$\text{id}_A =$ 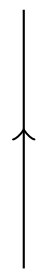 $\text{id}_{A^*} =$ 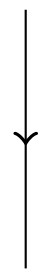 $g : A^* \to B^* =$ 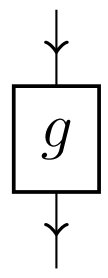 (3.12)

Thus the duality maps, commonly called "cups" and "caps", and their "snake equations" (3.11) have the following diagrammatic form:

$e_A =$ 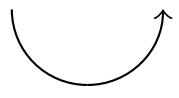 $d_A =$ 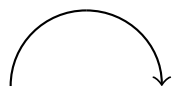 (3.13)

[2]These are sometimes called autonomous categories in the literature [63, 99].

[3]In general, there can be separate left and right duals for an object in any category, where the definition given here corresponds to a left dual. In a †-SMC (in fact in any braided monoidal category with left duals), these are necessarily equal, so we need only speak of one dual [63, Prop. 7.2].

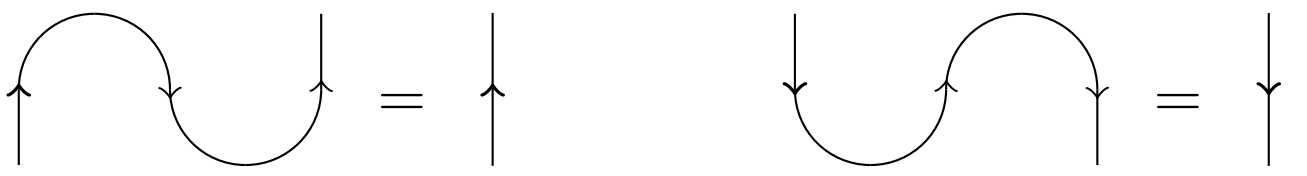

(3.14)

and are well behaved under the dagger functor, i.e.

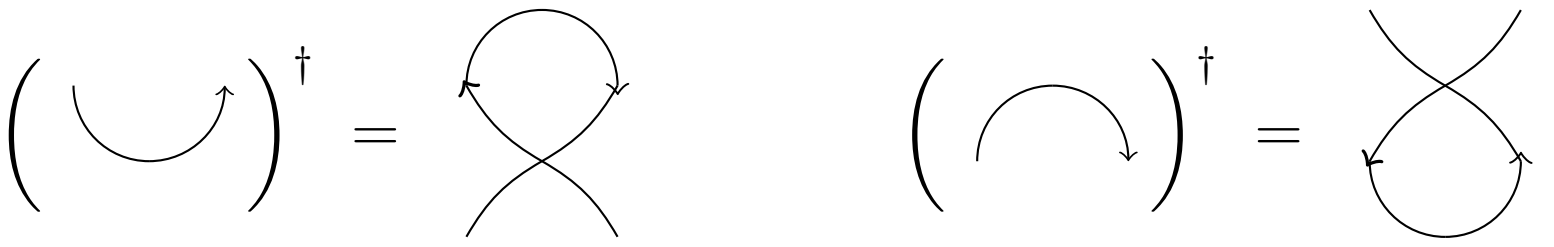

(3.15)

We can compose these caps and cups to define the dual of any process $f : A \to B$ as $f^* : B^* \to A^*$:

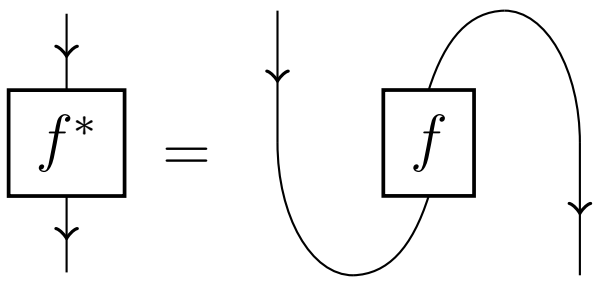

(3.16)

This is called the **upper-star** of $f$.

**Example 3.2.3.** To get a better handle on what duals are, we consider the example of **FHilb**. Given some finite dimensional Hilbert space $A \in \mathrm{Ob}(\mathbf{FHilb})$, its dual $A^*$ is the usual dual space, i.e. the space of linear functionals $A \to \mathbb{C}$. Note that in this case $A$ is isomorphic to $A^*$, and thus we can canonically map $\langle i| \mapsto |i\rangle$. Because of this isomorphism, we will often omit the arrows on wires when working in **FHilb**. Using this and given a basis $\{|i\rangle\}$ for $A$, the cups and the caps are maps:

$$e_A :: 1 \mapsto \sum_i \langle i| \otimes |i\rangle \cong \sum_i |i\rangle \otimes |i\rangle \tag{3.17}$$

$$d_A :: \sum_i |i\rangle \otimes \langle i| \cong \sum_i \langle i| \otimes \langle i| \mapsto 1 \tag{3.18}$$

We then immediately recognize $e_A$ as entangled state preparation. In particular, when $A$ is a two-dimensional Hilbert space, $e_A$ is a Bell state preparation of $|\psi_{00}\rangle = |00\rangle + |11\rangle$ and and $d_A$ is a post-selected Bell measurement for $|\psi_{00}\rangle$. For any process in **FHilb** its dual gives the transpose.

This provides a motivating example for the following, which acts as a generalized conjugation operation:

**Definition 3.2.4.** Given a process theory with duals $\mathbf{C}$, the **lower-star** is an involutive map

$$(-)_* : \mathrm{Arr}(\mathbf{C}) \to \mathrm{Arr}(\mathbf{C})$$
$$(f : A \to B) \mapsto f_* := (f^\dagger)^* = (f^*)^\dagger.$$

The lower-star operation introduces abstract probabilities for the generalized Born rule:

**Definition 3.2.5** (Generalized Born Rule)**.** For a state $X : I \to A$ and an effect $Y : A \to I$ in a †-SMC, the probability of outcome $Y$ given preparation $X$ is:

$$\mathrm{Prob}(Y|X) = (Y \circ X)_* \otimes Y \circ X : I \to I. \tag{3.19}$$

It is easy to see that this reduces to the usual Born rule in **FHilb**.

The cup and cap maps also provide a general definition for the trace of a process:

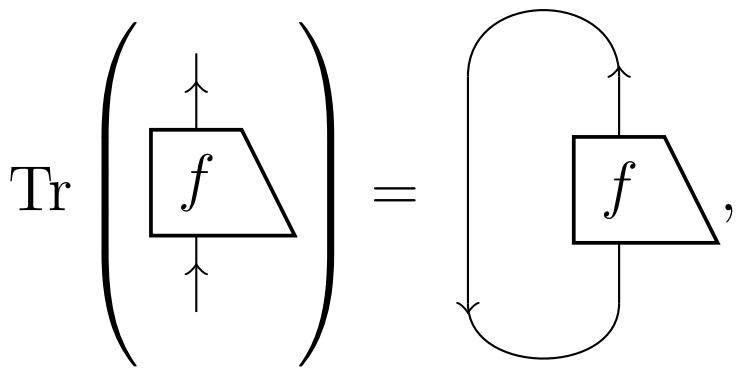

(3.20)

whose correspondence with the usual trace is easily shown [33]:

$$\mathrm{Tr}(f) = \left(\sum_i \langle ii|\right) \circ (\mathrm{id}_A \otimes f) \circ \left(\sum_j |jj\rangle\right) = \sum_i \langle i|f|i\rangle = \sum_i f_{ii} \tag{3.21}$$

### 3.2.1 Quantum-like process theories

The addition of states and a generalized Born rule (with further details in this section), give our process theories a notion of measurement and a quantum-like character.

**Definition 3.2.6.** (Quantum-like Process Theories) A †-SMC where all objects have duals is called a **†-compact category**. A **quantum-like process theory** (QPT) is a †-compact category where objects are interpreted as systems and morphisms are interpreted as processes.

Duals and their "wire-straightening" equations give diagrams in a quantum-like process theory a lot of power to encode topological equivalences. This is well expressed in the following graphical coherence theorem, whose mathematical details can be traced in Selinger's survey [99].

**Theorem 3.2.7** (Fundamental Theorem of Diagrams [33])**.** *Two diagrams in a QPT are considered equal if one can be smoothly transformed to another by bending, stretching, or crossing wires, and by moving boxes around. Given any two QPTs* **C** *and* **D** *and a functor* $F : \mathbf{C} \to \mathbf{D}$*, for any two diagrams* $d$ *and* $d'$ *in* **C***, if* $d = d'$ *as diagrams then* $F(d) = F(d')$ *in* **D***.*

To recap, we have so far introduced the structures of the dagger and duals, allowing us to compute measurement outcomes by manipulation of the diagrams alone. This contrasts with the usual quantum circuit formalism, which provides a representation of a protocol, but whose implementation must ultimately be understood through the matrix mechanics that accompany it. The use of this difference - the operational in operational process theories - can be illustrated with quantum teleportation, which provides a motivating example and was introduced in this form by Abramsky and Coecke [5].

**Example 3.2.8.** We saw in Example 3.2.3 that caps and cups represent preparation and post-selected measurement of one of the Bell states $|\psi_{00}\rangle$. The other Bell states can be prepared and/or measured by the application of a unitary to one of the entangled systems:

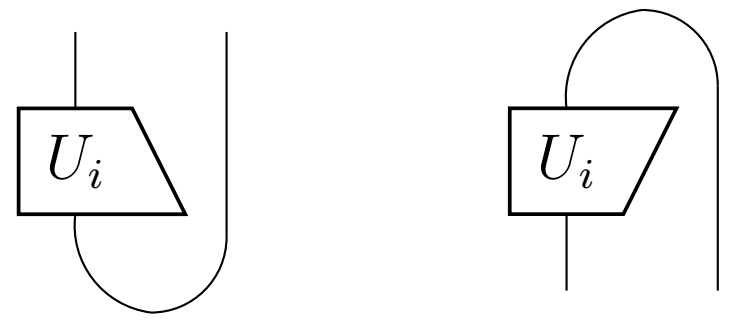

(3.22)

A teleportation protocol can then be represented by the process where Alice and Bob share an entangled system that will be used to teleport from Alice to Bob:

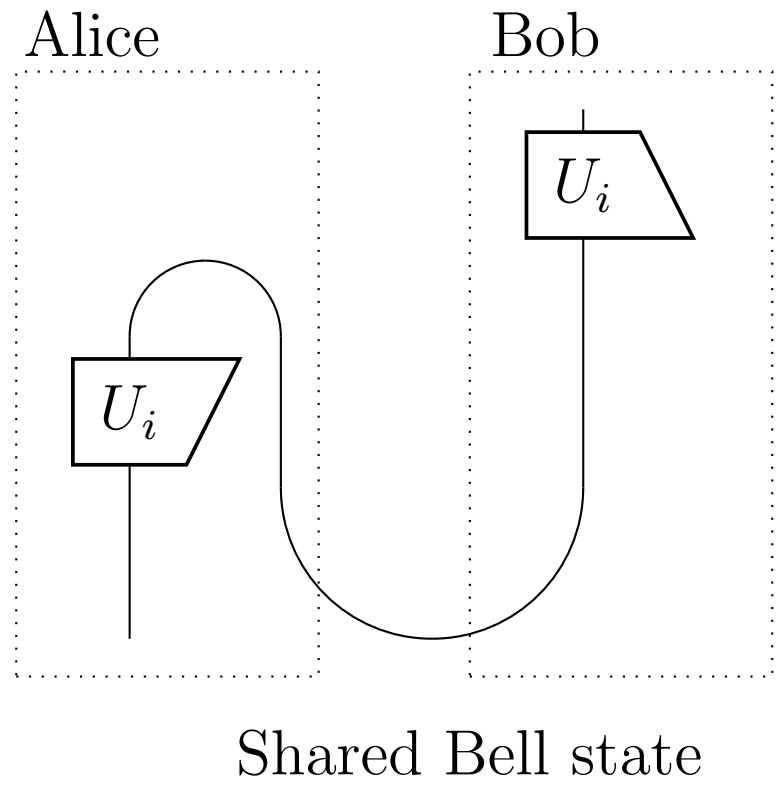

(3.23)

Here the cup map represents the shared Bell state. Alice performs a Bell measurement on the source system and her half of the entangled system. Bob then performs a unitary that matches Alice's Bell measurement. We can verify the effect of the protocol using purely diagrammatic equivalences:

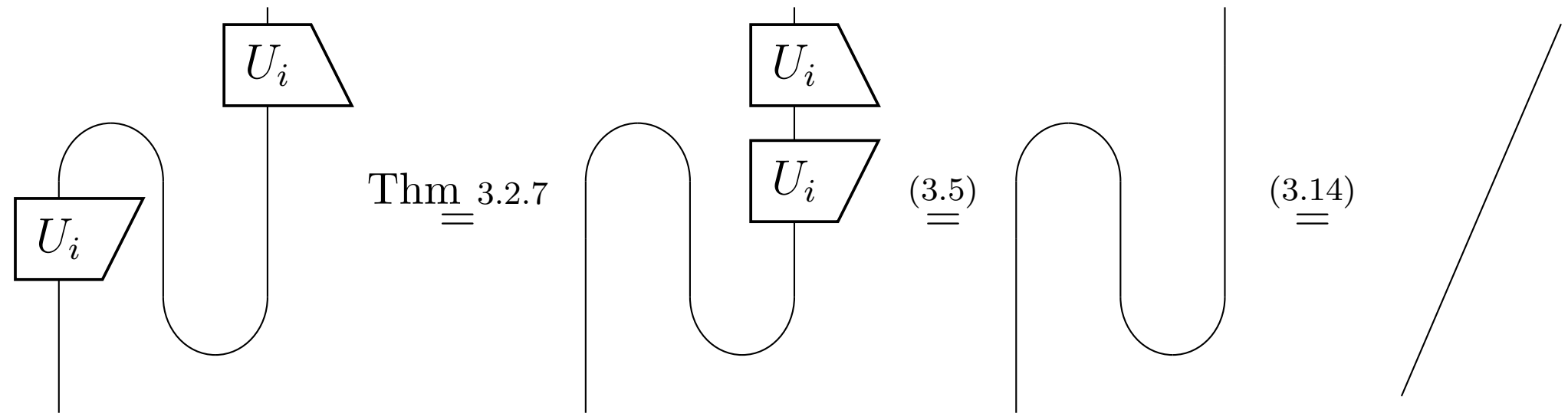

(3.24)

This protocol provides a good example where space-like and time-like intuition should be applied carefully. The application of the Bell effect and state connects Alice and Bob's systems by a single wire. Thus, the space-like separated pair of Alice and Bob's unitaries become equivalent to their time-like sequential composition, as show in the first step of (3.24). In fact, if one is reading the passage of time vertically in the diagrams, then it may be surprising to note that because

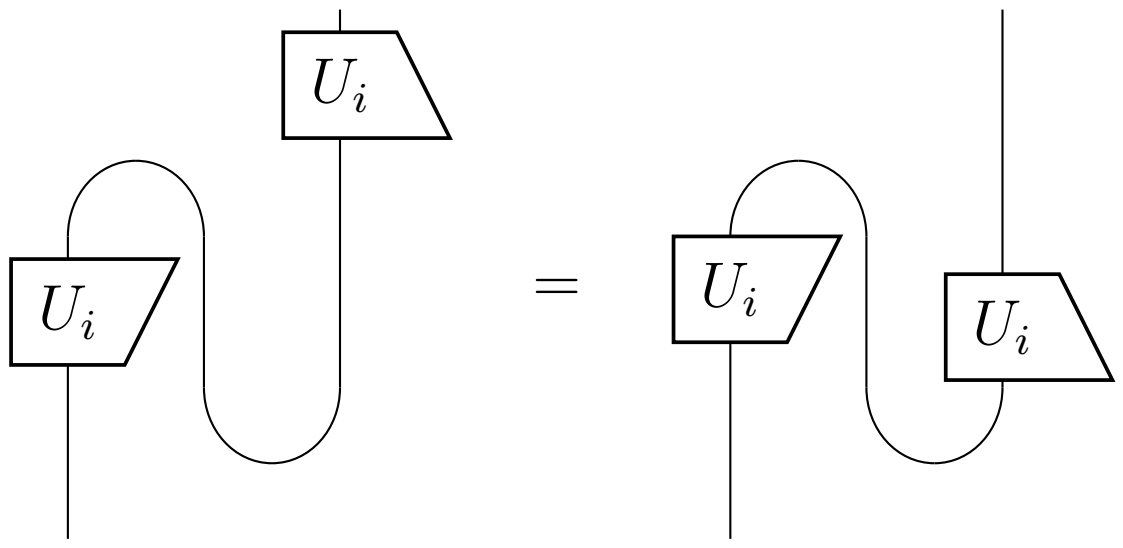

(3.25)

the time-ordering of Alice and Bob's unitaries appears to not affect the protocol. In some sense, the connectivity of the diagram means that Alice's unitary appears to happen first in the compositional sequence regardless of the time ordering in the protocol's implementation. It should be noted that this behavior is only present in the post-selected teleportation protocol that we present here. In general, when some classical communication is required, the time-ordering of course certainly matters.

This presentation of the teleportation protocol is useful in several ways:

1. The high level structure of the protocol (the teleportation) is immediately manifest without accompanying calculation.
2. It is obvious how to design equivalent teleportation protocols with multiple parties

and Bell measurements in more elaborate arrangements, e.g.

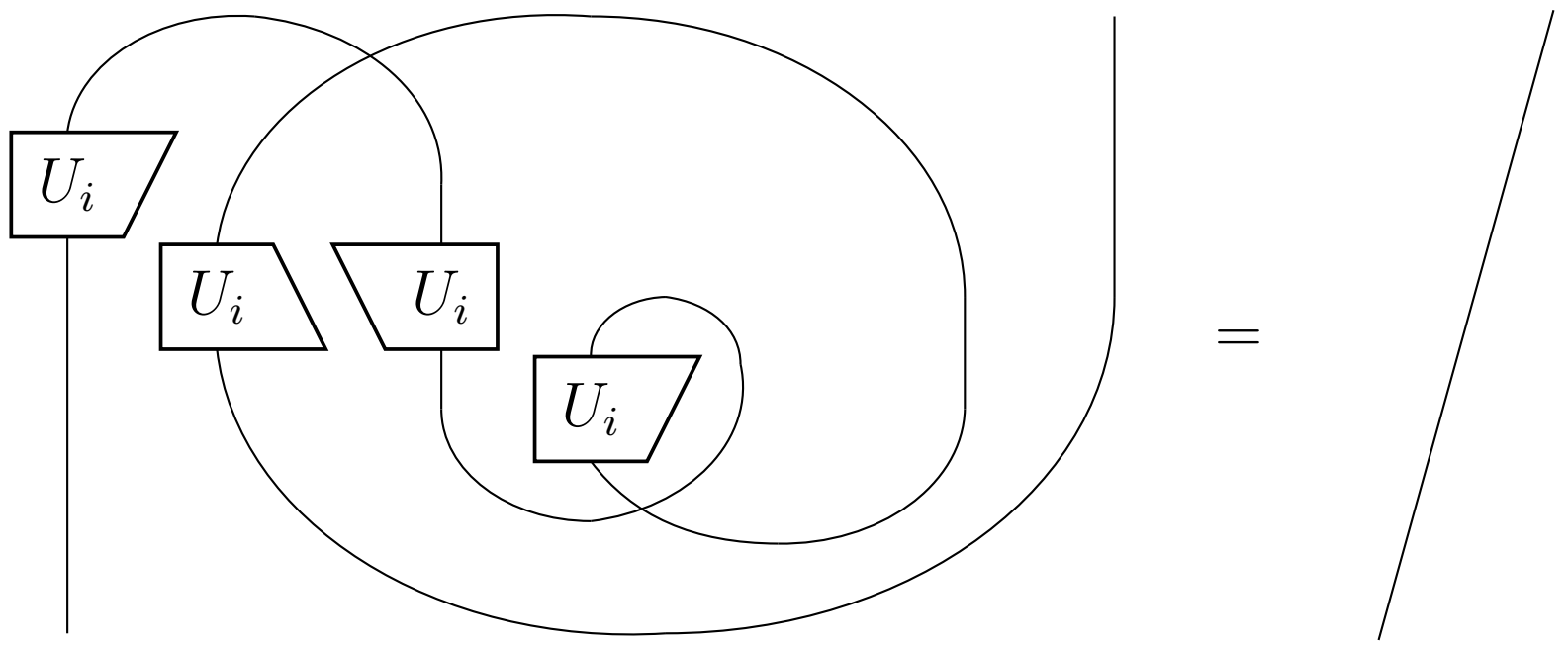

(3.26)

3. By specifying the teleportation protocol at the level of QPTs, we can consider models of teleportation in theories other than quantum theory, i.e. categories other than **FHilb**. As an example, the QPT **FRel** where systems are sets and processes are relations also has entangled states established by duals, where the dagger functor is the relational converse. All the protocol specifications given in this section apply equally well in the **FRel** setting, just as they will in any other QPT.

Quantum-like process theories already give us some power above and beyond the usual quantum circuit formalism, and we add more structures to the toolbox in the next sections. We have glossed over some of the details of how measurement should be represented in such theories, but the introduction of classical structures in the next section allows us to be more exact in Section 3.7.

## 3.3 Classical structures

Within a process theory we can construct objects that have the phenomenology of classical information: copying, deleting, etc. These structures become the observables of QPTs that allow us to extract classical information via measurements.

### 3.3.1 Monoids and comonoids

Monoids and comonoids on systems embody our notions of copying and comparing states. As they are particular kinds of processes, we draw them with distinct pictures. For $A \in \mathrm{Ob}(\mathbf{C})$, a monoid $(A, *, \mathbb{1})$ has states of $A$ as elements and the following maps:

$(- * -) :=$ 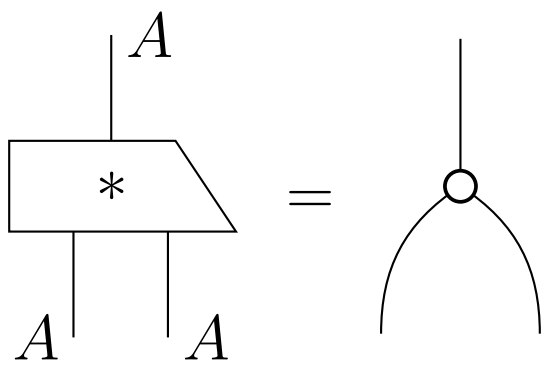

$\mathbb{1} :=$ 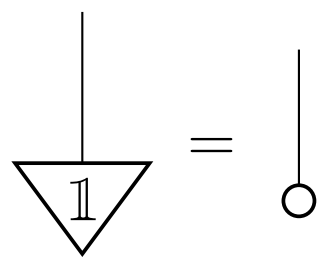

(3.27)

**Definition 3.3.1.** In a monoidal category, a **monoid** is a triple $(A, \ , \ )$ of an object $A$, a morphism $: A \otimes A \to A$ called the multiplication, and a state $: I \to A$ called the unit, satisfying associativity and unitality equations:

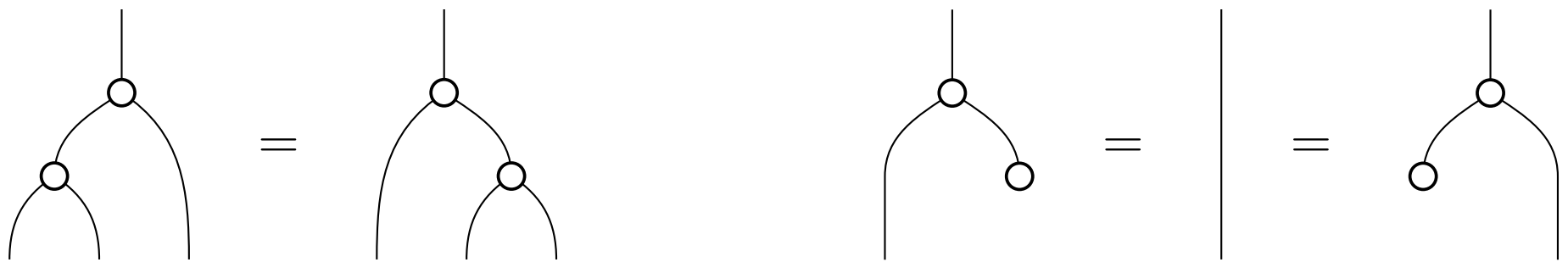

(3.28)

**Definition 3.3.2.** In a monoidal category, a **comonoid** is a triple $(A, \ , \ )$ of an object $A$, a morphism $: A \to A \otimes A$ called the comultiplication, and an effect $: A \to I$ called the counit, satisfying coassociativity and counitality equations:

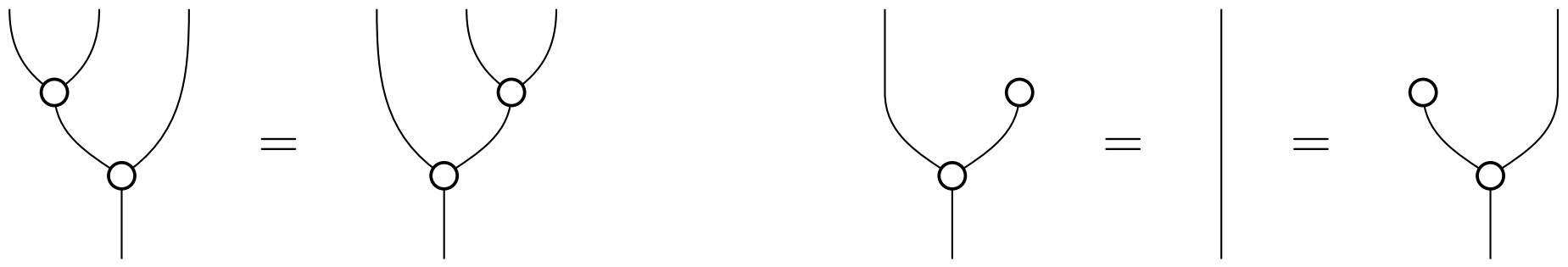

(3.29)

We use differently colored dots to represent different monoids on the same object. In a dagger monoidal category every monoid has a corresponding comonoid formed by applying the dagger functor, i.e. $(\ )^\dagger =$ and $(\ )^\dagger =$ . When a monoid and comonoid are the same color, we take this to mean that each is the dagger of the other.

**Definition 3.3.3.** In a monoidal category with object $A$, a **monoid homomorphism** $f : (A, \ , \ ) \to (A, \ , \ )$ is a map $f : A \to A$ such that

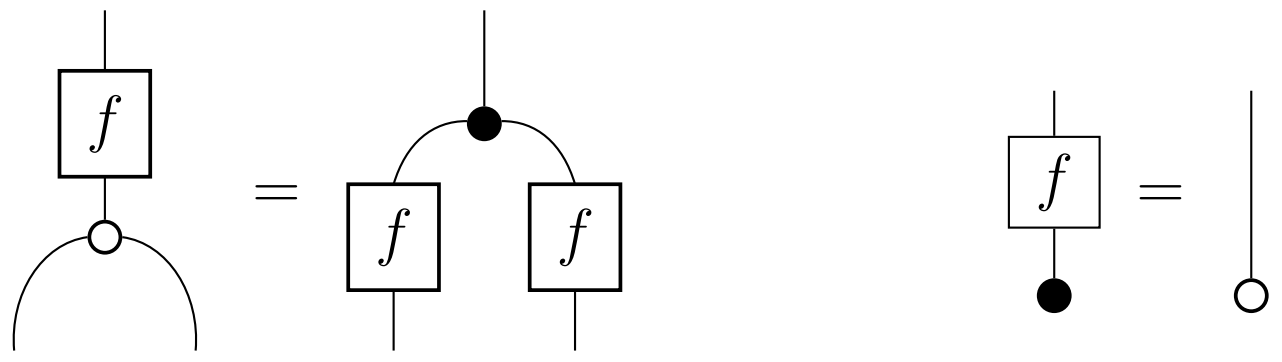

(3.30)

A **comonoid homomorphism** is defined similarly, but with the dagger of the conditions in (3.30).

In the next section we will ask for the comonoid and monoid to interact in various ways.

### 3.3.2 Generalized observables

When monoids and comonoids combine under certain rules, we obtain the structure of classical information on systems in a QPT. We can think of this as a way of embedding classical information into the systems of an arbitrary QPT.

**Definition 3.3.4.** In a †-SMC, the pair of a monoid $(A, \,, \,)$ and comonoid $(A, \,, \,)$ form a **dagger-Frobenius algebra** (†-FA) when the following equation holds:

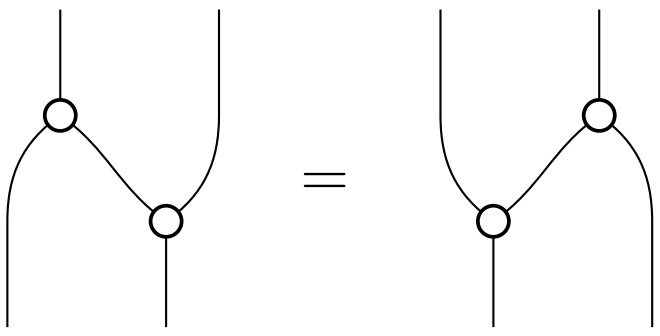

(3.31)

When  is commutative, the †-FA is commutative (is a †-CFA). The co-commutativity of  for a †-CFA follows [68, Thm 3.2.8].

**Definition 3.3.5.** A **classical structure** (○) is a dagger-Frobenius algebra $(A, \,, \,, \,, \,)$ satisfying the **specialness** (3.32) and **symmetry** (3.33) conditions:

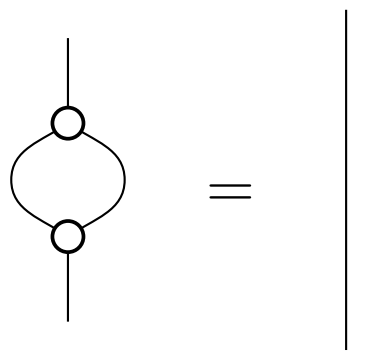

(3.32)

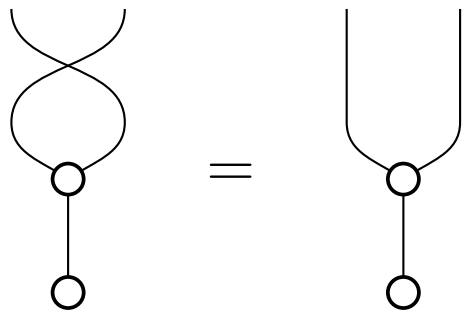

(3.33)

**Definition 3.3.6.** The set of **classical states** $K_\circ$ for a classical structure (○) are all states $j : I \to A$ such that:

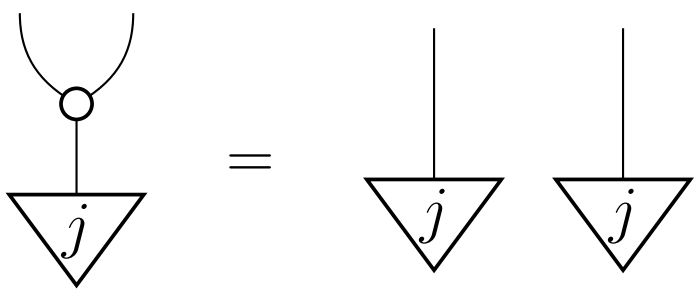

(3.34)

These classical states will typically be drawn as states that are the same color as the classical structure to which they correspond. It is also easy to determine the behavior of classical states under composition with the (co)unit:

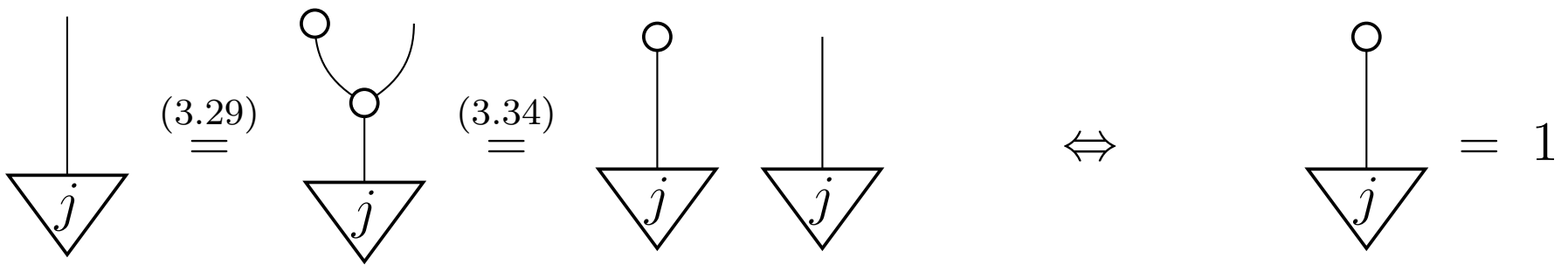

(3.35)

where 1 is the unit scalar, i.e. $1 \circ s = s = s \circ 1$ for all scalars $s$. In this sense the counit "erases" classical points.

The following theorems by Coecke et al. characterize the relationship between classical structures and observables.

**Theorem 3.3.7** ([37, Thm 5.1])**.** *Symmetric dagger Frobenius algebras in* **FHilb** *are orthogonal bases.*

The additional condition of specialness for classical structures acts as a normalizing condition so that:

**Theorem 3.3.8** ([37, Sec 6])**.** *Classical structures in* **FHilb** *are orthonormal bases.*

Classical structures are bases in **FHilb** and so are recognized as generalized **observables** in a QPT. The classical states of classical structures (Definition 3.3.6) form the elements of these bases, more specifically the eigenvectors of the observables, though we should be careful that in categories other than **FHilb**, they do not necessarily have all the familiar properties of bases for Hilbert spaces. For example, we usually expect to be able to distinguish maps by testing them on basis elements. In general this only holds for certain classical structures:

**Definition 3.3.9.** A classical structure ( $\circ$ ) has **enough classical states** when for all processes $f, g : A \to B$:

$$f = g \qquad \Leftrightarrow \qquad (\forall |j\rangle \in K_{\circ} : f \circ |j\rangle = g \circ |j\rangle) \tag{3.36}$$

**Remark 3.3.10.** The correspondence in Theorem 3.3.8 was modified for the infinite dimensional case in [7], where it still holds. We will only use the finite dimensional case in this thesis as we are concerned only with algorithms that run on computers of finite size.

**Definition 3.3.11.** In a QPT, the **dimension** $d(A)$ of an object $A$ equipped with a dagger-Frobenius algebra ($A$, 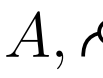, 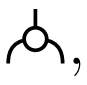, 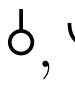, 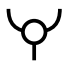), is given by the following composite:

$$\mathrm{d}(A) \quad := \tag{3.37}$$

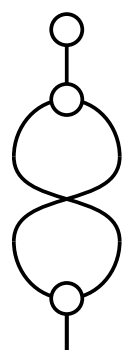

When the algebra is in fact a classical structure, (3.37) can be simplified to the composition of the unit and counit:

$$d(A) \; = \tag{3.38}$$

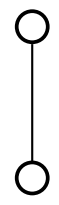

**Remark 3.3.12.** It is important to note that the dimension of a system does not always equal the number of classical states of the associated classical structure. One setting where this does not hold is **FRel**, where this fact plays an important role in its QPT characterization in Chapter 5.

An important result of Coecke and Duncan shows that the rules for classical structures can be summarized in a convenient normal form that is in keeping with the sorts of topological equivalences we are used to for diagrams of a QPT. Let the maps $\delta_n : A \to A^{\otimes n}$ be defined recursively by:

$$\delta_0 := \epsilon \qquad\qquad \delta_{n+1} := (\delta_n \otimes \mathrm{id}_A) \circ \delta \tag{3.39}$$

Let $\mu_n$ be similarly defined.

**Theorem 3.3.13** (Spider Theorem [31, Thm 6.11])**.** *Given a classical structure, let $f : A^{\otimes n} \to A^{\otimes m}$ be constructed from $\{\mu, \delta, \eta, \epsilon\}$ such that the diagram is connected. Then $f = \delta_m \circ \mu_n$.*

This normal form has a neat diagrammatic representation.

**Proposition 3.3.14** ([31, Thm 6.12])**.** *Given a classical structure on A, let $(\circ)_n^m$ denote the '$(n, m)$-legged spider':*

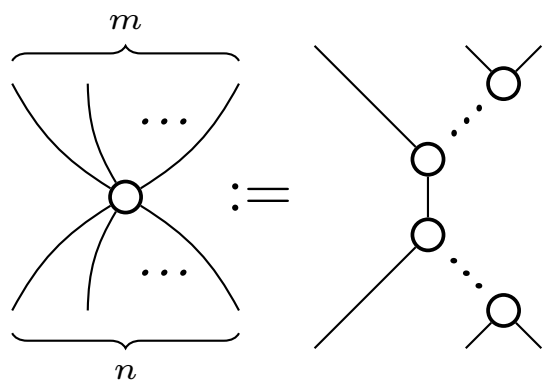

(3.40)

*then any process $A^{\otimes n} \to A^{\otimes m}$ built from $\{\mu, \delta, \eta, \epsilon\}$ which has a connected graph is equal to $(\circ)_n^m$. Spider composition is:*

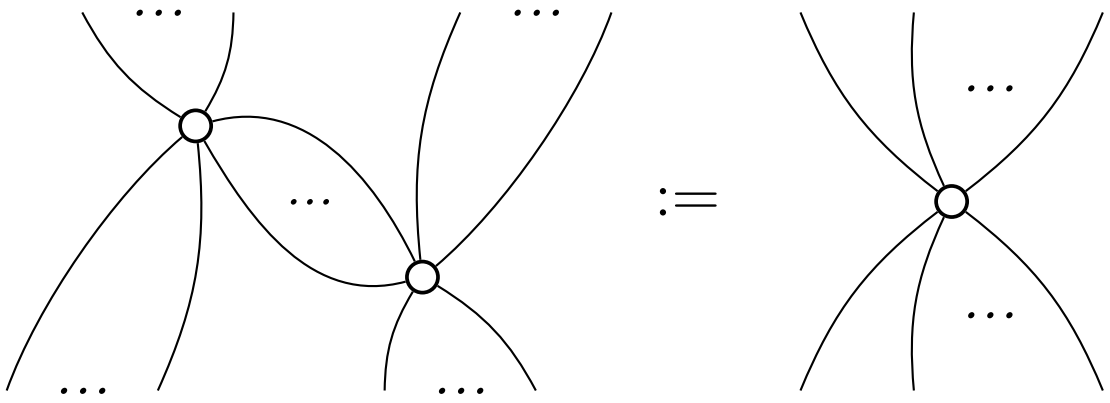

(3.41)

The spider rule makes it clear that every classical structure on $A$ can be used to make $A$ dual to itself (Definition 3.2.2) using caps ($\epsilon \circ \mu$) and cups ($\delta \circ \eta$) built from the white dot:

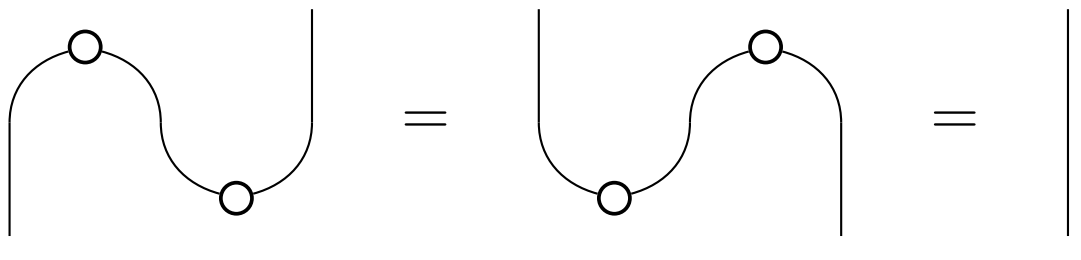

(3.42)

Kissinger provides direct proof of this fact [68, Thm 3.2.7].

The upper and lower star operations with respect to this white dot cup and cap corresponds in **FHilb** to transposition and conjugation in the white dot basis respectively. Further it can be shown that transposition with respect to a classical structure is equivalent to the dagger when applied to its classical states:

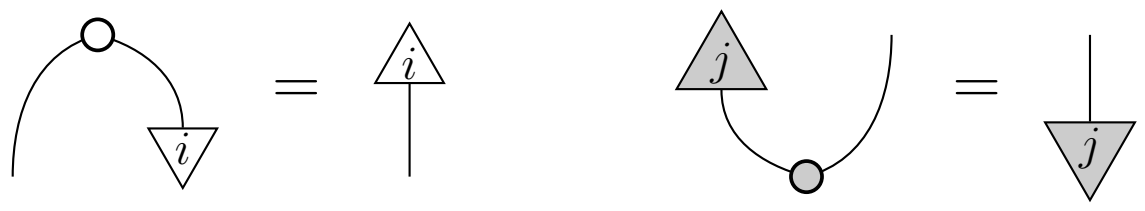

(3.43)

This is of course not true on general states.

## 3.4 Phases

Phases for QPTs were introduced by Coecke and Duncan [31]. They decorate classical structures with an abelian group in a particular way.

**Definition 3.4.1.** A **phase state** for a Frobenius algebra $(A, \curlyvee, \varphi)$ is a state $|\alpha\rangle$ such that:

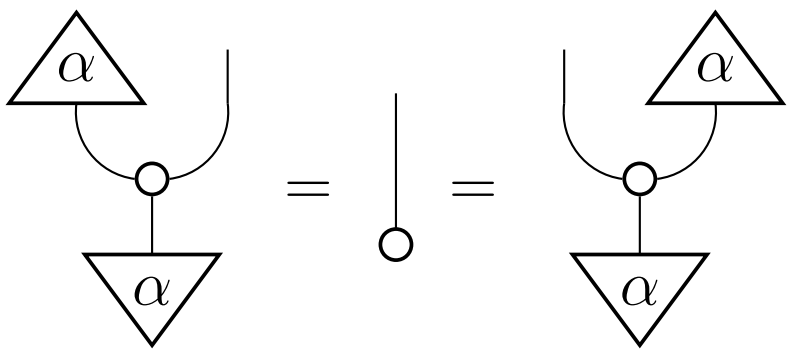

(3.44)

Each phase state corresponds to a **phase** that is a unitary map in the following form:

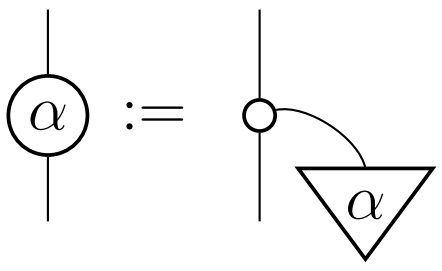

(3.45)

**Proposition 3.4.2** (Phase groups)**.** *Given a †-FA $(A, \curlywedge, \delta, \curlyvee, \varphi)$ in a QPT, its phases form a group under the following addition:*

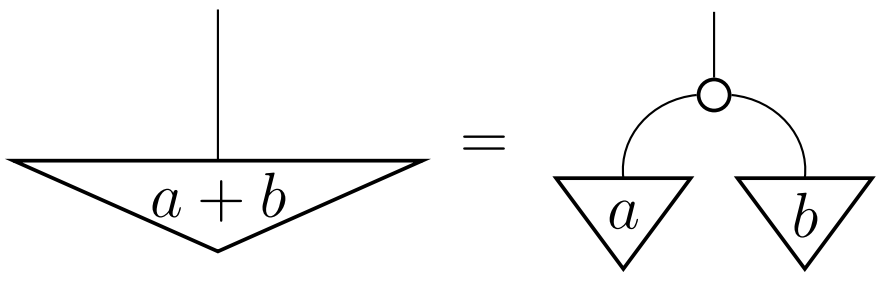

(3.46)

*with unit $\delta$. When the Frobenius algebra is part of a classical structure, then its phases form an abelian group.*

*Proof.* See [31, Sec. 7.4] for proofs. □

Phases can be added to the normal form from Theorem 3.3.13 to give a decorated spider rule [31, Thm 7.11]:

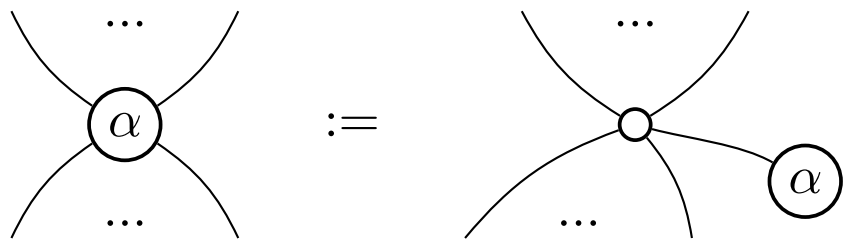

(3.47)

with composition

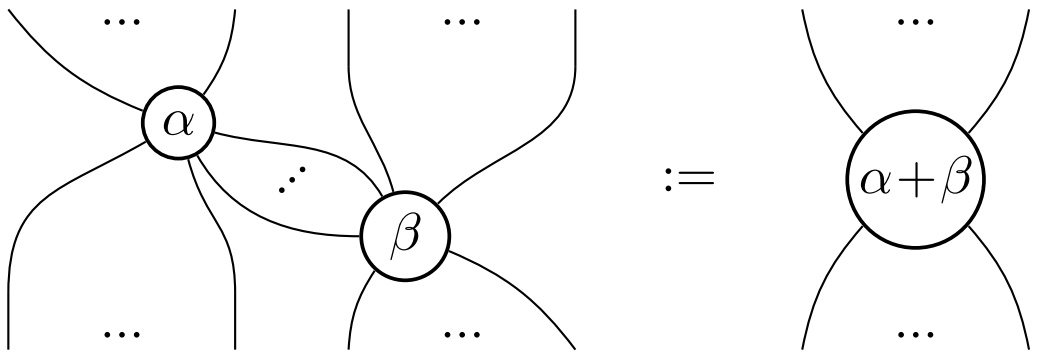

(3.48)

This normal form is a powerful simplifying tool and will often be used in the analysis of diagrams of processes in QPTs.

## 3.5 Complementarity

The notion of complementary bases can also be lifted to the general level of classical structures in QPTs [31]. This extension is perhaps best presented as emergent from a suitable generalization of mutual unbiasedness.[4] This means that for two bases $\{|i\rangle\}$ and $\{|j\rangle\}$ on a $D$-dimensional Hilbert space, $|\langle i|j\rangle|^2 = 1/D$ for all $i, j$. In diagrams we then have:

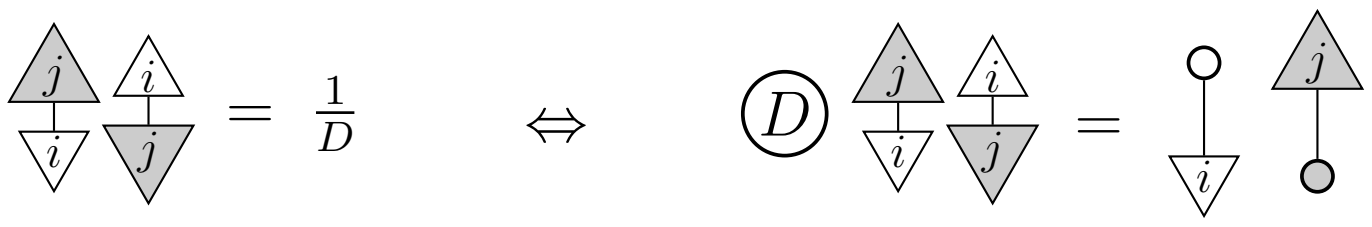

(3.49)

as the mutual unbiasedness condition, where we have assumed that $D$ is invertible (as all dimensions in **FHilb** are) and used (3.38) and (3.35) to obtain the right hand form.

**Definition 3.5.1** (Complementarity)**.** In a QPT, two classical structures ( ○ ) and ( ● ) on the same object are **complementary** when the following equation holds:

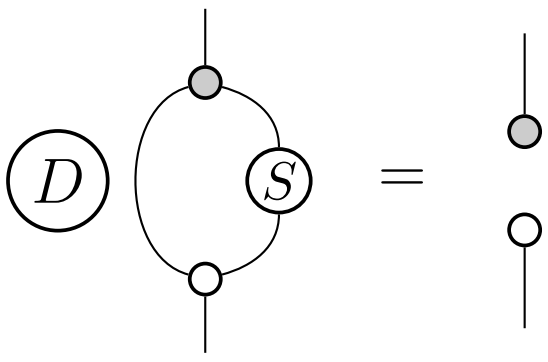

where

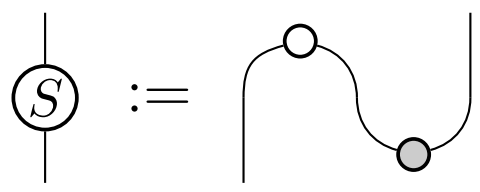

(3.50)

[4]Our presentation in this regard is heavily influenced by the approach taken by Coecke et al. in [33].

where $D$ is the dimension of the object and the map $S : A \to A$ is called the **antipode**.

The Equation (3.50) is equivalent to mutual unbiasedness on classical structures that have enough classical points by Definition 3.3.9. Then by a quick calculation:

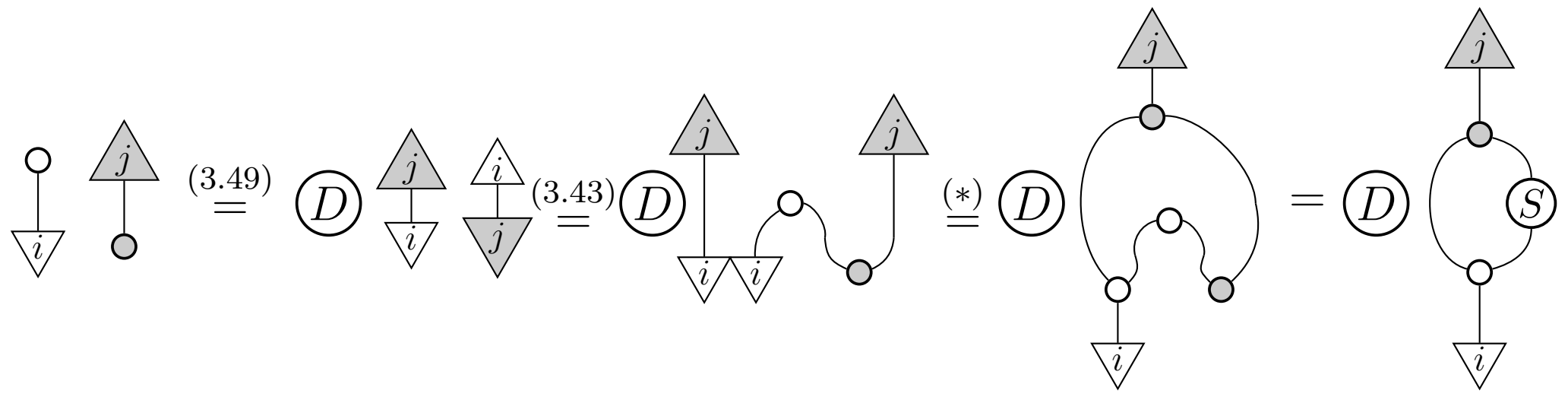

(3.51)

where $(*)$ is an application of Theorem 3.2.7, and the last step uses the fact that there are enough classical points.

Complementarity relates the classical states of classical structures to their phase groups.

**Lemma 3.5.2.** *If two classical structures in a QPT are complementary, then, up to an idempotent scalar, a state that is a classical point of one is a phase state for the other.*

*Proof.* We prove this using diagrams, following Coecke and Kissinger [35]:

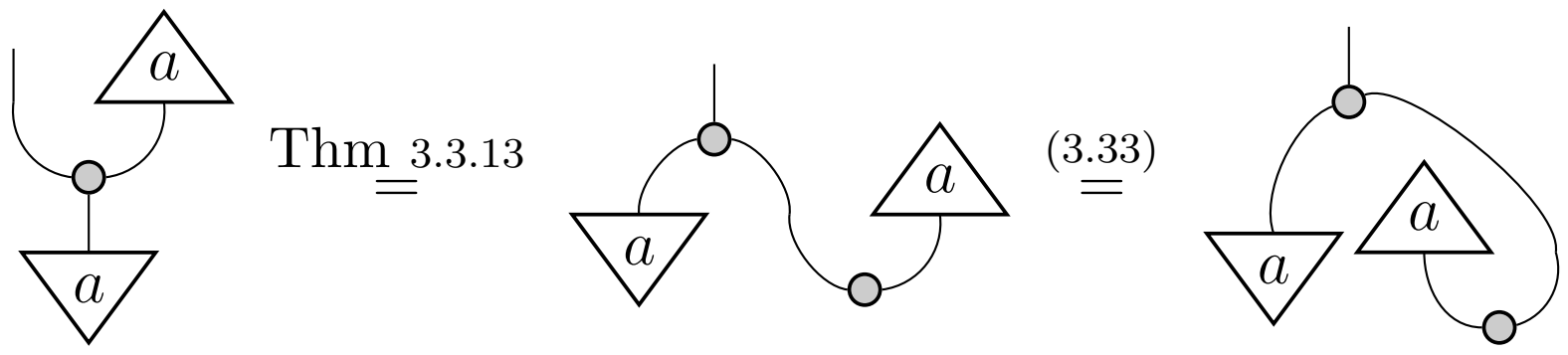

(3.52)

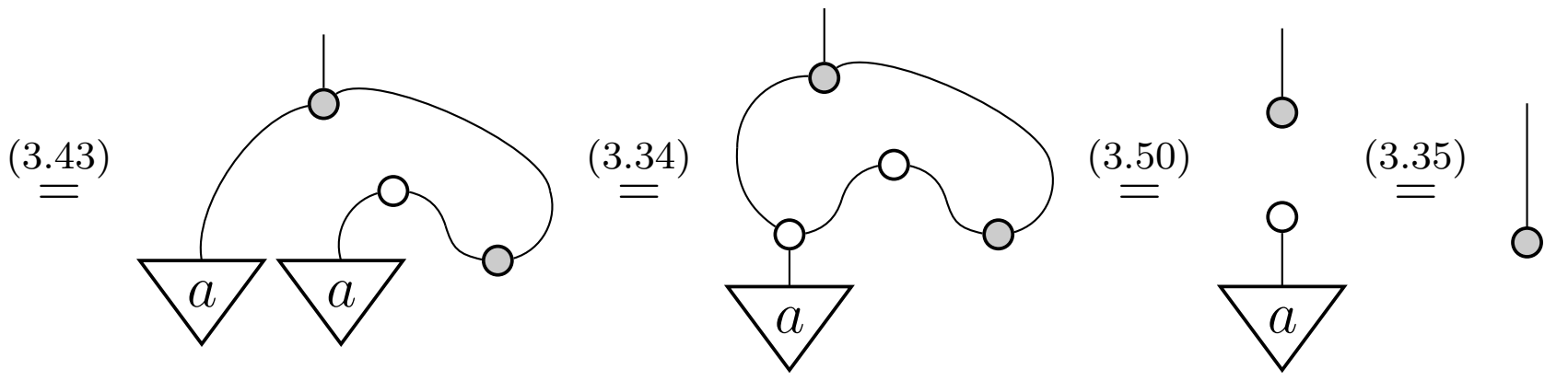

(3.53)

□

Though the converse of this lemma holds in **FHilb**, i.e. each phase is also a classical point of the complementary structure, the converse does not hold in general.

**Definition 3.5.3.** In a QPT, two classical structures (○) and (●) are **coherent** when the following equations hold:

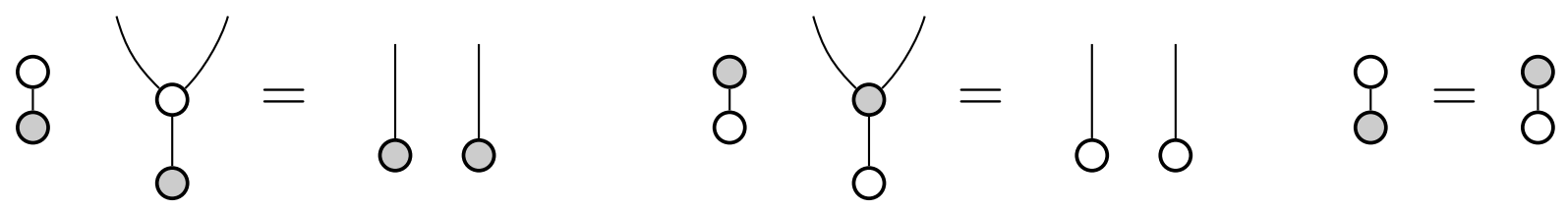

(3.54)

**Proposition 3.5.4** ([33, Prop. 3]). *In* **FHilb** *if we are given two self-adjoint operators corresponding to complementary classical structures, then we can always construct a pair of coherent classical structures with the same classical points.*

This means that in the QPT for quantum computation, we can take complementary classical structures to be coherent without loss of generality.

The terminology for the antipode comes from the fact that complementarity is almost enough to make the classical structures a Hopf algebra using this antipode. Some complementary classical structures do indeed form Hopf algebras and these ones are called strongly complementary. They will be discussed in Section 4.1 in detail.

**Example 3.5.5.** Complementary observables allow us to construct a generalized form of the controlled-not gate in any QPT. In **FHilb** we can choose classical structures on the two-dimensional Hilbert space that correspond with the usual Z and X observables:

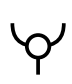 : $|0\rangle \mapsto |00\rangle$, $|1\rangle \mapsto |11\rangle$ 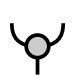 : $|+\rangle \mapsto |++\rangle$, $|-\rangle \mapsto |--\rangle$

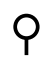 : $|0\rangle \mapsto 1$, $|1\rangle \mapsto 1$ 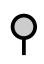 : $|+\rangle \mapsto 1$, $|-\rangle \mapsto 1$

$K_{○} = \{|0\rangle, |1\rangle\}$ $K_{●} = \{|+\rangle, |-\rangle\}$

These classical structures are coherent and complementary and each have $\mathbb{Z}_2$ as their phase groups, which we will write as $\{0, \pi\}$ where addition is modulo $2\pi$. By Lemma 3.5.2, we know that the classical points of the ○-structure are phases for the ●-structure. In particular we have:

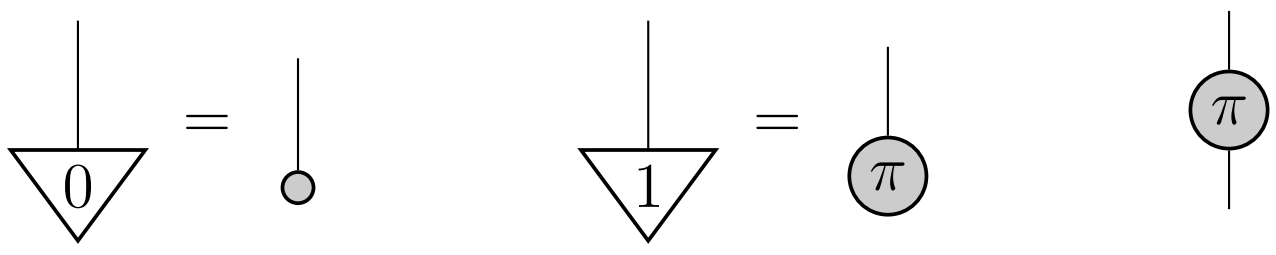 $= X\text{-gate} :: \begin{cases} |0\rangle \mapsto |1\rangle \\ |1\rangle \mapsto |0\rangle \end{cases}$ (3.55)

These structures can be used to define the controlled-not CNOT : $H_2 \otimes H_2 \to H_2 \otimes H_2$ whose diagram is:

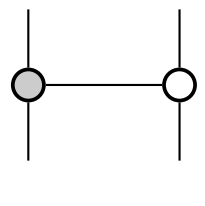 (3.56)

where the right system acts as the control. We can verify that this behaves as usual for qubits using Z (the ○-classical structure as their computational basis).

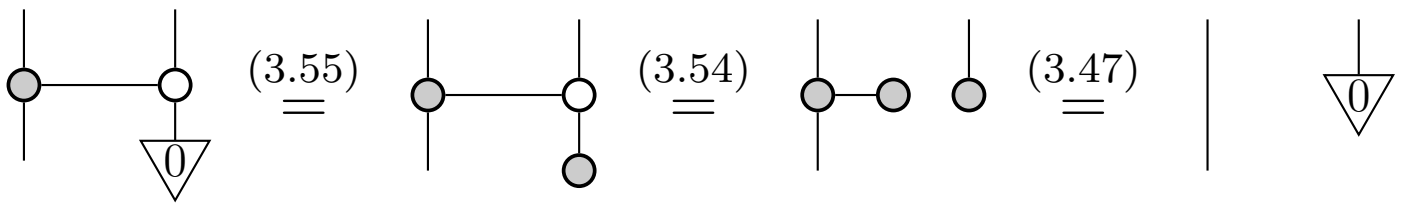

(3.57)

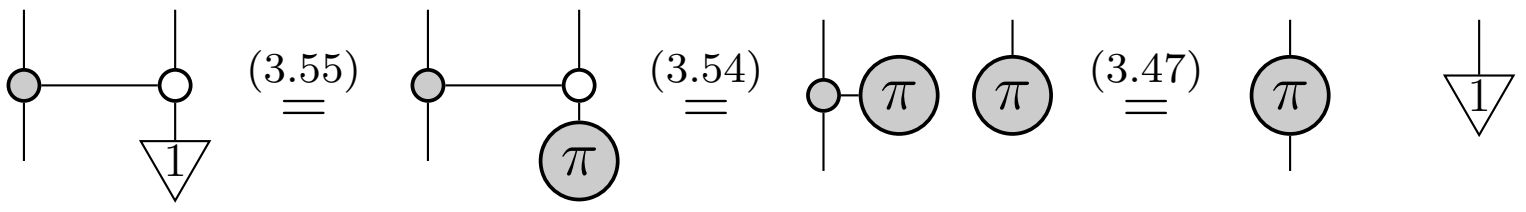

(3.58)

This shows that on computational basis elements, the control is left unchanged while a $X$-gate (a computational bit flip) is conditionally applied.

This example construction motivates the following definition on any kind of system in any QPT:

**Definition 3.5.6.** Given two classical structures (○) and (◉) that are complementary and coherent, the generalized **controlled-not** is the map:

$\sqrt{\mathrm{d}(A)}$ 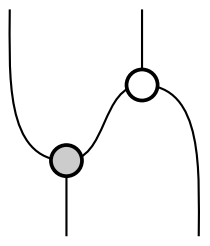 (3.59)

where we have assumed that a scalar equal to the square root of the dimension exists.

This scalar is needed to ensure the unitarity of the controlled-not gate, which we discuss in further details in Section 4.2.1.

## 3.6 Enriched QPTs

Some QPTs, such as **FHilb**, come equipped with linear structure on their processes. Abstractly these are enriched categories, i.e. categories where hom-sets are replaced with other objects.[5]

**Definition 3.6.1.** Given a monoidal category $K$, a $K$-enriched category $\mathbf{C}$ has objects $\mathrm{Ob}(\mathbf{C})$ such that:

- For every pair of $A, B \in \mathrm{Ob}(\mathbf{C})$ there is a hom-object:

$$\mathbf{C}(A, B) \in \mathrm{Ob}(\mathbf{K}). \tag{3.60}$$

[5] Kelly [67] can be used as a standard reference for enriched category theory.

- For all objects $A, B, C \in \text{Ob}(\mathbf{C})$ composition is given by a morphism in $\text{Arr}(\mathbf{K})$ of type:

$$\circ_{A,B,C} : \mathbf{C}(B, C) \otimes \mathbf{C}(A, B) \to \mathbf{C}(C, A). \tag{3.61}$$

Linear maps between Hilbert spaces themselves form a vector space, so **FHilb** is in fact a **FVect**-enriched category. Another way of saying this is that **FHilb** is enriched in **FVect**. As diagrams of a process theory are morphisms in a category, diagrams in an enriched category have additional operations. The **FVect** enrichment of **FHilb**, for example, allows us to sum diagrams, as in the following property. In the following definition **Mon** is the category of monoids and monoid homomorphisms.

**Definition 3.6.2.** In a **Mon**-enriched QPT, a set of states $\{|x\rangle\}$ on a system $A$ forms a **resolution of the identity** when:

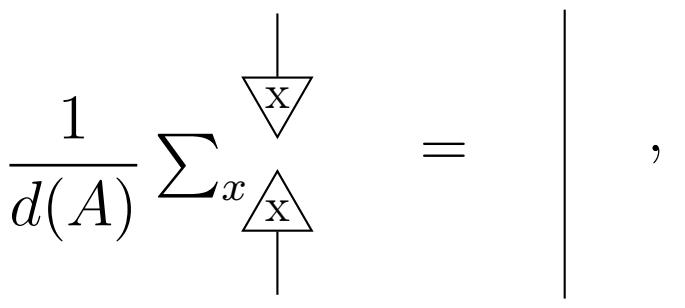

(3.62)

where the sum operation comes from a monoid structure of the hom-object **C**(A,A). In **FHilb** more specifically, this addition is inherited from the vector space enrichment. In fact, in Hilbert spaces all the classical states of any classical structure form a resolution of the identity. Further, in **FHilb**, we are able to link the classical structure maps to classical states in the following ways [33]:

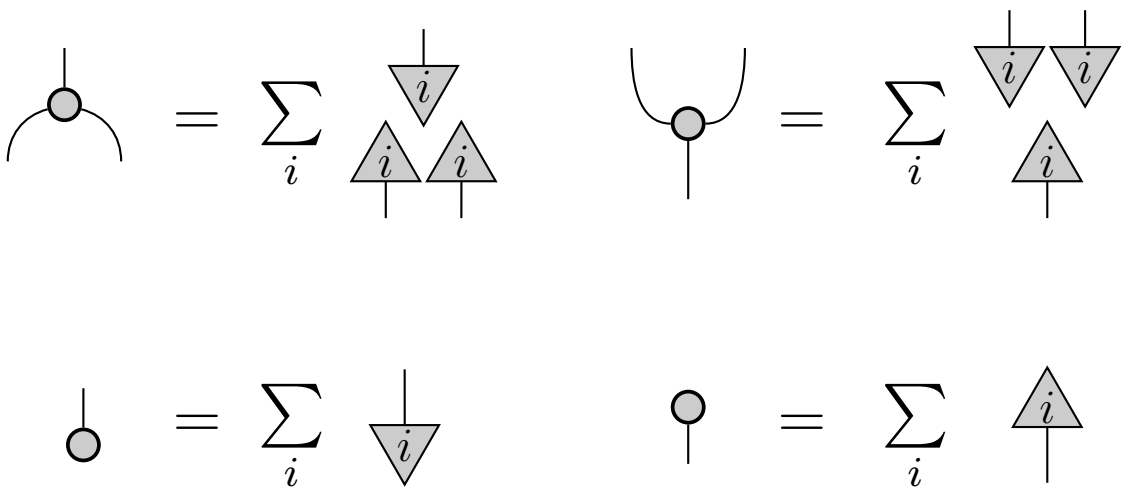

(3.63)

Arbitrary spiders can then be written as:

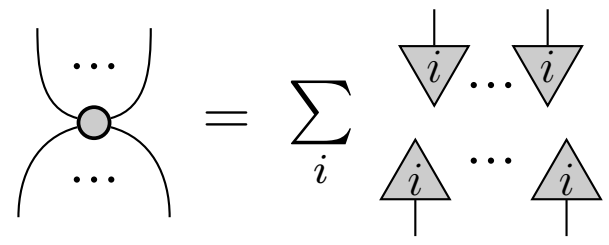

(3.64)

**Remark 3.6.3.** We emphasize that this connection between classical states and classical structures does not hold in general QPTs, but merely serves to further motivate abstract constructions that generalize from **FHilb**. We make use of this technique in the following section on measurements.

Gogioso provides more details on enriched categories, especially in regards to the results of Section 4.1 of this thesis, in [46, Sec. 6].

## 3.7 Measurements

We have already introduced a notion of post-selected measurement in Definition 3.2.5. This section uses Selinger's CPM (completely positive map) construction [98], to present measurement as a process that outputs classical information [32]. In particular, measurements are maps that decohere pure states into mixed states in a certain basis. We need to be more precise about systems and their duals here so will be more consistent about using wires decorated with the proper arrows.

Using caps and cups, we can construct the unique "name" of a process $\rho$ as:

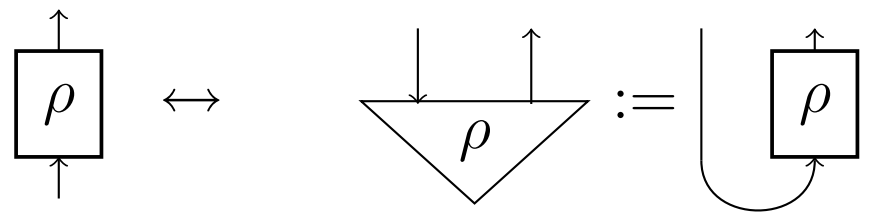

(3.65)

This is the usual Choi-Jamiolkowski isomorphism for map-state duality in quantum information. The "doubling" that occurs here, provides a natural way to represent measurements as completely positive maps. Suppose we wish to measure a system with respect to a classical structure ($\circ$), whose classical states form an orthonormal basis $\{|x_i\rangle\}$. The probability of getting the $i$-th measurement outcome is computed using the Born rule:

$$\mathrm{Prob}(i, \rho) = \mathrm{Tr}(|x_i\rangle\langle x_i|\,\rho) \tag{3.66}$$

We can write this probability distribution as a vector in the basis $\{|x_i\rangle\}$. That is, a vector whose $i$-th entry is the probability of the $i$-th outcome:

$$M_{\circ}(\rho) = \sum \mathrm{Tr}(|x_i\rangle\langle x_i|\,\rho)|x_i\rangle \tag{3.67}$$

So, $M$ defines a linear map from density matrices to probability distributions.

Expanding this graphically, we have:

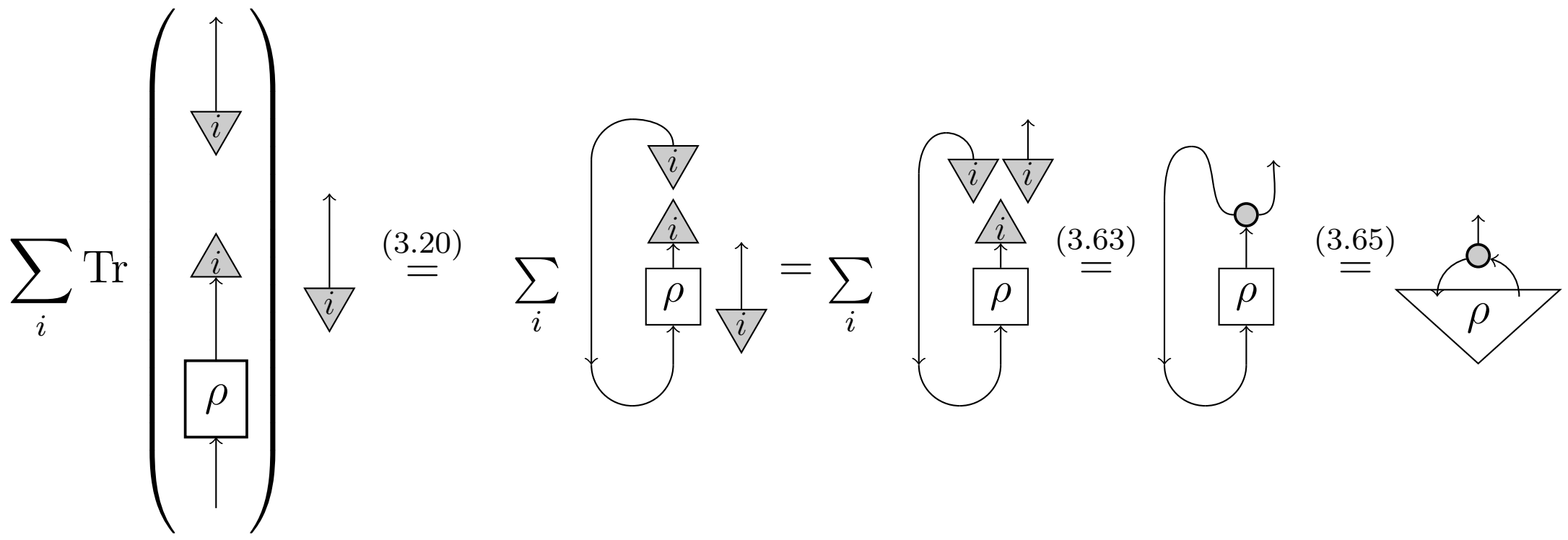

(3.68)

**Definition 3.7.1** ([32])**.** For a classical structure (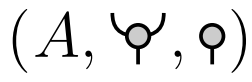), a measurement in that classical structure is defined as the following map:

$m_{\bigcirc}$ := 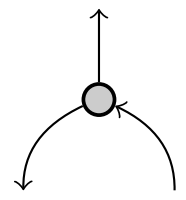 (3.69)

While the exact Born rule derivation does not apply in all QPTs (as traces and the linear structures are different in general), we can still consider Definition 3.7.1 as the abstract version of a QPT measurement.

**Example 3.7.2.** We will illustrate this measurement using an example on a qubit. Take the $Z$ and $X$ classical structures to be defined as they were in Example 3.5.5. Here $Z$ is the computational basis and the two classical points of $X$ are 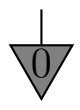 $= |+\rangle$ and 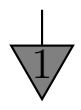 $= |-\rangle$. Thus a measurement of the state $|0\rangle$ in the $X$ (gray) basis is $m_{\bigcirc} \circ |0\rangle\langle 0|$, which is diagrammatically:

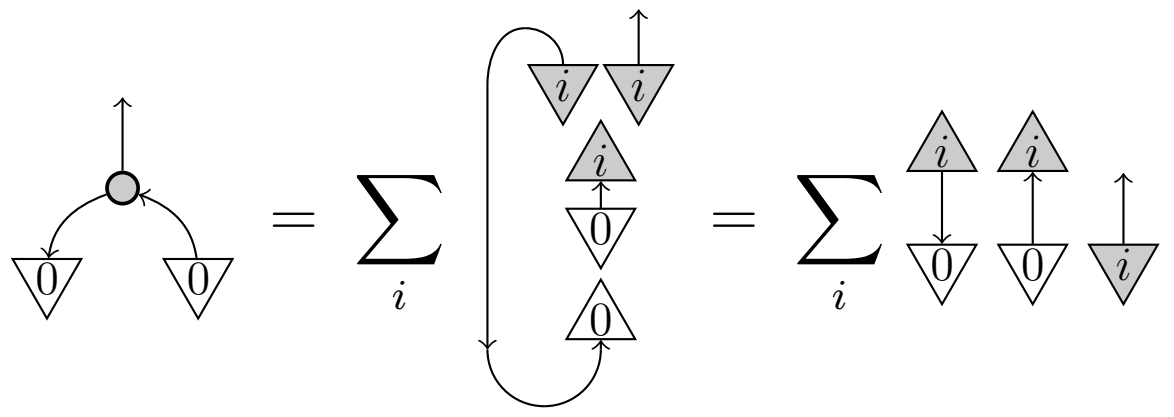 $= \frac{1}{2}|+\rangle + \frac{1}{2}|-\rangle.$ (3.70)

This gives the expected result that a measurement of $|0\rangle$ in the $X$ basis is a mixed state of $|+\rangle$ and $|-\rangle$.

We especially make use of this measurement presentation in Chapter 5, to describe Mermin non-locality tests.

## 3.8 Summary of QPTs

This chapter has introduced the main framework that is used in this thesis. This framework performs two functions:

1. our framework lifts the structure of quantum computation into the general setting of quantum-like process theories. This allows the study of protocols like teleportation, state transfer [57], quantum secret sharing (Section 5.4), quantum bit commitment [39], Mermin non-locality tests (Chapter 5), and other protocols (Chapters 4 and 6) to be handled at an abstract level and studied in different QPTs.

2. The categorical diagrams that accompany QPTs present a more powerful quantum circuit language. This is one that includes bases, classical information, complementarity, and phases explicitly and gives rules for manipulating these structures within the diagrams themselves. We present a summarizing table of these structures at the end of this chapter.

Both the generalization and diagrammatic formalism provide powerful tools throughout the results in this thesis.

Figure 3.1: A summary of the diagrammatic elements for QPTs above and beyond quantum circuit diagrams.

| | |
|---|---|
| The **dagger** $\dagger : \mathbf{C} \to \mathbf{C}$ generalizes the adjoint and vertically flips the whole diagram. See Definition 3.1.1. | 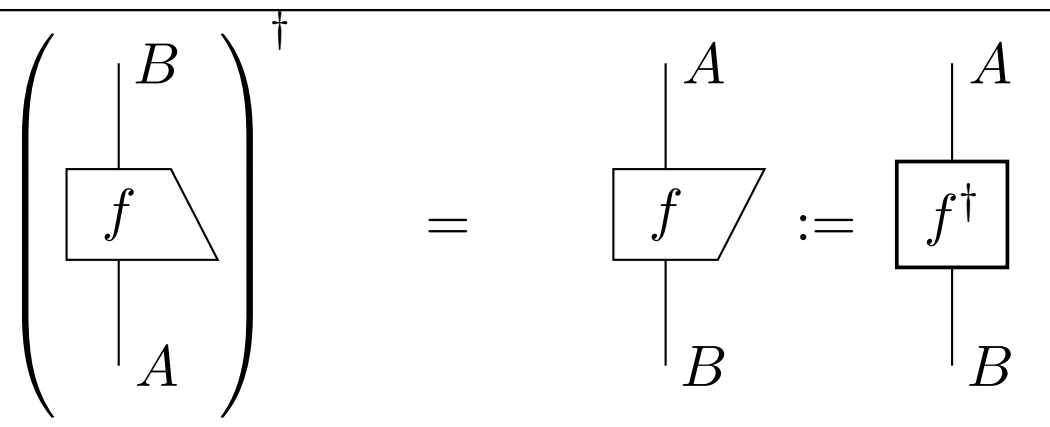  |
| **Scalars** $s : I \to I$. These float freely in diagrams. See Definition 3.1.3. | 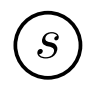  |
| **States** and **effects**. Definition 2.1.4 | 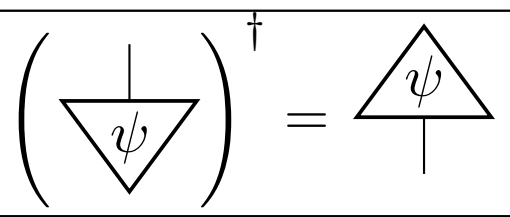  |
| **Post-selected measurement**. Section 3.2. | 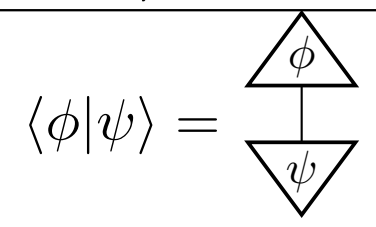  |
| **Bell states** and measurements, i.e. cups and caps. Definition 3.2.2. | 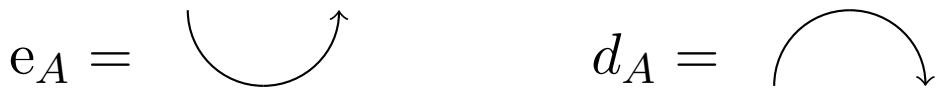  |
| **Duals**. See (3.16). | 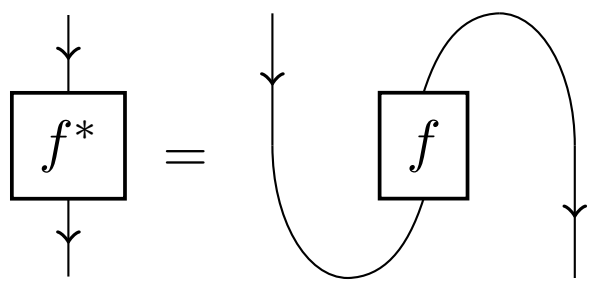  |
| **Classical structures** i.e. generalized observables (Def. 3.3.5). Their **classical states** (Def. 3.3.6) act as "basis elements". | 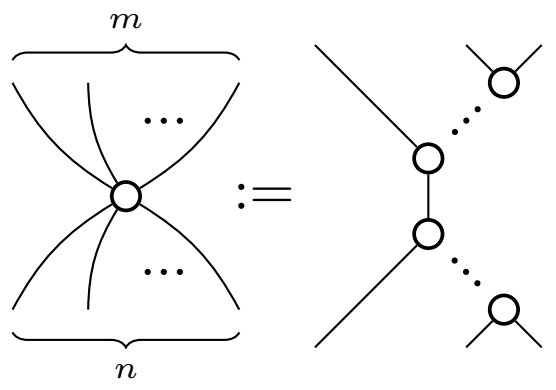  |
| **Dimension** of a system. Def. 3.3.11 | 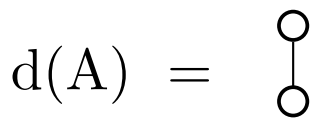  |
| **Phases**: A group of states for each classical structure. Section 3.4. | 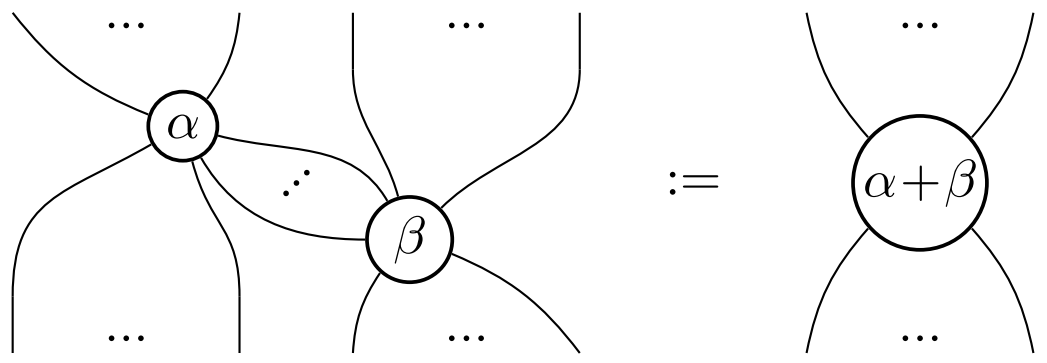  |
| **Complementarity** between classical structures. Definition 3.5.1. | 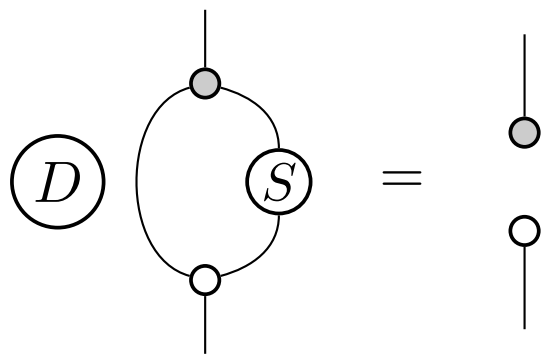  |
| **Measurement** by a classical structure. Definition 3.7.1. | 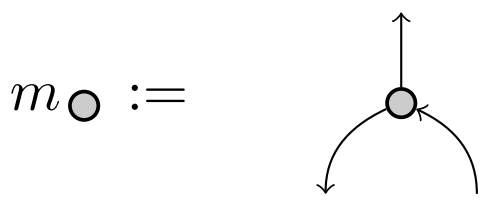  |

# Chapter 4

# Quantum Algorithms

**Chapter Abstract**

This chapter applies quantum-like process theories to the study of quantum algorithms in three ways. The first section connects the Fourier transform to strongly complementary observables in a QPT. This provides a new mathematical setting for Fourier theory and pinpoints the reason why the Fourier transform appears naturally in quantum computation. These results are ongoing collaborative work by the author and Gogioso that updates our preprint [46].

The second section applies QPTs to the verification and generalization of quantum blackbox algorithms. We provide a general construction of unitary oracles in arbitrary process theories and prove an equivalence between the unitarity of these oracles and the existence of complementary observables. We then extend work by Vicary [104] to present a quantum blackbox algorithm for a new problem that identifies group homomorphisms. These results are extensions of collaborative work by the author and Vicary that were published in [115].

The third section of this chapter presents a new toy model for quantum blackbox algorithms in the QPT of sets and relations along with several mathematical results. This setting is usually interpreted as a model for nondeterministic classical computation and so demonstrates a technique for modelling quantum algorithms classically. Results of this section are adapted from work by the author in the preprint [116].

## 4.1 The Fourier transform in QPTs

Ongoing work in quantum algorithms emphasizes the need for a structural understanding of quantum speedups [1]. In this section we focus on the quantum

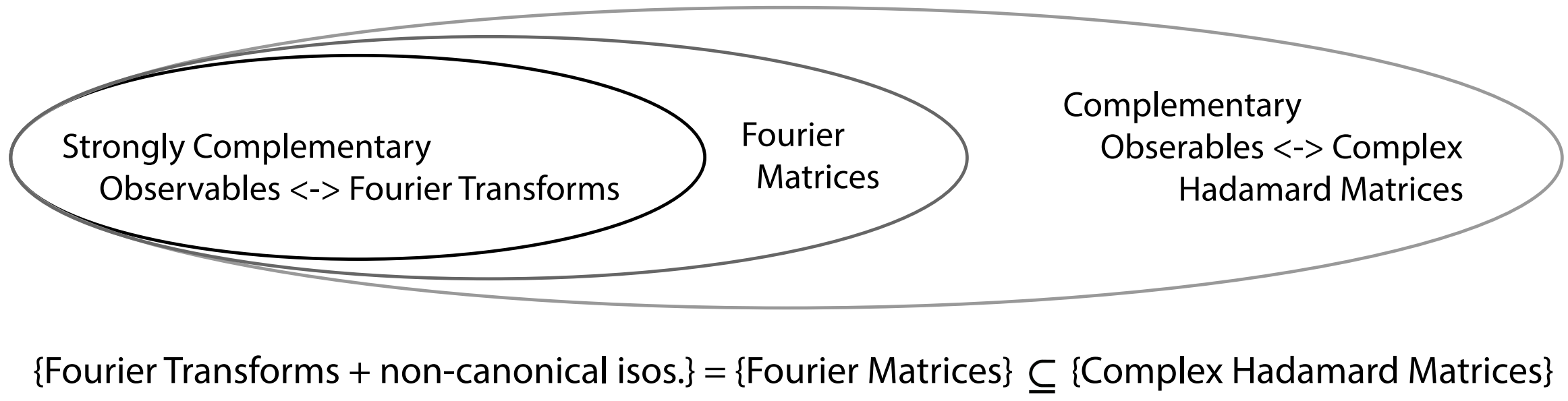

Figure 4.1: Schematic of the relationship between Fourier transforms, Fourier matrices, Hadamard matrices, strongly complementary observables, and complementary observables.

Fourier transform and the structure in quantum theory that enables it. We elucidate a general connection in any QPT between the Fourier transform and strongly complementary observables, i.e. Hopf algebras in dagger symmetric monoidal categories. We emphasize that, while they happen to coincide for qubits and systems composed of qubits, the Fourier transform of a general system is mathematically distinct from a Fourier matrix. In particular, a Fourier matrix is a Fourier transform equipped with the choice of an isomorphism that is, in general, non-canonical. These Fourier matrices then correspond to strongly complementary observables (with a choice of isomorphism) in the same way that complex Hadamard matrices correspond to complementary ones. The relationship between these concepts is illustrated in Figure 4.1. The section proceeds along the following outline:

Section 4.1.1 gives background for the traditional notion of the Fourier transform as is relevant for our construction. We also emphasize the relationship between several different, but related concepts: the Fourier transform, Fourier matrices, and (complex) Hadamard matrices.

In Section 4.1.2 we cover the definition of strong complementarity [31], which has been used in the foundations of quantum mechanics to study non-locality [32, 47] (Chapter 5), quantum secret sharing [47, 114] (Section 5.4), and blackbox quantum algorithms [104, 115, 116] (Section 4.2). This allows a generalization beyond **FHilb** to strongly complementarity pairs of a quasi-Special †-Frobenius Algebra (†-qSFA or †-qSCFA if commutative) and a †-SCFA. We use this generalization to embed finite groups in arbitrary dagger symmetric monoidal categories.

In Section 4.1.3 the usual Fourier transform concepts from Section 4.1.1 are lifted to general symmetric monoidal categories. We construct the accompanying general definitions for multiplicative characters and the abelian Fourier transform in this setting. These results allow us to provide categorical versions, with abstract proofs, of the Fourier inversion theorem, the convolution theorem, and Pontryagin duality that are all based on a strongly complementary pair of observables.

In Section 4.1.4, we study **Rel** as an example setting for our categorical Fourier transform. This example is of particular interest as it often acts as a toy model for quantum theory [44, 57, 85, 116]. We find that while a generalized Fourier matrix is not suitably defined, a Fourier transform can be.

In Section 4.1.5 we review extensions of these results to the non-abelian case, with accompanying Fourier transform. Then, in Section 4.1.6, we summarize how these results relate to measurements in the "representation basis."

These results both move Fourier theory into a new mathematical setting and capture the structural connection between quantum theory and the Fourier transform. Though this connection has been much exploited in quantum algorithms, this work is the first abstract presentation that shows its place in the structure of quantum theory, i.e. alongside strongly complementary observables.

### 4.1.1 The Fourier transform

We begin with a quick review of Pontryagin duality and the Fourier transform as it relates to quantum computation. A number of different notions related to the Fourier transform on finite abelian groups can be found in mathematics, physics, computer science and quantum computation, so it is useful to clarify them:

1. In mathematics, the Fourier transform is understood through Pontryagin duality.

2. In physics and signal processing, the Fourier transform is understood as a transformation of fields/signals from time/space domain to energy[1]/momentum domain.

3. In quantum computing, we have Fourier matrices and (complex) Hadamard matrices that correspond to unitary quantum processes.

This section is ultimately concerned with the first notion, where the Fourier transform is defined on locally compact groups. Still, the other notions are relevant, as our work situates this abstract definition in the context of QPTs, a structure inherited from quantum information.

[1] Or frequency.

We begin by explaining the relationship of Notion 1 with the others listed above. In what follows, $(G, \cdot, 0)$ is a finite abelian group of order $N$, and $\mathbb{C}^\times = \mathbb{C} \setminus \{0\}$ is the multiplicative group of non-zero complex numbers.

A **(multiplicative) character** of $G$ is a group homomorphism $\chi : G \to \mathbb{C}^\times$.[2] If $G = \prod_j \mathbb{Z}_{n_j}$,[3] a multiplicative character $\chi_h$ takes the following form, for any $g, h \in G$ s.t. $h = \prod_j h_j$ and $g = \prod_j g_j$:

$$g \mapsto \exp\left[ i \sum_j \frac{2\pi}{n_j} \left( g_j h_j \pmod{n_j} \right) \right] \tag{4.1}$$

The set of characters, with pointwise multiplication defined as $(\chi \cdot \psi)(x) := \chi(x)\psi(x)$, forms a group; this is called the **Pontryagin dual** (or **dual group**) of $G$, and is denoted $G^\wedge$. In fact, the Pontryagin construction can be made (contravariantly) functorial on the category **Grp** of groups and group homomorphisms. Define $f^\wedge : G^\wedge \to H^\wedge$, for any $f : H \to G$ morphism of abelian groups, as follows:

$$f^\wedge(\chi) = \chi \circ f.$$

From (4.1) it is not hard to see that $G^\wedge \cong G$.[4] However, this isomorphism is *not canonical*. This means that there is no natural way of identifying the multiplicative characters with group elements, and we must keep track of our choice of isomorphism $G^\wedge \cong G$. Remarkably though, there is a canonical isomorphism $G \cong (G^\wedge)^\wedge$ given as follows, making the functor $(-)^\wedge : \mathbf{Grp} \to \mathbf{Grp}$ is its own (weak) inverse:

$$g \mapsto (\chi \mapsto \chi(g)).$$

We are now ready to introduce the Fourier transform in the context of Pontryagin duality: this is the most abstract among the notions, and the others are derived from it. Let $\mathrm{L}^2[G]$ denote the space of functions $f : G \to \mathbb{C}$. These functions are necessarily square-integrable (as $G$ is finite), and thus $\mathrm{L}^2[G]$ is an $N$-dimensional complex Hilbert space (and lives in the category **FHilb** of finite-dimensional complex Hilbert spaces and linear maps).

**Definition 4.1.1.** The **Fourier transform** for a finite abelian group $G$ is a bijection $\mathcal{F}_G : \mathrm{L}^2[G] \to \mathrm{L}^2[G^\wedge]$, sending $f : G \to \mathbb{C}$ to the $\bar{f} := \mathcal{F}_G[f] : G^\wedge \to \mathbb{C}$ defined as follows:

$$\mathcal{F}_G[f](\chi) := \frac{1}{N} \sum_{g \in G} \chi^{-1}(g) f(g). \tag{4.2}$$

[2] For finite $G$, $\chi$ maps into the subgroup $S^1 \subseteq C^\times$ of unit complex numbers.

[3] Which is always true when $G$ is finite, for some family $(n_j)_j$ of positive integers.

[4] Note that if $G \cong H$, then there are always exactly as many isomorphisms $G \cong H$ as there are automorphisms $G \cong G$.

The **Inverse Fourier transform** is the inverse bijection $\mathcal{F}_G^{-1} : \mathrm{L}^2[G^\wedge] \to \mathrm{L}^2[G]$ and is defined as follows:

$$\mathcal{F}_G^{-1}[\bar{f}](g) := \sum_{\chi \in G^\wedge} \chi(g)\bar{f}(\chi). \tag{4.3}$$

The Fourier transform is natural. This means it is invariant under automorphisms of abelian groups (note that the isomorphism $G \cong G^\wedge$ was not). Let $\Psi : G \to H$ be some isomorphism where $M_\Psi = \mathrm{L}^2[G] \to \mathrm{L}^2[H]$ is the corresponding unitary isomorphism that takes $f \mapsto f \circ \Psi$. We then always have:

$$M_{\Psi^\wedge} \circ \mathcal{F}_H \circ M_\Psi = \mathcal{F}_G, \tag{4.4}$$

There are a number of properties of interest for the Fourier transform, some rather straightforward and others more complicated to prove. One of specific interest to this work, because of its wide application and relationship with structures in QPTs, is the Convolution Theorem. The space $\mathrm{L}^2[G]$ comes with a distinguished orthonormal basis, given by the **delta functions** $(\delta_g)_{g\in G}$ defined as follows.

$$\delta_g(h) := \begin{cases} 1, & \text{if } h = g. \\ 0, & \text{otherwise.} \end{cases} \tag{4.5}$$

We sometimes refer to this as the **computational basis**, the name usually given to it in the context of (group-theoretic) quantum algorithms.

The computational basis comes with a monoid structure, defined below and with unit $\delta_0$:

$$(\delta_g * \delta_h) := \delta_{g+h}. \tag{4.6}$$

Linearly extended to $\mathrm{L}^2[G]$, this structure yields the **convolution operation** $\big(\mathrm{L}^2[G], *, \delta_0\big)$.

$$(f * f') = \left(\sum_{g\in G} f(g)\delta_g\right) * \left(\sum_{g'\in G} f'(g')\delta_{g'}\right) = \sum_{g\in G}\sum_{g'\in G'} f(g)f'(g')\delta_{g+g'} \tag{4.7}$$

$$= \sum_{h\in G}\left(\sum_{g'\in G} f(h-g')f'(g')\right)\delta_h \tag{4.8}$$

The Fourier transforms of the delta functions yield the following orthogonal basis for $\mathrm{L}^2[G^\wedge]$, which we refer to as the **basis of evaluation functions**:

$$\xi_g := \sqrt{N}\mathcal{F}[\delta_{-g}] = \left(\chi \mapsto \sum_{h\in G} \chi^{-1}(h)\delta_{-g}(h)\right) = (\chi \mapsto \chi(g)).$$

The basis of evaluation functions also comes with a monoid structure, with unit $\xi_0 : \chi \mapsto 1$:

$$(\xi_g \cdot \xi_h) := \chi \mapsto \xi_g(\chi)\xi_h(\chi) = \chi \mapsto \chi(g)\chi(h).$$

Functions $F \in \mathrm{L}^2[G^\wedge]$ on the dual group have the following expansion in terms of evaluation functions:

$$F = \sum_{g \in G} \left( \frac{1}{N} \sum_{\chi \in G^\wedge} F(\chi)\chi^{-1}(g) \right) \xi_g$$

Linearly extended to $\mathrm{L}^2[G^\wedge]$, the monoid structure above yields the **pointwise multiplication** $\big(\mathrm{L}^2[G^\wedge], \cdot, \xi_0\big)$:

$$(F \cdot F') = \tau \mapsto \sum_{\chi,\kappa \in G^\wedge} F(\chi)F'(\kappa) \left( \frac{1}{N} \sum_{g \in G} \chi^{-1}(g)\tau(g) \right) \left( \frac{1}{N} \sum_{g' \in G} \kappa^{-1}(g')\tau(g') \right) \tag{4.9}$$

$$= \tau \mapsto F(\tau)F'(\tau) \tag{4.10}$$

We use the (easy to check) fact that, for any $\chi, \tau \in G^\wedge$, the expression $\frac{1}{N}\sum_{g \in G} \chi^{-1}(g)\tau(g)$ yields 1 if $\tau = \chi$ and 0 otherwise (this is usually referred to as **orthogonality of (multiplicative) characters**).

**Theorem 4.1.2** (Convolution Theorem)**.** *The Fourier transform is a monoid isomorphism in* **FHilb**, *from the convolution monoid* $\big(\mathrm{L}^2[G], *, \delta_0\big)$ *to the pointwise multiplication monoid* $\big(\mathrm{L}^2[G^\wedge], \cdot, \xi_0\big)$. *This statement amounts exactly to the following expression (for every* $f \in \mathrm{L}^2[G]$*), which is the usual formulation of the Convolution Theorem:*

$$\mathcal{F}_G(f') \cdot \mathcal{F}_G(f) = \mathcal{F}_G(f * f'). \tag{4.11}$$

This concludes our presentation of the Fourier transform in the context of Pontryagin duality. A further reference for details of the topics in this presentation is [93]. The Fourier transform finds wide applicability in signal processing, physics, engineering and the applied sciences, but the full formulation based on Pontryagin duality is rarely used, if mentioned at all. In the engineering context, one usually considers periodic real-valued or complex-valued functions on a $D$-dimensional space, discretized in a rectangular $D$-dimensional lattice, and defines the (Discrete) Fourier transform as a transformation on them. Due to the periodicity conditions, complex-valued functions on a rectangular $D$-dimensional lattice can be equivalently seen as living in $\mathrm{L}^2[G]$, where $G = \prod_{j=1}^{D} \mathbb{Z}_{n_j}$ and $n_j$ is the number of lattice sites along the $j$-th dimension. The Fourier transform $\mathcal{G} : \mathrm{L}^2[G] \to \mathrm{L}^2[G^\wedge]$ defined above sends these functions onto functions on another, isomorphic $D$-dimensional lattice corresponding to $G^\wedge$. In order to obtain functions living back on the original lattice, one *fixes an isomorphism* $\Psi : G \to G^\wedge$ (traditionally the one from Equation 4.1), and defines the Discrete Fourier transform as the following transformation on $\mathrm{L}^2[G]$:

$$\mathbf{F} := f \mapsto \mathcal{F}_G(f) \circ \Psi. \tag{4.12}$$

This definition has the advantage of working with functions on the same lattice, but the disadvantage of implicitly depending on the choice $\Psi$ of isomorphism.[5] The transformation $\mathbf{F}$ from Equation 4.12 is in fact a unitary automorphism of $\mathrm{L}^2[G]$. Its matrix $(\mathbf{F}_{hg})_{h,g\in G}$ in the computational basis is:

$$\mathbf{F}_{hg} = \exp\left[ i \sum_j \frac{2\pi}{n_j} \left( g_j h_j \pmod{n_j} \right) \right] \tag{4.13}$$

and it is called a **Fourier matrix** in the context of quantum computing.

Fourier matrices correspond to a Fourier transform along with a choice of the isomorphism. Thus the Fourier matrices, exactly like the definition of the Discrete Fourier transform above, are non-canonical, and depend on an implicit choice of isomorphism $\Psi$. This contrasts with the Fourier transform, which is itself canonical.

Fourier matrices are a subclass of more general **complex Hadamard matrices**: orthogonal matrices[6] whose complex entries are unimodular, in particular (real) **Hadamard matrices** are orthogonal matrices with entries $\pm 1$. Having defined these four different terms (the Fourier transform, Fourier matrices, Hadamard matrices, and complex Hadamard matrices, see Figure 4.1) we will clarify a few ways that they appear in quantum computation.

There is a particularly interesting reason the lack of canonicity of Fourier matrices is not usually an issue in quantum computing. Most of the algorithms are traditionally formulated for qubits, and the state-space of a $D$-qubit system is isomorphic to $\mathrm{L}^2[G]$ for $G = \prod_{j=1}^{D} \mathbb{Z}_2$. The group $\mathbb{Z}_2$ has a unique automorphism (the identity), and thus a unique isomorphism $\mathbb{Z}_2 \to \mathbb{Z}_2^\wedge$, resulting in the familiar matrix

$$\begin{pmatrix} 1 & 1 \\ 1 & -1 \end{pmatrix}, \tag{4.14}$$

which is both the only Fourier matrix on a two dimensional system and, in fact, a Hadamard matrix. There is then a unique isomorphism $\Psi : \mathbb{Z}_2^N \to (\mathbb{Z}_2^N)^\wedge$ which can be obtained by local qubit operations only, namely the $N$-fold tensor product of the isomorphism in 4.14; if multi-qubit operations are allowed, however, the isomorphism is not unique. We stress that for general groups, i.e. for combinations of quantum systems where some have dimensions larger than two, the Fourier transform in terms of Pontryagin duality does not fix a unique Fourier matrix (not even requiring that it is obtained by local operations only). Furthermore, not all complex Hadamard matrices correspond to a Fourier matrix. We'll return to these ideas in Section 4.1.6, but until then we use Fourier transform to refer explicitly to the canonical one defined in terms of Pontryagin duality.

[5]This is a common issue in signal processing and physics, where it is related to the symmetry group of the underlying space and the choice of units of measure for energy/frequency. We will not discuss this further.

[6]Here an orthogonal matrix $H$ is a square matrix such that $H^T H = H H^T = \mathbb{1}$.

There are a number of existing generalizations in the literature of the Fourier transform presented here that we make contact with to varying degrees.

1. Pontryagin theory can be extended from finite abelian groups to arbitrary locally compact abelian groups equipped with the Haar measure: the groups $G$ and $G^\wedge$ are not necessarily isomorphic (e.g. $\mathbb{R}^\wedge = \mathbb{R}$ but $\mathbb{Z}^\wedge = S^1$), but the Fourier transform is still a canonical isomorphism between $\mathrm{L}^2[G]$ and $\mathrm{L}^2[G^\wedge]$, and it's still true that $(G^\wedge)^\wedge = G$.

2. The representation theory can be extended from abelian to arbitrary locally compact groups by observing that $\mathrm{L}^2[G]$ is always a $C^\star$ algebra, and considering the Gelfand-Naimark representation. In the abelian case, this representation coincides with the Fourier transform. This connection is elaborated on in [46].

3. Tannaka-Krein duality provides a different generalisation from compact abelian groups to arbitrary compact groups that we saw previously in Example 2.2.2: the finite-dimensional linear representations of a compact group $G$ form a symmetric monoidal category $\Pi(G)$, generalising $G^\wedge$, with representations $R : G \to \mathrm{End}\,[V_R]$ as objects, intertwiners (linear maps $f : V_R \to V_S$ s.t. $f \circ R(g) = S(g) \circ f$ for all $g \in G$) as morphisms and tensor product of representations as monoidal tensor. The category $\Pi(G)$ comes with a complex conjugation operation on morphisms, and a theorem of Tannaka shows that the set $\Gamma(\Pi(G))$ of all self-conjugate monoidal natural transformations $\mathrm{id}_{\Pi(G)} \to \mathrm{id}_{\Pi(G)}$ forms (once equipped with composition of natural transformations and an appropriate topology) a compact group isomorphic to $G$. A generalisation of Tannaka-Krein duality to braided monoidal categories appears in the representation theory of Drinfeld-Jimbo quantum groups. In this work, we do not deal with either Tannaka-Krein theory or Drinfeld-Jimbo quantum groups. For more on quantum groups and their connection to Hopf algebras see, e.g. [24, 101].

### 4.1.2 Strong Complementarity

We will eventually show that the Fourier transform is related to a special type of complementarity called strong complementarity. In this section we introduce this strongly complementary notion. Definition 3.3.5 introduced classical structures as special and symmetric Frobenius algebras. Here we operate with the slightly weaker notion of a quasi-special commutative Frobenius algebras. This is for convenience and allows us to lump scalars together rather than having to keep track of them at every step.

**Definition 4.1.3.** A **quasi-special** †-Frobenius algebra 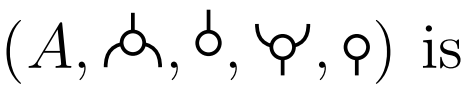 is one that satisfies the following equation for some invertible scalar $N$:

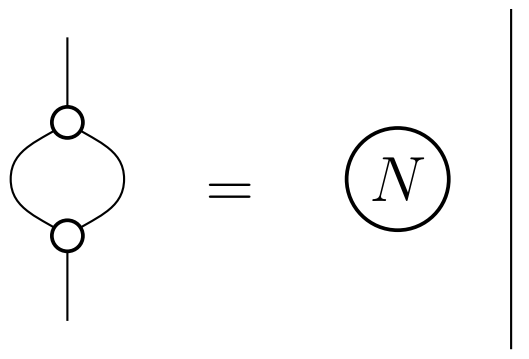

(4.15)

We will use the shorthand †-qSFA, and refer to $N$ as the **normalisation factor** for the †-qSFA.

These †-qSFA's can be thought of as generalized orthogonal bases that are normalize-able (as long as the square root of the scalar $\sqrt{N}$ is invertible) even if they are not normalized. While classical states can be defined in any †-SMC, the following definition for matching families requires an appropriate *zero scalar*. For this and other reasons, we consider categories enriched over commutative monoids,[7] i.e. where homsets come with a commutative monoid structure $(\mathbf{C}(A, B), +, 0)$, and we require the appropriate distributivity laws between the tensor product and the monoidal structure:

$$(f + g) \otimes h = (f \otimes h) + (g \otimes h) \tag{4.16}$$
$$f \otimes (g + h) = (f \otimes g) + (f \otimes h) \tag{4.17}$$
$$0 \otimes f = 0 \tag{4.18}$$
$$f \otimes 0 = 0 \tag{4.19}$$

We will refer to these as **distributively CMon-enriched** †-SMCs.

[7]Refer to Section 3.6 for more detail on enriched QPTs.

**Definition 4.1.4.** Let $|x\rangle_{x\in X}$ be a finite family of states $I \to \mathcal{G}$ in a †-SMC which is distributively **CMon**-enriched. A **matchable family** $|x\rangle_{x\in X}$ for a monoid $(\mathcal{G}, \mu_\bullet, \eta_\bullet)$ are those for which the following holds for all $x, y \in X$:

$$\mu_\bullet \circ (|x\rangle \otimes |y\rangle) = \begin{cases} |x\rangle \text{ if } |x\rangle = |y\rangle \\ 0 \text{ otherwise} \end{cases} \tag{4.20}$$

We re-emphasize that while †-qSCFA's correspond to bases in **FHilb** by Theorem 3.3.8, the general notion of a (orthogonal) basis is somewhat different.

**Definition 4.1.5.** A finite family of states $|x\rangle_{x\in X} : I \to \mathcal{H}$ is a **(orthogonal) basis** (for $\mathcal{H}$) if it satisfies the following conditions:

(i) Orthogonality, i.e. $\langle y|x\rangle = 0$ if $x \neq y$ (where $\langle y|$ stands for $|y\rangle^\dagger$).

(ii) Completeness, i.e. for every $f, g : \mathcal{H} \to \mathcal{H}'$ we have that $\forall x : X\, f|x\rangle = g|x\rangle$ implies $f = g$.

A finite family of co-states $\langle x|_{x\in X} : \mathcal{H} \to I$ is a **(orthogonal) cobasis** (for $\mathcal{H}$) if the family of states $|x\rangle_{x\in X} : I \to \mathcal{H}$ is a basis.

When the classical states for a classical structure form a basis in this manner, the algebra has "enough classical points" (Definition 3.3.9). In **FHilb**, this is the usual linear-algebraic notion of orthogonal basis.

Strong complementarity was originally introduced by Coecke and Duncan in [31] as the additional rule that makes classical structures into a Hopf algebra.[8]

**Definition 4.1.6.** A pair of †-qSFAs $(A, \mu_\circ, \eta_\circ, \delta_\circ, \epsilon_\circ)$ and $(A, \mu_\bullet, \eta_\bullet, \delta_\bullet, \epsilon_\bullet)$, henceforth written as $(\circ, \bullet)$, is **strongly complementary** if they are coherent (Definition 3.5.3) and satisfy the following **bialgebra equation** (4.21):

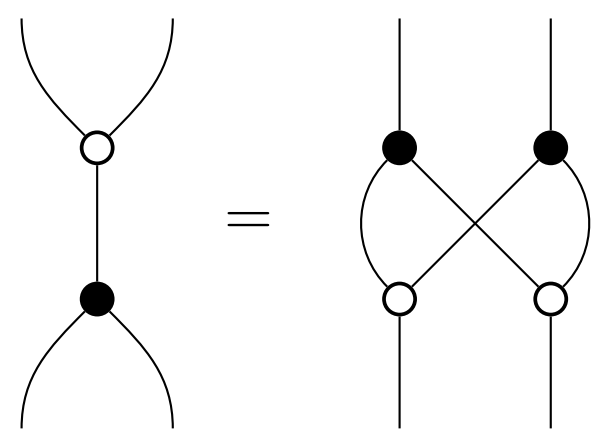

(4.21)

Though this definition is usually given for classical structures, we generalise to †-qSFAs to include non-commutative algebras and, hence, our later construction of a generalized non-abelian Fourier transform.

[8]They are also studied in this form, though separately from the process theoretic framework, as a foundation for graphical linear algebra by Bonchi et al. [21].

**Remark 4.1.7.** Recall that under certain assumptions on the †-FAs that are common in process theories, the antipode is self-adjoint [68, Lem. 7.2.6], though we will work in the more general setting.

It is easy to see that the name is an apt one, i.e. that strongly complementarity classical structures are also complementary in the sense of Definition 3.5.1:

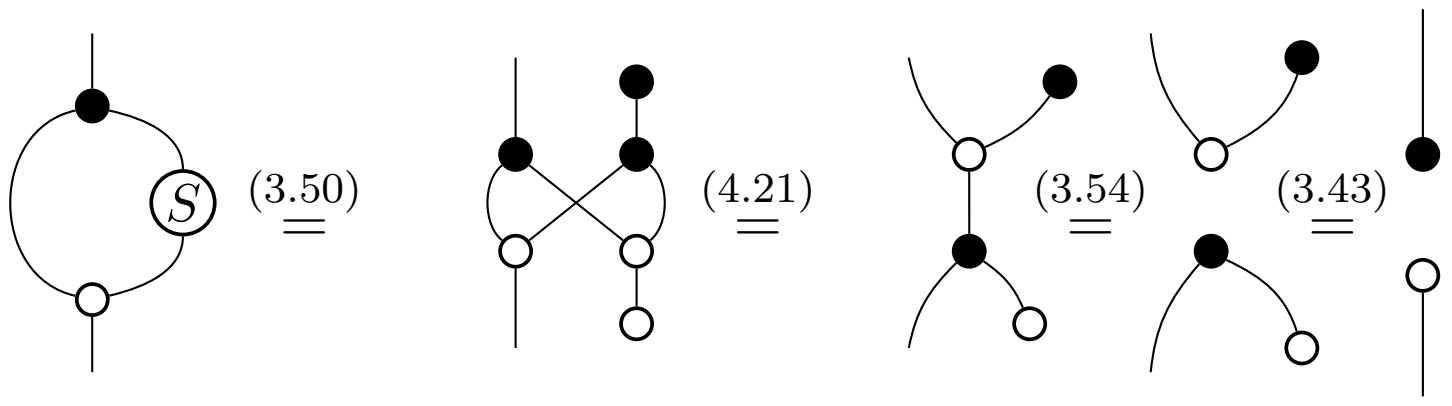

(4.22)

where we have also assumed a self-adjoint antipode in the first step.

As we have slightly generalized the definition of strong complementarity, we also wish to present a slightly more general concept of an antipode that is not self-inverse. Note the slight difference between this definition and Definition 3.5.1. Our results will, of course, still hold in the case of a self-adjoint antipode.

**Definition 4.1.8.** Given a strongly complementary pair of †-FAs (○, ●) on some object $\mathcal{G}$ in a †-SMC, the **antipode** ◇ : $\mathcal{G} \to \mathcal{G}$ is defined to be the following map:

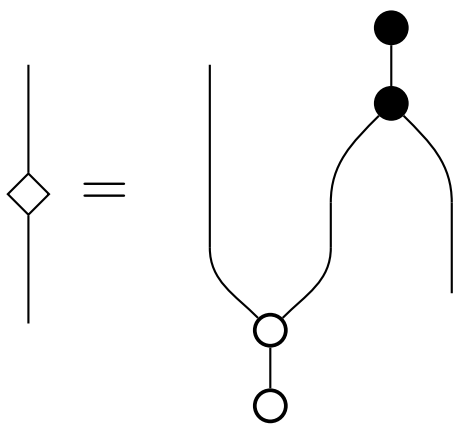

(4.23)

**Lemma 4.1.9.** *Given a strongly complementary pair of †-FAs (○, ●) on some object $\mathcal{G}$ in a †-SMC, the* ***antipode inverse*** ◇$^{-1}$ : $\mathcal{G} \to \mathcal{G}$ *is the following map:*

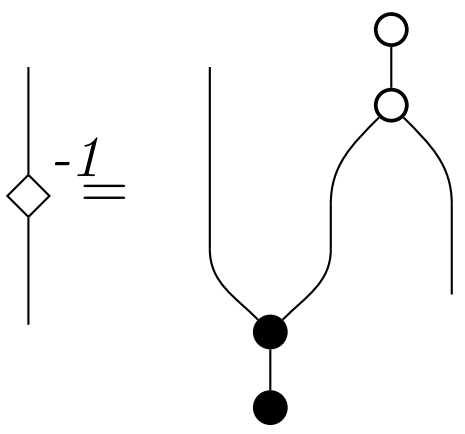

(4.24)

*Furthermore, if at least one of the two †-FAs has a finite matchable family that forms a basis, then the antipode is self-adjoint and unitary, i.e. antipode and antipode inverse coincide.*

*Proof.* The fact that $\diamond^{-1}$ as defined is indeed the inverse of $\diamond$ is an immediate consequence of the Frobenius law (one application per colour). Now suppose without loss of generality that the matchable states $|g\rangle_{g\in G}$ of $\bullet$ form a basis, and remember that $\circ$ acts as some (possibly non-abelian) group $(G, \cdot, 1)$ on them (Lemma 3.5.2). Then $\langle h|\diamond|g\rangle = \langle h\cdot g|1\rangle$ and $\langle h|\,\diamond^{-1}|g\rangle = \langle 1|h\cdot g\rangle$ and $\langle h|\,\diamond^{\dagger}|g\rangle = \langle 1|g\cdot h\rangle$ and $\langle h|\,(\diamond^{-1})^{\dagger}|g\rangle = \langle g\cdot h|1\rangle$ coincide for all $g, h \in G$, proving that $\diamond = \diamond^{-1} = \diamond^{\dagger} = |g\rangle \mapsto |g^{-1}\rangle$ for all $g \in G$. □

Coecke and Duncan showed that strongly complementary classical structures have a specific relationship between their phase groups and classical states.

**Theorem 4.1.10** ([31]). *Let $(A, \delta_\circ, \epsilon_\circ)$ and $(A, \delta_\bullet, \epsilon_\bullet)$ be a pair of strong complementary classical structures with finite numbers of classical states. Then $K_{\bullet} \subseteq P_{\circ}$, i.e. the classical states of the black classical structure form a subgroup of the phase group of the white classical structure. The converse is true when $(A, \delta_\bullet, \epsilon_\bullet)$ has enough classical points.*

This leads to Kissinger's motivating classification of strongly complementary classical structures:

**Corollary 4.1.11** ([32, Cor. 3.10]). *Every pair of strongly complementary classical structures in* **FHilb** *is of the following form:*

$$\begin{cases} \delta_\bullet :: |g\rangle \mapsto |g\rangle \otimes |g\rangle \\ \epsilon_\bullet :: |g\rangle \mapsto 1 \end{cases} \qquad \begin{cases} \mu_\circ :: |g\rangle \otimes |h\rangle \mapsto \frac{1}{\sqrt{D}}|g+h\rangle \\ \eta_\circ :: 1 \mapsto \sqrt{D}|0\rangle \end{cases} \tag{4.25}$$

*where $(G = \{g, h, \ldots\}, +, 0)$ is a finite Abelian group. Conversely, each such pair is always strongly complementary.*

Inspired by this classification, we use strongly complementary structures to embed groups into an arbitrary QPT.

**Definition 4.1.12.** An **internal group**, denoted by $(\mathcal{G}, \mu_\circ, \eta_\circ, \delta_\bullet, \epsilon_\bullet)$ or $(\mathcal{G}, \circ, \bullet)$ when no confusion should arise), consists of a strongly complementary pair on the same object $\mathcal{G}$ of a †-SMC and

1. A †-qSFA $(\mathcal{G}, \mu_\circ, \eta_\circ, \delta_\circ, \epsilon_\circ)$, the **group structure**, denoted by $\circ$.
2. A †-qSCFA $(\mathcal{G}, \mu_\bullet, \eta_\bullet, \delta_\bullet, \epsilon_\bullet)$, the **point structure**, denoted by $\bullet$.

The multiplication and unit for the group structure are called **group multiplication** and **group unit**, and the antipode $\diamond$ for the pair is called the **group inverse**. An **abelian internal group** is one where the group structure is commutative.

The internal groups in a QPT form a category **Grp**[**C**], with objects given by the strongly complementary pairs $(\mathcal{G}, \circ, \bullet)$, and morphisms $(\mathcal{G}, \circ, \bullet) \to (\mathcal{G}', \circ, \bullet)$ given by $f : \mathcal{G} \to \mathcal{G}'$ in **C** that are co-monoid homomorphisms $f : (\curlyvee_\bullet, \bullet) \to (\curlyvee_\bullet, \bullet)$ and monoid homomorphisms $f : (\curlywedge_\circ, \circ) \to (\curlywedge_\circ, \circ)$; the abelian internal groups form a full subcategory **AbGrp**[**C**]. We will refer to these morphisms as **internal group homomorphisms**, both when seen as morphisms in **Grp**[**C**] and in **C**.

**Theorem 4.1.13.** *If $(\mathcal{G}, \circ, \bullet)$ is an (abelian) internal group in any †-SMC, then $(\curlywedge_\circ, \circ)$ acts as an (abelian) group $G$ on the classical points of $(\curlyvee_\bullet, \bullet)$, henceforth the **group elements**. Furthermore, this correspondence yields an equivalence between the category of (abelian) internal groups in* **FHilb** *and the category of finite (abelian) groups.*

In **FHilb**, the point structure $(\bullet)$ characterises the group elements $|g\rangle_{g \in G}$ as an orthonormal basis for $\mathcal{G}$. This is the basis of delta functions from Equation 4.5, with $|g\rangle := \delta_g$. The corresponding isomorphism $\mathrm{L}^2[G] \cong \mathcal{G}$ sends any square-integrable $f : G \to \mathbb{C}$ to the vector $|f\rangle \in \mathcal{G}$ defined by $|f\rangle = \sum_{g \in G} f(g)|g\rangle$. Also under this isomorphism, the multiplicative fragment $(\mathcal{G}, \curlywedge_\circ, \circ)$ of the internal group structure acts as the convolution operation from Equation 4.7. Simply put, an internal group $\mathbb{G} = (\mathcal{G}, \circ, \bullet)$ in **FHilb** consists of:

(i) a space $\mathcal{G}$

(ii) a distinguished orthogonal basis, encoded by the †-qSCFA $\bullet$

(iii) a group structure on that basis, encoded by the †-qSFA $\circ$

From the point of view of the category **Grp**[**C**], $\mathbb{G}$ should be understood as the group $G$ encoded by $\circ$, while from the point of view of the category **FHilb** it should be considered as endowing $\mathcal{G}$ with the structure of $\mathrm{L}^2[G]$.[9] As we abstract away from Hilbert spaces, we will take this conceptual standpoint. Sometimes, when talking about an internal group $\mathbb{G} = (\mathcal{G}, \circ, \bullet)$, we will refer to states $|f\rangle : I \to \mathcal{G}$ as **states of** $\mathbb{G}$, generalising square-integrable functions $f \in \mathrm{L}^2[G]$.

[9] In this correspondence, the orthogonal basis in $\mathcal{G}$ corresponds to the basis of delta functions in $\mathrm{L}^2[G]$, as given in Equation 4.5. The groups structure given by $\circ$ corresponds to the convolution operation from Equation 4.7.

### 4.1.3 Abelian Fourier transform

The previous section provides us with the basic tools to do group theory in arbitrary †-SMCs. We can now connect it to the more traditional theory from in Section 4.1.1. We begin by introducing multiplicative characters as co-states, building on ideas from Vicary [104] and [115].

The use of the $L^2[G]$ notation in this section is consistent with the fact that $L^2$-spaces over finite groups are exactly finite-dimensional Hilbert spaces that come with a canonical choice of basis (the group elements) and a group operation over them. Throughout, we have identified $L^2[\hat{G}] \cong L^2[G]^\star$ as the multiplicative characters are a basis of $L^2[G]^\star$.

**Definition 4.1.14.** A **multiplicative character** for a pair $(\circ, \bullet)$ in a †-SMC is a monoid homomorphism from ( , ) to the canonical monoid on the trivial object $I$ induced by the unitors, or equivalently it is a co-state $\blacktriangle$ : $\mathcal{G} \rightarrow I$ satisfying the following equations:

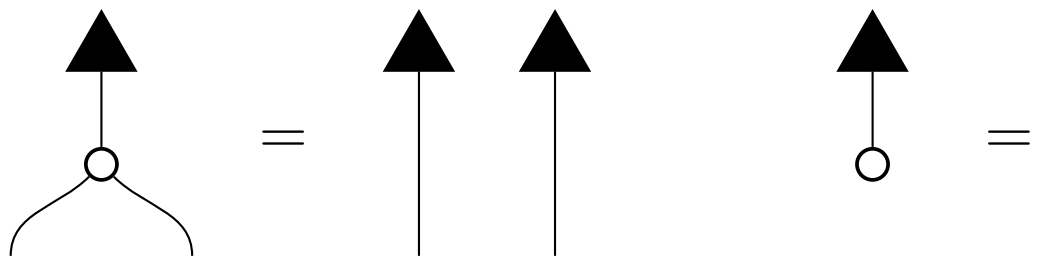
(4.26)

**Lemma 4.1.15.** *If $(\mathcal{G}, \circ, \bullet)$ is an internal group in a †-SMC, then the classical states of the group structure are exactly the (adjoints of its) multiplicative characters. In the case of* **FHilb***, the group structure of an abelian internal group thus characterises the (group theoretic) multiplicative characters of $G$ as a co-basis for $\mathcal{G}$.*

*Proof.* The first part is immediate, the second follows from the equivalence of classical structures and bases in Theorem 3.3.8. □

In **FHilb**, the multiplicative characters of an internal group $(\mathcal{G}, \circ, \bullet)$ are co-states $\mathcal{G} \rightarrow \mathbb{C}$, while the multiplicative characters defined in Section 4.1.1 are group homomorphisms $G \rightarrow \mathbb{C}^\times$. Under the isomorphism $L^2[G] \cong \mathcal{G}$ given by the point structure, the multiplicative characters of the internal group are exactly the linear extensions to $L^2[G]$ of the multiplicative characters of $G$. If the internal group is abelian, then the multiplicative characters are exactly the adjoints of the unique orthogonal basis associated with the $\bullet$ structure.

**Theorem 4.1.16.** *Let $\mathbb{G} = (\mathcal{G}, \circ, \bullet)$ be an abelian internal group in a †-SMC* $\mathbf{C}$*. Then $\mathbb{G}^\wedge := (\mathcal{G}, \bullet, \circ)$ is an abelian internal group in the †-SMC* $\mathbf{C}^{\mathbf{op}}$*, and we shall refer to it as the* ***Pontryagin dual*** *of $(\mathcal{G}, \circ, \bullet)$. The group elements of $(\mathcal{G}, \bullet, \circ)$ are exactly the multiplicative characters of $(\mathcal{G}, \circ, \bullet)$ – this is to say that $\Upsilon_\bullet$ acts as a group, the* ***pointwise multiplication*** *group, on the multiplicative characters, with the* ***trivial character*** *$\varphi_\bullet$ as unit. The antipode acts again as group inverse.*

It is worth clarifying that the pointwise multiplication of Equation 4.9 is different from the pointwise multiplication of Theorem 4.1.16: the former is a pointwise product of functions of characters, and would correspond to the co-monoid $(\Upsilon_\circ, \varphi_\circ)$ (because the group multiplication duplicates multiplicative characters), while the latter is a pointwise product of functions of group elements, and thus corresponds to the co-monoid $(\Upsilon_\bullet, \varphi_\bullet)$ (which duplicates group elements). Also, note that $(\mathbb{G}^\wedge)^\wedge = \mathbb{G}$, as in the traditional formulation of Pontryagin duality.

The usual formulation of the Fourier transform, and of its properties, involves several summations, but a careful analysis shows that they boil down to appropriate resolutions of the identity (Definition 3.6.2), like $\frac{1}{N}\sum_{g\in G} |g\rangle\langle g| = \mathrm{id}_{\mathrm{L}^2[G]}$, and to various formulations of orthogonality of characters. The following lemma shows that, from a categorical perspective, the two are equivalent.

**Lemma 4.1.17.** *Let $\bullet$ be a †-qSFA with normalization $N$ on an object $\mathcal{G}$ in a †-SMC which is distributively* $\mathbf{CMon}$*-enriched, and let $|x\rangle_{x\in X}$ be a finite set of classical states for the co-monoid $(\Upsilon_\bullet, \varphi_\bullet)$ such that*

*(a.) the family is* ***orthogonal****, i.e. $\langle x'|x\rangle = 0$ (the zero scalar) for all $x \neq x'$*

*(b.) the family is* ***normalisable****, i.e. $\langle x|x\rangle$ is an invertible scalar for all $x$.*

*Then the following are equivalent:*

*(i) The classical states $|x\rangle_{x\in X}$ form a (orthogonal) basis, as per Definition 4.3.11.*

*(ii) The classical states $|x\rangle_{x\in X}$ form a* ***resolution of the identity****, as per Definition 3.6.2.*

*(iii) The adjoints of the classical states form an* ***(orthogonal) partition of the counit****, i.e. they satisfy the following equation:*

$\frac{1}{N}\sum_x$ 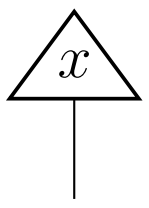 $= \varphi_\bullet$ (4.27)

*Proof.* Since we have assumed that $\langle\chi|\chi\rangle$ is invertible, then $\langle\chi|\chi\rangle = N$.

- $(i) \implies (ii)$ Suppose that the classical states form a basis, i.e. suppose that $\forall\chi, f \circ |\chi\rangle = g \circ |\chi\rangle$ implies $f = g$ (completeness). Then we get, for all $\chi'$:

$$\left(\frac{1}{N}\sum_{\chi} |\chi\rangle\langle\chi|\right) \circ |\chi'\rangle = \frac{1}{N}|\chi'\rangle\langle\chi'|\chi'\rangle = |\chi'\rangle = \mathrm{id}_{\mathcal{G}} \circ |\chi'\rangle$$

where we have used orthogonality. We conclude $(ii)$ by completeness.

- $(ii) \implies (iii)$ By using the fact that $\text{⫯} \circ |\chi\rangle = 1$ for all $\chi \in X$ we immediately get $(iii)$:

$$\text{⫯} \circ \left(\frac{1}{N}\sum_{\chi} |\chi\rangle\langle\chi|\right) = \frac{1}{N}\sum_{\chi} (\text{⫯} \circ |\chi\rangle)\langle\chi| = \frac{1}{N}\sum_{\chi}\langle\chi|$$

- $(ii) \implies (i)$ All we have to prove is completeness, as orthogonality of the family $|\chi\rangle_{\chi\in X}$ was assumed as a hypothesis of the lemma. Assume $f \circ |\chi\rangle = g \circ |\chi\rangle$ for all $\chi$, then we get:

$$\frac{1}{N}\sum_{\chi} f \circ |\chi\rangle\langle\chi| = \frac{1}{N}\sum_{\chi} g \circ |\chi\rangle\langle\chi|$$

But the LHS is $f \circ \mathrm{id}_{\mathcal{G}}$, i.e. $f$, and the RHS is $g \circ \mathrm{id}_{\mathcal{G}}$, i.e. $g$.

- $(iii) \implies (ii)$ Assume that $\text{⫯} = \frac{1}{N}\sum_{\chi}\langle\chi|$, then we get (using Frobenius law in the first equality):

$$\begin{aligned}
\mathrm{id}_{\mathcal{G}} &= (\text{⫯} \otimes \mathrm{id}_{\mathcal{G}}) \circ (\curlywedge \otimes \mathrm{id}_{\mathcal{G}}) \circ (\mathrm{id}_{\mathcal{G}} \otimes \curlyvee) \circ (\mathrm{id}_{\mathcal{G}} \otimes \text{⫰}) \\
&= \frac{1}{N}\sum_{\chi}\frac{1}{N}\sum_{\chi'} |\chi'\rangle\langle\chi'|\chi\rangle\langle\chi| \\
&= \frac{1}{N}\sum_{\chi} |\chi\rangle\langle\chi|
\end{aligned}$$

As a final remark, note that orthogonality, assumed separately, is already included in the definition of basis used in point $(i)$; it is, however, necessary to assume it explicitly in points $(ii)$ and $(iii)$. As for point $(iii)$, a counterexample can be found in **FRel**, by replacing an orthogonal family with the one obtained by repeating some element $\langle\chi|$, and using the fact that $\langle\chi| + \langle\chi| = \langle\chi|$ (since the enriched monoidal operation $\sum$ in **FRel** is just the set union $\cup$). As for point $(ii)$, one can consider the category of finite-dimensional vector spaces over the field with 2 elements (where $1 + 1 = 0$): if $|\chi\rangle$ is a norm-1 vector in a 1-dimensional space $\mathcal{G}$, the family $(|\chi\rangle, |\chi\rangle, |\chi\rangle)$ is non-orthogonal, and yet a resolution of the identity as $|\chi\rangle\langle\chi| + |\chi\rangle\langle\chi| + |\chi\rangle\langle\chi| = |\chi\rangle\langle\chi| = \mathrm{id}_{\mathcal{G}}$ (this cannot happen in **FHilb**). $\square$

The formulation in terms of orthogonal partition of the counit is related to the orthogonality of (multiplicative) characters traditionally mentioned in the context of Fourier transform (e.g. used here in Equation 4.9), as the following lemma shows.

**Theorem 4.1.18** (Orthogonality of Multiplicative Characters)**.** *Let* $(\mathcal{G}, \circ, \bullet)$ *be an internal group in a* †*-SMC* $\mathbf{C}$*, and* $N$ *be the normalisation factor for the quasi-special condition of* ○*. Assume that the characters are all orthogonal, in the sense that* $\langle\chi|\chi'\rangle = 0$ *for* $\chi \neq \chi'$*, and that* $\langle 1|1\rangle = N$*, where* $\langle 1| :=$ ⚲ *is the trivial character. Then if* $|\chi\rangle, |\chi'\rangle$ *are (not necessarily distinct) multiplicative characters of the internal group, the following* ***orthogonality of multiplicative characters*** *holds:*

$\frac{1}{N}$ 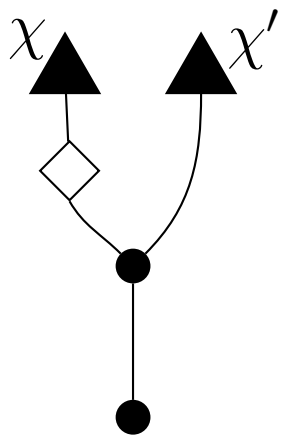

$= \quad \delta_{\chi\chi'}$ (4.28)

*Now assume that* $\mathbf{C}$ *is distributively* $\mathbf{CMon}$*-enriched. If the family* $\langle g|_{g\in G}$ *of (adjoints of) group elements is normalisable and forms an orthogonal partition of the counit, then Equation 4.28 can be re-written in the following, more familiar form (where we have set* $|\chi^{-1}\rangle := |\chi\rangle \circ$◇*):*

$$\frac{1}{N}\sum_{g\in G}\langle\chi^{-1}|g\rangle\langle\chi'|g\rangle = \delta_{\chi\chi'} \tag{4.29}$$

*Proof.* By Theorem 4.1.16, the comultiplication ⟟ acts as a group on the multiplicative characters, and ◇ as the group inverse. The LHS of Equation 4.28 can be re-written as $\langle\chi^{-1}\cdot\tau|1\rangle$, and $\cdot$ is the pointwise multiplication: since we assumed that the multiplicative characters are orthogonal, the result follows. In order to obtain Equation 4.29 from Equation 4.28, all we have to do is observe that the group elements $|g\rangle_{g\in G}$ form an orthogonal partition of the unit ⫯ (by taking adjoints), and that they are classical points of ●. □

Note that, by Definition 4.1.8 and Frobenius law for ●, Equation 4.28 can equivalently be written as the following, stating that the multiplicative characters are a matchable family (Definition 4.1.4) for (⟟, ⚲):

$\frac{1}{N}$ 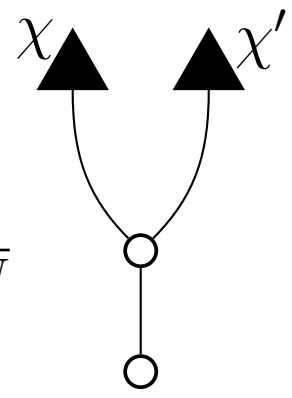

$= \quad \delta_{\chi\chi'}$ (4.30)

Equations 4.28 and 4.30 provide a summation-free version of the orthogonality of multiplicative characters of Equation 4.29 (under appropriate conditions). This leads us to the following summation-free definition of the Fourier transform, valid for any internal group in any †-SMC.

**Definition 4.1.19.** Let $\mathbb{G} = (\mathcal{G}, \circ, \bullet)$ be an internal group in any †-SMC. The **Fourier transform** is defined to be the following mapping $\mathcal{F}_{\mathbb{G}}$ of states of $\mathcal{G}$ to co-states of $\mathcal{G}$:

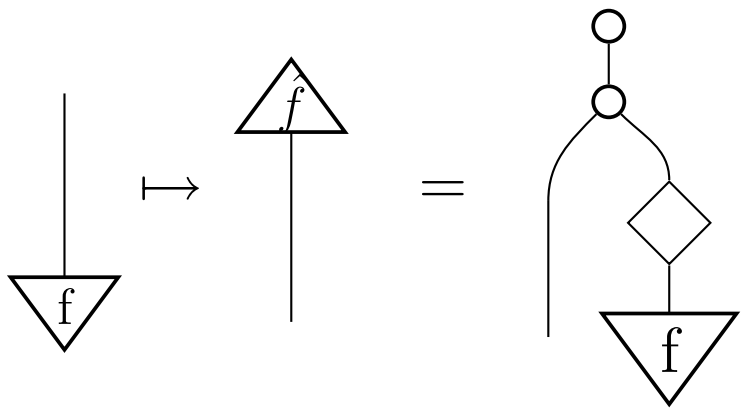

(4.31)

The **inverse Fourier transform** is defined to be the following mapping $\mathcal{F}_{\mathbb{G}}^{-1}$ of co-states of $\mathcal{G}$ to states of $\mathcal{G}$:

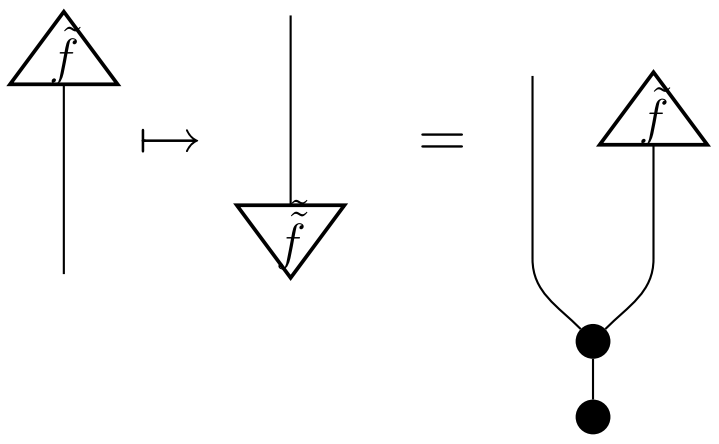

(4.32)

Under appropriate circumstances, the Fourier transform of Definition 4.1.19 takes the more familiar form of Equation 4.2, as shown by the following lemma and its subsequent application to **FHilb**.

**Lemma 4.1.20.** *Let $\mathbb{G} = (\mathcal{G}, \circ, \bullet)$ be an internal group in a †-SMC which is distributively* **CMon**-*enriched. Further assume that the multiplicative characters and the group elements of $\mathbb{G}$ are both finite, normalisable families, which form an orthogonal partition of the counits* ⚲ *and* ⚲ *respectively. Then the Fourier transform of Definition 4.31 can be written in the following way:*

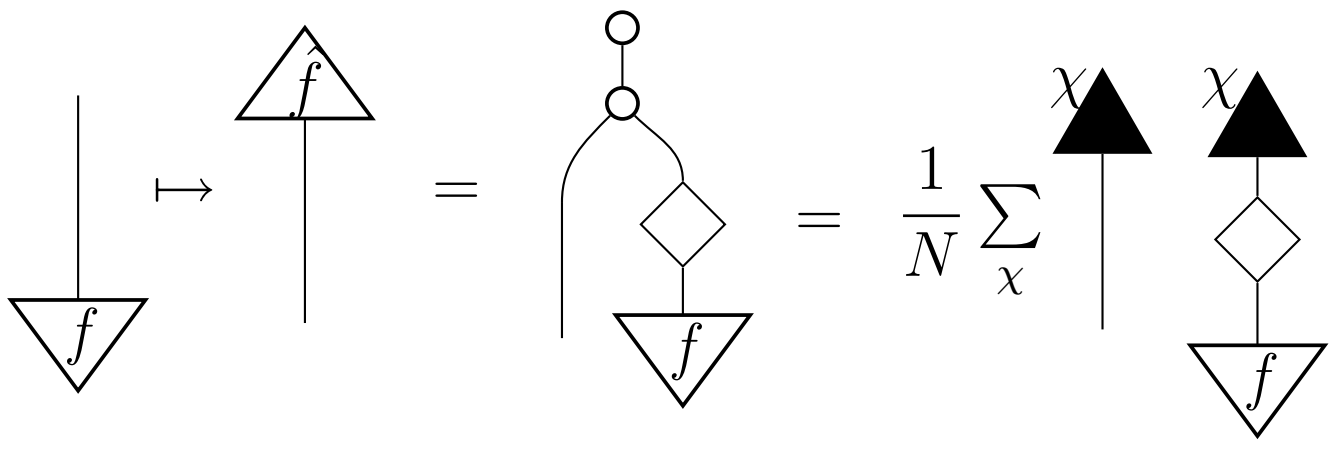

(4.33)

*Furthermore, the Inverse Fourier transform of Definition 4.32 can be written in the following way:*

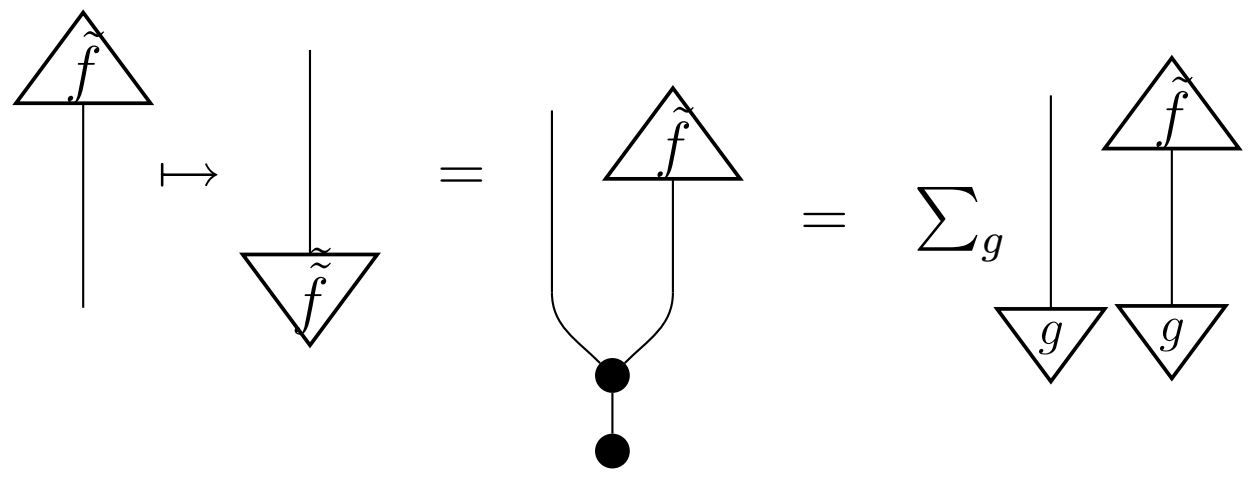

(4.34)

*Proof.* For the Fourier transform, use the fact that the multiplicative characters form an orthogonal partition of the counit ⫯, as per Equation 4.27, and that they are classical states of ⩛, as per Equation 4.26. Similar reasoning is used for the Inverse Fourier transform. □

In **FHilb**, the conditions of Lemma 4.1.20 hold for abelian internal groups (but not for non-abelian ones, as the multiplicative characters fail to form a basis). The rightmost expression in equation 4.33 can be written as follows, where we have $|\chi^{-1}\rangle = |\chi\rangle \circ \diamond$ (as in Lemma 4.1.18):

$$\frac{1}{N}\sum_{\chi}\langle\chi|\langle\chi^{-1}|f\rangle$$

The vector $|f\rangle$ can be expanded by using a resolution of the identity in terms of the group elements, courtesy of Lemma 4.1.17:

$$\sum_{\chi}\langle\chi|\frac{1}{N}\sum_{g}\langle\chi^{-1}|g\rangle\langle g|f\rangle$$

Now we use the isomorphism $\mathcal{G} \cong \mathrm{L}^2[G]$ induced by the point structure, under which $f : \mathrm{L}^2[G]$ gets mapped to $|f\rangle := \sum_g |g\rangle f(g)$, to obtain:

$$\sum_{\chi}\langle\chi|\frac{1}{N}\sum_{g}\langle\chi^{-1}|g\rangle f(g)$$

Furthermore, the multiplicative characters of $\mathbb{G}$ are, in **FHilb** and under the isomorphism $\mathcal{G} \cong \mathrm{L}^2[G]$ above, the linear extensions of the multiplicative characters of the $G$, and we can re-write the above as:

$$\sum_{\chi}\langle\chi|\frac{1}{N}\sum_{g}\chi^{-1}(g)f(g)$$

Finally, we use the isomorphism $\mathcal{G}^\star \cong \mathrm{L}^2[G^\wedge]$ induced by the group structure,[10] under which $\tilde{f} : \mathrm{L}^2[G^\wedge]$ gets mapped to $\langle \tilde{f}| := \sum_\chi \langle \chi | \tilde{f}(\chi)$, to finally obtain:

$$\tilde{f}(\chi) = \frac{1}{N} \sum_g \chi^{-1}(g) f(g)$$

This is exactly the same as Equation 4.2, and a similar reasoning shows that in **FHilb** Equation 4.34 coincides with Equation 4.3. Therefore Definition 4.1.19 matches the traditional definition in the case of abelian internal groups of **FHilb**, but it remains to be seen under which circumstances and in which form its usual properties extend to internal groups in arbitrary †-SMCs. Here we will focus on three particularly important results: the Fourier Inversion Theorem, the Convolution Theorem and Pontryagin Duality. In order to clarify their categorical formulation, we summarize the role played by each structure:

(i) When states $\mathbb{C} \to \mathcal{G}$ are identified as functions in $\mathrm{L}^2[G]$ via the basis of group elements, the monoid (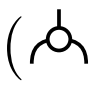, 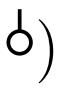) acts as the *convolution* operation on $\mathrm{L}^2[G]$:

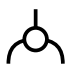

$$\circ \left( \sum_g f(g) \,|g\rangle \otimes \sum_g f'(g) \,|g\rangle \right) = \sum_g \left( \sum_h f(h) f'(g-h) \right) |g\rangle \tag{4.35}$$

$$= \sum_g (f \star_G f')(g) \,|g\rangle \tag{4.36}$$

(ii) When states $\mathbb{C} \to \mathcal{G}$ are identified with functions in $\mathrm{L}^2[G]$ via the basis of group elements, the monoid (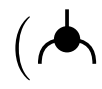, 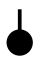) acts as the *pointwise multiplication* operation on $\mathrm{L}^2[G]$:

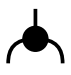

$$\circ \left( \sum_g f(g) \,|g\rangle \otimes \sum_g f'(g) \,|g\rangle \right) = \sum_g f(g) f'(g) \,|g\rangle \tag{4.37}$$

(iii) When co-states $\mathcal{G} \to \mathbb{C}$ are identified with functions in $\mathrm{L}^2[G^\wedge]$ via the co-basis of multiplicative characters, the monoid[11] (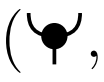, 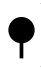) acts as the *convolution* operation on $\mathrm{L}^2[G^\wedge]$:

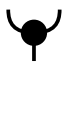

$$\circ \left( \sum_\chi f(\chi) \,\langle\chi| \otimes \sum_\chi f'(\chi) \,\langle\chi| \right) = \sum_\chi \left( \sum_\sigma f(\sigma) f'(\chi - \sigma) \right) \langle\chi| \tag{4.38}$$

$$= \sum_\chi (f \star_{G^\wedge} f')(\chi) \,\langle\chi| \tag{4.39}$$

---

[10]Where we have used the fact that **FHilb** can be **FHilb**-enriched, and thus that the homset $\mathbf{FHilb}(\mathcal{G}, \mathbb{C})$ can be canonically endowed with the finite-dimensional Hilbert space structure of the space $\mathcal{G}^\star$ of linear functionals $\mathcal{G} \to \mathbb{C}$.

[11]It is a co-monoid in **FHilb**, but when acting on co-states it is a monoid.

(iv) When co-states $\mathcal{G} \to \mathbb{C}$ are identified with functions in $L^2[G^\wedge]$ via the co-basis of multiplicative characters, the monoid. ($\curlyvee'$, $\circ\!\!\mid$) acts as the *pointwise multiplication* operation on $L^2[G^\wedge]$:

$$\curlyvee' \circ \left( \sum_{\chi} f(\chi) \langle \chi | \otimes \sum_{\chi} f'(\chi) \langle \chi | \right) = \sum_{\chi} f(\chi) f'(\chi) \langle \chi | \qquad (4.40)$$

**Theorem 4.1.21** (Categorical Fourier Inversion Theorem). *Let* $\mathbb{G} = (\mathcal{G}, \circ, \bullet)$ *be an internal group in any* †*-SMC* **C**. *Then the Fourier transform and the Inverse Fourier transform from Definition 4.1.19 are mutually inverse bijections between states and co-states of* $\mathcal{G}$.

*Proof.* The following shows that $\mathcal{F}_{\mathbb{G}}^{-1} \circ \mathcal{F}_{\mathbb{G}} = \mathrm{id}_{\mathcal{G}}$, by using the Definition 4.1.8 for the antipode and Lemma 4.1.9:

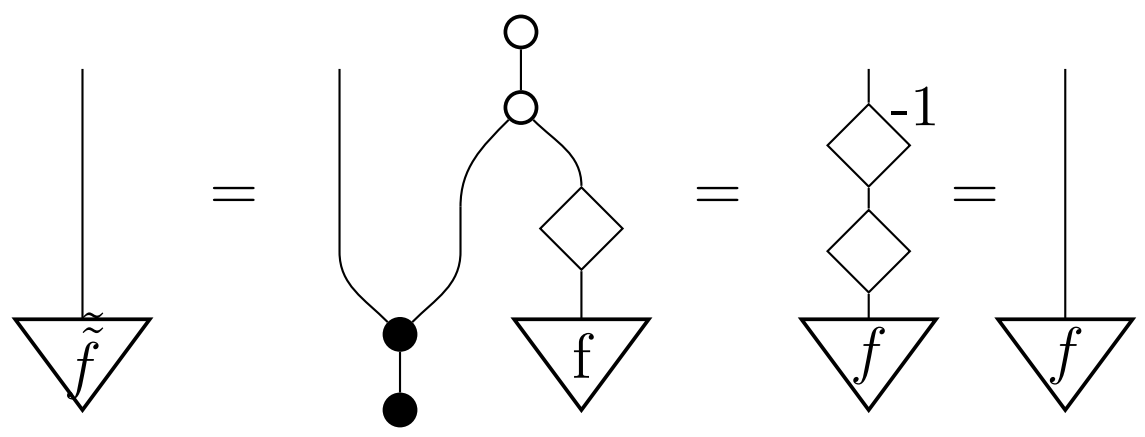

(4.41)

A similar proof holds for $\mathcal{F}_{\mathbb{G}} \circ \mathcal{F}_{\mathbb{G}}^{-1} = \mathrm{id}_{\mathcal{G}}$, by expanding the antipode in terms of its definition and using Frobenius law (once per colour, exactly like in the proof of Lemma 4.1.9). □

**Theorem 4.1.22** (Categorical Convolution Theorem). *Let* $\mathbb{G} = (\mathcal{G}, \circ, \bullet)$ *be an internal group in any* †*-SMC* **C**. *Then the Fourier transform is a monoid homomorphism:*

$$\mathcal{F}_{\mathbb{G}} : (\mathcal{G}, \curlywedge, \mathrm{o\!\!\mid}) \longrightarrow (\mathcal{G}, \curlyvee', \circ\!\!\mid) \qquad (4.42)$$

*Proof.* That $\mathcal{F}_{\mathbb{G}}$ preserves the unit $\mathrm{o\!\!\mid}$ is obvious by the unitality of $\curlywedge$. Using associativity, followed by one application of Frobenius Law, we get the desired:

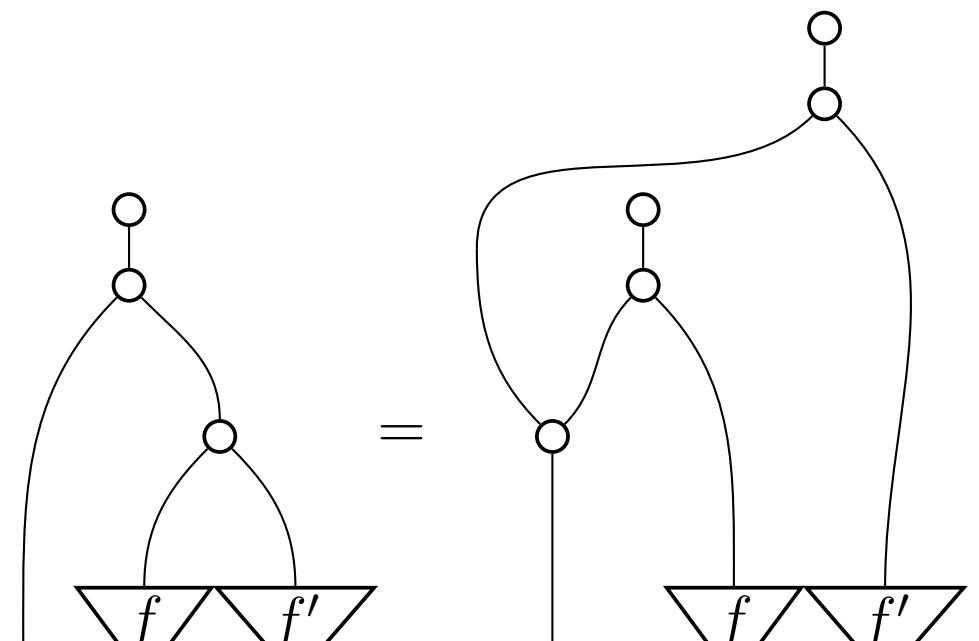

(4.43)

□

**Theorem 4.1.23** (Categorical Pontryagin Duality)**.** *Let* $\mathbb{G} = (\mathcal{G}, \circ, \bullet)$ *be an internal group in a †-SMC* $\mathbf{C}$*. Then the Fourier transform* $\mathcal{F}_{\mathbb{G}}$ *is a bijection between states of* $\mathbb{G}$ *and states of* $\mathbb{G}^\wedge$*, which is furthermore canonical in the sense that:*

$$ {}^{(\varphi^\wedge)}M \cdot \mathcal{F}_{\mathbb{H}} \cdot {}_{\varphi}M = \mathcal{F}_{\mathbb{G}} \tag{4.44} $$

*where* $\varphi : \mathbb{G} \to \mathbb{H}$ *is any unitary isomorphism of internal groups in* $\mathbf{C}$*, for* $\mathbb{H} = (\mathcal{H}, \circ, \bullet)$ *any other internal group of* $\mathbf{C}$*. We have defined the following:*

*(i)* $\varphi^\wedge := \varphi$ *is an isomorphism of internal groups* $\mathbb{H}^\wedge \to \mathbb{G}^\wedge$ *in* $\mathbf{C}^{\mathbf{op}}$

*(ii)* ${}_{\varphi}M$ *as the map sending state* $|f\rangle : I \to \mathcal{G}$ *to state* $\varphi \circ |f\rangle : I \to \mathcal{H}$

*(iii)* ${}^{(\varphi^\wedge)}M$ *as the map sending co-state* $\langle f| : \mathcal{H} \to I$ *(a state in* $\mathbf{C}^{\mathbf{op}}$*) to co-state* $\langle f| \circ \varphi : \mathcal{G} \to I$

*Proof.* The bijection is proven by Theorem 4.1.21, so all we have to show is canonicity:

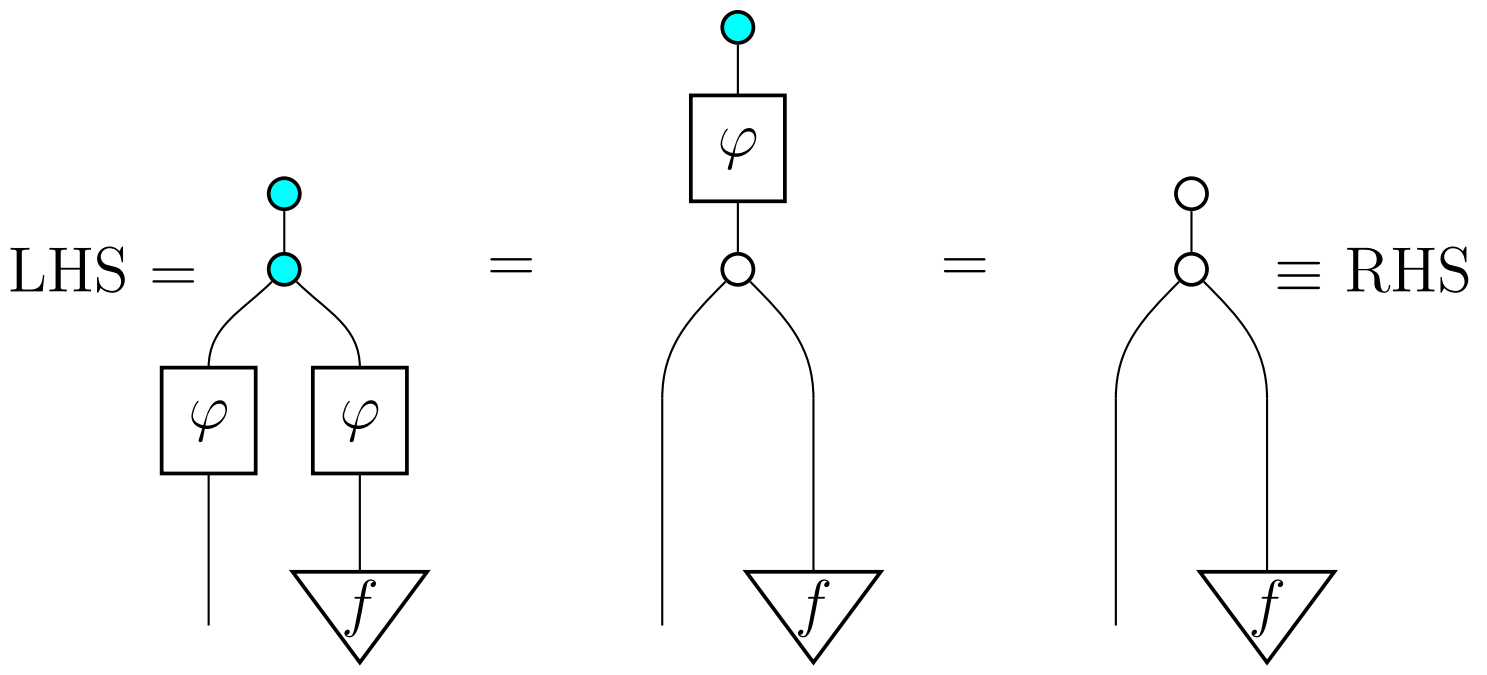

(4.45)

The first equality follows from the fact that $\varphi$ is a morphism of internal groups. The second equality follows from the (easy to check) fact that, if $\varphi$ is a unitary isomorphism $\varphi : \mathbb{G} \to \mathbb{H}$, then $\varphi^\dagger$ is a unitary isomorphism $\varphi^\dagger : \mathbb{H} \to \mathbb{G}$, and hence we have:

$$ ⚲ \circ \varphi = \left(\varphi^\dagger \circ ⫯\right)^\dagger = (⫯)^\dagger = ⚲ \tag{4.46} $$

□

In **FHilb** (with abelian internal groups), (4.45) takes the form of (4.4). To conclude, we provide a categorical definition of Fourier matrices, which helps to frame the difference between them and the Fourier transform.

**Definition 4.1.24.** Let $\mathbb{G} = (\mathcal{G}, \circ, \bullet)$ be an internal group in a †-SMC $\mathbf{C}$. A **Fourier matrix** is defined to be a co-monoid isomorphism $F : (\Upsilon, ⚲) \to (\Upsilon, ⚲)$ which is furthermore a monoid isomorphism $F : (\curlywedge, ⫯) \to (\curlywedge, ⫯)$. We write it as:

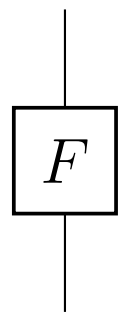

(4.47)

Definition 4.1.24 may look cryptic at first, but it is in fact quite natural as we clarify in the following remark.

**Remark 4.1.25.** A co-monoid isomorphism $F : (\delta_\bullet, \epsilon_\bullet) \to (\delta_\circ, \epsilon_\circ)$ is an isomorphism $F : \mathcal{G} \to \mathcal{G}$ which satisfies:

$$\delta_\circ \cdot F = (F \otimes F) \cdot \delta_\bullet \tag{4.48}$$

$$\epsilon_\circ \cdot F = \epsilon_\bullet \tag{4.49}$$

and in particular it maps $\bullet$-classical states to $\circ$-classical states (since it is an isomorphism, it is a bijection between the classical states of the two comonoids). The requirement that $F$ is furthermore a monoid isomorphism $F : (\mu_\circ, \eta_\circ) \to (\mu_\bullet, \eta_\bullet)$ amounts to the requirement that:

$$F \circ \mu_\circ = \mu_\bullet \circ (F \otimes F) \tag{4.50}$$

$$F \circ \eta_\circ = \eta_\bullet \tag{4.51}$$

and in particular, as a bijection of classical states, $F$ is a group isomorphism from the group given by $(\mu_\circ, \eta_\circ)$ acting on $\circ$-classical states to the group given by $(\mu_\bullet, \eta_\bullet)$ acting on $\circ$-classical states.

In **FHilb** (with abelian internal groups), Definition 4.1.24 yields the usual Fourier matrices. Indeed $F$ corresponds to an isomorphism $\Psi : G \to G^\wedge$ (by considering its action on classical states): the linear isomorphism $F$ is itself the Discrete Fourier transform of Equation 4.12 (where $\mathcal{G}$ is identified with $L^2[G]$ and $L^2[G^\wedge]$ using $\bullet$ and $\circ$ respectively), and the matrix of $F$ in the basis defined by $\bullet$ is the Fourier matrix of Equation 4.13.

### 4.1.4 The Fourier transform in the category FRel

The abstract correspondence between strongly complementary observables and Fourier transforms in Section 4.1.3 means that a characterization of strongly complementary observables in any †-SMC allows for the definition of a Fourier transform in that category. In this section, we apply this idea to **FRel**, the category of finite sets and relations (Example 2.2.2). One can find more details on **FRel** as a QPT in Section 4.3. In this setting, the relevant classical structures are classified by abelian groupoids, in a sense made clear by Theorem 4.1.27 below. Also recall that the monoidal identity is given by the singleton, i.e. $I = \{\star\}$. Please note that in **FRel** the scalars are the booleans $\{\bot, \top\}$, and $\top = \mathrm{id}_I : I \to I$ is the only invertible scalar.

As a consequence, all †-qSFAs are automatically †-SFAs (and thus all †-qSCFAs are in fact classical structures).

**Definition 4.1.26.** A **(finite) abelian groupoid** on some finite set $A$ is any finite family $(G_\lambda, +_\lambda, 0_\lambda)_{\lambda\in\Lambda}$ of finite abelian groups such that $(G_\lambda)_{\lambda\in\Lambda}$ is a finite partition of $A$ into disjoint subsets. We denote one such groupoid by $\oplus_{\lambda\in\Lambda} G_\lambda$, leaving the groups' structures understood.

**Theorem 4.1.27.** *Let $\circ$ be a classical structure in* **FRel** *on a finite set $X$. Then there exists a (unique) abelian groupoid $\oplus_{\lambda\in\Lambda} G_\lambda$ on $A$ such that, for all $a, b \in A$:*

$$\curlywedge \circ (\{a\} \times \{b\}) = \begin{cases} \{a +_\lambda b\} \text{ if for some } \lambda \in \Lambda \text{ we have } a, b \in G_\lambda \\ \emptyset \text{ otherwise} \end{cases} \tag{4.52}$$

*Furthermore, each abelian groupoid defines a unique classical structure in this way. The classical states of the co-monoid fragment $(\curlyvee, \bot)$ are the family $(G_\lambda)_{\lambda\in\Lambda}$, which is also a matching family for the monoid fragment $(\curlywedge, \top)$.*[12]

*Proof.* Proven by Pavlovic [85], and extended to the case of non-commutative †-SFAs / non-abelian groupoids by Heunen et al. [56]. □

Evans et al. show that the groupoids corresponding to complementary / strongly complementary classical structures take a particularly nice form:

**Theorem 4.1.28** ([44])**.** *Let $\oplus_{\gamma\in\Gamma} H_\gamma$ and $\oplus_{\lambda\in\Lambda} G_\lambda$ be abelian groupoids on some finite set $A$, and let $\bullet$ and $\circ$ be the corresponding classical structures on $A$ in* **FRel**. *Then $\bullet$ and $\circ$ are strongly complementary if and only if:*

*(i)* *there is a finite abelian group $H$ such that for all $\gamma \in \Gamma$ we have $H_\gamma \cong H$ as groups.*

*(ii)* *there is a finite abelian group $G$ such that for all $\lambda \in \Lambda$ we have $G_\lambda \cong G$ as groups.*

*(iii)* *for each $(\lambda, \gamma) \in \Lambda \times \Gamma$, the intersection $G_\lambda \cap H_\gamma$ is a singleton.*

In particular, this means that $\Lambda \cong H$ and $\Gamma \cong G$ as sets: as a consequence, we will write abelian groupoids corresponding to strongly complementary classical structures as $\oplus^{|G|} H$ and $\oplus^{|H|} G$, leaving the indexing of the partitions as understood. We will also implicitly label the elements of the underlying set $A$ as:

$$A \cong \{(h, g) \text{ s.t. } h \in H \text{ and } g \in G\} \tag{4.53}$$

[12] Recall that the states $I \to A$ of a finite set $A$ in **FRel** are exactly the subsets of $A$.

In **FHilb**, strongly complementary pairs of classical structures have the same number of classical states, and their monoid fragments act on each other's classical states as isomorphic groups $G$ and $G^\wedge$. As a consequence, it is possible to fix isomorphisms between the two groups and construct Fourier matrices as per Definition 4.1.24. In **FRel**, on the other hand, Fourier matrices only exist in very special cases.

**Theorem 4.1.29.** *Let $\mathbb{G} = (\mathcal{G}, \circ, \bullet)$ be an abelian internal group in* **FRel***, and let $Z = \oplus^{|G|} H$ and $X = \oplus^{|H|} G$ be the groupoids corresponding to the $\bullet$ and $\circ$ classical structures respectively. Then $(\mu_\circ, \eta_\circ)$ acts on the $\bullet$-classical states $(H_g)_{g \in G}$ as the finite abelian group $G$, and $(\mu_\bullet, \eta_\bullet)$ acts on the $\circ$-classical states $(G_h)_{h \in H}$ as the finite abelian group $H$.*

*Proof.* It is sufficient to prove for $(\mu_\circ, \eta_\circ)$ acting on the $\bullet$-classical states. Indeed we have that $\eta_\circ = H_0$ is a $\bullet$-classical state by strong complementarity, and that:

$$\mu_\circ \circ (|H_g\rangle \times |H_{g'}\rangle) = \bigcup_{h,h' \in H} \mu_\circ \circ (\{(h,g)\} \times \{(h',g')\}) \tag{4.54}$$

$$= \bigcup_{h \in H} \{(h, g+g')\} = |H_{g+g'}\rangle \tag{4.55}$$

□

**Example 4.1.30.** The groupoids $Z = \mathbb{Z}_2 \oplus \mathbb{Z}_2 \oplus \mathbb{Z}_2$ and $X = \mathbb{Z}_3 \oplus \mathbb{Z}_3$ correspond to strongly complementary structures, call them $\bullet$ and $\circ$ respectively, on a 6-element set $A$. We can label the elements of $A$ as:

$$A \cong \{(h,g) \text{ s.t. } h \in \mathbb{Z}_2 \text{ and } g \in \mathbb{Z}_3\} \tag{4.56}$$

The classical states of $\bullet$ are the 3 subsets $(\mathbb{Z}_2, g)$ for $g \in \mathbb{Z}_3$, while the classical states of $\circ$ are the 2 subsets $(h, \mathbb{Z}_3)$ for $h \in \mathbb{Z}_2$. The monoid $(\mu_\circ, \eta_\circ)$ acts on the $\bullet$-classical states as the group $\mathbb{Z}_3$, while the monoid $(\mu_\bullet, \eta_\bullet)$ acts on the $\circ$-classical states as the group $\mathbb{Z}_2$.

**Corollary 4.1.31.** *Let $\mathbb{G} = (A, \circ, \bullet)$ be an abelian internal group in* **FRel***. Fourier matrices for $\mathbb{G}$ exist if and only if the two groups $G$ and $H$ are isomorphic.*

*Proof.* By Remark 4.1.25, a Fourier matrix gives a group isomorphism between the groups given by the action of the two monoid fragments on each other's classical states: in this case, the existence of a Fourier matrix forces $G$ and $H$ to be isomorphic. On the other hand, if $\Psi : G \to H$ is a group isomorphism, one could define a map $M_\Psi : A \to A$ as follows:

$$M_\Psi := \bigcup_{g \in G} |G_{\Psi(g)}\rangle\langle H_g| \tag{4.57}$$

This satisfies the monoid and comonoid homomorphism requirements from Definition 4.1.24, but is not an isomorphism in **FRel**. To get an isomorphism, and prove the existence of a relevant Fourier matrix in **FRel**, we consider a new map $t : A \to A$, which we give in the form of a relation $t \subseteq A \times A$:

$$t := \big\{ \big((h, g), (\Psi g, \Psi^{-1} h)\big) \;\text{ s.t. }\; h \in H \text{ and } g \in G \big\} \tag{4.58}$$

□

The Fourier transform as given by Definition 4.1.19 is valid in any †-SMC with a pair of strongly complementary classical structures, and in particular it holds in **FRel**. The more traditional formulation given by Lemma 4.1.20, which allows one to see the Fourier transform as a canonical isomorphism $\mathrm{L}^2[G] \cong \mathrm{L}^2[G^\wedge]$, is based on the assumption that the group elements of an abelian internal group $\mathbb{G} = (\mathcal{G}, \circ, \bullet)$ in **FHilb** form a basis, giving $\mathrm{Hom}_{\mathbf{FHilb}}(\mathbb{C}, \mathcal{G})$ the Hilbert space structure of $\mathrm{L}^2[G]$ in a natural way, and that the multiplicative characters form a co-basis, giving $\mathrm{Hom}_{\mathbf{FHilb}}(\mathbb{C}, \mathcal{G})$ the Hilbert space structure of $\mathrm{L}^2[G^\wedge]$ in a natural way.

In **FRel**, however, the assumptions of Lemma 4.1.20 only hold in one, somewhat trivial case. The classical states of a classical structure in **FRel** are always a finite, orthogonal and normalisable family, and **FRel** is distributively **CMon**-enriched as required. However, as Lemma 4.1.32 below notes, there is a unique classical structure on any finite set $A$ with classical states forming the resolution of the identity, and the only abelian internal group in **FRel** satisfying the assumptions of Lemma 4.1.20 is the one on the tensor unit $\{\star\}$ of **FRel**.

**Lemma 4.1.32.** *Let $\circ$ be a classical structure on a finite set $A$ in* **FRel***, and let $Z$ be the associated abelian groupoid. The classical points of $\bullet$ form an orthogonal resolution of the identity if and only if the abelian groupoid is discrete, i.e. $Z = \bigoplus^A \mathbb{Z}_1$. Furthermore, if $(\bullet, \circ)$ is a strongly complementary pair on $A$, with $\bullet$ associated to an abelian groupoid $Z = \bigoplus^A \mathbb{Z}_1$, then the abelian groupoid associated with $\circ$ is in fact a group, in the form $X = G$ for some finite abelian group $G = (A, +, 0)$ on the element set $A$. As a consequence, the only abelian internal group in* **FRel** *satisfying the assumptions of Lemma 4.1.20 is the (unique) abelian internal group on the tensor unit $\{\star\}$.*

*Proof.* In **Rel** the scalars are 0 or 1 and summation is given by set union. Thus a resolution of the identity must satisfy the following equation, where each $\chi$ is a classical state:

$\bigcup_\chi$ 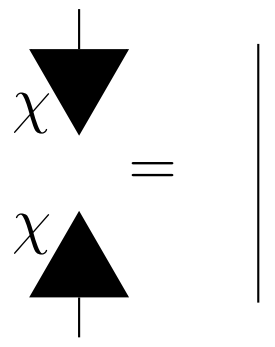 (4.59)

In the specific case of $Z = \bigoplus^A \mathbb{Z}_1$, we have that classical points are in the form $\chi = \{(\star, a)\}$ and (4.59) reads

$$\bigcup_{a\in A}\{(\star, a)\} \circ \{(a, \star)\} = \bigcup_{a\in A}\{(a, a)\} = \mathrm{id}_A. \tag{4.60}$$

When $Z$ is of the generic form $Z = \bigoplus_{\lambda\in\Lambda} G_\lambda$, the classical points are in the form $\chi = G_\lambda$, and (4.59) reads

$$\bigcup_{\lambda\in\Lambda}\{(\star, g')|g' : G_\lambda\} \circ \{(g, \star)|g : G_\lambda\} = \bigcup_{\lambda\in\Lambda}\{(g, g')|g, g' : G_\lambda\}, \tag{4.61}$$

which cannot be the identity if $|G_h| > 1$. Therefore the unique classical structures ○ with classical states forming an orthogonal resolution of the identity are the ones associated to discrete abelian groupoids, in the form $Z = \bigoplus^A \mathbb{Z}_1$. If ● is one such classical structure, and ○ is strongly complementary to it, then it follows immediately from Theorem 4.1.28, and subsequent remarks, that the abelian groupoid $X$ associated to ○ must be in fact an abelian group, in the form $X = \oplus^{\mathbb{Z}_1} G$, where $G$ is some finite abelian group with element set $A$. But to satisfy the assumptions of Lemma 4.1.20, we must have that $X$ is also a discrete groupoid, which forces $G \cong \mathbb{Z}_1$ and $A \cong \{\star\}$. □

So there is no direct parallel in **FRel** of the **FHilb** view that the Fourier transform is a canonical isomorphism $\mathrm{L}^2[G] \cong \mathrm{L}^2[G^\wedge]$, or of its traditional formulation from Equation 4.2. At first sight, this seems to be because the classical states of the two structures of a generic abelian internal group need not form a basis. The question then naturally arises whether restricting our attention to some subclass of states (e.g. those which are linear combinations of the classical states) would lead to a canonical isomorphism similar to the **FHilb** one. Theorem 4.1.33 below answers this question negatively.

Define the **span** $\langle s_j \rangle_{j\in J}$ of a family of states $s_j \subseteq A$ to be the set of all states $r \subseteq A$ which can be obtained as the union $r = \cup_{j\in I} s_j$ of some subfamily $(s_j)_{j\in I\subseteq J}$.

If the $(s_j)_{j\in J}$ are pairwise disjoint, then the states $r \in \langle s_j\rangle_{j\in J}$ are exactly those that descend to boolean functions over the set $\{s_j \text{ s.t. } j \in J\}$:

$$r \in \langle s_j\rangle_{j\in J} \quad \mapsto \quad (j \mapsto \langle s_j|r\rangle) \in \mathbb{B}[J] \tag{4.62}$$

where we denoted by $|s\rangle : I \to A$ the state corresponding to subset $s \subseteq A$. Now consider an abelian internal group $\mathbb{G} = (A, \bullet, \circ)$ in **FRel**, and let $Z = \bigoplus^{|G|} H$ and $X = \bigoplus^{|H|} G$ be the abelian groupoids corresponding to $\bullet$ and $\circ$ respectively. Then $\langle H_g\rangle_{g\in G}$, seen as the set $\{0,1\}^G$ of boolean functions on $\{H_g \text{ s.t. } g \in G\}$, plays the role in **Rel** that $\mathrm{L}^2[G]$ played in **FHilb**, and similarly $\langle G_h\rangle_{h\in H}$, seen as $\{0,1\}^H$, is the analogue of $\mathrm{L}^2[H]$ (and takes the place $\mathrm{L}^2[G^\wedge]$ had in **FHilb**).

**Theorem 4.1.33.** *Let $\mathbb{G} = (A, \bullet, \circ)$ be an abelian internal group in* **FRel**, *and let $Z = \bigoplus^{|G|} H$ and $X = \bigoplus^{|H|} G$ be the abelian groupoids corresponding to $\bullet$ and $\circ$ respectively. Then $\langle H_g\rangle_{g\in G} \cap \langle G_h\rangle_{h\in H} = \{\emptyset, A\}$. In particular, under the correspondence of* (4.62)*, the Fourier transform of* **FRel** *does not restrict to an isomorphism $\{\bot,\top\}^G \cong \{\bot,\top\}^H$, nor can it be restricted to an isomorphism $S_G \cong S_H$ for any $S_G \subseteq \{\bot,\top\}^G$ and $S_H \subseteq \{\bot,\top\}^H$ containing non-constant functions.*

*Proof.* Let $r \in \langle H_g\rangle_{g\in G} \cap \langle G_h\rangle_{h\in H}$, then $r$ is either the empty set or it contains some $(h', g') \in A$. But if $(h', g') \in r$, then for all $g \in G$ we have $(h', g) \in r$ (because $r \in \langle G_h\rangle_{h\in H}$), and then for any $g \in G$ we have that for all $h \in H$ $(h, g) \in r$ (because $r \in \langle H_g\rangle_{g\in G}$). Thus for all $g \in G$ and $h \in H$ we have $(h, g) \in r$, i.e. $r = A$. The state $r = \emptyset$ in $\langle H_g\rangle_{g\in G} \cap \langle G_h\rangle_{h\in H}$ corresponds the constant $\bot$ function in $\{\bot,\top\}^H$ and in $\{\bot,\top\}^G$ (under Equation 4.62), while the state $r = A$ corresponds to the constant $\top$ function. The Fourier transform maps the constant $\bot$ function of $\{\bot,\top\}^G$ to the constant $\bot$ function of $\{\bot,\top\}^H$, and similarly with the constant $\top$ functions. However, if $r \in \langle H_g\rangle_{g\in G}$ is neither empty nor the whole of $A$, i.e. if it corresponds to a non-constant function in $\{\bot,\top\}^G$, then its Fourier transform does not lie in the span $\langle G_h\rangle_{h\in H}$, and thus cannot be seen as a function in $\{\bot,\top\}^H$. □

As a consequence of Theorem 4.1.33, the best analogue in **FRel** of the statement that the Fourier transform is a canonical isomorphism $\mathrm{L}^2[G] \cong \mathrm{L}^2[G^\wedge]$ in **FHilb** is the trivial statement that the Fourier transform in **FRel** is an isomorphism $\{\bot_G, \top_G\} \cong \{\bot_H, \top_H\}$ between the constant functions of $\{\bot,\top\}^G$ and of $\{\bot,\top\}^H$.

To summarise, in **FHilb** we have the following views of the Fourier transform for abelian internal groups:

1. As quantum Fourier transform, implemented by application of a Fourier matrix (subject to a non-canonical choice of isomorphism $G \cong G^\wedge$) followed by operations in the computational basis.

2. In the sense of Pontryagin duality, as a canonical isomorphism $\mathrm{L}^2[G] \cong \mathrm{L}^2[G^\wedge]$.

3. Again as quantum Fourier transform, but implemented by measuring in a basis that is strongly complementary to the computational basis.

In **FRel**, on the other hand, things are very different:

1. Except in the case where $Z = \bigoplus^{|G|} G$ and $X = \bigoplus^{|G|} G$ (isomorphic, but different), no Fourier matrix can exist in **Rel**.

2. The Fourier transform does not give, in **Rel**, an isomorphism $\{0,1\}^G \cong \{0,1\}^H$ between the spaces of boolean-valued functions on the group elements / multiplicative characters.

3. The operational definition based on strong complementarity, however, is still valid in **FRel**.

To conclude, the following examples give explicit examples of quantum Fourier transforms in **Rel**.

**Example 4.1.34.** Take $G = \mathbb{Z}_2 = \{0,1\}$, $H = \mathbb{Z}_1 = \{\star\}$, $Z = G = \{0_\star, 1_\star\}$ and $X = H \oplus H = \{\star_0, \star_1\}$. The computational basis is the family $(H_g)_{g:G}$ of copyable points for $X$, i.e. $H_0 = \{(\star, 0)\}$ and $H_1 = \{(\star, 1)\}$. The character family used for the quantum Fourier transform consists a single classical state $G_\star = \{(\star, 0), (\star, 1)\}$ for $Z$. In this case all states can be prepared in the computational basis, but the measurement in the character family will be trivial.

**Example 4.1.35.** Take $G = \mathbb{Z}_2 = \{0,1\}$, $H = \mathbb{Z}_2 = \{a,b\}$, $Z = G \oplus G = \{0_a, 1_a, 0_b, 1_b\}$ and $X = H \oplus H = \{a_0, b_0, a_1, b_1\}$. The computational basis is the family $(H_g)_{g:G}$ of classical states for $X$, i.e. $H_0 = \{(a,0),(b,0)\}$ and $H_1 = \{(a,1),(b,1)\}$. The character family used for the quantum Fourier transform is the family $(G_h)_{h:H}$ of classical states for $Z$, i.e. $G_a = \{(a,0),(a,1)\}$ and $G_b = \{(b,0),(b,1)\}$.

It is part of the process theoretic programme that the operational features of quantum theory should be modelled categorically, and that any category sharing features with **FHilb** should be considered, at least in principle, as a potential (toy?) model of quantum mechanics. The category **FRel** is an example of one such model. We have shown that Fourier matrices do not generalise well outside **FHilb**, and certainly fail to implement a quantum Fourier transform in **Rel**, but that our treatment of Fourier theory based on strong complementarity goes through unharmed. As a consequence, categorical quantum algorithms where the quantum Fourier transform is formulated using strong complementarity will straightforwardly generalise to **Rel** and other categories. We take this to be an indication that our perspective on the quantum Fourier transform is conceptually sound and operationally advantageous.

### 4.1.5 Non-abelian Fourier transform

In **FHilb**, abelian internal groups satisfy the assumptions of Lemma 4.1.20, and the corresponding Fourier transform can be seen, via enrichment, as a canonical isomorphism of $\mathrm{L}^2$-spaces $\mathrm{L}^2[G] \cong \mathrm{L}^2[G^\wedge]$. Non-abelian internal groups in **FHilb**, however, fail those assumptions, as the classical states of the group structure never form a basis. However, it is a consequence of the Peter-Weyl theorem that the irreducible representations can be used to obtain a resolution of the identity. We will introduce matrix algebras in a QPT in order to handle these multi-dimensional representations. This then allows us, the rest of the section, to review the generalisation of our treatment in the previous sections as is presented by Gogioso in [46]. The work presented there allows full-blown representation theory and concludes with a formulation of non-abelian Fourier transform connected with the Gelfand–Naimark–Segal construction.

First, let's see why non-abelian internal groups in **FHilb** cannot satisfy the assumptions of Lemma 4.1.20. The classical states of the point structure form an orthogonal basis, and are the elements of some non-abelian group $G$. The classical states of the group structures are the multiplicative characters of $G$: they are always orthogonal and normalisable, but as long as we show that there are less of them than the number of elements of $G$, we can conclude that they won't form a co-basis as would be required by Lemma 4.1.20. But the multiplicative characters of a group $G$ are the same as the multiplicative character of its abelianization $G'$, which is always strictly smaller than $G$ (at least by a factor of 2). Therefore there are always strictly less multiplicative characters than group elements for a non-abelian internal group in **FHilb**, and hence the assumptions of Lemma 4.1.20 always fail. It turns out that this is not restricted to **FHilb**:

**Theorem 4.1.36.** *Let $\mathbb{G} = (\mathcal{G}, \circ, \bullet)$ be an internal group in a †-SMC* **C**. *If the classical states of $\circ$ form a basis, then $\mathbb{G}$ is necessarily an abelian internal group.*

*Proof.* All we have to show is that $\curlywedge = \curlywedge \circ s_{\mathcal{GG}}$, where $s_{\mathcal{GG}}$ is the braiding operator. Equivalently, it is enough to show that $\curlyvee = s_{\mathcal{GG}} \circ \curlyvee$. But indeed for any classical state $|\chi\rangle$ of $\circ$ we have $\curlyvee \circ |\chi\rangle = |\chi\rangle \otimes |\chi\rangle = s_{\mathcal{GG}} \circ |\chi\rangle \otimes |\chi\rangle = s_{\mathcal{GG}} \circ \curlyvee \circ |\chi\rangle$. As we have assumed that the classical states form a basis, this completes the proof. □

To deal with the non-abelian case, we will introduce matrix algebras into QPTs. Recall that Hilbert spaces $\mathcal{H}_n$ of dimension $n$ are isomorphic to $\mathbb{C}^n$. This means that in **FHilb** the name (3.65) of a morphism is an $n \times n$ matrix:

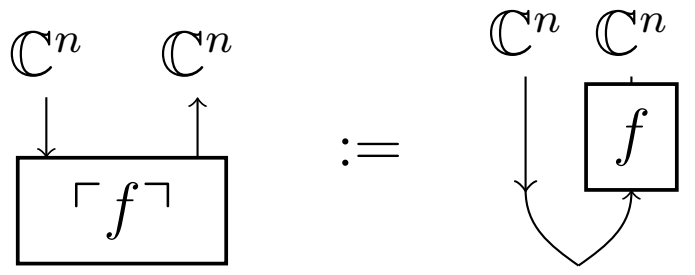

(4.63)

This motivates us, following Vicary [104], to build up a †-Frobenius algebra (a matrix algebra) in an arbitrary QPT using caps and cups. The monoid, unit, comonoid, and counit of this algebra are given as follows:

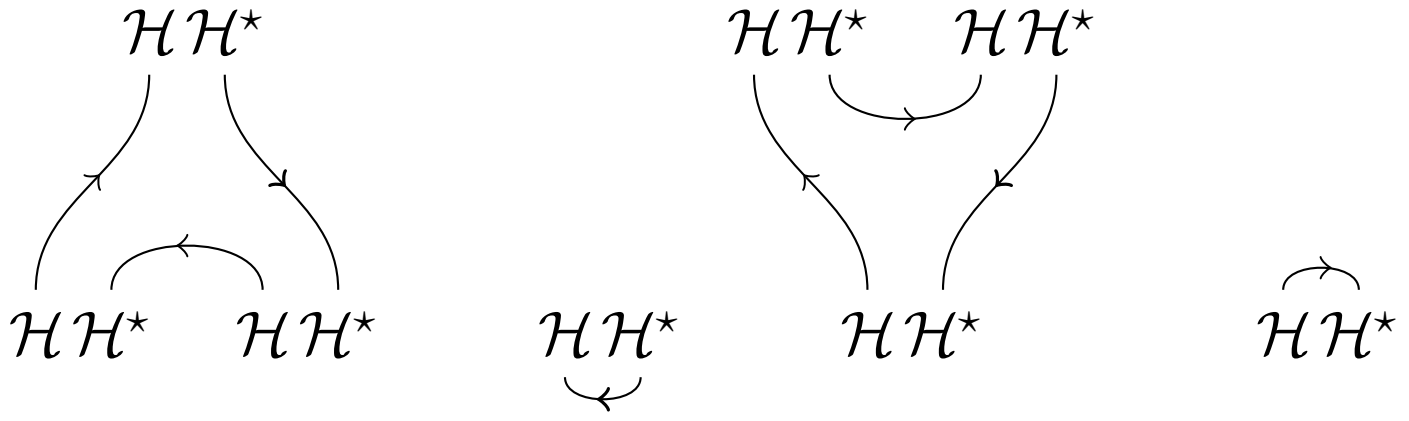

(4.64)

One-dimensional representations of a monoid $(\mathcal{G}, \curlywedge, \bot)$ are morphisms into the trivial monoid on $\mathcal{H} = I$. In general though, multidimensional representations are morphisms into the monoid from (4.64).

**Definition 4.1.37.** The **representations** of a monoid $(\mathcal{G}, \curlywedge, \bot)$ in a compact-closed †-SMC are the morphisms $\rho : \mathcal{G} \to \mathcal{H} \otimes \mathcal{H}^\star$ satisfying the first two equations in (4.65). The representations of an internal groups $(\mathcal{G}, \circ, \bullet)$ are the representations of the monoid $(\mathcal{G}, \curlywedge, \bot)$ part of the internal group. A representation of an internal group is **unitary** if it satisfies the third as well. A representation is **isometric** if it is an isometry.

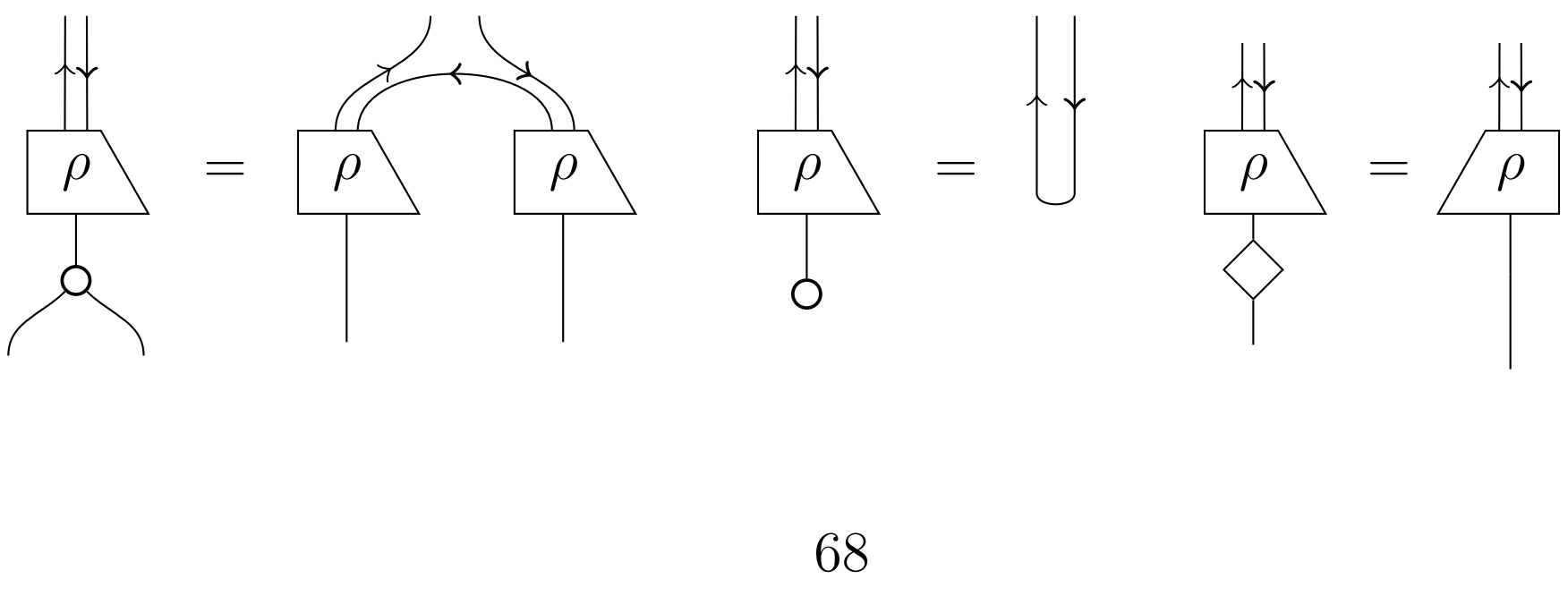

(4.65)

**Definition 4.1.38.** The **character** associated with a representation $\rho : \mathcal{G} \to \mathcal{H} \otimes \mathcal{H}^\star$ of a monoid / internal group in a compact-closed †-SMC is the morphism $\chi_\rho : \mathcal{G} \to I$ defined by Equation 4.66. In the case of internal groups, a character $\chi_\rho$ is **unitary** if the representation $\rho$ is.

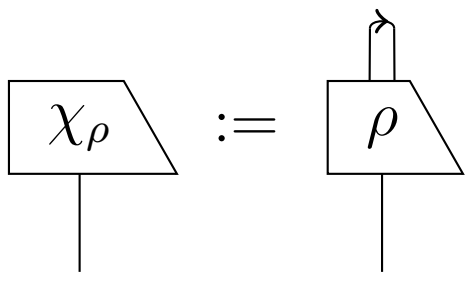

(4.66)

These notions clearly generalize those given in Section 4.1.3. The orthogonality of representations and characters by these abstract definitions is categorically proven by Gogioso in [46]. This allows a generalization of the Abelian Fourier transform from the previous section

**Lemma 4.1.39.** *Let* $\mathbb{G} = (\mathcal{G}, \circ, \bullet)$ *be an internal group in a compact-closed †-SMC which is distributively* **CMon**-*enriched. Further assume that* $(\rho)_{\rho \in \mathcal{R}}$ *is a finite, normalisable family of representations of* $\mathbb{G}$ *(with normalisation factors* $(N_\rho)_{\rho \in \mathcal{R}}$*), which forms an orthogonal resolution of the identity. Then the Fourier transform of Definition 4.31 can be written in the following way:*

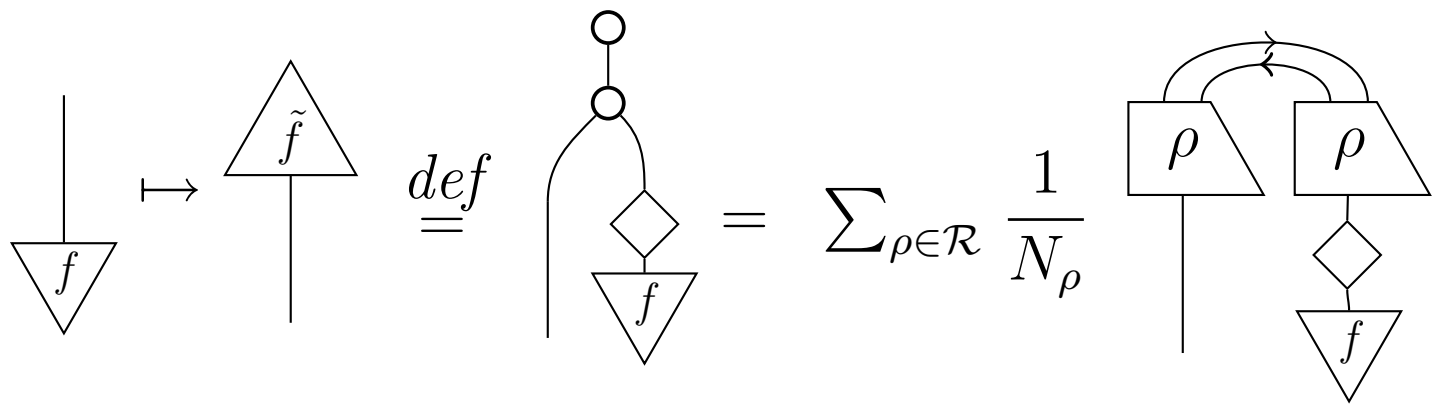

(4.67)

*Proof.* Given in [46]. It proceeds along the same lines as that of the abelian version given here. ☐

Gogioso extends this perspective to provide a categorical version of the Gelfand-Naimark theorem [46].

### 4.1.6 Measurements and representation theory

We have seen that the representations of an abelian internal group can form a basis, and, in particular, that they are the dagger of classical states of one half of a strongly complementary pair of classical structures $(\circ, \bullet)$. Indeed, even in the non-abelian case, representations of larger than one dimension will also be orthogonal and form a resolution of the identity [46]. If we consider the group elements as embedded in classical states of $\circ$ then the group's representations are effects for the $\bullet$ classical structure. By the QPT Born rule (Definition 3.2.5) we now have a full and abstract grasp on what it means to measure in the "representation basis" or "Fourier basis", i.e. composing with a representation's effect acts as a post-selected measurement of that outcome in the "representation basis." We clarify the results in this section that will appear in our characterization of quantum algorithms with Figure 4.2. The generalized Fourier matrix will play a particularly important role in characterizing quantum-like algorithms in process theories. Indeed the quantum Fourier transform algorithm, when implemented in a QPT, acts like a Fourier matrix (Figure 4.2).

Figure 4.2: Summary of the results of Section 4.1 as they pertain to the analysis of quantum algorithms in the following sections. These constructions are with reference to an internal group $\mathbb{G} = (\mathcal{G}, \circ, \bullet)$ in a QPT (Definition 4.1.12).

| | |
|---|---|
| **Group structure**; acts as a group on its elements and as pointwise multiplication on representations. The unit is the identity element. | 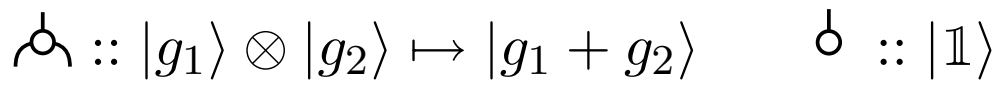  |
| **Group elements**; these are classical states of $\bullet$. | 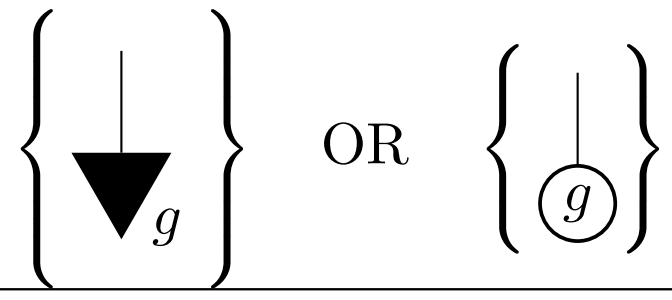  |
| **Representation structure**; acts as pointwise multiplication on group elements and as the representation group on representations. The unit is the trivial representation. | 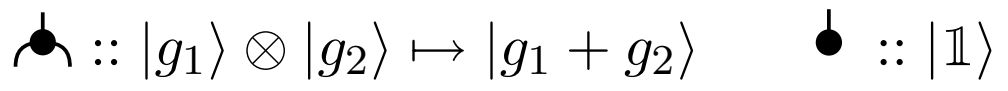  |
| **Representations** and **representation states** (abelian $\mathbb{G}$); these are daggers of each other and are classical co-states and states of $\circ$ respectively. | 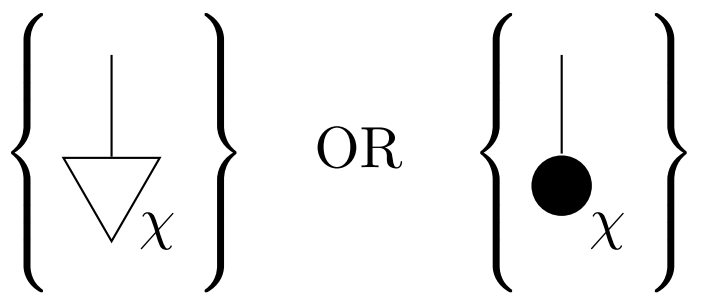  |
| **Representations** and **representation states** (non-abelian $\mathbb{G}$); these are daggers of each other and act as generalized classical co-states and states of $\circ$ respectively. | 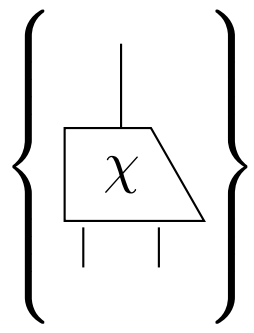  |
| **Fourier transform**; For arbitrary state $f$; Definition 4.1.19 and Lemma 4.1.39 (any $\mathbb{G}$). | 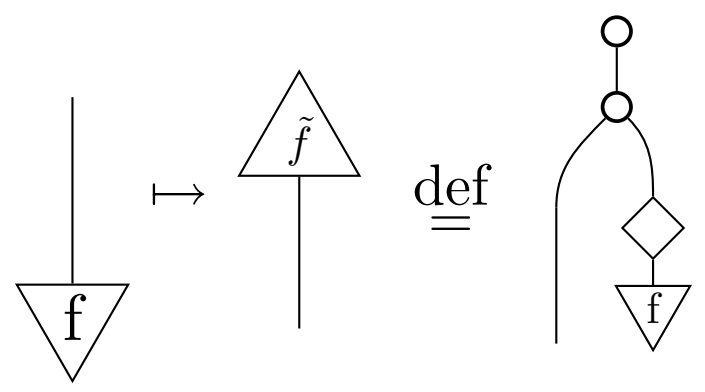  |
| **Fourier matrix**; Definition 4.1.24. This can be thought of as a color-change operation. | 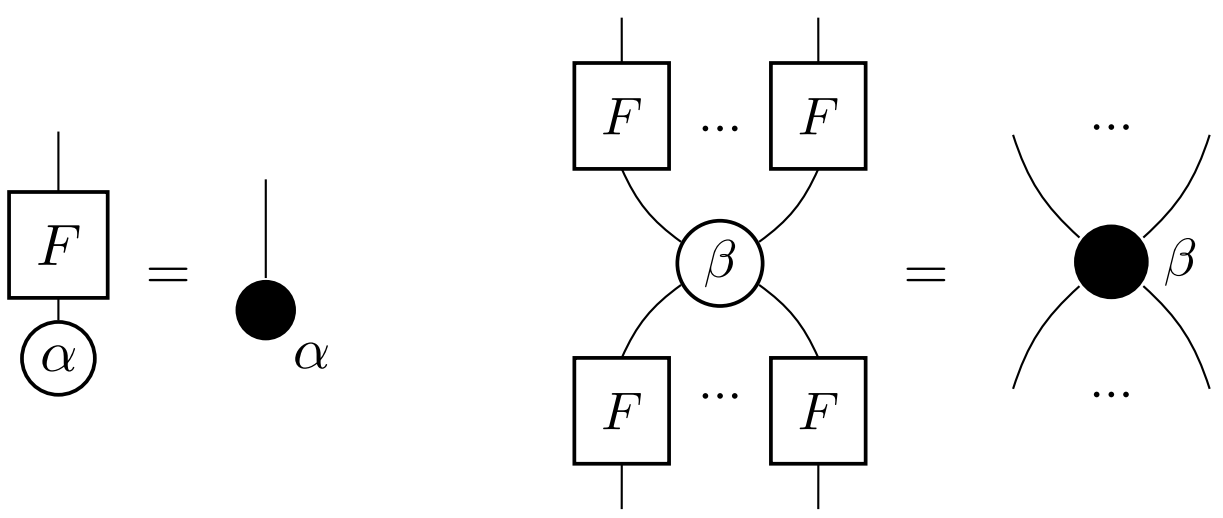  |

## 4.2 Quantum Blackbox Algorithms

The Fourier transform is a powerful tool in most quantum algorithms for algebraic problems. Now that we have a structural handle on this tool, we use it to verify, generalize, and construct quantum algorithms.

In this section we first consider the structure of unitary oracles in general and then review the Deutsch-Jozsa, Grover's single shot, and hidden subgroup algorithms as examples of the approach. These three initial algorithms were analyzed by Vicary in [104], though the structure of the underlying oracles was assumed there. We then expand on this base to develop a new quantum algorithm for the GROUPHOMID problem. The analysis of this new algorithm is presented in Section 4.2.4.

### 4.2.1 The abstract structure of unitary oracles

When we program an abstract problem into an oPT, we have a choice of embeddings, but typically assign classical information to the classical states of some classical structure. This was, for example, the case in our treatment of the Fourier transform. Of course, we will also want to consider how to program functions between those classical states. An equivalent definition for classical states is that they are self-conjugate comonoid homomorphisms from the trivial comonoid on the identity object to the classical structure monoid on a system. Classical functions will then be self-conjugate comonoid homomorphisms in general.

**Definition 4.2.1.** In a monoidal dagger-category, a comonoid homomorphism $f : (A, \blacktriangledown, \bullet) \to (B, \triangledown, \circ)$ is **self-conjugate** when the following property holds:

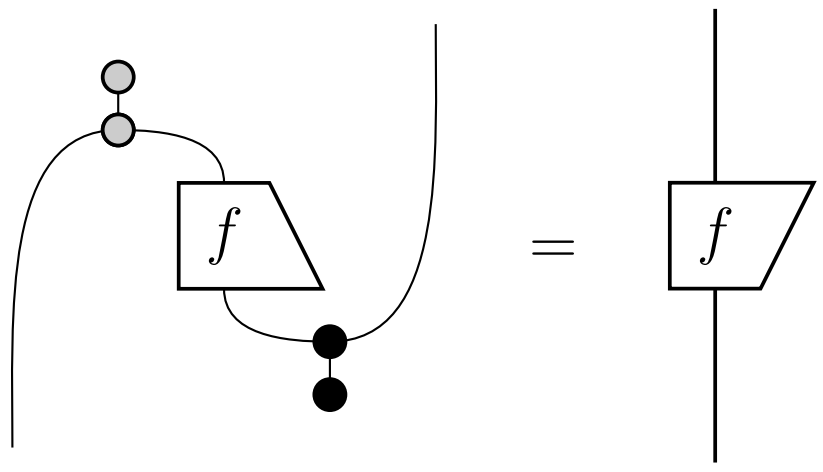

(4.68)

**Lemma 4.2.2.** *In* **Hilb***, comonoid homomorphisms* $f : (A, \blacktriangledown, \bullet) \to (B, \triangledown, \circ)$ *of classical structures are self-conjugate.*

*Proof.* Recall that comonoid homomorphisms between classical structures in **Hilb** are exactly classical functions between the classical states [37]. The linear maps on either side of (4.68) will be the same if and only if their matrix elements are the same, obtained by composing with $|i\rangle$ at the bottom and $\langle j|$ at the top. On the left-hand side, this gives the following result:

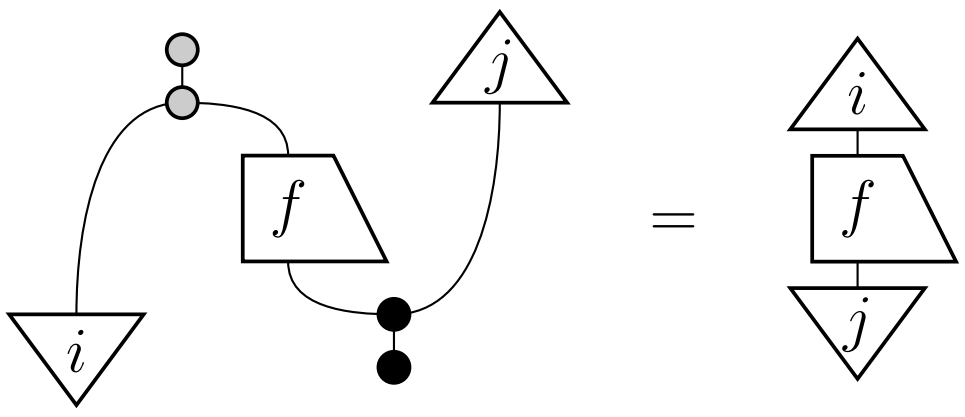

$$= \begin{cases} 1 & \text{if } i = f(j), \\ 0 & \text{if } i \neq f(j). \end{cases} \tag{4.69}$$

On the right we can do this calculation:

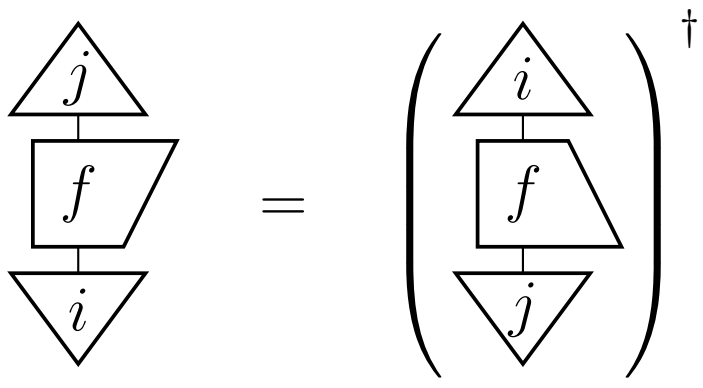

$$= \begin{cases} 1 & \text{if } i = f(j) \\ 0 & \text{if } i \neq f(j) \end{cases}^{\dagger} = \begin{cases} 1 & \text{if } i = f(j), \\ 0 & \text{if } i \neq f(j). \end{cases} \tag{4.70}$$

This is the same result as for the left-hand side, and so Equation (4.68) holds. □

The oracle that we will build here is similar to the CNOT from Section 3.5.5. Recall that a pair of symmetric dagger-Frobenius algebras can be used to build a linear map in the following way:

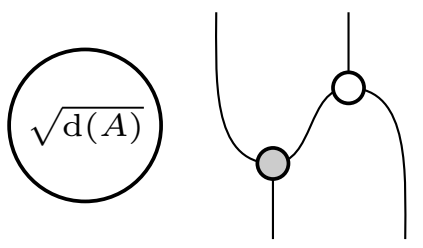

(4.71)

Here we have assumed that we operate in a suitably enriched category where square roots of scalars exist. In the rest of this section we will suppress drawing circles around scalars. The difference between scalars and labels of the diagram will be clear from context. Two classical structures are complementary exactly when the composite (4.71) is unitary, as we show in the following theorem.

**Theorem 4.2.3** (Complementarity via a unitary)**.** *In a dagger symmetric monoidal category, two classical structures are complementary if and only if the composite* (4.71) *is unitary.*

*Proof.* Composing (3.59) with its adjoint in one order, we obtain the following:

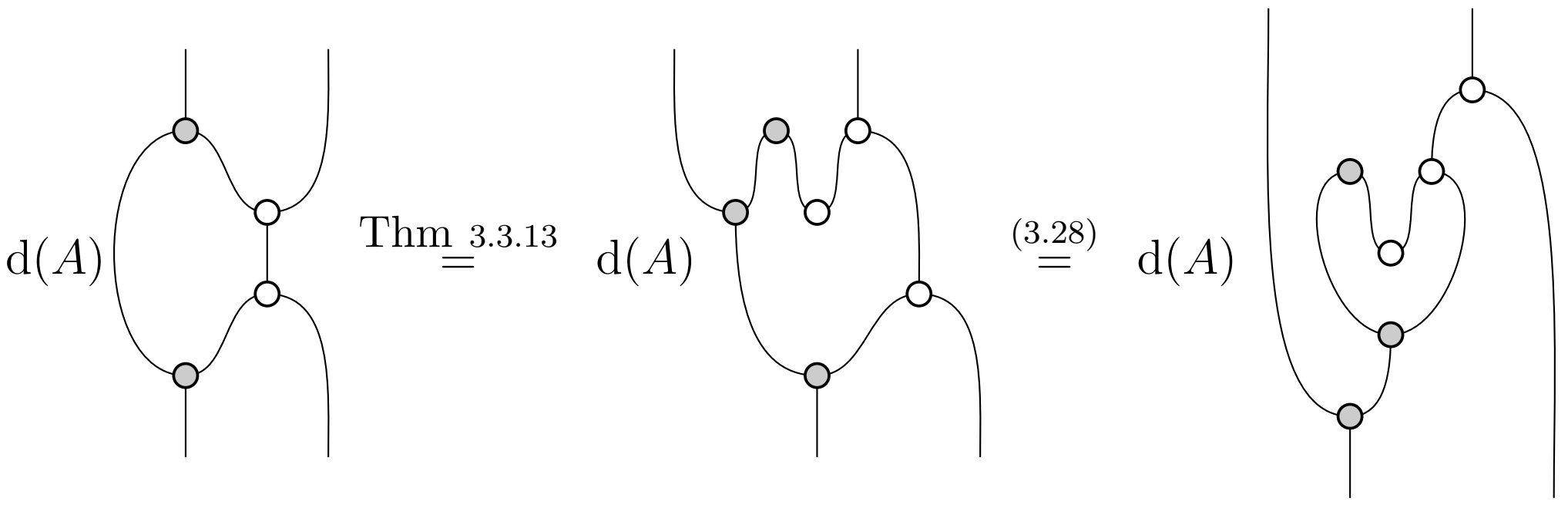

(4.72)

If the complementarity condition (3.50) holds then this is clearly the identity on $A \otimes A$. The other composite can be shown to be the identity in a similar way, and so (3.59) is unitary.

Conversely, suppose (3.59) is unitary. Then the final expression of (4.72) certainly equals the identity on $A \otimes A$:

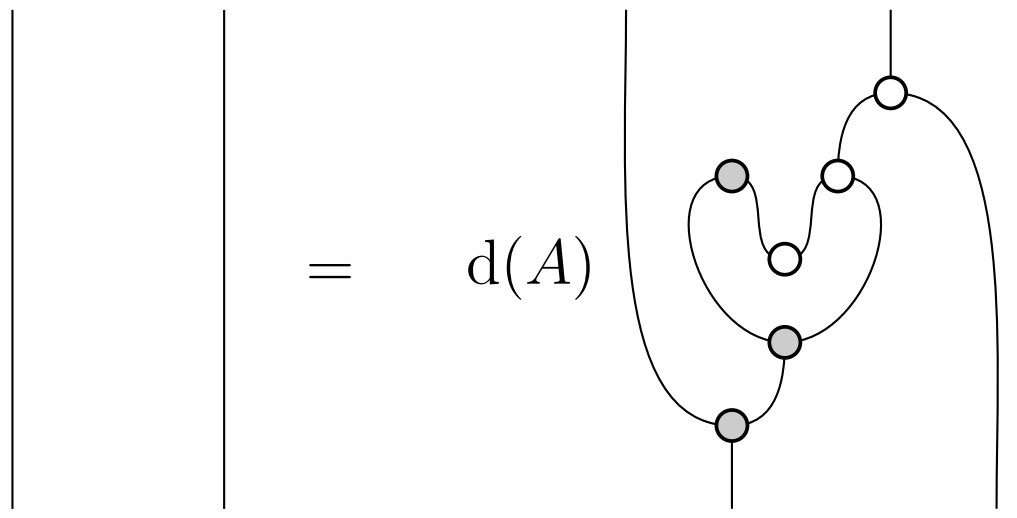

(4.73)

Composing with the black counit at the top-left and the white unit at the bottom-right then gives back complementarity condition (3.50) as required:

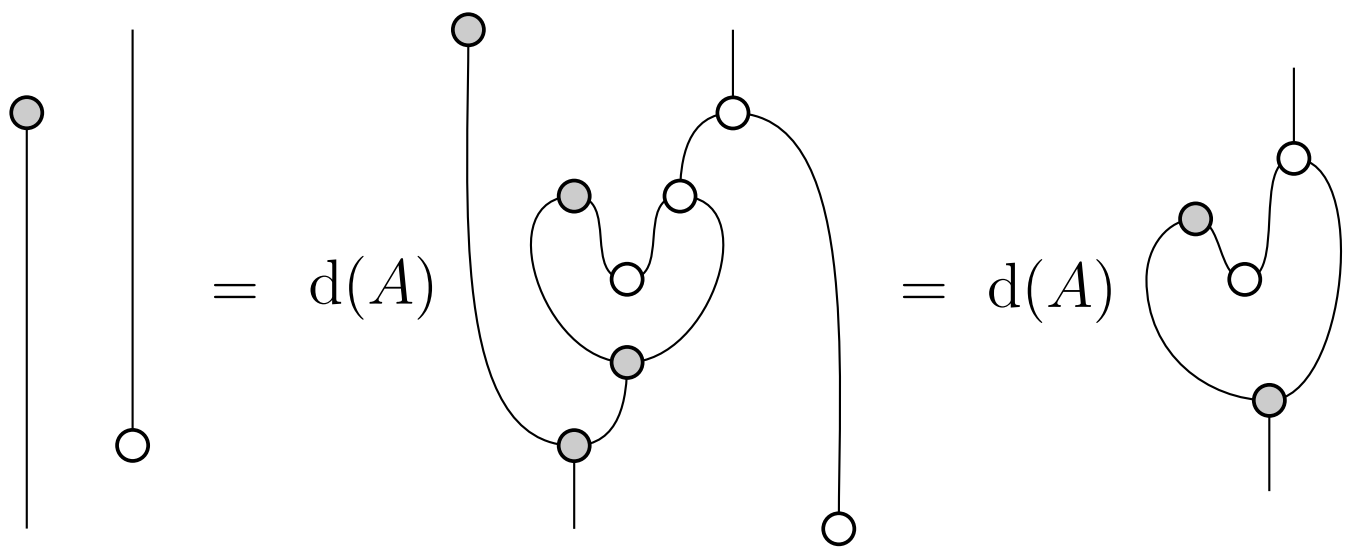

(4.74)

This completes the proof. □

This pair of complementary observables automatically gives rise to a much larger family of unitaries, one for each self-conjugate comonoid homomorphism onto one of the classical structures in the pair. Lemma 4.2.2 demonstrated that in **FHilb**, every comonoid homomorphism of classical structures is self-conjugate.

**Definition 4.2.4** (Oracle)**.** In a symmetric monoidal dagger-category, given a dagger-Frobenius comonoid $(A, \blacktriangledown, \bullet)$, a pair of complementary symmetric dagger-Frobenius comonoids $(B, \triangledown, \circ)$ and $(B, \triangledown', \circ')$, and a self-conjugate comonoid homomorphism $f : (A, \blacktriangledown, \bullet) \to (B, \triangledown', \circ')$, the **oracle** is defined to be the following endomorphism of $A \otimes B$:

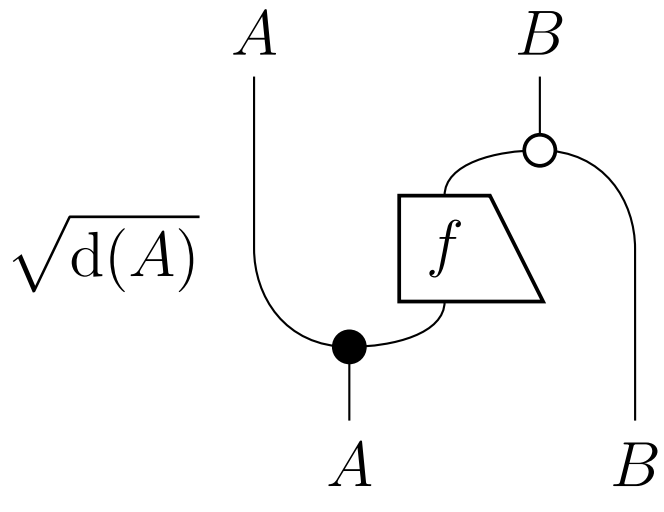

(4.75)

**Theorem 4.2.5.** *Oracles are unitary.*

*Proof.* To demonstrate that the oracle (4.75) is unitary, we must compose it with its adjoint on both sides and show that we get the identity in each case. In one case, we obtain the following, making use of the Frobenius laws, self-conjugacy of $f$, associativity and coassociativity, the fact that $f$ preserves comultiplication, the complementarity condition, the fact that $f$ preserves the counit, and the unit and counit laws:

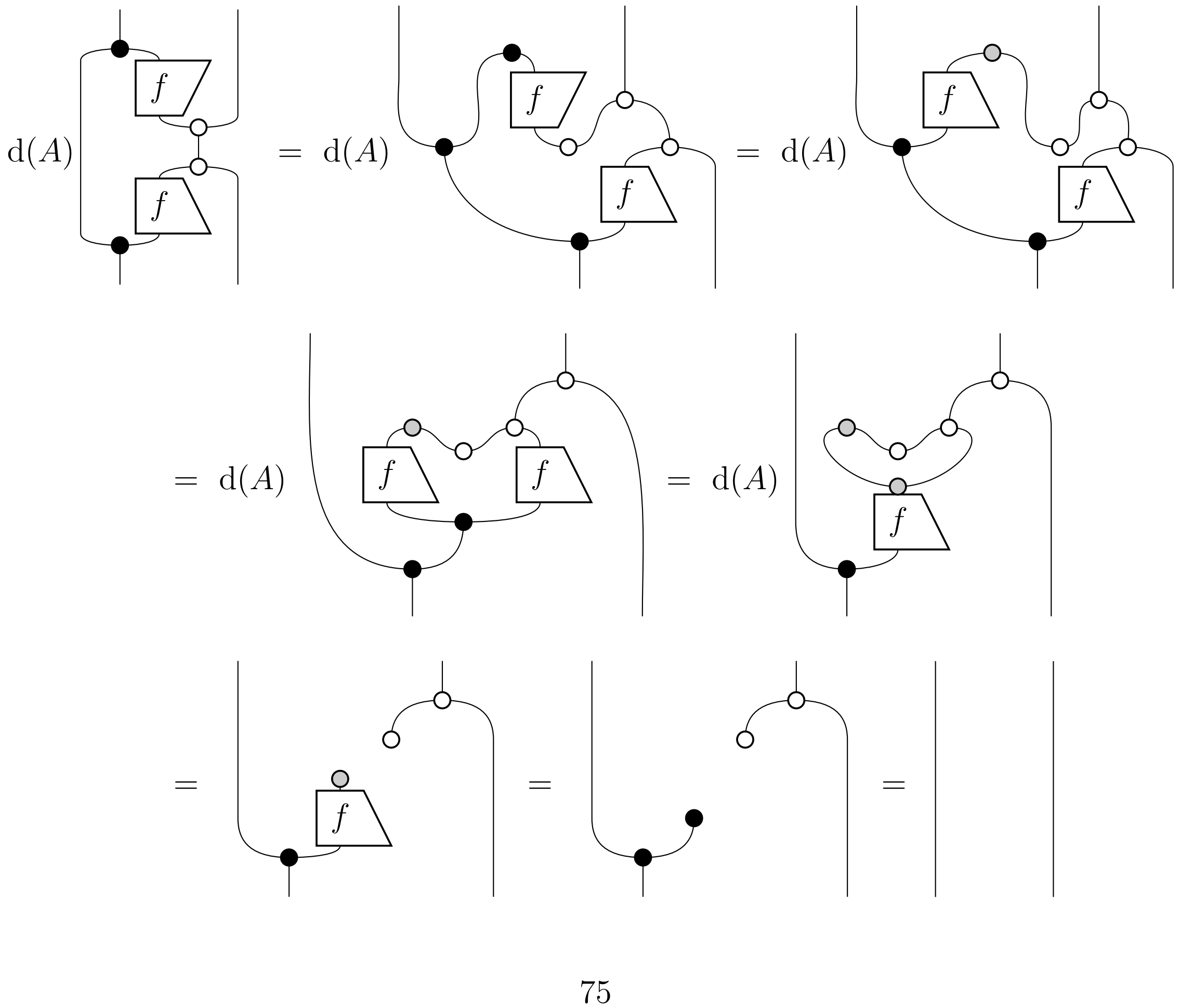

There is a similar argument that the other composite also gives the identity. □

Note that this construction works for pairs of observables that a complementary, but not necessarily strongly complementary. Still, in the examples we consider, we will use strongly complementary ( $\circ$, $\circ$ ) as we wish to embed an internal group onto them.

### 4.2.2 The Deutsch-Jozsa algorithm

In this section, the abstract structure of the Deutsch-Jozsa (DJ) quantum algorithm is presented and its function is verified abstractly, following Vicary [104]. We recall that the Deutsch-Jozsa algorithm presents a quantum algorithm with an exponential speedup over exact classical computation [40]. In its original formulation, given a function $f : \{0,1\}^n \to \{0,1\}$ that is promised to be either constant or balanced (i.e. outputs the same number of 0's as 1's), the algorithm determines which of the two classes $f$ is in. As we will see, this formulation can be easily verified and generalized by the QPT approach.

First we define the promise at the level of processes in an oPT.

**Definition 4.2.6.** Let $A$ and $B$ be systems in a QPT such that $A$ has an internal group ($\mathbb{G}$, $\circ$, $\bullet$) and $B$ has an internal group ($\mathbb{H}$, $\circ$, $\circ$). A process $f : A \to B$ is

(i) **constant** when for $\circ$-classical state $c : I \to B$

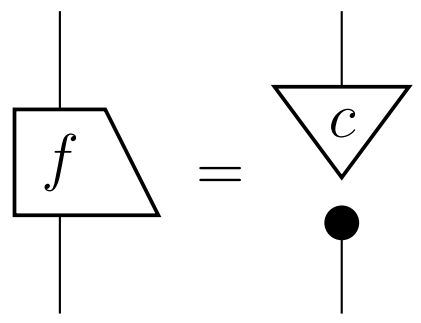

(4.76)

(ii) **balanced** when for non-trivial representation $\sigma : B \to I$

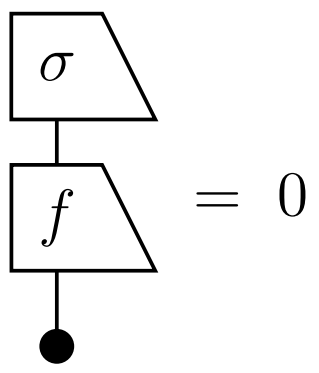

(4.77)

It is easy to see that this definition coincides with the original formulation in **FHilb** for internal groups $\mathbb{G} = \mathbb{Z}_2^n$ and $\mathbb{H} = \mathbb{Z}_2$. In this case the $\circ$-classical states of $B$ are the group elements $\{|0\rangle, |1\rangle\}$, so a constant function simply ignores all input (as $\bullet$ sends all input to 1) and outputs one group element. In the balanced case,

there is only one non-trivial representation of $\mathbb{Z}_2$, namely $\sigma(0) = 1$ and $\sigma(1) = -1$. Thus

$$\sigma \circ f \circ \bullet = \sigma\left(\sum_{x\in\mathbb{G}} |f(x)\rangle\right) = \sum_{x\in\mathbb{G}} \sigma\left(|f(x)\rangle\right), \tag{4.78}$$

which is clearly 0 if and only if half the number of inputs for which $f(x) = 0$ is the same as the number of inputs for which $f(x) = 1$. Picking different internal groups and representations gives different behaviors for balanced functions.

**Example 4.2.7.** Let $\mathbb{G} = \mathbb{Z}_2^n$ and $\mathbb{H} = \mathbb{Z}_2 \times \mathbb{Z}_2$. We then have a choice of different non-trivial representations. Choose $\sigma = (1, -1, 1, -1)$. Denoting $\# f_0$ as the number of inputs for which $f(x) = 0$, balanced functions are now $f$ such that:

$$\# f_{00} + \# f_{10} = \# f_{01} + \# f_{11}$$

Directly mapping the (rotated) quantum circuit for the DJ algorithm into a QPT process gives the following, where the internal groups $(\mathbb{G}, \circ, \bullet)$ and $(\mathbb{H}, \circ, \circ)$ are inherited from the promise of constant and balanced:

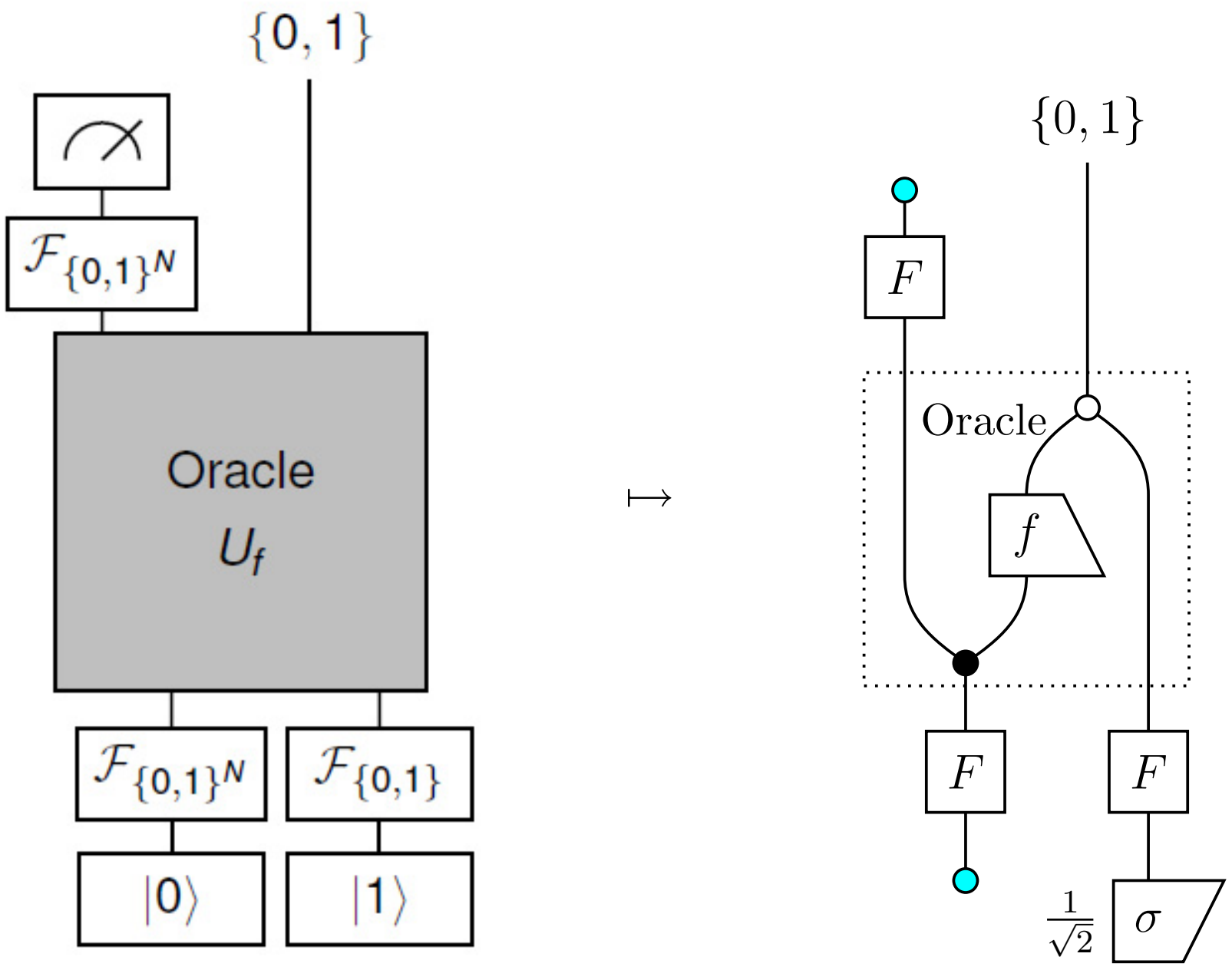

(4.79)

In Section 4.1, we saw that the quantum Fourier transform takes classical states of one observable to representations that correspond to classical states of its strongly complementary partner. Simplifying the above using the Fourier transform rules we obtain the annotated diagram, where the drawing of circles around normalizing scalars is suppressed:

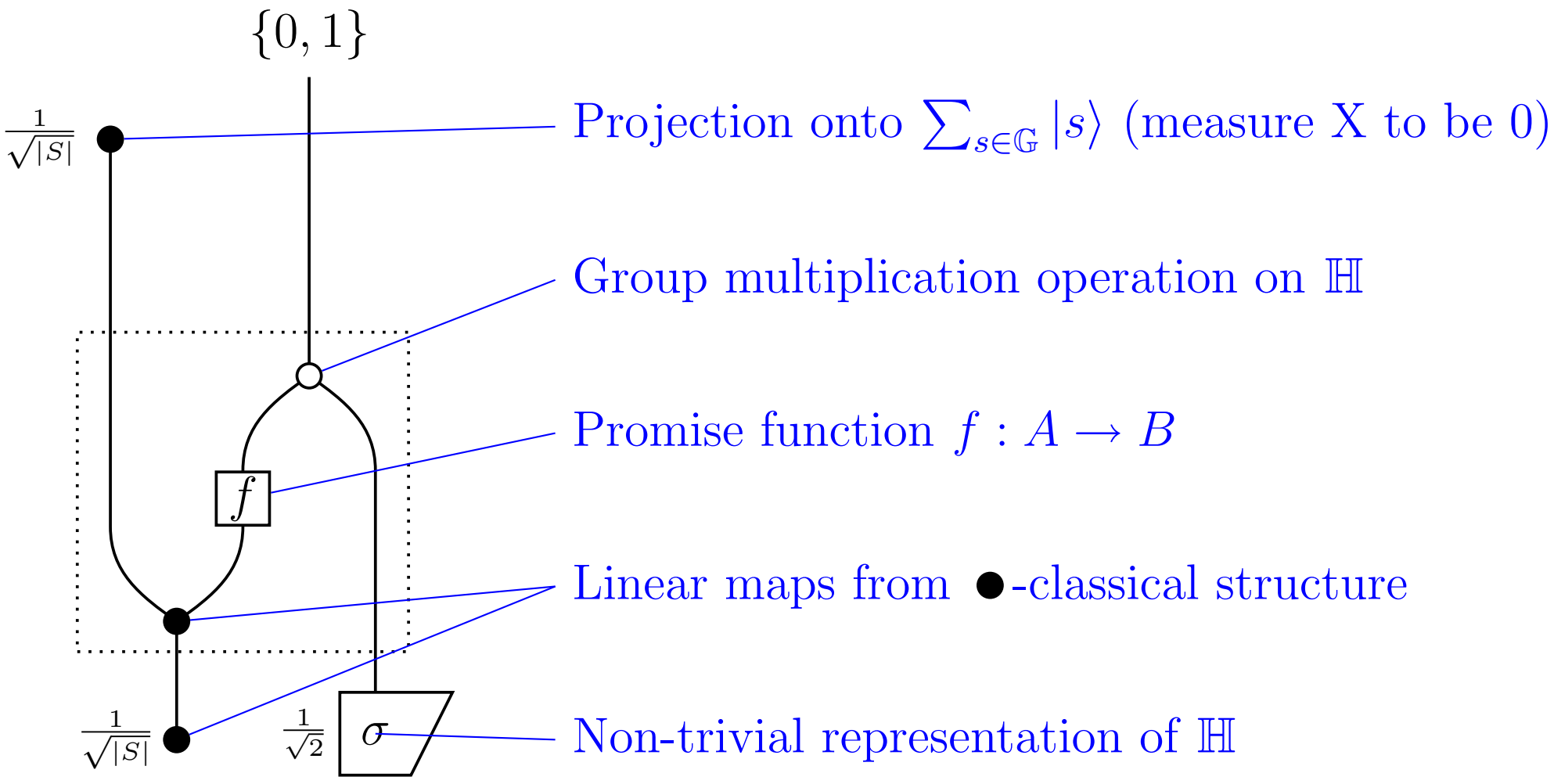

(4.80)

Using the rules of QPTs we obtain:

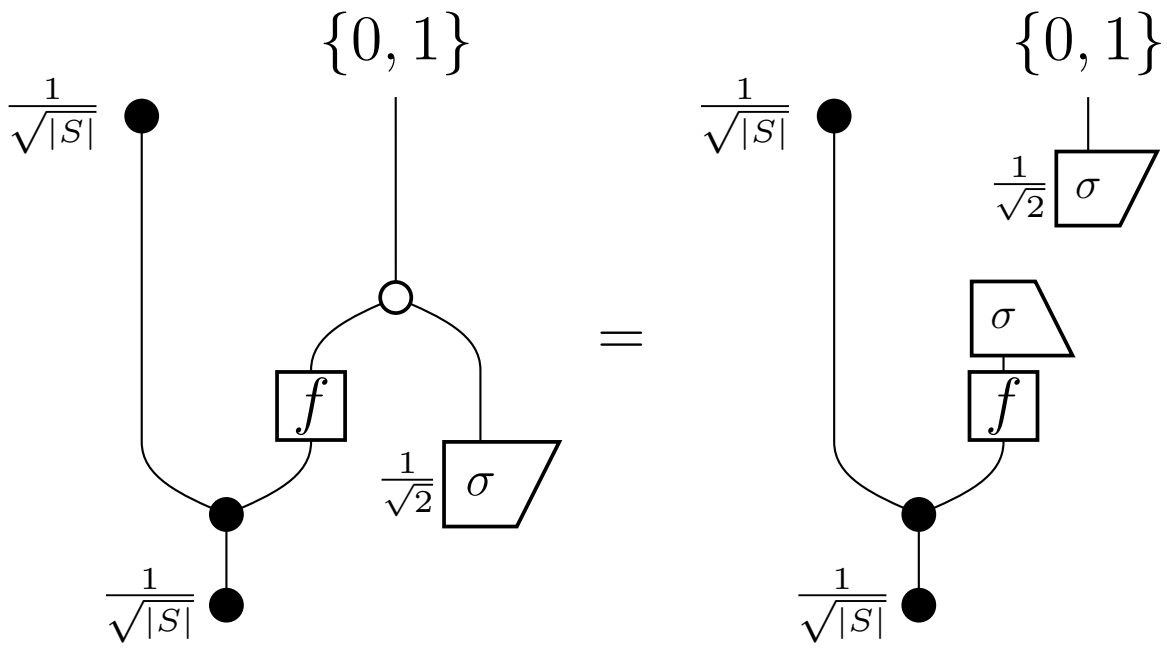

(4.81)

If we now ignore the right hand system, this becomes:

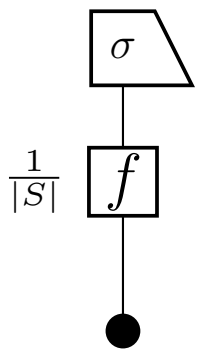

$$= \begin{cases} 0 \text{ if } f \text{ is balanced} \\ \sigma(c) \text{ if } f \text{ is constant s.t. } f(x) = c \end{cases} \qquad (4.82)$$

Thus, by the QPT Born rule (Definition 3.2.5), there is no probability that an outcome $\sigma$ would be measured for a balanced function. Conversely, should the coefficients of $\sigma$ be square roots of the identity, then $\sigma$ will always be measured for constant $f$. This allows us to deterministically distinguish large generalized classes of constant and balanced $f$ with a single oracle query.

In a single proof, this method and its extension to arbitrary finite internal groups (c.f. [104]) captures the original formulation of the problem as well as the generalizations of Høyer [58] and Batty et al. [18]. Further, we can now model the algorithm in QPTs besides quantum theory that have strongly complementary observables. We expand on this idea in Section 4.3 with a toy model in **FRel**.

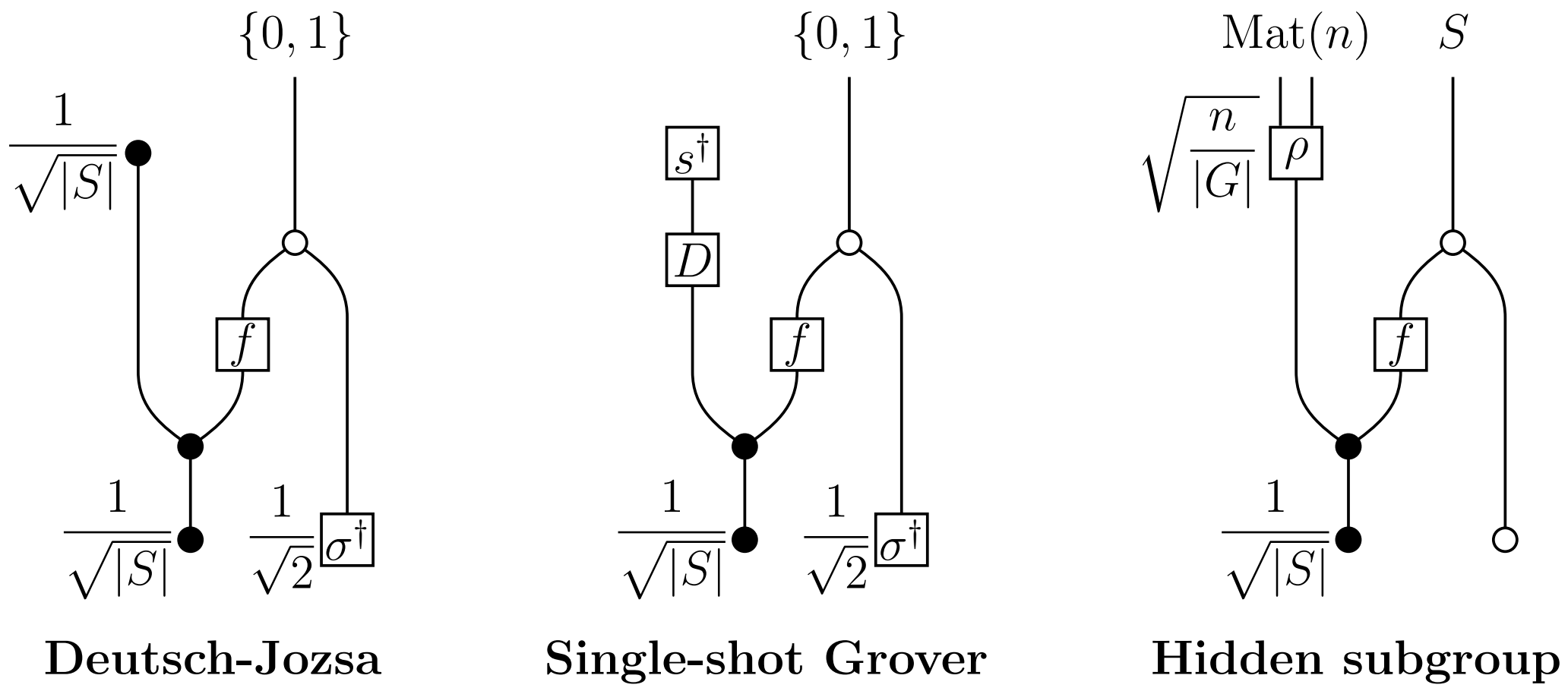

Figure 4.3: Three blackbox quantum algorithms presented as processes in a quantum-like process theory [104].

### 4.2.3 The Grover's and hidden subgroup algorithms

Through similar methods, the single-shot Grover's and hidden subgroup algorithms can be analyzed using categorical techniques [104]. These algorithms have the diagrammatic forms given in Figure 4.3. The single-shot Grover's generalization produces a variant where the indicator function maps into a group, rather than a set. When the target group is, for example, $\mathbb{Z}_3$ the indicator function has three marking "colors." Should the elements be marked by indicator function in a $4 : 1 : 1$ ratio, a single query can be used to return one of the rarer elements with certainty. An interesting path for the future development of this work would be to extend this to a generalization of amplitude amplification. We conjecture that in certain cases, where the input amplitudes follow a known ratio, a speedup over the usual amplitude amplification algorithm can be found.

The function of the quantum part of the hidden subgroup algorithm can also be demonstrated using QPT techniques. Vicary [104] abstractly verifies that the algorithm returns a uniform sampling of representations that factor through the hidden subgroup. In considering the structure of the hidden subgroup algorithm, we notice that the input system in the lower right is set to the unit for the $\circ$-classical structure. This unit, via strong complementarity, is equivalent to a uniform superposition over all possible input representations. This leads Vicary and the author to make the following conjecture:

**Conjecture 4.2.8.** *The performance of the hidden subgroup algorithm can be improved by a particular chosen distribution of input representation in the second system.*

Though in small examples it is easy to see the advantage, we have so far been unable to find a general rule beyond using a uniform distribution over representations excepting the trivial representation, which provides no information.

### 4.2.4 The group homomorphism identification algorithm

In this section we use the framework of QPTs to construct a new deterministic quantum algorithm to identify group homomorphisms.

**Definition 4.2.9** (Group homomorphism identification problem)**.** Given finite groups $G$ and $A$ where $A$ is abelian, and a blackbox function $f : G \to A$ that is promised to be a group homomorphism, identify the homomorphism $f$.

We demonstrate a quantum algorithm that solves the group homomorphism identification problem with a number of queries equal to the number of simple factors of the abelian group $A$.

Høyer [58] gives a bound on classical algorithms for a similar problem that we can easily extend.

**Lemma 4.2.10.** *Given finite groups $G$ and $A$, where $A$ is abelian and $G$ has a generating set of order $m$, and a blackbox function $f : G \to A$ that is promised to be a group homomorphism, a classical algorithm can determine $f$ if and only if we have made $m$ oracle queries.*

*Proof.* Once we have evaluated $f$ classically on the generating set of $G$, we have fully characterized $f$. The other direction is [58, Lem. 9] where $G$ is abelian. This extends to non-abelian $G$ as the pre-image of some homomorphism $f : G \to A$ for abelian $A$ and non-abelian $G$ must be an abelian subgroup $X \subseteq G$ (or $f$ is trivial). We can then consider the GROUPHOMID problem for abelian groups and $f : X \to A$. □

We will demonstrate that the query complexities of quantum and classical algorithms for this problem depend on different and unrelated parameters. Instances where the order of the generating set of $G$ is larger than the number of factors in the target group $A$ give a quantum advantage.

In the simpler case where $G$ is an abelian group this quantum algorithm was previously described by Høyer [58], though his algebraic presentation differs significantly from ours. Høyer also notes that the algorithm by Bernstein and Vazirani in [20] is an instance of the abelian group identification problem where $G = \mathbb{Z}_n^n$ and $A = \mathbb{Z}_2$. Independently, Cleve et. al. [29] also presented an algorithm for the abelian case where $G = \mathbb{Z}_2^n$ and $A = \mathbb{Z}_2^m$.

Our results give a new approach to the solution of the group homomorphism identification problem that both extends the existing results to the case where $G$ is non-abelian, and clearly connects the structure of our algorithm to that of other black-box quantum algorithms, such as the Deutsch-Jozsa and hidden subgroup algorithms.[13]

We will proceed using the abstract structure defined earlier, but will now work in the QPT **FHilb**. Recall that any choice of orthonormal basis for an object $A$ in **FHilb** endows it with a dagger-Frobenius algebra $(A, ⋏, ●, ⋎, ●)$, whose copying map $d : A \to A \otimes A$ is defined as the linear extension of $d(|i\rangle) = |i\rangle \otimes |i\rangle$. Any finite group $G$ induces a different dagger-Frobenius algebra on an object $A = \mathbb{C}[G]$, the Hilbert space with orthonormal basis given by the elements $G$, with multiplication given by linear extension of the group multiplication; we represent this structure as $(A, ⋏, ○, ⋎, ○)$. These two Frobenius algebras are strongly complementary.

Recall from Definition 4.1.37 that, for finite $G$, its representations can be characterized as the homomorphisms $G \xrightarrow{\rho} \mathrm{Mat}(n)$. The homomorphism conditions take the following form [104, Section A.7]:

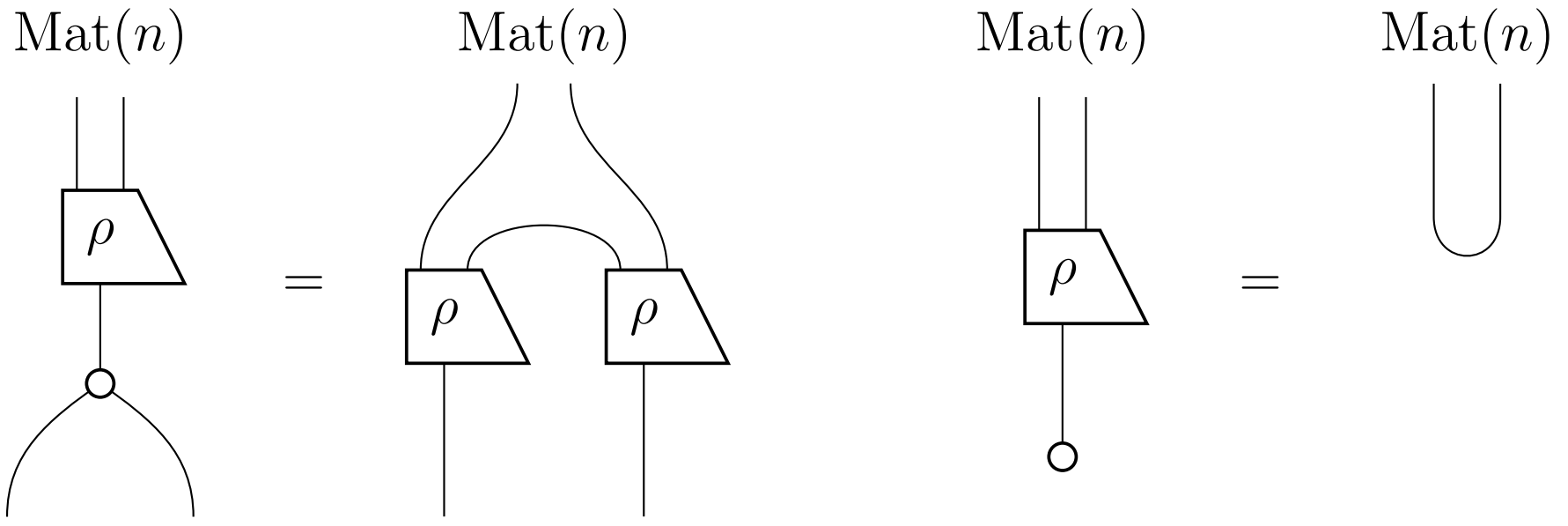

(4.83)

As we are currently focused on **FHilb**, the arrows delineating the difference between an object and its dual can be suppressed, as Hilbert spaces are isomorphic to their dual. These equations will be essential for our proofs below.

[13] This connection at the QPT level motivated our construction and we discovered Høyer's work afterwards.

### The algorithm

The structure of the quantum algorithm that solves the group homomorphism identification problem is given by the topological diagram (4.84) below. Here $\sigma : G \to \mathbb{C}$ is a normalized irreducible representation of $G$, representing the result of the measurement, and $\rho : A \to \mathbb{C}$ is a normalized irreducible representation of $A$. The representation $\rho$ is one-dimensional as $A$ is an abelian group. Physically, we are able to produce the input state $\rho$ efficiently, using $O(\log n)$ time steps, via the quantum Fourier transform for any finite abelian group [30]. The measurement result $\sigma$ arises from a measurement in the Fourier basis, which can, by a similar procedure for any finite group [25], also be implemented efficiently.

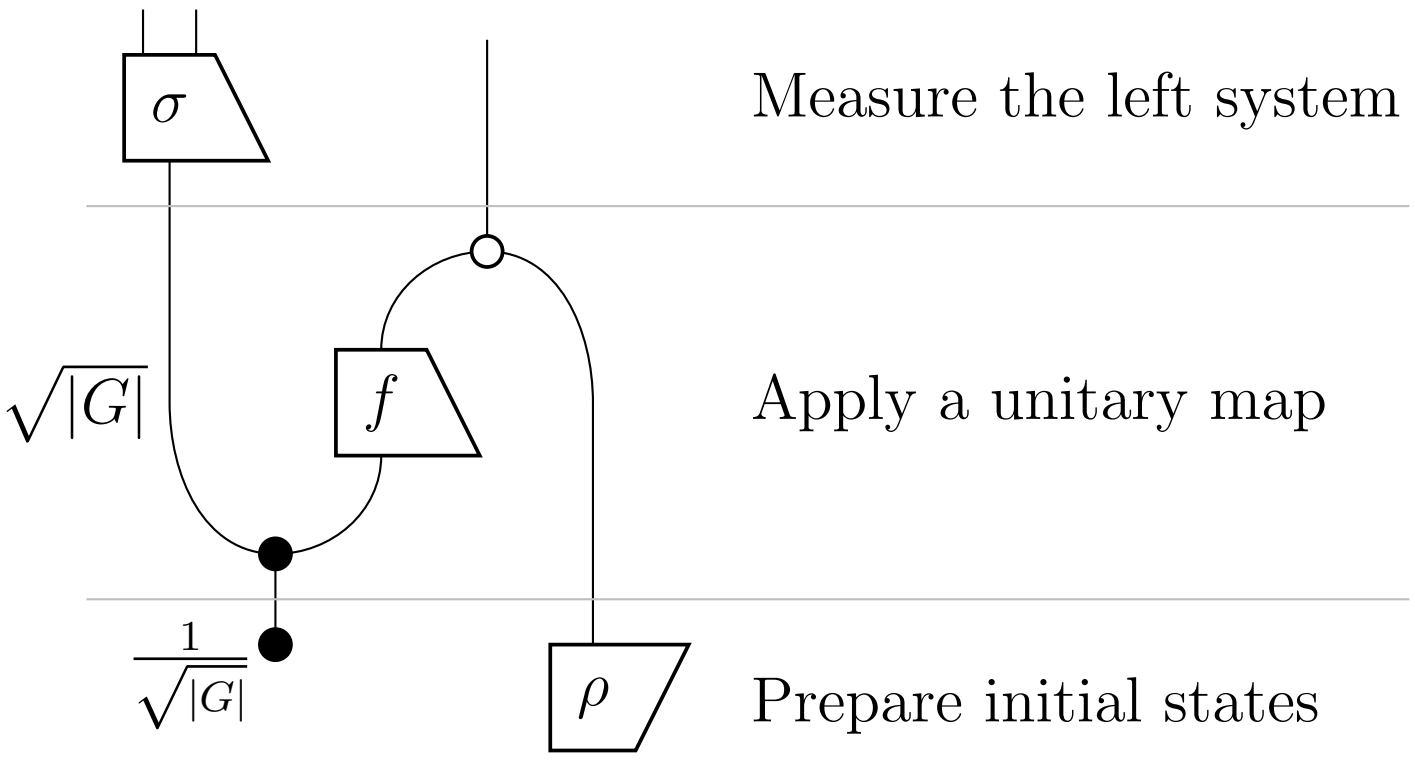

(4.84)

We can compare the structure of this algorithm to that of the standard quantum algorithm for the hidden subgroup problem. There, the second system is prepared in a state given by the identity element of the group, corresponding to a uniform linear combination of the irreducible representations. A later measurement of this second system—which is not a part of the standard hidden subgroup algorithm, but can be done without changing the result of the procedure—would collapse this combination to a classical mixture of these representations. The hidden subgroup algorithm therefore contains an amount of classical nondeterminism in its initial setup. In principle removing this, and selecting the input representation strategically, can only improve performance, and we take advantage of this here.

We analyze the effect of our new algorithm as follows.

**Lemma 4.2.11.** *The algorithm defined by* (4.84) *gives output* $\sigma$ *with probability given by the square norm of* $\sigma \circ f^* \circ \rho^*$.

*Proof.* Using that $\rho$ is a group homomorphism and simple diagrammatic rewrites defined in [104, Section A.9], we show the following, making use of the fact that representations are copyable points for group multiplication:

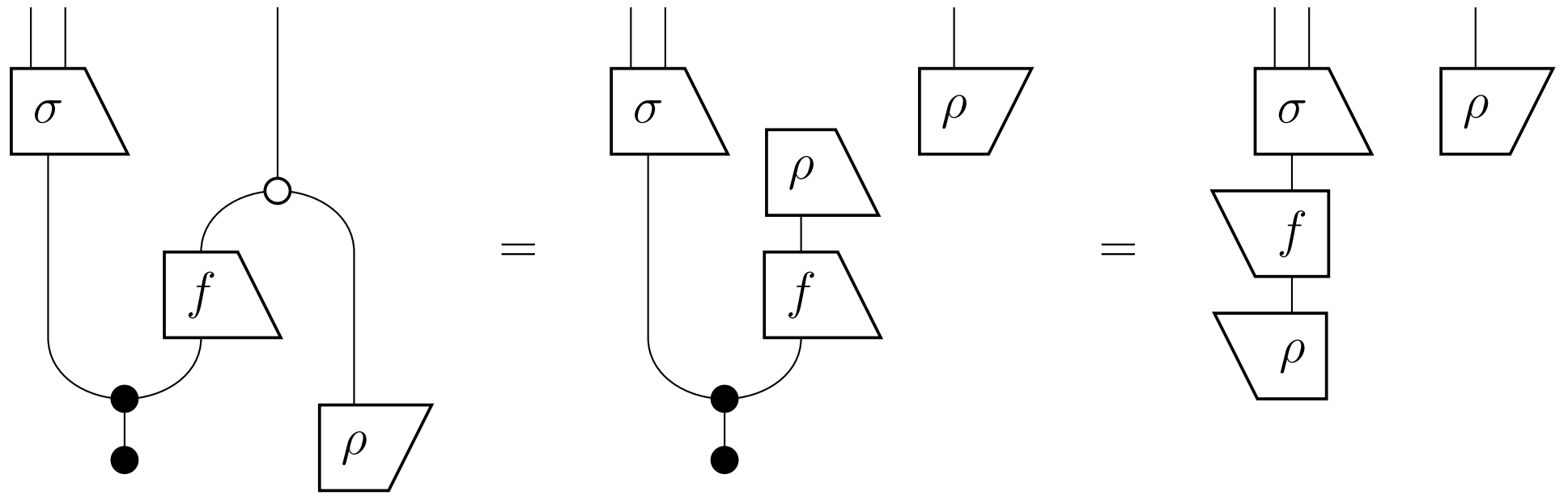

(4.85)

The left hand system is thus in the state $\sigma \circ f^* \circ \rho^*$, and using the Born rule, the squared norm of this state gives the probability of this experimental outcome. □

**Lemma 4.2.12.** *The composite $\rho \circ f$ is an irreducible representation of $G$.*

*Proof.* The map $f$ is a homomorphism, so $\rho \circ f : G \to \mathbb{C}$ is a one-dimensional representation of $G$. All one-dimensional representations are irreducible, so $\rho \circ f$ is an irreducible representation. □

**Lemma 4.2.13.** *One-dimensional representations are equivalent only if they are equal.*

*Proof.* Let $\rho_1, \rho_2 : G \to \mathbb{C}$ be irreducible representations of $G$. If they are isomorphic, then there exists a linear map $\mathcal{L} : \mathbb{C} \to \mathbb{C}$, i.e. some complex number, such that $\forall g \in G$

$$\mathcal{L}\rho_1(g) = \rho_2(g)\mathcal{L}.$$

Hence we see that $\forall g \in G$, $\rho_1(g) = \rho_2(g)$. □

**Theorem 4.2.14** (Structure theorem for finite abelian groups)**.** *Every finite abelian group is isomorphic to a direct product of cyclic groups of prime power order.*

*Proof.* See [9, Theorem 6.4] for a proof of this standard result. □

**Theorem 4.2.15.** *For a finite group $G$ and cyclic group of prime power order $\mathbb{Z}_{p^n}$, the algorithm* (4.84) *identifies a group homomorphism $f : G \to \mathbb{Z}_{p^n}$ in a single query.*

*Proof.* Choose the input representation $\rho$ to be the fundamental representation of $\mathbb{Z}_{p^n}$. This representation is faithful. This means exactly that

$$\rho \circ f = \rho \circ f' \qquad \Leftrightarrow \qquad f = f'.$$

Thus $\rho \circ f$ and $\rho \circ f'$ are different irreducible representations if and only if $f$ and $f'$ are different group homomorphisms. The single measurement on the state $(\rho \circ f)^*$ is performed by the algorithm in the representation basis of $G$, allowing us to determine $\rho \circ f$ up to isomorphism. Due to Lemma 4.2.13 we know that each equivalence class contains only one representative, and thus we can determine $f$ with a single query. □

**Theorem 4.2.16.** *For any two finite groups $G$ and $A$, where $A$ is abelian with $n$ simple factors, the quantum algorithm* (4.84) *can identify a group homomorphism $f : G \to A$ with $n$ oracle queries.*

*Proof.* We prove the result by induction.

**Base case.** When $A = \mathbb{Z}_{p^n}$ is simple, then by Theorem 4.2.15 we can identify the homomorphism with a single query.

**Inductive step.** If $A$ is not simple, then we must have $A = H_1 \times H_2$ by Theorem 4.2.14, where the following hold:

1. The product $\times$ is the direct product whose projectors $(p_1, p_2)$ are homomorphisms.
2. $H_1$ and $H_2$ are groups with $n_1$ and $n_2$ factors respectively such that the theorem holds, i.e. homomorphisms of the type $f_1 : G \to H_1$ and $f_2 : G \to H_2$ can be identified in $n_1$ and $n_2$ queries respectively.

Since $p_1 \circ f$ and $p_2 \circ f$ are homomorphisms, we can run subroutines of the algorithm to determine them. Hence we recover $f$ as

$$f(x) = ((p_1 \circ f)(x), (p_2 \circ f)(x)).$$

The first subroutine will require $n_1$ queries and the second will require $n_2$ queries, so the total number of queries will be $n_1 + n_2$, which is the number of factors of $H_1 \times H_2$. □

### Optimality

In this subsection we investigate the information theoretic bounds solutions to the GROUPHOMID algorithm.[14] While we are able to construct bounds on both the number of quantum and the number of classical queries required, these bounds turn out to be overtly weak lower bounds, that, in particular, do not show an asymptotic separation between quantum and classical. Recall, though, that the result of Lemma 4.2.10 does give a separation for our algorithm. Still, it is hoped that future work can build on these techniques to construct better bounds to analyze the optimality of our quantum algorithm.

[14]The author would like to thank Ronald de Wolf for suggesting this analysis.

We begin with a few useful group theoretic lemmas. Recall that the homset $\mathbf{Grp}(G, A)$ denotes the set of all homomorphisms between groups $G$ and $A$.

**Lemma 4.2.17.** *For $Z_{p^a}$, $Z_{p^b}$ cyclic groups of prime power order, the number of group homomorphisms between $Z_{p^a}$ and $Z_{p^b}$ is given by $p^{\min(a,b)}$.*

*Proof.* Consider representatives $x \in Z_{p^a}$ and $y \in Z_{p^b}$. For cyclic groups a homomorphism is exactly defined by its image on the generator of that group. We will now consider two cases.

For the first, where $a \geq b$, we are able to map the single generator of $Z_{p^a}$ to any element of $Z_{p^b}$ as it clear by Lagrange's theorem that the order of each $y$ divides $p^b$ which in this case also divides $p^a$. Thus there is one homomorphism for each $y$, i.e. $p^b$ homomorphisms.

For the second we have $a < b$. The only available images for the generator of $Z_{p^a}$ are those with orders that divide $p^a$, i.e. where $p^a/|y|$ is an integer. Since $|y| = |1^y| = \frac{p^b}{\gcd y, p^b}$ this condition requires

$$\frac{p^a}{|y|} = \gcd(y, p^b)p^{a-b}$$

to be an integer. We argue that necessary and sufficient condition for this is that $p^{b-a}$ divides $y$. Sufficient is obvious and necessary is shown by checking that $y/\gcd(y, p^b) \cdot \gcd(y, p^b)/p^{a-b}$ is an integer as it is a product of two integers. The number of such $y$ is $p^a$, with one homomorphism for each.

Taking these two cases together we find that the number of group homomorphisms is given by $p^{\min(a,b)}$. □

**Lemma 4.2.18.** *Let abelian groups $G$ and $A$ have structure theorem factorizations as $G \cong \mathbb{Z}_{p_1^{a_1}} \times ... \times \mathbb{Z}_{p_n^{a_n}}$ and $A \cong \mathbb{Z}_{p_1^{b_1}} \times ... \times \mathbb{Z}_{p_n^{b_n}}$ for prime $p_i$ and $a_i, b_i \in \mathbb{N} \cup \{0\}$. Write the set of factor sizes for $G$ as $G^f = \{p_i^{a_i}\}_{1 \leq i \leq n}$ and similarly define $A^f$. Let $\Lambda = G^f \cap A^f$. The largest possible number of homomorphisms between $G$ and $A$ is given by:*

$$|\mathbf{Grp}(G, A)| = \prod_{i=1}^{|\Lambda|} \Lambda_i \tag{4.86}$$

*Proof.* Recall that $|\mathbf{Grp}\left(\mathbb{Z}_{p_i^{a_i}}, \mathbb{Z}_{p_j^{b_i}}\right)| = 1$ when $p_i \neq p_j$. In that case we can only count the trivial homomorphism. Now, using their factorizations, we split the hom functor over the factors of the groups. This gives:

$$|\mathbf{Grp}(G, A)| = \prod_{i=1}^{n} \prod_{j=1}^{n} |\mathbf{Grp}(\mathbb{Z}_{p_i^{a_i}}, \mathbb{Z}_{p_j^{b_j}})| \tag{4.87}$$

$$= \prod_{i=1}^{n} |\mathbf{Grp}(\mathbb{Z}_{p_i^{a_i}}, \mathbb{Z}_{p_i^{b_i}})| \tag{4.88}$$

$$= \prod_{i=1}^{n} p_i^{\min(a_i, b_i)} \qquad \text{by Lemma 4.2.17} \tag{4.89}$$

Clearly this number will be maximized when $a_i = b_i$, which are exactly the factors in the intersection set $\Lambda$. □

This lemma allows us to consider several specific cases:

1. Case $G = A = \Lambda$. Then $|\mathbf{Grp}(G, A)| = |G| = |A|$.
2. Case $G = A \times A'$. Then $A^f \subseteq G^f \Rightarrow |\mathbf{Grp}(G, A)| = |A| \leq |G|$.
3. Case $A = G \times G'$. Then $G^f \subseteq A^f \Rightarrow |\mathbf{Grp}(G, A)| = |G| \leq |A|$.

We can attempt to use these cases to analyze the information theoretic lower bounds on classical and quantum queries for this algorithm. Such a lower bound is given by a ratio of the total information ($\log |\mathbf{Grp}(G, A)|$) and the maximum information that can be obtained per query. In the classical case we have access to $\log |G|$ bits, which gives a maximum bound on the information per query of $\log |G|$. In the quantum case, we have qubits for both the source and target groups, so the maximum information per query is $\log |G| + \log |A|$. We can now consider the lower bounds that emerge for each of the cases given above, where we write $\#CQ$ for the number of classical queries and $\#QQ$ for the number of quantum queries. These will, unfortunately, turn out not to be particularly enlightening lower bounds.

1. Case $G = A = \Lambda$. Thus $\#CQ \geq \log |G| / \log |G| = 1$ and $\#QQ \geq \log |G| / (2 \log |G|) = 1/2$.
2. Case $G = A \times A'$. Thus $\#CQ \geq \log |A| / \log |G| \leq 1$, as $A$ is smaller than $G$. Further $\#QQ \geq \log |A| / (\log |G| + \log |A|) \leq 1$.
3. Case $A = G \times G'$. Thus $\#CQ \geq \log |G| / \log |G| = 1$ and $\#QQ \geq \log |G| / (\log |G| + \log |A|) \leq 1$.

To conclude, it seems that despite the encouraging fact that the query complexities of the classical and quantum algorithms depend on seemingly independent variables, the structure of the promise space of group homomorphisms limits our ability to say something meaningful about the optimal quantum queries required.

Let $G$ have $m$ generators and $A$ have $n$ simple factors. Our best separation is then to use Lemma 4.2.10 and our algorithm to conclude that the classical query complexity is $\Theta(m)$ and the quantum query complexity is $\mathcal{O}(n)$.

### Extension to the non-abelian case

We now consider the more general case where the target group $A$ is non-abelian. We do not know how to extend the algorithm described above to this case. Nevertheless, it is instructive to analyze this scenario in our process theoretic approach.

Irreducible representations of a non-abelian group $A$ are not necessarily one dimensional, though we are still able to compute them via the Fourier transform efficiently [25]. In this case the algorithm has the following structure, where $\psi$ represents the initial state of the right-hand system in the representation space:

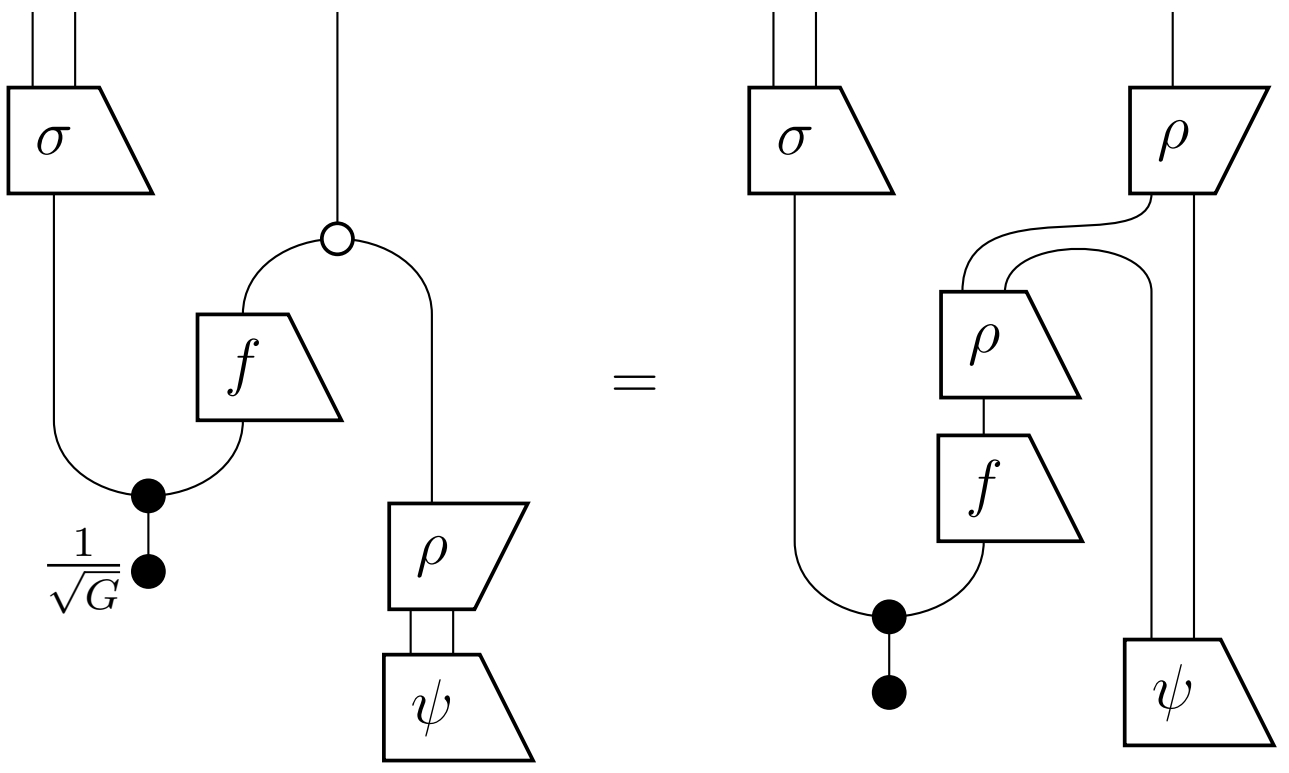

(4.90)

Recall the notation for multi-dimensional characters from Section 4.1. We notice two additional features in this case. First, it is clear that the left and right systems are no longer in a product state at the end of the protocol, as they were in the final diagram of (4.85). Second, we now have an additional choice when preparing the input representation $\rho$; in order to construct a state from a representation $\rho$ we also must choose the state $\psi$.

While this provides a clear description of the algorithm in this more general setting, it is not clear that it would identify homomorphisms into non-abelian groups. Complications include the lack of a structure theorem that satisfies the conditions for Theorem 4.2.16, and that Lemma 4.2.12 no longer applies. In this setting it may be useful to make the problem easier by restricting to the identification of homomorphisms up to *natural isomorphism*, i.e. where two homomorphisms $f_1, f_2 : G \to H$ are considered equivalent when there exists some $\eta \in H$ such that, for all $g \in G$, we have $\eta f_1(g)\eta^{-1} = f_2(g)$.

## 4.3 Models of quantum algorithms in sets and relations

In this section, we construct abstract models of blackbox quantum algorithms using a model of quantum computation in sets and relations, a setting that is usually considered for nondeterministic classical computation. This alternative model of quantum computation (QCRel), though unphysical, nevertheless faithfully models its computational structure. Our main results are models of the Deutsch-Jozsa, single-shot Grover's, and GroupHomID algorithms in QCRel. These results provide new tools to analyze the protocols from quantum computation and improve our understanding of the relationship between computational speedups and the structure of physical theories. They also exemplify a method of extending physical/computational intuition into new mathematical settings.

### 4.3.1 Introduction

Having grasped the abstract structure at play in the protocols and algorithms of quantum computation, we can conceive of modelling quantum computation in QPTs other than Hilbert spaces and linear maps. There are two main thrusts that make this investigation interesting. The first is to further analyze the structure of quantum computation, focusing on the relationship between computational speedups and the structure of physical theories. We use the QCRel model defined here to analyze some example quantum algorithms as non-deterministic classical algorithms while preserving their query-complexity (and, in fact, all their abstract structure). The second thrust regards the insights that become available by extending physical/computational intuition into new areas of mathematics. While other toy models of a relational flavor for quantum mechanics have been proposed by several authors [43, 53, 96, 100], and some even discuss protocols [59], these works have not developed the structures necessary to model quantum algorithms.

First we construct our chosen model of quantum information. This is the QPT in **FRel**, rather than **FHilb**, and it is introduced by rephrasing the axioms of quantum mechanics. Next we present our novel contributions: relational models of unitary oracles, the Deutsch-Jozsa algorithm, the single-shot Grover's algorithm, and the group homomorphism identification algorithm.

### 4.3.2 The model of quantum computation in relations

We begin with definitions of the key components of quantum computation in this new setting, e.g. systems, states, bases, observables, etc. The following definitions are motivated by Chapter 3, whose general theorems prove useful. To avoid distracting repetition of notation, we use generic terminology to refer to the relational setting within this section. For example **system** is intended to mean **relational system**, i.e. a set. When we wish to refer to the quantum setting we explicitly denote this e.g. **quantum system** refers to a finite dimensional Hilbert space.

**Axiom 4.3.1.** *A **system** is a set $H$ with **states** $|\psi\rangle$ given by subsets $\psi \subseteq H$.*

Each state in our notation is a boolean column vector written as a labelled ket, to follow the convention in quantum mechanics where states are complex valued column vectors as in the following example.

**Example 4.3.2.** Consider a three element system $\{0, 1, 2\}$, the relation $R = \{(0, 0), (0, 2), (1, 1)\}$ and the state $|\psi\rangle = \{0\}$. In terms of boolean matrices and vectors the composition $R \circ |\psi\rangle$ is written as:

$$\begin{pmatrix} 1 & 0 & 0 \\ 0 & 1 & 0 \\ 1 & 0 & 0 \end{pmatrix} \begin{pmatrix} 1 \\ 0 \\ 0 \end{pmatrix} = \begin{pmatrix} 1 \\ 0 \\ 1 \end{pmatrix}. \tag{4.91}$$

The state $|\psi \vee \phi\rangle$ has elements in the union of sets $\psi$ and $\phi$. We often use $|\psi\rangle$ to mean the relation $\{\star\} \to H$ that relates the singleton set to all the elements in $\psi$.

**Axiom 4.3.3.** *A **composite system** $H$ of $n$ subsystems is given by the Cartesian product so that $H = H_1 \times ... \times H_n$. **Composite states** are any subset of $H$.*

**Definition 4.3.4.** For a relation $R : A \to B$ from set $A$ to $B$, the **converse relation** is denoted $R^{-1} : B \to A$ where for $x \in A$ and $y \in B$, $xRy$ if and only if $yR^{-1}x$.

The converse replaces the †-adjoint in quantum mechanics. This leads to:

**Definition 4.3.5.** A relation $R : H_1 \to H_2$ is **unitary** if and only if $R \circ R^{-1} = \mathrm{id}_{H_1}$ and $R^{-1} \circ R = \mathrm{id}_{H_2}$, where $Q \circ R$ means $Q$ after $R$.

This is the relational analog to the usual unitarity of linear maps in quantum mechanics and has an obvious interpretation:

**Theorem 4.3.6.** *Relations are unitary if and only if they are bijections.*

**Axiom 4.3.7.** ***Evolution*** *of systems is given by unitary relations.*

This means that states of system $A$ can evolve to states of system $B$ if and only if there is a bijection between them. Note that this implies that there do not exist physical evolutions between systems of different cardinality.

**Definition 4.3.8.** For a state $|\psi\rangle : \{\star\} \to H$, denote its relational converse as $\langle\psi| : H \to \{\star\}$ called its **effect**.

A state preparation followed by an effect amounts to an experiment with a post-selected outcome. Effects are maps to $\{\star\}$ that return whether the outcome state $|\psi\rangle$ is possible. We give an example to illustrate:

**Example 4.3.9.** The preparation of the state $|\phi\rangle$ followed by a post-selected measurement of the effect $\langle\psi|$ is given by the relation

$$\langle\psi|\phi\rangle := \langle\psi| \circ |\phi\rangle : \{\star\} \to H \to \{\star\}$$

This is either the identity relation that we interpret to mean a measurement outcome of $\langle\psi|$ is *possible*, or it is the empty relation that we interpret to mean the measurement outcome $\langle\psi|$ is *impossible*. It is clear that the outcome $\langle\psi|$ is possible if there exists some element of $H$ in both $\psi$ and $\phi$. Otherwise it is impossible. In this sense our relational quantum computation is a deterministic model of quantum computation.

This interpretation allows us to define a generalized version of the Born rule[15] to describe measurement in our model.

**Axiom 4.3.10** (Relational Born Rule)**.** *The possibility of measuring the state $|\psi\rangle$, having prepared state $|\phi\rangle$, is given by the image of:*

$$\langle\psi|\phi\rangle : \{\star\} \to \{\star\} \tag{4.92}$$

In the relational model, bases are characterized as particular generalizations of groups known as *groupoids* [57, 85]. Groupoids can be viewed as groups where multiplication is relaxed to be a partial function.

[15]In quantum theory, the Born rule gives the probability of measuring the outcome state $|\psi\rangle$ following preparation in state $|\phi\rangle$ as $|\langle\psi|\phi\rangle|^2$ where $\langle\psi|\phi\rangle : \mathbb{C} \to \mathbb{C}$ is the inner product of the two state vectors [84]. The QPT generalization is given in Definition 3.2.5.

**Definition 4.3.11.** For a system $H$, a **basis** $Z$ is a direct sum (disjoint union) of abelian groups $Z = G_0 \oplus G_1 \oplus ...$ where $|Z| = |H|$. Multiplication with respect to this list of groups will be written as $\bullet_Z$ and is defined in the following way. For elements $x, y \in Z$ such that $x \in G_i$ and $y \in G_j$ we have the partial function:

$$x \bullet_Z y = \begin{cases} x +_{G_i} y & i = j \\ \text{undefined} & \text{otherwise} \end{cases} \tag{4.93}$$

This makes $Z$ an **abelian groupoid** with groupoid multiplication $\bullet_Z$.

We will sometimes take a categorical perspective on groupoids. A groupoid $Z = \bigoplus^N G_i$ made up $N$ groups is a category whose set of objects is isomorphic to the set of groups $\{G_i\}$ and whose morphisms are elements of $Z$, e.g. $x \in Z$ such that $x \in G_1$ is a morphism $x : G_1 \to G_1$.

At first guess, one might be motivated by the intuition that a basis for a system breaks it up into parts, and so a basis would be a partition of $H$. This is not a bad start, however, bases have additional structure: namely that we can copy, delete and combine them at will. This idea is used to motivate Definition 4.3.11 by abstracting bases to called classical structures (Definition 3.3.5).

Recall that classical structures' properties, allowing the copying, deleting, and combining that accompany classical (as opposed to quantum) information, give them this name. We interpret them in the relational model of quantum computation through the following theorem:

**Lemma 4.3.12** ([44, 85])**.** *The classical structures in the category of sets and relations are exactly the abelian groupoids.*[16]

### Complementarity

Complementary bases are important features of quantum theory. In the general setting, complementary bases are understood as mutually unbiased bases in a certain sense (Definition 3.5.1). In relations Evans et al. gives a more direct characterization:[17]

**Theorem 4.3.13** ([44])**.** *Two bases $Z$ and $X$ are complementary if and only if they are of the following form. Basis $Z = \bigoplus^{|H|} G$ and basis $X = \bigoplus^{|G|} H$ given by copies of abelian groups $G$ and $H$ respectively.*

[16] In [56] this connection is extended to the non-abelian case where it is shown that all relative Frobenius algebras are groupoids.

[17] Theorem 4.3.13 holds as long as we consider bases to be the same if their lists of groups are isomorphic.

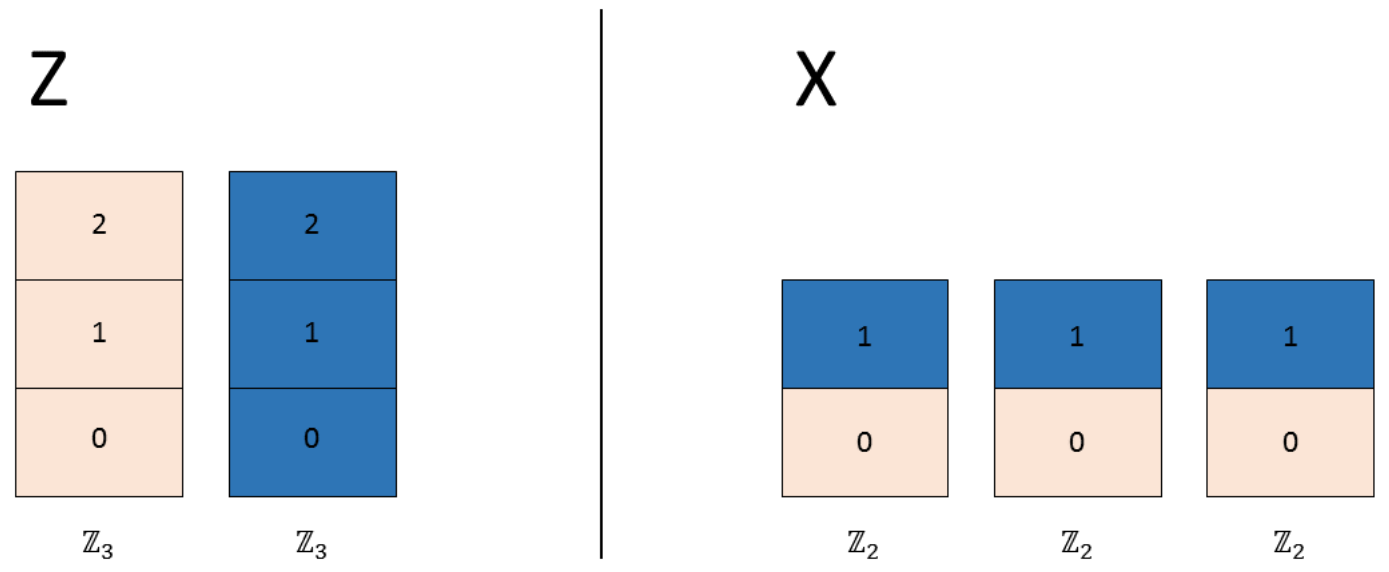

Figure 4.4: An example of two complementary bases on the system of six elements. Here $Z = \mathbb{Z}_3 \oplus \mathbb{Z}_3$ and $X = \mathbb{Z}_2 \oplus \mathbb{Z}_2 \oplus \mathbb{Z}_2$. The two classical states of $Z$ are each three element subsets and are colored in pink and blue. The unbiased states of $X$ to which they correspond are colored to match.

This theorem follows from the requirement that the classical states of one basis must be isomorphic to the unbiased states of its complement. We will return to this idea in the Section 4.1.1 when we address the quantum Fourier transform in relations. Classical and unbiased states of bases in the relational model are specified by Evans et al. [44] in the following corollaries that again correspond to abstract definitions from Chapter 3. An example on the six element system is illustrated with Figure 4.4.

**Corollary 4.3.14** ([44])**.** *The* ***classical states*** *of a basis $Z = \bigoplus^N G$ are the subsets corresponding to the groups $G_0, G_1, \ldots$ where we forget the group structure. They will often be denoted $|Z_i\rangle$.*

**Corollary 4.3.15** ([44])**.** *The* ***unbiased states*** *for a basis $Z = \bigoplus^N G$ are subsets $U$ such that for a fixed $g \in G$, $|U\rangle = \bigoplus^N \{g\}$. Thus there is exactly one element in each unbiased $U$ from each component $G_i$ of $Z$.*

**Example 4.3.16.** Take $Z = \mathbb{Z}_2 \oplus \mathbb{Z}_2 = \{0_a, 1_a, 0_b, 1_b\}$. The classical states of $Z$ are $|G_a\rangle = |0_a \vee 1_a\rangle$ and $|G_b\rangle = |0_b \vee 1_b\rangle$. The unbiased states of $Z$ are $|U_0\rangle = |0_a \vee 0_b\rangle$ and $|U_1\rangle = |1_a \vee 1_b\rangle$.

It is easy to check that bases as specified by Theorem 4.3.13 have the property that each classical state $|Z_i\rangle$ of the basis $Z$ corresponds to one unbiased state of $X$ and vice versa. This allows us to call these bases mutually unbiased, i.e. complementary [44].

### Phases

Phases are also defined in this relational setting. In Hilbert space quantum mechanics a quantum phase for an n-dimensional system is given by the vector

$$\begin{pmatrix} e^{i\phi_1} \\ \vdots \\ e^{i\phi_n} \end{pmatrix}.$$

These quantum phases form an abelian group and can be applied as phase gates. Their relational counterparts are described by the following lemma from [57]:

**Lemma 4.3.17.** *For a basis $Z = \bigoplus_i^N G_i$, the* ***phase group*** *$B(Z)$ is given by $\prod_i^N G_i$.*

**Example 4.3.18.** Consider the basis $\mathbb{Z}_2 \oplus \mathbb{Z}_2$ for the four element system $\{00, 01, 10, 11\}$. Let $|\psi\rangle$ be the state $|00 \vee 10\rangle$. Application of the phase 11 results in

$$11|00 \vee 10\rangle = |11 \vee 01\rangle.$$

We are also able to interpret GHZ states and density matrices in sets and relations.

**Definition 4.3.19.** For a basis $Z = \bigoplus_i^N G_i$, a **GHZ** state is given by

$$GHZ_Z \ := \{ \ (a, b, c) \mid \ \forall \, a, b, c \in Z, \ a \bullet_Z b \bullet_Z c = \mathrm{id}_{G_i} \text{ for some } i \ \}.$$

**Definition 4.3.20.** For a state $|\psi\rangle$, the **density matrix** $|\psi\rangle\langle\psi|$ is given by the relation $xRy$ s.t. $x, y \in \psi$.

### The Model QCRel

**Definition 4.3.21.** Axioms 1-4, and subsequent definitions, specify the quantum-like process theory for **quantum computation in relations: QCRel**.

By the QPT construction, this clearly makes QCRel a model of quantum computation in sets and unitary relations. It is worth noting that QCRel can be simply viewed as a local hidden variable theory. Consider the set $H$ to be the set of ontic states such that for $\phi \subseteq H$ the state $|\phi\rangle$ is non-deterministically in any of the ontic states in the subset $\phi$. This perspective on **Rel**, discussed further by Abramsky and Heunen [8], shows that QCRel provides a non-deterministic local hidden variable model for computational aspects of quantum mechanics. This means that protocols exist for entanglement, teleportation, and, as we show in this section, some familiar blackbox algorithms.

### 4.3.3 Unitary Oracles

In order to model blackbox quantum algorithms in this setting, we must define the oracles themselves. We do this by building up from an abstract definition of the controlled-not gate. Let the gray classical structure on a system $A$ be given by a basis $Z = \bigoplus^{|H|} G$ and the white classical structure be a basis $X = \bigoplus^{|G|} H$. The comonoid for the gray dot is then the relation 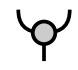 $: A \to A \times A$ that for $x, a, b \in H$ is given by

$$\{(x, (a, b)) \mid a \bullet_Z b = x\}.$$

**Definition 4.3.22.** The abstract controlled-not is given by a composition of the comonoid for Z and the monoid for X:

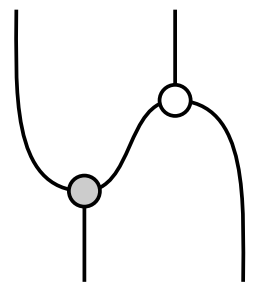

$$\begin{aligned}&\text{CNOT: } H \times H \to H \times H :: \\ &\{((x, y), (a, b \circ_X y)) \mid a \bullet_Z b = x\}.\end{aligned} \tag{4.94}$$

Example 3.5.5 showed that in the traditional quantum setting of Hilbert spaces and linear maps, this exactly corresponds to the usual controlled-not. Recall that Theorem 4.2.3 showed that two classical structures are complementary if and only if their corresponding controlled-not is unitary. This leads to the following corollary for complementary bases in QCRel:

**Corollary 4.3.23.** *Two bases (Z and X) in QCRel are complementary, in the sense of Theorem 4.3.13, if and only if the relation in* (4.94) *is a bijection.*

*Proof.* The relevant relation can clearly be seen to be the composite in Definition 4.3.22 as:

$$\{((a, b, y), (a, b \circ_X y))\} \circ \{((x, y), (a, b, y)) \mid a \bullet_Z b = x\}. \tag{4.95}$$

Thus the abstract proof of Theorem 4.2.3 goes through unchanged. □

An oracle is then introduced as a controlled-not where we have embedded a particular kind of relation that abstractly must be a self-conjugate comonoid homomorphism. We construct such relations in the following lemmas.

**Definition 4.3.24.** Let $G$ and $H$ be groupoids with with groupoid multiplications $\bullet_G$ and $\bullet_H$ respectively. Let $\mathrm{id}_G = \bigcup_{X \in \mathrm{Ob}(G)} \mathrm{id}_X$ and similarly define $\mathrm{id}_H$. A **groupoid homomorphism relation** $R : G \to H$ obeys the following condition for $g_1, g_2 \in G$:

$$R(g_1 \bullet_G g_2) = R(g_1) \bullet_H R(g_2) \tag{4.96}$$

Note that while this in many ways resembles a groupoid homomorphism, it is actually a weakening of this notion, in that groupoid homomorphism relations are not required to be total functions and have no explicit requirement on their identity morphisms. From another perspective, this definition is the groupoid generalization of many-to-many group homomorphisms [107].

**Definition 4.3.25.** A **monoid homomorphism relation** is a monoid homomorphism in **Rel**. Specifically, let $A$ and $B$ be sets equipped with monoids $(A, \overset{\circ}{\wedge}, \overset{|}{\circ})$ and $(B, \overset{\bullet}{\wedge}, \overset{|}{\bullet})$ respectively. A relation $r : A \to B$ is a monoid homomorphism when is obeys the following two conditions:

$$r \circ \overset{\circ}{\wedge} = \overset{\bullet}{\wedge} \circ (r \times r) \tag{4.97}$$

$$r \circ \overset{|}{\circ} = \overset{|}{\bullet} \tag{4.98}$$

A **comonoid homomorphism relation** is defined similarly, using duals of the above conditions.

**Lemma 4.3.26.** *A groupoid homomorphism relation that is surjective on objects is a monoid homomorphism relation.*

*Proof.* Throughout this proof we refer to a groupoid as a category where the elements of the groupoid are the morphisms. From this perspective a group is a groupoid with a single object. Consider a groupoid homomorphism relation $R : G \to H$ on objects $X, A, B$ of $G$ and morphisms $f$ of $G$. In order to show that $R$ is a monoid homomorphism relation we first show that it preserves the unit (4.98). We have $R(\bigcup_{X \in \mathrm{Ob}(G)} \mathrm{id}_X) = \bigcup_{Y \in \mathrm{Ob}(H)} \mathrm{id}_Y$. Recall that for a set $A$, $R(A) = \bigcup_{a \in A} R(a)$. It is that case that

$$R(\bigcup_{X \in \mathrm{Ob}(G)} \mathrm{id}_X) = \bigcup_{X \in \mathrm{Ob}(G)} R(\mathrm{id}_X) = \bigcup_{X \in \mathrm{Ob}(G)} \mathrm{id}_{R(X)} \quad \text{def. of group hom. rel.} \tag{4.99}$$

$$= \bigcup_{R(X) \in \mathrm{Ob}(G)} \mathrm{id}_{R(X)} = \bigcup_{Y \in \mathrm{Ob}(H)} \mathrm{id}_Y \quad \text{surjective on objects} \tag{4.100}$$

where we have used the fact that $R$ is surjective on the groupoid objects, which implies that every object of $H$ is in the image of $R$ and that $|\mathrm{Ob}(G)| \geq |\mathrm{Ob}(H)|$.

The second monoid homomorphism condition (4.97) is to preserve multiplication, i.e. that for subsets $K$ and $J$ of $G$ we have

$$R(K +_G J) = R(K) +_H R(J). \tag{4.101}$$

Here we recall that for two sets $A$ and $B$, $A + B = \{a + b | a \in A \text{ and } b \in B\}$. Thus,

$$R(K +_G J) = R(\bigcup_{k \in K, j \in J} k +_G j) = \bigcup_{k \in K, j \in J} R(k +_G j) \tag{4.102}$$

$$= \bigcup_{k \in K, j \in J} R(k) +_H R(j) \qquad \text{def. of group hom. rel.} \tag{4.103}$$

$$= R(K) +_H R(J). \tag{4.104}$$

This completes the proof. □

We then dualize the proof of Lemma 4.3.26 to conclude that:

**Lemma 4.3.27.** *Let $F : H \to G$ be a functor such that $F^{op}$ is a groupoid homomorphism relation that is surjective on objects. $F$ is a comonoid homomorphism relation.*

We call these comonoid homomorphism relations **classical relations**. These are relations that properly preserve the structure of the bases where classical data is embedded. In the quantum case they take basis elements to basis elements. Some examples in QCRel are listed in Appendix A. In order to define unitary oracles, we also need these relations to be self-conjugate (Definition 4.2.1). Luckily, this is always the case in relations:

**Lemma 4.3.28.** *All classical relations $f : Z^A \to Z^B$ between groupoids $Z^A = \bigoplus^N G$ and $Z^B = \bigoplus^{N'} H$ are self-conjugate.*

*Proof.* In QCRel, our dagger-Frobenius structures are groupoids and, if they are complementary to some other groupoid, then they are of the form $Z^A = \bigoplus^N G$ and $Z^B = \bigoplus^{N'} H$. We annotate the definition of self-conjugacy for some arbitrary element $(g, n)$, the element $g$ from the $n$-th group. Recall that $f^\dagger = f^{-1}$ in QCRel.

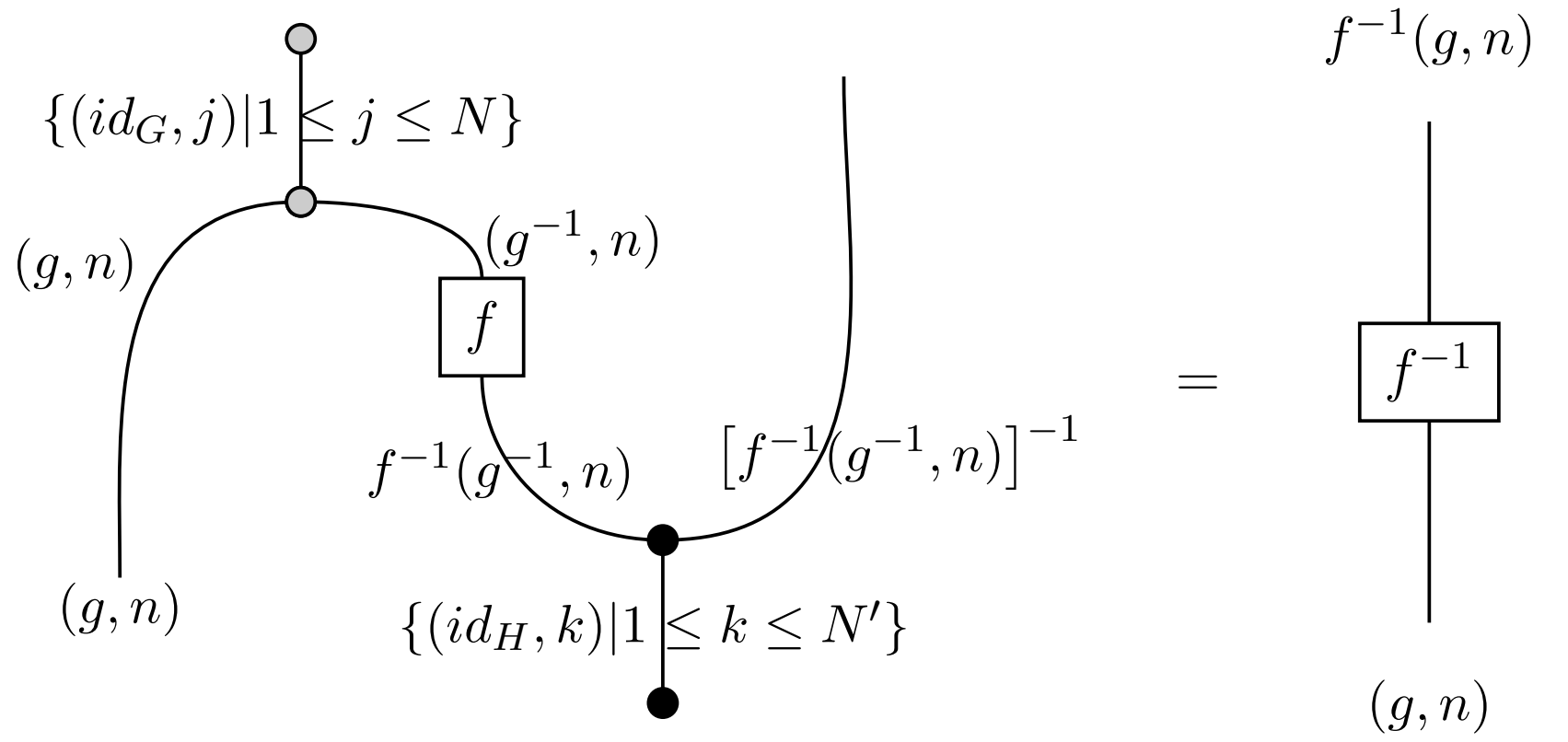

(4.105)

Thus, a relation $f$ is self-conjugate if and only if for all elements $(g, n)$ it is the case that $[f^{-1}(g^{-1}, n)]^{-1} = f^{-1}(g, n)$. From Lemma 4.3.27 the converse of the classical relation $f$ is a monoid homomorphism relation whose multiplication is the groupoid operation, so this condition will hold. □

Classical relations, as self-conjugate comonoid homomorphisms, lead to unitary oracles.

**Definition 4.3.29** (Relational Oracle)**.** Given a groupoid $Z^A : (A, \curlywedge^{\bullet}, \bullet)$, a pair of complementary groupoids $Z^B : (B, \curlywedge^{\circ}, \circ)$ and $X^B : (B, \curlywedge^{\circ}, \circ)$, and a classical relation $R : (A, \curlywedge^{\bullet}, \bullet) \to (B, \curlywedge^{\circ}, \circ)$, an **oracle** is defined to be the following endomorphism of $A \times B$:

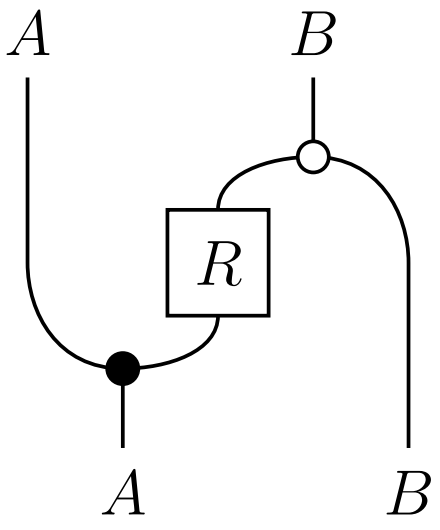

$$\begin{aligned} &\text{OracleRel(R)} : A \times B \to A \times B :: \\ &\{((x, y), (a, c \circ_X y)) \mid \\ &\qquad \exists b \in A, s.t.\ a \bullet_{Z^A} b = x \ and\ bRc\}. \end{aligned} \tag{4.106}$$

**Theorem 4.3.30.** *Oracles are unitary.*

*Proof.* Proved in the abstract setting by Theorem 4.2.3, when $R$ is a self-conjugate comonoid homomorphism. Though there are others, classical relations $R$ are necessary and sufficient in our cases as the algorithms that follow additionally require that the comonoids be part of classical structures. □

**Corollary 4.3.31.** *OracleRel is a bijection.*

*Proof.* This follows directly from Theorem 4.3.30 and Theoremm 4.3.6. □

## 4.3.4 The Fourier transform in relations

In Section 4.1.4, we saw that there are several perspectives on the Fourier transform in **FRel**, only some of which are nontrivial. In this section we take the operational perspective on the generalized quantum Fourier transform whose definition is motivated through the relationship between classical and unbiased states of two bases. For abelian groups $G$ and $H$, consider two groupoids $Z = \bigoplus^{|H|} G$ and $X = \bigoplus^{|G|} H$ to be complementary bases of the same system.

**Definition 4.3.32.** The **quantum Fourier transform in relations** corresponds to preparing classical states of $Z$ and measuring them against classical states of $X$. is an isomorphism from the classical states of $Z$ to the unbiased states of $X$, i.e.

$$\{G_h\} \mapsto \{h_g | \forall g \in G\}$$

**Example 4.3.33.** Take $G = \mathbb{Z}_2 = \{0,1\}$, $H = \mathbb{Z}_1 = \{\star\}$, $Z = G$ and $X = H \oplus H = \{(\star,0),(\star,1)\}$. The computational basis is the family $|H_g\rangle_{g\in G}$ of classical states for $X$, i.e. $H_0 = \{(\star,0)\}$ and $H_1 = \{(\star,1)\}$. The quantum Fourier basis is a single classical state $G_\star = \{(\star,0),(\star,1)\}$ for $Z$. In this case all states can be prepared in the computational basis, but measurement in the quantum Fourier basis is trivial.

**Example 4.3.34.** Take $G = \mathbb{Z}_2 = \{0,1\}$, $H = \mathbb{Z}_2 = \{a,b\}$, $Z = G \oplus G = \{(0,0),(1,0),(0,1),(1,0)\}$ and $X = H \oplus H = \{(a,a),(b,a),(a,b),(b,b)\}$. The computational basis is the family $|H_g\rangle_{g\in G}$ of classical states for $X$, i.e. $H_0 = \{(a,a),(b,a)\}$ and $H_1 = \{(a,b),(b,b)\}$. The quantum Fourier basis is the family $|G_h\rangle_{h\in H}$ of classical states for $Z$, i.e. $G_a = \{(0,0),(0,1)\}$ and $G_b = \{(1,0),(1,1)\}$.

See Section 4.1 and [46] to fully motivate this definition of the Fourier transform in QCRel and for its relationship to the usual Hadamard and Fourier transforms for Hilbert spaces and linear maps.

### 4.3.5 The Deutsch-Jozsa algorithm in QCRel

The well known Deutsch-Jozsa algorithm is an early quantum algorithm that demonstrates a speedup over exact classical computation [40]. It takes as input a function promised to be either constant or balanced and returns which, deterministically using only a single oracle query. In this section, we model the algorithm's steps in QCRel just as it is implemented with Hilbert spaces and linear maps. This approach is somewhat dual to the usual one where different algorithms are compared on the same problem. Here we run the abstract protocol from Section 4.2.2 (implemented in a different model) with the same query complexity and compare the different problems that it solves.

To run this algorithm in QCRel we use two systems. System $A$ has cardinality $n$ and system $B$ has cardinality $\geq 2$. Take $Z^A = \bigoplus^{|H^A|} G^A$ and $X^A = \bigoplus^{|G^A|} H^A$ to be complementary bases of $A$. Take $Z^B = \bigoplus^{|H^B|} G^B$ and $X^B = \bigoplus^{|G^B|} H^B$ to be complementary bases of $B$, such that $X^B$ has at least two classical states. In analogy with the usual specification, the algorithm proceeds with the following steps.

1. Prepare $A$ in the zero state $|Z_0^A\rangle$. Prepare $B$ in the state given by the second classical state of $Z^B$, i.e. $|Z_1^B\rangle$.

2. Apply the Fourier transform, as given by Definition 4.3.32, to each system, resulting in states $|X_0^A\rangle$ and $|X_1^B\rangle$ respectively.

3. Apply an oracle (4.106), built from a classical relation $f : Z^A \to Z^B$.

4. Again apply the Fourier transform to system $A$ and then measure it in the $Z$ basis.

This sequence of steps is an instance in sets and relations of the abstract Deutsch-Jozsa algorithm from [104], which translates to the following relation, where we have already applied the Fourier map to the input and output. See (4.80).

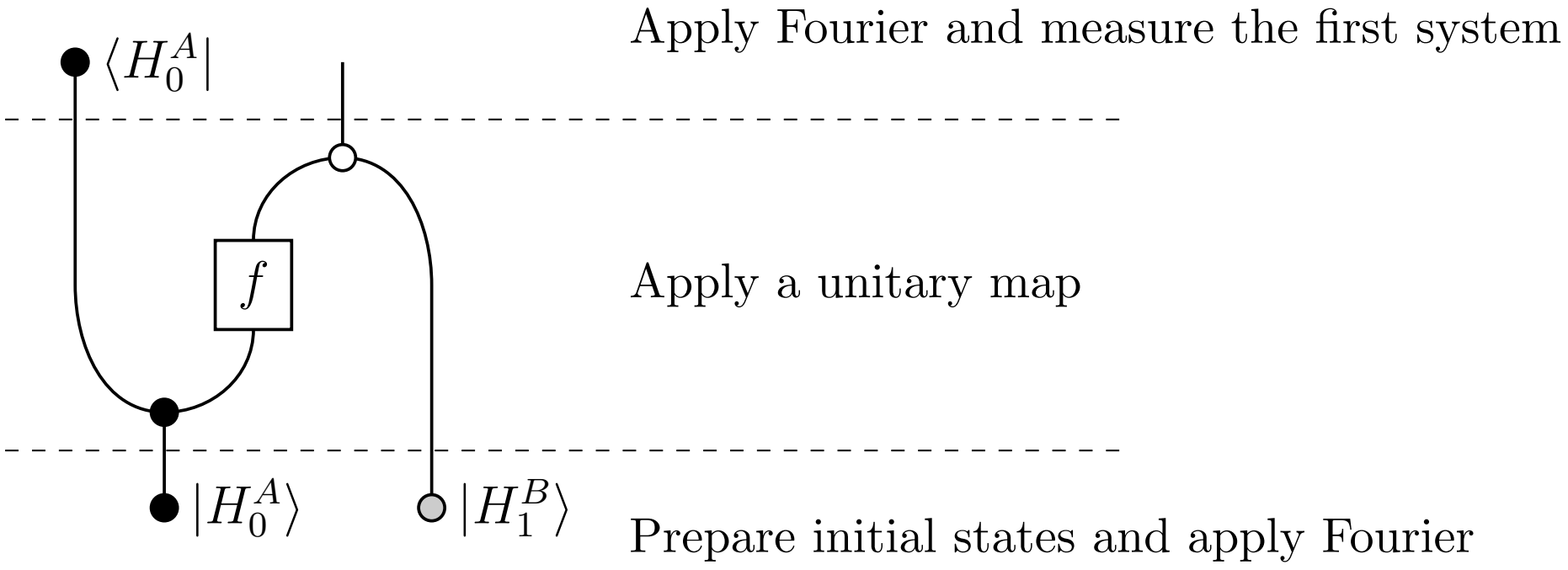

(4.107)

that is explicitly written as:

$$
\begin{aligned}
\mathrm{DJAlg}(f) &:: \{\star\} \times \{\star\} \to \{\star\} \times B \\
&= \langle H_0^A| \times \mathrm{id}_B \circ \mathrm{OracleRel}(f) \circ |H_0^A\rangle \times |H_1^B\rangle \\
&= \{((\star,\star),(\star,z)) \mid z \in H_1^B \text{ and } \exists y \in H_0^A, \text{ s.t. } yfz\},
\end{aligned}
$$

Note the correspondence between (4.107) and (4.80). Thus by the derivation in Section 4.2.2 we have:

**Theorem 4.3.35.** *In any dagger compact category with complementary bases, the algorithm in Equation 4.107 will, with a single oracle query, distinguish* ***constant*** *and* ***balanced*** *classical relations $f : Z^A \to Z^B$ according to the following abstract definitions. Here $|x\rangle$ is a classical state of $Z^A$ and the zero scalar $0$ is, in* **Rel**, *the empty relation:*

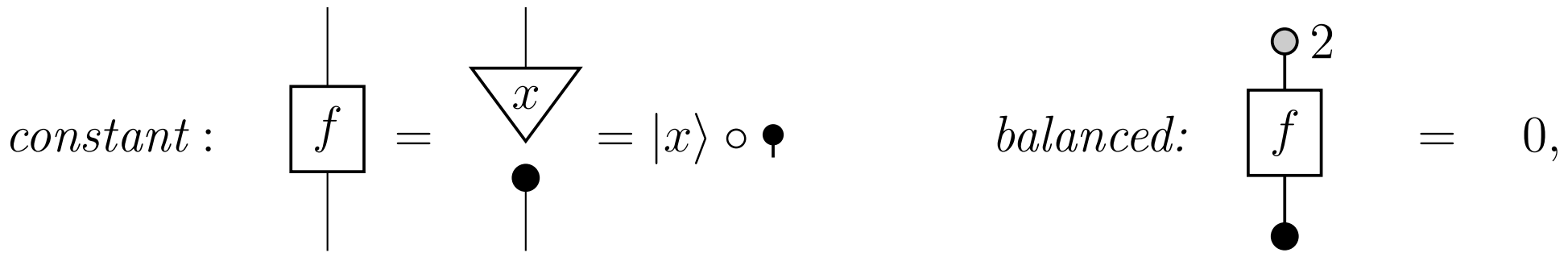

(4.108)

*where* 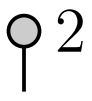 *is the dagger adjoint of the second classical state of $X^B$.*

That these definitions coincide with the usual ones for constant and balanced functions is shown in Section 4.2.2. In fact, it is easy to check that any constant relation will be a classical relation. In QCRel, the effect 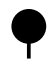 is $\langle H_1^A|$, which acts as a measurement of system $A$ after applying the oracle. We illustrate the details of the QCRel model of this algorithm by example and then with general definitions.

**Example 4.3.36.** Take $A = \{0, 1, 2, 3\}$ and $B = \{a, b, c, d\}$ to be four element systems. We define complementary bases on these systems as the following:

| System $A$ | System $B$ |
|---|---|
| $Z^A = \mathbb{Z}_2 \oplus \mathbb{Z}_2$ s.t.<br>$Z_0^A = \{0, 1\}, Z_1^A = \{2, 3\}$ | $Z^B = \mathbb{Z}_2 \oplus \mathbb{Z}_2$ s.t.<br>$Z_0^B = \{a, b\}, Z_1^B = \{c, d\}$ |
| $X^A = \mathbb{Z}_2 \oplus \mathbb{Z}_2$ s.t.<br>$X_0^A = \{0, 2\}, X_1^A = \{1, 3\}$ | $X^B = \mathbb{Z}_2 \oplus \mathbb{Z}_2$ s.t.<br>$X_0^B = \{a, c\}, X_1^B = \{b, d\}$ |

From Equation 4.108, we then define constant and balanced classical relations using the following dictionary:

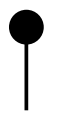 $= \{(0, \star), (2, \star)\}$, the classical state adjoint $\langle X_0^A|$ (4.109)

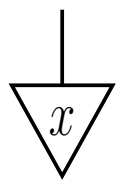 $= \{(\star, a), (\star, b)\}$ OR $\{(\star, c), (\star, d)\}$, a classical state of $Z^B$ (4.110)

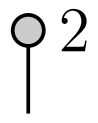 $= \{(b, \star), (d, \star)\}$, the classical state adjoint $|X_1^B\rangle$ (4.111)

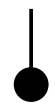 $= \{(\star, 0), (\star, 2)\}$, the classical state $|X_0^A\rangle$ (4.112)

Thus there are two constant classical relations[18] $f : Z^A \to Z^B$, one for each classical state of $Z^B$. They are:

$$\{(0, a), (0, b), (2, a), (2, b)\} \quad \text{and} \quad \{(0, c), (0, d), (2, c), (2, d)\}.$$

By Definition 4.3.35, balanced classical relations are those which do not relate 0 or 2 to either $b$ or $d$. There are four balanced classical relations for this example:

$$\{(0, c), (2, c), (1, d), (3, d)\} \qquad \{(0, a), (1, b), (2, c), (3, d)\}$$
$$\{(2, a), (3, b), (0, c), (1, d)\} \qquad \{(0, a), (2, a), (1, b), (3, b)\}$$

For a classical relation promised to be in one of these two classes, we can distinguish which with a single oracle query.

[18] A list of more example classical relations is given in Appendix A.

We generalize these definitions of constant and balanced classical relations to the following:

**Definition 4.3.37.** Let $Z^A = \oplus^N G_i$. A **constant relation** $f : Z^A \to Z^B$ relates all $\mathrm{id}_{G_i}$ to a single classical state of $Z^B$.

**Definition 4.3.38.** A relation $f : Z^A \to Z^B$ is **balanced** when no element in $X_0^A$ is related to an element in $X_1^B$.

**Theorem 4.3.39.** *The Deutsch-Jozsa algorithm defined above distinguishes constant relations from balanced relations in a single oracle query.*

*Proof.* This follows immediately from Vicary's abstract proof of the Deutsch-Jozsa algorithm from Section 4.2.2. □

**Remark 4.3.40.** It is important to note that the constant or balanced classes of classical relations in Example 4.3.36 can, as non-deterministic classical relations, be distinguished by a single query with input $|1\rangle, |3\rangle$, or $|1 \vee 3\rangle$. This observation might lead us to conclude that constant and balanced classical relations $f : Z^A \to Z^B$ can be distinguished by a single query of $f$ on any classical state of $X^A$ except the first one. In fact this is the case, as we now show.

**Lemma 4.3.41.** *There does not exist a balanced classical $f : Z^A \to Z^B$ relation whose pre-image is given only by elements of $X_0^A$.*

*Proof.* If the pre-image of $f$ is given only by elements of $X_0^A$, then the image of $f^{-1}$ must be the this same set $X_0^A$. However, balanced classical relations by definition have $\langle X_0^A| \circ f^{-1} \circ |X_1^B\rangle = 0$. Thus no element of $X_1^B$ can be in the pre-image of $f^{-1}$ (in the image of $f$). Let $\delta_{Z^A} : A \to A \otimes A$ be the relational converse of the groupoid multiplication for $Z^A$ (and likewise for $\delta_{Z^B}$). Write $-X_1^B = \{x^{-1} \mid x \in X_1^B\}$. Then if no element of $X_1^B$ is in the image of $f$, we have:

$$0 = \langle X_1^B| \otimes \langle -X_1^B| \circ (f \otimes f) \circ \delta_{Z^A} \tag{4.113}$$

$$0 = \langle X_1^B| \otimes \langle -X_1^B| \circ \delta_{Z^B} \circ f \qquad \text{by the dual of (4.97)} \tag{4.114}$$

$$0 = \langle X_0^B| \circ f, \tag{4.115}$$

As classical relations are comonoid homomorphisms they must obey the duals of (4.97) and (4.98), but this last equation contradicts the dual of (4.97). Thus $f$ cannot both be a balanced classical relation and have a pre-image of only $X_0^A$. □

This allows us to show that we will always be able to construct a simple query of constant and balanced classical relations in order to distinguish them.

**Theorem 4.3.42.** *Constant and balanced classical relations can be distinguished in a single query whose input state has non-zero intersection with any classical state other than* $|Z_0^A\rangle$.

*Proof.* Let $|Z_i^B\rangle$ be a classical state of $Z^B$ and ♦ be the classical co-state $\langle Z_0^A|$. Any constant classical relation has the form $f : |Z_i^B\rangle \circ$ ♦, and so the pre-image of $f$ is only the elements in the first classical state of $A$. Thus, a query of a constant classical relation with input from any other classical state will return the empty set. Further, Lemma 4.3.41 shows that it is not possible to construct balanced relations which will have the same behavior. Thus a query with any element not in ♦ will suffice to identify the class of the classical relation. □

Theorem 4.3.42 shows that the model of the Deutsch-Jozsa algorithm does not result in a classical algorithm for a non-trivial problem. While this would have been a nice side effect of our analysis, it is not the main goal. The more important takeaway is that the operation of the quantum algorithm can be accurately modeled in QCRel and that it does optimally solve a problem in that context. We will discuss this further after two more examples.

### 4.3.6 Single-shot Grover's Algorithm

The usual Grover's algorithm [52] takes as input a set $S$ and an indicator function $f : S \to \{0, 1\}$ and outputs an element $s \in S$ such that $f(s) = 1$. Though the algorithm is usually probabilistic and runs a repeated series of "Grover steps", here we consider the deterministic version that runs with a single step. In this section we will consider Vicary's generalization of the single-shot Grover algorithm where the codomain of the indicator function is allowed to be an arbitrary group [104]. Our setup requires the set $S$, as one system, as well as another system $B$. We define the basis $Z^S = \bigoplus^{|H^S|} G^S$ and $X^S = \bigoplus^{|G^S|} H^S$ on the $S$ system. System $B$ has complementary bases $Z^B = \bigoplus^{|H^B|} G^B$ and $X^B = \bigoplus^{|G^B|} H^B$. Let $|X_0^B\rangle$ be the first classical state of $X^B$, e.g. is $X^B = \mathbb{Z}_2 \oplus \mathbb{Z}_2$ then $|X_0^B\rangle = \{(\star, 1), (\star, 3)\}$, where 1 and 3 are the non-identity elements of that factors of $X^B$. Let $\langle X_\rho^B|$ be the converse of a classical state of $X^S$. Recall that $|Z_0^S\rangle = \{(\star, g) \,|\, g \in G^S \text{ is the first factor group of } Z^S\}$ is a classical point of $Z^S$, and that, by the complementary relationship of classical and unbiased points (Section 4.3.4), $|X_0^S\rangle \cong \{(\star, \mathrm{id}_{G_i^S}) \,|\, G_i^S \text{ is a factor group of } Z^S\}$.

In QCRel, the algorithm proceeds by the following steps:

1. Prepare system $S$ in the state $|Z_0^S\rangle$ and system $B$ in the state $|X_0^B\rangle$.
2. Apply the Fourier transform to system $S$, resulting in state $|X_0^S\rangle$.
3. Apply the oracle for a classical indicator relation $f : Z^S \to Z^B$.
4. Apply a diffusion relation $D : S \to S$ (defined below) to system $S$.
5. Measure system $S$ in the $X^S$ basis.

Vicary's [104] presentation for this procedure is:

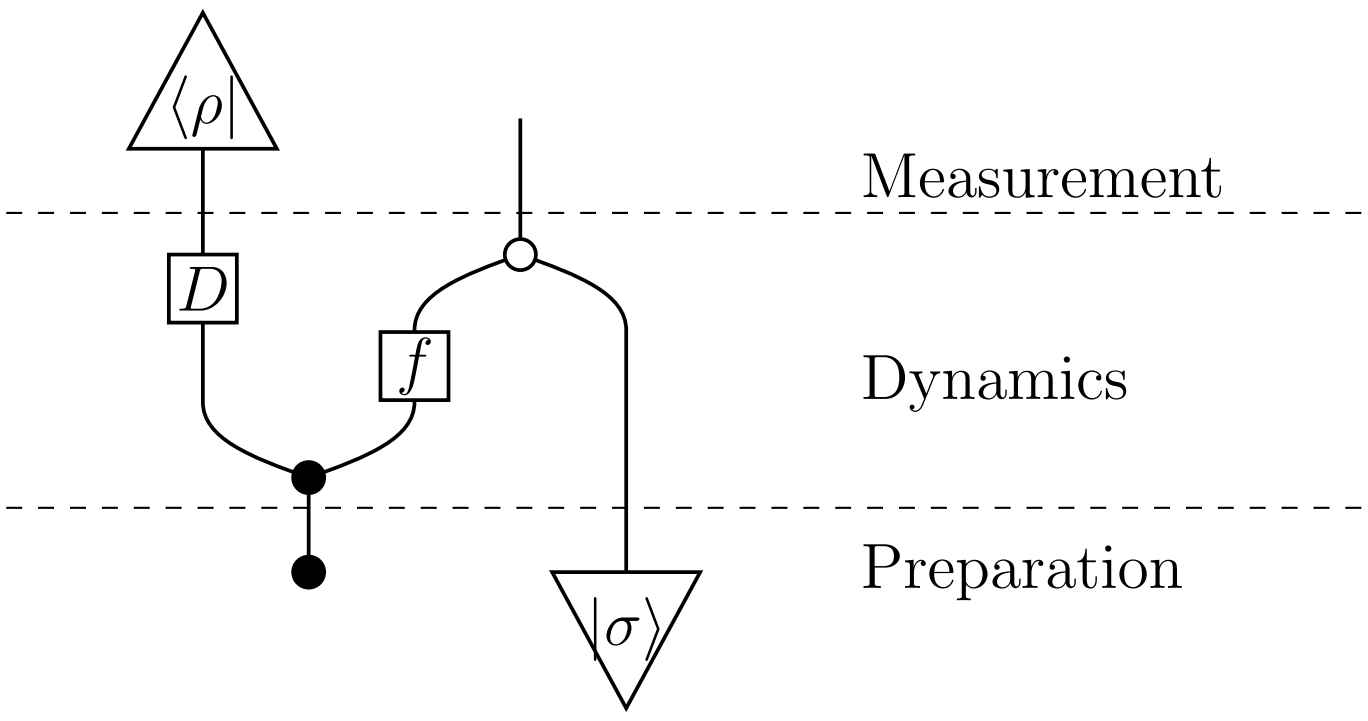

(4.116)

where numerical scalars have been dropped relative to that reference as there is only one non-zero scalar in QCRel.[19] Compare (4.116) to Figure 4.3. Recall that $\bullet : \{\star\} \to S$ relates the singleton to the elements of $H_0$ and that $\bullet$ is its relational converse. We will use the map $\bullet \circ \bullet : S \to S$ in the following definition. Here there is a special relation $D : S \to S$ called the diffusion operator:

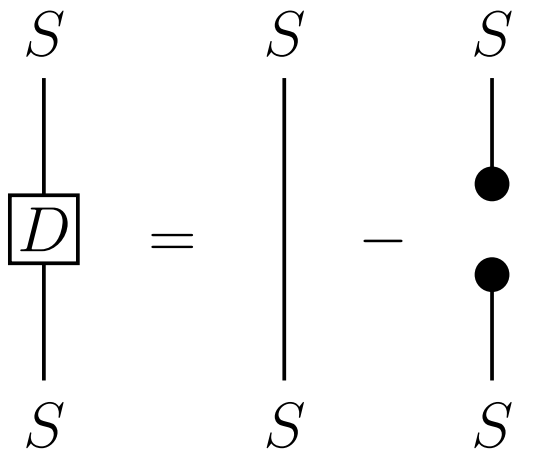

$$D := \{(x, x) \,|\, x \in S\} \triangle (H_0 \times H_0) \tag{4.117}$$

where the subtraction of two relations is given by the symmetric difference of their images. Explicitly then, the relational model for Grover's algorithm is:

$$\begin{aligned}
\mathrm{Grover}(f) : \{\bullet\} \times \{\bullet\} &\to \{\bullet\} \times B \\
&= \langle X_\rho^S| \times \mathrm{id}_B \circ D \times \mathrm{id}_B \circ \mathrm{OracleRel}(f) \circ |X_0^S\rangle \times |\sigma\rangle \\
&= \{((\bullet, \bullet), (\bullet, c \circ_X x)) \,| \\
&\qquad x \in X_0^B, y \in X_\rho^S, \mathrm{id}_{G_n}, b, z \in S \text{ s.t. } z \bullet_{Z^S} b = \mathrm{id}_{G_n} \text{ and } bfc, zDy\}
\end{aligned}$$

[19] Recall that scalars in a monoidal category with identity object $I$ are maps $s : I \to I$. Thus in **Rel** $I = \{\star\}$, so the only scalars are the empty relation and the identity relation on the singleton set.

**Theorem 4.3.43.** *Equation 4.116 is zero only for classical states of $X^S$ denoted $|X^S_\rho\rangle$ that satisfy the following equation:*

$$\langle X^B_0| \circ f \circ |X^S_\rho\rangle = \langle X^B_0| \circ f \circ |X^S_0\rangle \tag{4.118}$$

*Proof.* Proven in [104]. See Section 3.2, Equation (34). $\square$

Here $|X^B_0\rangle$ can be generalized to any fixed classical state $|X^B_\sigma\rangle$. This allows a generalization of the single-shot Grover's algorithm where the cardinality of system $B$ is increased [104]. Consequently, the LHS of Equation 4.118 tests if any element in the classical state $|X^S_\rho\rangle$ is related to any of the elements in $|X^B_\sigma\rangle$. The RHS tests if any of the elements of $X^S_0$ are related to $|X^B_\sigma\rangle$.

**Proposition 4.3.44.** *The QCRel single-shot Grover algorithm only returns states $|X^S_\rho\rangle$ such that for all $h \in X^S_0$, $s \in X^S_\rho$ and $x \in X^B_\sigma$*

$$hfx \quad = \quad \neg(sfx).$$

*In other words, the only elements that can be possibilistically measured under the QCRel Born rule (Axiom 4.3.10) are elements of $S$ that have the opposite mapping to $X^B_\sigma$, under the relation $f$, than elements of $X^S_0$.*

*Proof.* By Theorem 4.3.43 and definitions. $\square$

**Example 4.3.45.** Let $S = \{0, 1, 2, 3\}$ and choose $Z^S = \mathbb{Z}_2 \oplus \mathbb{Z}_2$ and $X^S = \mathbb{Z}_2 \oplus \mathbb{Z}_2$ as $G$ (black) and $H$ (white) bases respectively, so that $Z^S_0 = |0 \vee 1\rangle$ and $X^S_0 = |0 \vee 2\rangle$. Let $B$ be the same four element system with the bases $Z^B = \mathbb{Z}_2 \oplus \mathbb{Z}_2$ and $X^B = \mathbb{Z}_2 \oplus \mathbb{Z}_2$ and choose $|X^B_\sigma\rangle = |1 \vee 3\rangle$. The diffusion operator is then given by

$$\begin{aligned} D &:= \{(0,0), (1,1), (2,2), (3,3)\} - \{(0,0), (0,2), (2,0), (2,2)\} \\ &= \{(1,1), (3,3), (0,2), (2,0)\}. \end{aligned}$$

In this case, $D$ happens to be a bijection, it is a unitary relation and thus a possible evolution in QCRel.[20] Let $f$ be the classical relation[21] $\{(0, 2), (2, 2), (1, 3), (3, 3)\}$, where elements of $X^S_0$ are not related to elements of $X^B_\sigma$. Thus the above algorithm will only return classical states of $X^S$ that *are* related, under $f$, to $|X^B_\sigma\rangle$. The only possible outcome state is $|1 \vee 3\rangle$.

[20] This will not be the case whenever $S$ has more than two factor groups. Unitarity is a stringent condition on processes in QCRel.

[21] See Appendix A for a list of classical relations $\mathbb{Z}_2 \oplus \mathbb{Z}_2 \to \mathbb{Z}_2 \oplus \mathbb{Z}_2$.

**Example 4.3.46.** This is the same as the above example, but take $f$ to be the classical relation $\{(0,0),(2,0),(0,1),(2,1)\}$. As an element of $X_0^S$ is related to $|X_\sigma^B\rangle$, the algorithm will return classical states of $X^S$ which are *not* mapped to $|X_\sigma^B\rangle$, i.e. the state $|1 \vee 3\rangle$.

If we loosen the QCRel model slightly, and do not require $f$ to be a classical relation, then it is easy to see that this algorithm solves a non-trivial classical search problem. Consider an $f$ that maps all classical states of $X^S$ to $|X_\sigma^B\rangle$ except exactly one other *marked* classical state $X_i^S$, where $i \neq 0$. As in Example 4.3.46, we know that the QCRel Grover's algorithm will output only $X_i^S$ with a single query of $f$. We can easily imagine that there are a large number of classical states of $X^S$ that might be marked and, further, that the image of the marked classical state under $f$ is indistinguishable from that of any other classical state (other than not being related to $|X_\sigma^B\rangle$). A deterministic classical approach to this problem would certainly require many queries of $f$, and it is not immediate that there is an obvious non-deterministic solution with a single query; though the QCRel Grover's algorithm we describe here does achieve exactly that.

### 4.3.7 The Groupoid homomorphism promise algorithm

This section models the group homomorphism algorithm from Section 4.2.4 in QCRel. The quantum version of the algorithm takes as input a blackbox function $f : G \to A$ promised to be one of the homomorphisms between group $G$ and abelian group $A$. It then outputs the identity of the homomorphism. In that section the full identification algorithm is built up by multiple calls to an instance of the problem for cyclic groups.[22] It is this cyclic group subroutine that we consider here. In the relational setting we will move from groups to groupoids. Let groupoid $H$ be complementary to groupoid $G$ and groupoid $B$ be complementary to groupoid $A$. The QCRel GroupHomID algorithm then takes as input a groupoid isomorphism $f : G \to A$. Let $|\rho\rangle$ be a classical state of $H$, and $|\sigma\rangle$ be a classical state of $B$.

[22] Making use of the structure theorem for abelian groups to complete the general case.

The algorithm has the following abstract specification [115]:

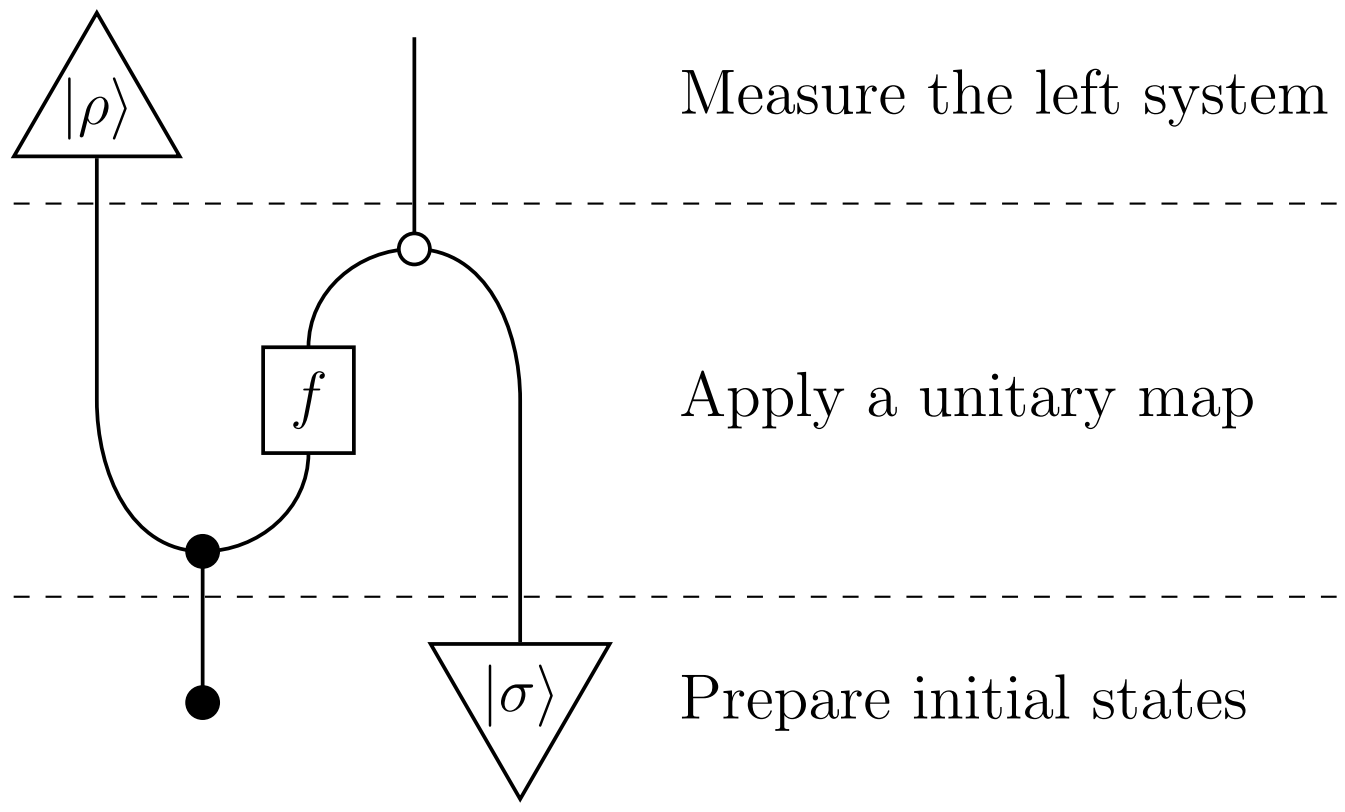

(4.119)

Let the factor groups of a groupoid $G$ be denoted $G_n$. This gives the following relational model for the algorithm:

$$\begin{aligned}
\text{GroupHomID}(f) : \{\bullet\} \times \{\bullet\} &\to \{\bullet\} \times B \\
&= \langle\rho| \times \text{id}_B \circ \text{OracleRel}(f) \circ |H_0\rangle \times |\sigma\rangle \\
&= \{((\bullet, \bullet), (\bullet, c \circ_X x)) \mid \\
&\qquad x \in \sigma, y \in \rho, \text{id}_{G_n}, b \in A \text{ s.t. } y \bullet_G b = \text{id}_{G_n} \text{ and } bfc\}
\end{aligned}$$

**Theorem 4.3.47.** *The algorithm defined by* (4.119) *has output state* $|\rho\rangle$ *only when for some* $x \in \rho$ *and some* $y \in \sigma$ *we have* $(y, x) \in f$.

*Proof.* The verification in Section 4.2.4 simplifies the algorithm in Equation 4.119 to:

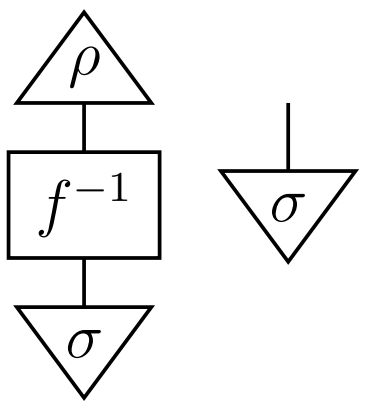

$$= \quad \{(\bullet, (\bullet, x)) \mid xf^{-1}y \text{ for some } y \in \rho\}, \qquad (4.120)$$

where we see that post-selection on the left hand system implies the theorem's condition via the QCRel Born rule (Axiom 4.3.10). □

**Theorem 4.3.48.** *If* $f$ *is a groupoid isomorphism then the algorithm in Equation 4.119 returns all states.*

*Proof.* Groupoid isomorphisms relate every element of the domain to some element in the codomain and relate every element of the codomain to some element of the domain. □

Still, we can imagine running the algorithm from (4.119) where any classical relation $f$ is allowed as input to obtain non-trivial outcomes. The investigation of the usefulness of this algorithm is part of future work.

## 4.4 Conclusion

In this chapter we have investigated the structure of quantum algorithms using the QPT perspective. We have contributed the following new results:

1. A definition of the Fourier transform in general QPTs along with abstract proofs of the Fourier Inversion Theorem, Pontryagin Duality, and the Convolution Theorem.

2. A definition of the Fourier Transforms and an operational quantum Fourier transform in general oPTs.

3. We use complementary observables to construct a unitary oracle in a general QPT and provide an abstract proof of its unitarity. We show an equivalence between the complementarity of those observables and the unitarity of this oracle.

4. We pose a new problem, the GROUPHOMID problem, and give a quantum algorithm for its solution with favorable query complexity over classical algorithms.

5. We provide models for the operation of quantum algorithms in non-deterministic classical computation by explicitly constructing the Deutsch-Jozsa algorithm, the single-shot Grover's algorithm, and the GROUPHOMID algorithm in the QPT QCRel. Along the way we introduce an operational notion of a QPT Fourier transform in **FRel** and characterize self-conjugate comonoid homomorphisms and unitary oracles in this category.

Models of quantum algorithms in other QPTs (such as QCRel), perform two functions. Firstly, they help us further understand the relationship between different kinds of computation. Our main question is:

*How are different models of computation related to each other?*

Typically, computer scientists address this question by keeping computational problems fixed and writing different algorithms for the same problem that take advantage of the computational model. This view is the basis of the field of computational complexity and has given many insights into the differences between computational models. The approach in this thesis aims to extend this view. In a sense, the QPT framework allows us to think about running the same algorithms on different QPTs as we would think about running them on different kinds of hardware. We "compile" algorithms into different QPTs and analyzes the results to address our main questions. It is an algorithm-centric vs. problems-centric approach to computation.

As compared to the foundational interest of the first function, the second function of this investigation is more practical. Mapping algorithms between computational models allows us to generate new problems and algorithms. Some of these will be trivial (as in the QCRel Deutsch-Jozsa algorithm of Section 4.3.5) and some will not be (as in QCRel Grover's of Section 4.3.6). An important thread of future work is to understand the conditions that allow for one case and not that other.

In terms of quantum algorithms specifically, the successes of this approach open up new avenues to study other blackbox quantum algorithms at this level of abstraction. We discuss more details of the outlook for future work in Chapter 7.

# Chapter 5

# Mermin Non-locality

**Chapter Abstract**

The study of non-locality is fundamental to the understanding of quantum mechanics. The past 50 years have seen a number of non-locality proofs, but its fundamental building blocks, and the exact role it plays in quantum protocols, has remained elusive. In this paper, we focus on a particular flavour of non-locality, generalising Mermin's argument on the GHZ state. Using strongly complementary observables, we provide necessary and sufficient conditions for Mermin non-locality in quantum-like process theories. We show that the existence of more phases than classical points (aka eigenstates) is not sufficient, and that the key to Mermin non-locality lies in the presence of certain algebraically non-trivial phases. This allows us to show that **FRel**, a favourite toy model for categorical quantum mechanics, is Mermin local. By considering the role it plays in the security of the HBB CQ (N,N) family of Quantum Secret Sharing protocols, we argue that Mermin non-locality should be seen as a resource in quantum protocols. Finally, we challenge the unspoken assumption that the measurements involved in Mermin-type scenarios should be complementary, opening the doors to a much wider class of potential experimental setups than currently employed.

In short, we give conditions for Mermin non-locality tests on any number of systems, where each party has an arbitrary number of measurement choices, where each measurement has an arbitrary number of outcomes and further, that is in any quantum-like process theory.

## 5.1 Introduction

Non-locality is a fundamental property of quantum mechanics. It impacts both foundations and application, ruling out the existence of **local hidden variable theories** consistent with quantum theory [19], and underpinning protocols like quantum key distribution [42] and quantum secret sharing [75]. The importance of this property pushed the development of methods to characterise it both in general (e.g. Abramsky and Brandenburger's sheaf-theoretic methods) and in specific extensions of quantum theory (e.g. Barrett's [17] generalized probabilistic theories).

We focus on a particular possibilistic class of non-locality arguments generalized from Mermin's argument [80] and related to the recent work on All-versus-Nothing arguments by Abramsky et al. [4]. These experiments produce possibilistic evidence for quantum mechanical non-locality, i.e. certain measurement outcomes that can only be realized by non-local theories. Mermin scenarios are typically described by triples $(N, M, D)$ for $N$ parties with $M$ measurement choices for each party, each having $D$ classical outcomes. Current literature generalises from the original $(3, 2, 2)$ scenario [80] to derive non-locality proofs for the $(3, 3, 2)$[94], $(D + 1, 2, D)$[117], $(N > D,\ 2,\ D \text{ even})$[83], and $(\text{odd } N, 2, \text{even } D)$[61]. One contribution of our work is to extend the work of [32] to cover all $(N, M, D)$ scenarios in quantum theory.

In [32], Coecke et al. used strong complementarity to formulate Mermin arguments within the the framework of Categorical Quantum Mechanics [6]. Not only does this approach help generalize non-locality arguments within quantum theory, but it also paved the way towards an understanding of Mermin non-locality in quantum-like process theories. As a corollary, they are able to identify the difference between qubit stabilizer quantum mechanics (which is non-local) and Spekken's toy theory (which is local) in the structure of the respective phase groups [32, 34].

In Sections 5.2 and 5.3, we remove implicit assumptions about phase groups and classical points from [32] and use strongly complementary structures to generalise Mermin measurements to any QPT, defining Mermin non-locality as the existence of a Mermin measurement scenario not admitting a local hidden variable model.

In Section 5.3, we show that strong complementarity is not sufficient to characterise Mermin non-locality. The phase group structure is shown to provide necessary algebraic conditions in abstract process theories, as summarised by our first main result:

**Theorem. 5.3.8.** Let **C** be a †-SMC, and ( ○ , ● ) be a strongly complementary pair of †-qSCFAs. If the group of ○-phases is a trivial algebraic extension of the subgroup of $\bullet$-classical points, i.e. if there exist no algebraically non-trivial $\circ$-phases, then **C** is Mermin local.

Thus $\circ$-phase groups which are trivial algebraic extensions of the respective subgroups of $\bullet$-classical points always lead to local hidden variable models, regardless of whether there are enough $\bullet$-classical points to form a basis and/or strictly more $\circ$-phases than $\bullet$-classical points. Indeed, we show that the category **FRel** of finite sets and relations is Mermin local (despite it having arbitrarily many more $\circ$-phases than $\bullet$-classical points), and confirm that Spekkens' toy theory is Mermin local (despite it having enough $\bullet$-classical points to form a basis).

Also in Section 5.3, we show that the existence of algebraically non-trivial $\circ$-phases is sufficient, under mild additional assumptions, to formulate a non-locality argument. This leads to our second main result:

**Theorem. 5.3.7.** Let **C** be a †-SMC, and $(\circ, \bullet)$ be a strongly complementary pair of †-qSCFAs. Suppose further that the $\bullet$-classical points form a basis. If the group of $\circ$-phases is a non-trivial algebraic extension of the subgroup of $\bullet$-classical points, then **C** is Mermin non-local.

As a consequence, we confirm that Stabilizer Quantum Mechanics is Mermin non-local.

In Section 5.4, we argue that our concrete characterisation as the existence of algebraically non-trivial phases can be used to see Mermin non-locality as a resource in the construction of quantum protocols. We exemplify this by showing how the security of the HBB CQ (N,N) family of Quantum Secret Sharing protocols from [75, 72] directly relates to the flavour of non-locality explored in this work.

In Section 5.5, we use our general framework to investigate Mermin non-locality in **FHilb**, the historical arena of quantum mechanics. The traditional formulation of Mermin arguments relies on sets of complementary measurements, such as the $X$ ($\bullet$ measurement with $\circ$-phase 0) and $Y$ ($\bullet$ measurement with $\circ$-phase $\frac{\pi}{2}$) measurements of the qubit in the original $(3, 2, 2)$ Mermin argument. We show how, even in the case of $(N, 2, D)$ scenarios, many more possible measurements exist than complementary ones. This result opens a wealth of novel experimental configurations for tests of Mermin non-locality and, through results of Section 5.4, new configurations for quantum secret sharing protocols as well.

## 5.2 Mermin measurements

Unlike Bell tests, which produce outcomes with probabilities that are forbidden to local hidden variable theories, the Mermin argument produces outcomes which are impossible to observe in a local hidden variable theory [80]. This section introduces the definitions necessary to generalise the Mermin argument to process theories. We make use of the standard definitions for strongly complementary observables, phase states and phases. We often refer to quasi-special †-Frobenius algebras as **non-degenerate observables** and use the shorthand †-qSFA. The acronym †-qSCFA refers to a commutative †-qSFA. Definitions of these concepts were given in Chapter 3 and Section 4.1.

We make particular use of the notion of classical structures with enough classical points to form a basis (Definitions 3.3.9 and 4.3.11). Recall that in **FHilb**, the objects are vector spaces and any vector space basis clearly all obey the abstract conditions of Definition 4.3.11. In general QPTs this will not be the case.

We also recall the following fact about phases (Section 3.4) for strong complementarity from Section 4.1 and [32], where it is clarified.

**Theorem 5.2.1.** *Let* $\circ$ *and* $\bullet$ *be strongly complementary †-qSFAs. Phase states (resp. phases) of* $\circ$ *form a group under the action of* $(\curlywedge, \dot{\circ})$. *This group of phase states is denoted the* ***phase group*** $P_{\circ}$. *The classical points (resp. the induced phases) of* $\bullet$ *are a subgroup, i.e.* $K_{\bullet} \subseteq P_{\circ}$.

As the phase group of a †-qSCFA is commutative, we use additive notation: given two $\circ$-phase states $|\alpha\rangle$ and $|\beta\rangle$, we denote by $|\alpha + \beta\rangle$ their addition in the phase group. From now on, we interchangeably use phase states and phases, leaving disambiguation to context.

The GHZ states and Mermin measurements are the main ingredients needed in our argument. GHZ states appear in the ZX calculus fragment of our framework in [31] and are generalized to the definition that we use in [32].

**Definition 5.2.2.** Given a †-qSFA $\circ$ in a †-SMC, an $N$**-partite GHZ state** for $\circ$ is:

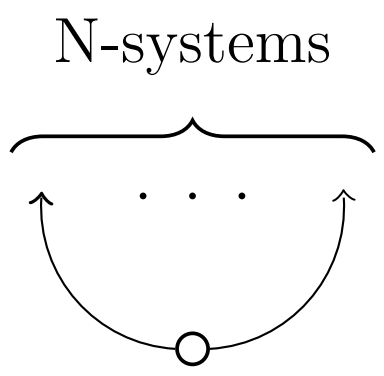

(5.1)

We follow [32] to build Mermin type scenarios out of them.

**Definition 5.2.3.** Let ○ and ● be a pair of strongly complementary †-qSFAs. An $N$-partite **Mermin measurement** is obtained by applying $N$ ○-phases $\alpha_1, ..., \alpha_N$ to the $N$ components of an $N$-partite GHZ state, and then measuring each component in the ● structure:

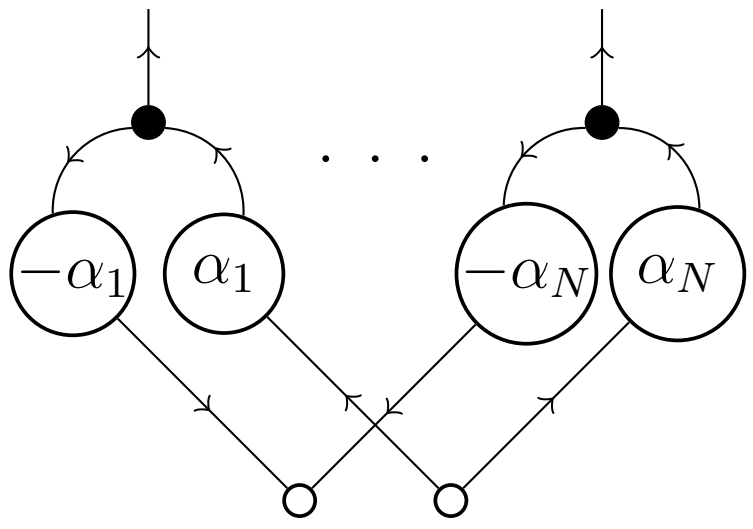

(5.2)

We further require that $\sum_i \alpha_i$, where the sum is taken in the group of phases, be a ●-classical point.

**Lemma 5.2.4.** *The $N$-partite Mermin measurement shown in Equation 5.2 is equivalent to the following state:*

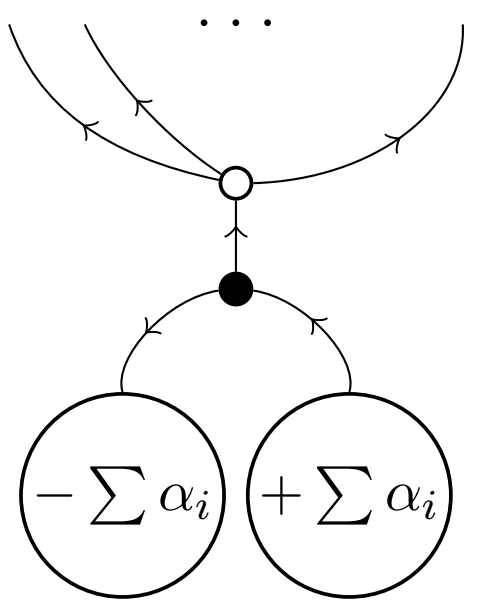

(5.3)

*Proof.* Pushing the phases down through the ○ nodes and using strong complementarity. See [32]. □

While this defines a single Mermin experiment, the full non-locality argument requires the joint outcomes of several Mermin measurements.

**Definition 5.2.5.** Let ○ and ● be strongly complementary †-qSCFAs on a space $\mathcal{H}$ in a †-SMC. An $N$-partite **Mermin measurement scenario** (for ○ and ●) is any non-empty, finite collection of Mermin measurements $\underline{\alpha}^s = (\alpha_1^s, ..., \alpha_N^s)_{s=1,...,S}$ of the $N$-partite GHZ state in the form of (5.1).

In the category **FHilb** of finite-dimensional Hilbert spaces, an $N$-partite Mermin measurement scenario where $a_1, ..., a_M$ are the distinct ○-phases appearing in the scenario and $\mathcal{H}$ is $D$-dimensional is exactly the usual $(N, M, D)$ Mermin scenario. This correspondence is clarified in Section 5.3, where we derive our generalized Mermin non-locality argument.

## 5.3 Mermin locality and non-locality

The last definitions we need for our main results, Theorems 5.3.7 and 5.3.8, are those of local hidden variable models (following the construction of [32]) and non-trivial algebraic extensions.

**Definition 5.3.1.** Let $\circ$ and $\bullet$ be strongly complementary †-qSCFAs on some system $\mathcal{H}$. Consider an $N$-partite Mermin measurement scenario $(\underline{\alpha}^s)_{s=1,...,S}$, and let $a_1, ..., a_M$ be the distinct $\circ$-phases appearing in it. The **local map** for the scenario is the map $\mathcal{H}^{\otimes(M\cdot N)} \to \mathcal{H}^{\otimes(N\cdot S)}$ defined as follows:

a. we group the input wires in $N$ groups of $M$ wires: we say that the $r$-th wire of $i$-th group is the $a_r$ **input wire for system** $i$

b. we group the output wires in $S$ groups of $N$ wires: we say that the $j$-th wire of $r$-th group is the $j$-th **output wire for measurement** $s$

c. each input wire is connected to a $\bullet$ node

d. for all $r, i, j$ and $s$, the $\bullet$ node of each $a_r$ input wire for system $i$ is connected to the $j$-th output wire for measurement $s$ if and only if $i = j$ and $\alpha^s_j = a_r$

The following diagram details the procedure:

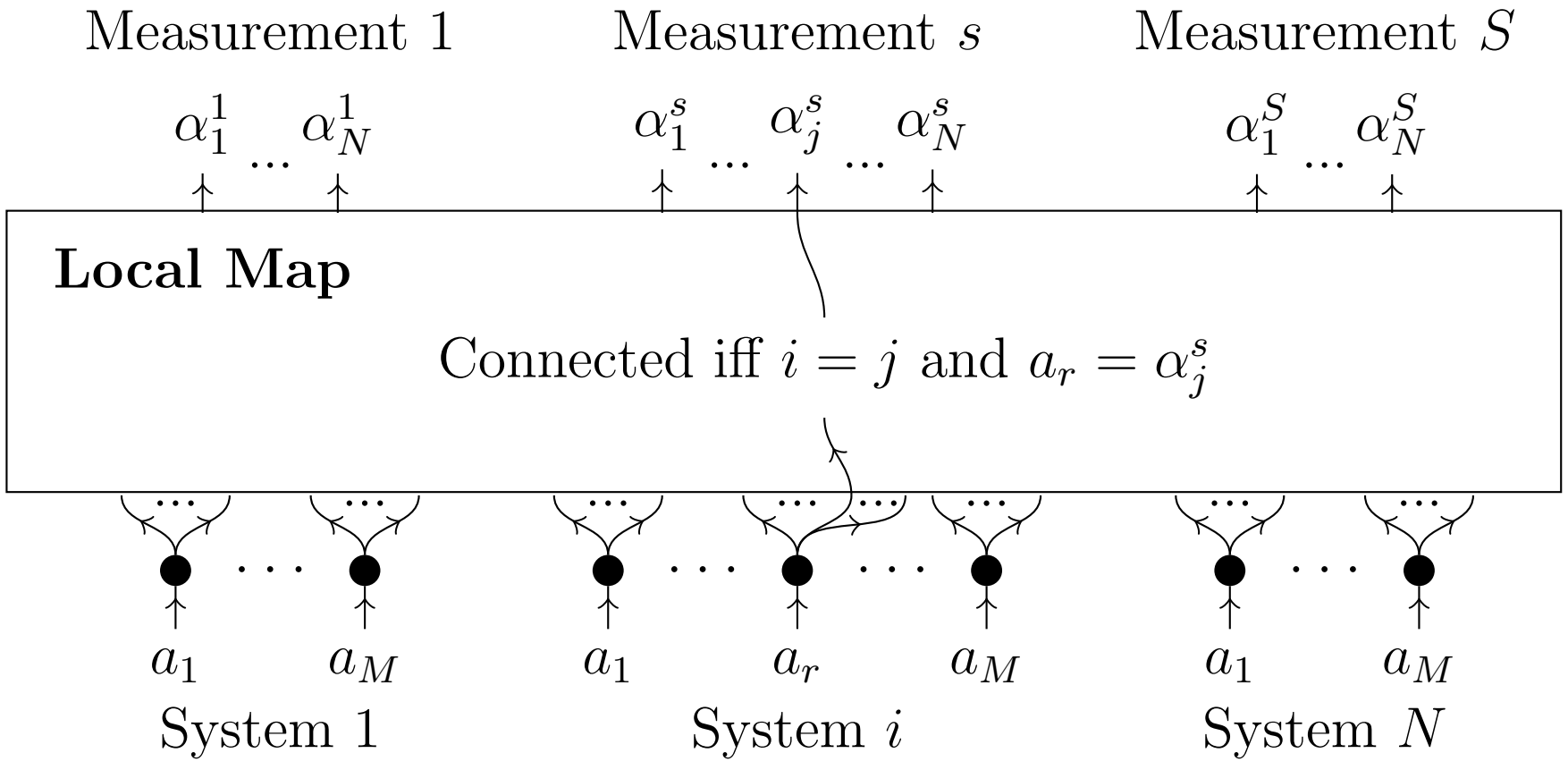

(5.4)

A **local hidden variable model** for an $N$-partite Mermin measurement scenario is a state $\Lambda$ of $\mathcal{H}^{\otimes(N\cdot S)}$, obtained by applying the local map for the scenario to some state $\Lambda'$ of $\mathcal{H}^{\otimes(M\cdot N)}$. We further require that for each $s = 1, ..., S$, the Mermin measurement $\underline{\alpha}^s$ is the same as the state obtained from $\Lambda$ by composing an $\bullet$ with each output wires of each measurement $t$ with $t \neq s$:

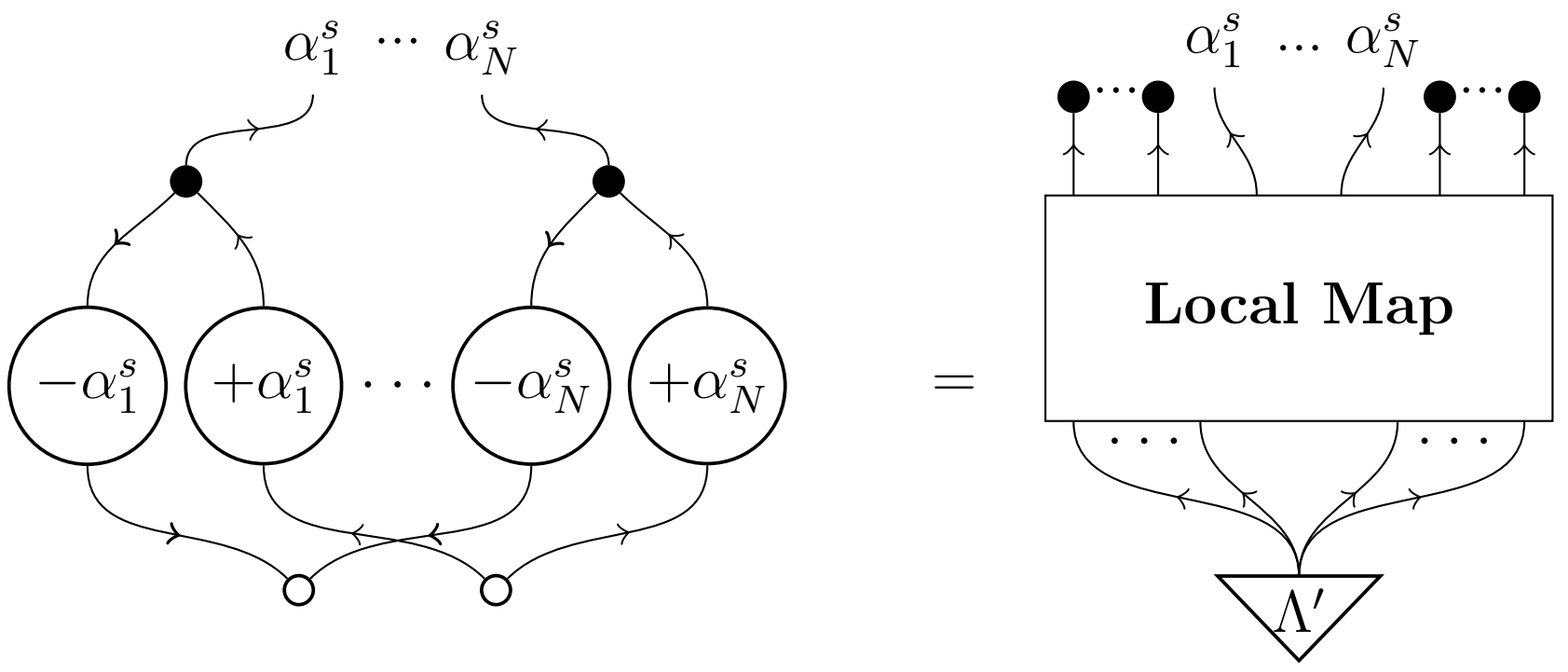

(5.5)

The definition of local hidden variables finally allows us to formulate our generalised notion of Mermin non-locality.

**Definition 5.3.2.** We say a †-SMC **C** is **Mermin non-local** if there exists a Mermin scenario for some strongly complementary pair ( ○, ● ) of †-qSCFAs which has no local hidden variable model. If for all strongly complementary pairs no such measurement exists, then we say that **C** is **Mermin local**.

Mermin non-locality will shortly be shown to be equivalent to the following algebraic property of the group of ○-phases. The following examples will be used later on to investigate some abstract process theories of interest.

**Definition 5.3.3.** Let $(G, +, 0)$ be an abelian group and $(H, +, 0)$ be a subgroup. We say that $G$ is a **non-trivial algebraic extension** of $H$ if there exists a finite system of equations $(\sum_{j=1}^{l} n_j^p \cdot x_j = h^p)_p$, with $h^f \in H$ and $n_j^p \in \mathbb{Z}$, which has solutions in $G$ but not in $H$. Otherwise, we say $G$ is a **trivial algebraic extension** of $H$.

If $G = P_{○}$ is a non-trivial algebraic extension of $H = K_{●}$, then the ○-phases involved in any solution $x_j := \alpha_j$ to a system unsolvable in $K_{●}$ will be called **algebraically non-trivial phases**.

**Example 5.3.4.** Let $G = \{0, \pi/2, \pi, -\pi/2\} < \mathbb{R}/2\pi\mathbb{Z}$ and $H = \{0, \pi\} < G$. Then $G$ is a non-trivial algebraic extension of $H$, because the single equation $2x = \pi$ has no solution in $H$ but has solution(s) $\pm\pi/2$ in $G$. It is in fact this example that yields the original argument in **FHilb** from [32].

**Lemma 5.3.5.** *Let $(G, +, 0)$ be an abelian group and $(H, +, 0)$ be a subgroup. Suppose that there is a function $\Phi : G \to H$ such that for any equation $\sum_{j=1}^{l} n_j \cdot x_j = h$ with $h \in H$ and $n_j \in \mathbb{Z}$, if $x_j := g_j$ is a solution in $G$, $x_j := \Phi(g_j)$ is also a solution (in $H$). Then $G$ is a trivial algebraic extension of $H$.*

*Proof.* Consider a system with solution $x_j := g_j$ in $G$. Then $x_j := \Phi(g_j)$ solves each individual equation in $H$, and thus also the system. □

**Example 5.3.6.** Let $(K, +, 0)$ be any finite abelian group, and $G = K \times K'$ for some finite non-trivial abelian group $(K', +, 0)$. Let $H < G$ be the subgroup $K \times \{0\}$. If $h = (k, 0) \in H$, then any equation $\sum_{j=1}^{N} n_j \cdot x_j = h$ is equivalent to the following pair of equations, where $\pi_K$ and $\pi_{K'}$ are the quotient projections onto $K \cong G/K'$ and $K' \cong G/K$ respectively:

a. $\sum_{j=1}^{N} n_j \cdot \pi_K x_j = k$ in $K$

b. $\sum_{j=1}^{N} n_j \cdot \pi_{K'} x_j = 0$ in $K'$

If $x_j := g_j = (\pi_K g_j, \pi_{K'} g_j)$ is a solution in $G$, then $x_j := (\pi_K g_j, 0)$ is a solution in $H$. Define $\Phi$ to be the map $g_j : G \mapsto (\pi_K g_j, 0) \in H$ and use Lemma 5.3.5 to conclude that $G$ is a trivial algebraic extension of $H$.

We are now able to introduce our first main result:

**Theorem 5.3.7** (Mermin Non-Locality). *Let* $\mathbf{C}$ *be a* †*-SMC, and* ($\circ$, $\bullet$) *be a strongly complementary pair of* †*-qSCFAs. Suppose further that the* $\bullet$*-classical points form a basis. If the group of* $\circ$*-phases is a non-trivial algebraic extension of the subgroup of* $\bullet$*-classical points, then* $\mathbf{C}$ *is Mermin non-local.*

*Proof.* For clarity, we present a proof where the system of equations that defines the phase group as a non-trivial algebraic extension is composed of a single equation. The construction for general systems of $l$ equations consists of $l$ copies of the construction we explicitly give.

Let $a_1, ..., a_M$ be $\circ$-phases and $a \neq 0$ be (the phase induced by) a $\bullet$-classical point such that the following equation (in additive $\mathbb{Z}$-module notation, for $n_r \in \mathbb{Z}$) has solution $(x_r := a_r)_{r=1,...,M}$ in the group of $\circ$-phases, but has no solution in the subgroup of (phases induced by) $\bullet$-classical points:

$$\sum_{r=1}^{M} n_r \cdot a_r = a \tag{5.6}$$

This means that we are assuming the group of $\circ$-phases are a non-trivial algebraic extension of the subgroup of $\bullet$-classical points. Without loss of generality, assume that $n_r \neq 0$ and $a_r \neq 0$ for all $r = 1, ..., M$.

Let $k$ be the exponent of the group of $\bullet$-classical points, and define the following Mermin measurement, where each $a_r$ appears $n_r$ times and 0 appears $n_0$ times, for some $n_0$ such that $V := \sum_{r=0}^{M} n_r \equiv 1 \pmod{k}$

$$\underline{\alpha} = (a_1, ..., a_1, ..., a_M, ..., a_M, 0, ..., 0) \tag{5.7}$$

Define a $V$-partite Mermin measurement scenario with $S := n_0 + V$ and:

$$\underline{\alpha}^s := (0, 0, ..., 0, 0) \text{ for } s = 1, ..., n_0 \tag{5.8}$$

$$\underline{\alpha}_i^{n_0+v} := \underline{\alpha}_{i+v \pmod{V}} \text{ for } v = 1, ..., V \tag{5.9}$$

The scenario has $n_0$ measurements with only 0 phases (the **controls**) and $V$ measurements with cyclic permutations of $\underline{\alpha}$ (the **variations**). The following diagram depicts the scenario:

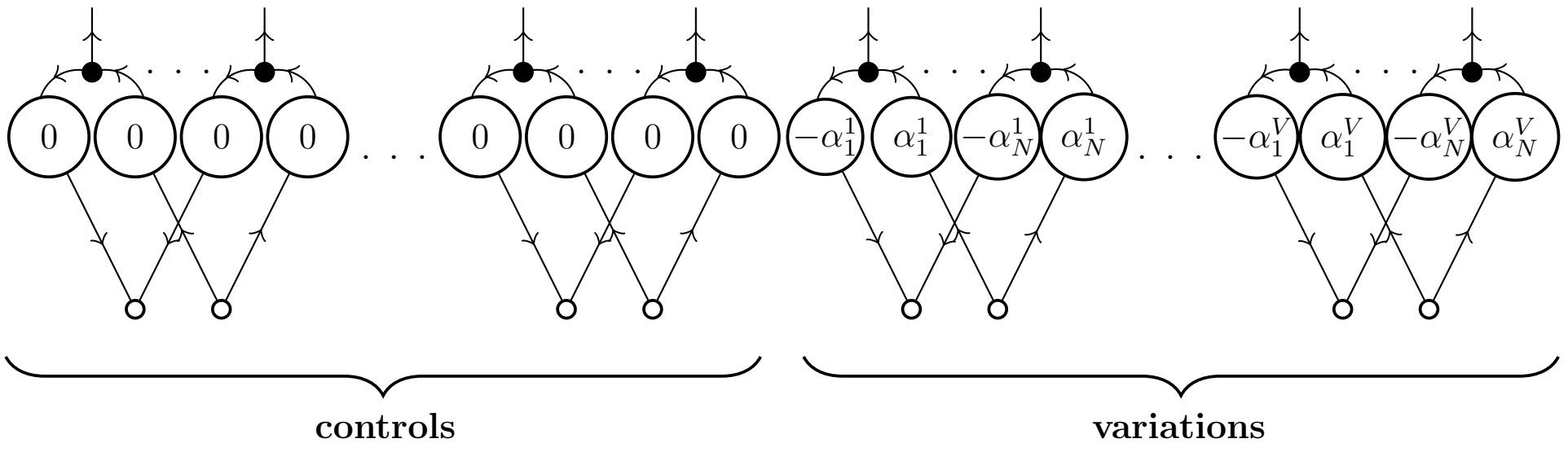

(5.10)

To show that the scenario from Equation 5.10 does not admit a local hidden variable:

1a. we add up (in the group of $\circ$-phases) all the components of each control, using Lemma 5.2.4, and obtain 0 from each control

1b. we add up all the components of each variation, again using Lemma 5.2.4, and obtain $a$ from each variation

2a. we add up the result from all the controls, and obtain $\Sigma_C := n_0 \cdot 0 = 0$

2b. we add up the result from all variations, and obtain $\Sigma_V := V \cdot a = a$ , using the fact that $a$ is in the subgroup of (phases induced by) $\bullet$-classical points and $V$ is congruent to 1 modulo the exponent of the subgroup

3. we subtract $\Sigma_C$ from $\Sigma_V$, using the antipode $\diamond$ of the strongly complementary pair $(\circ, \bullet)$, and obtain $a - 0 = a$

4. we test the result against the $\bullet$-classical point $\langle a|$, and obtain the non-zero scalar $\langle a|a\rangle$

The procedure is summarised by the following diagram:

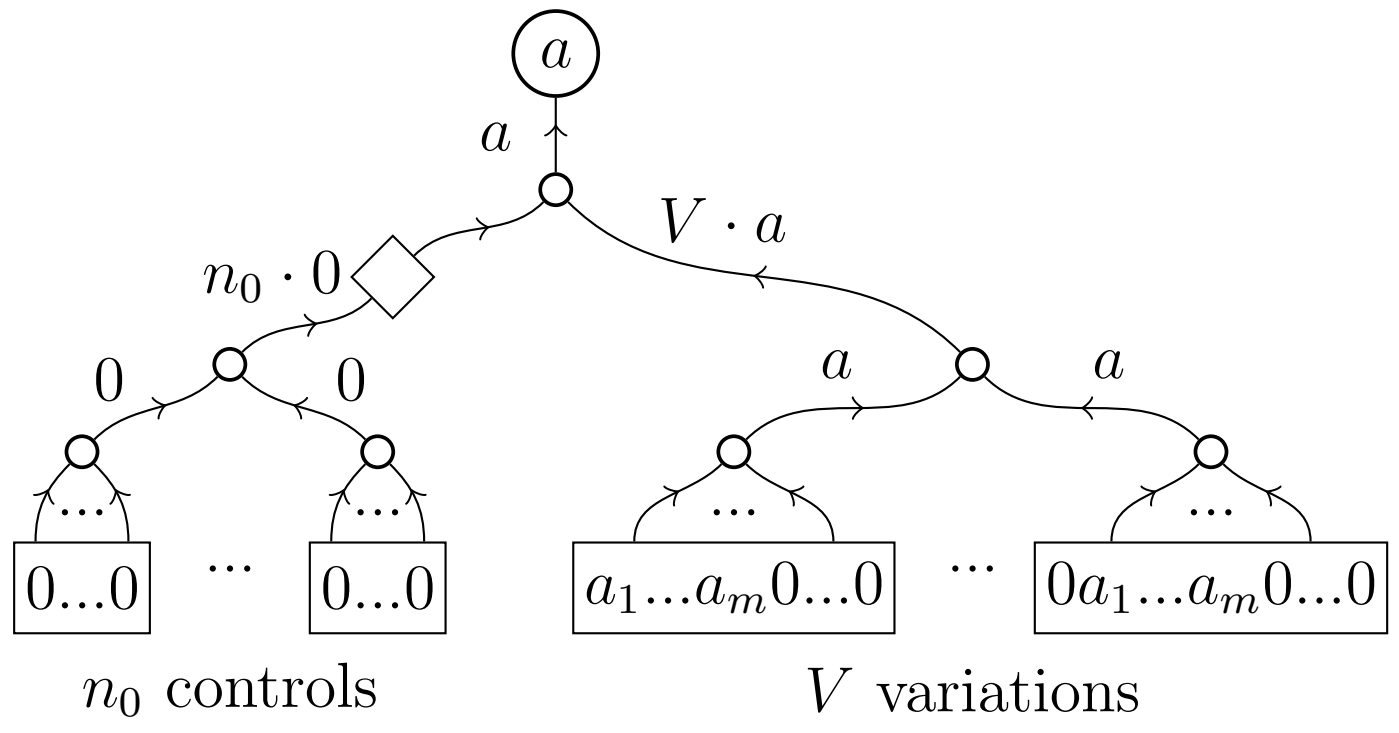

(5.11)

The same procedure applied to any local hidden variable model always yields the 0 scalar. A local hidden variable model is nothing but the local map for the scenario applied to some state, so it is enough to show that the above procedure yields the constant 0 function when composed with the local map:

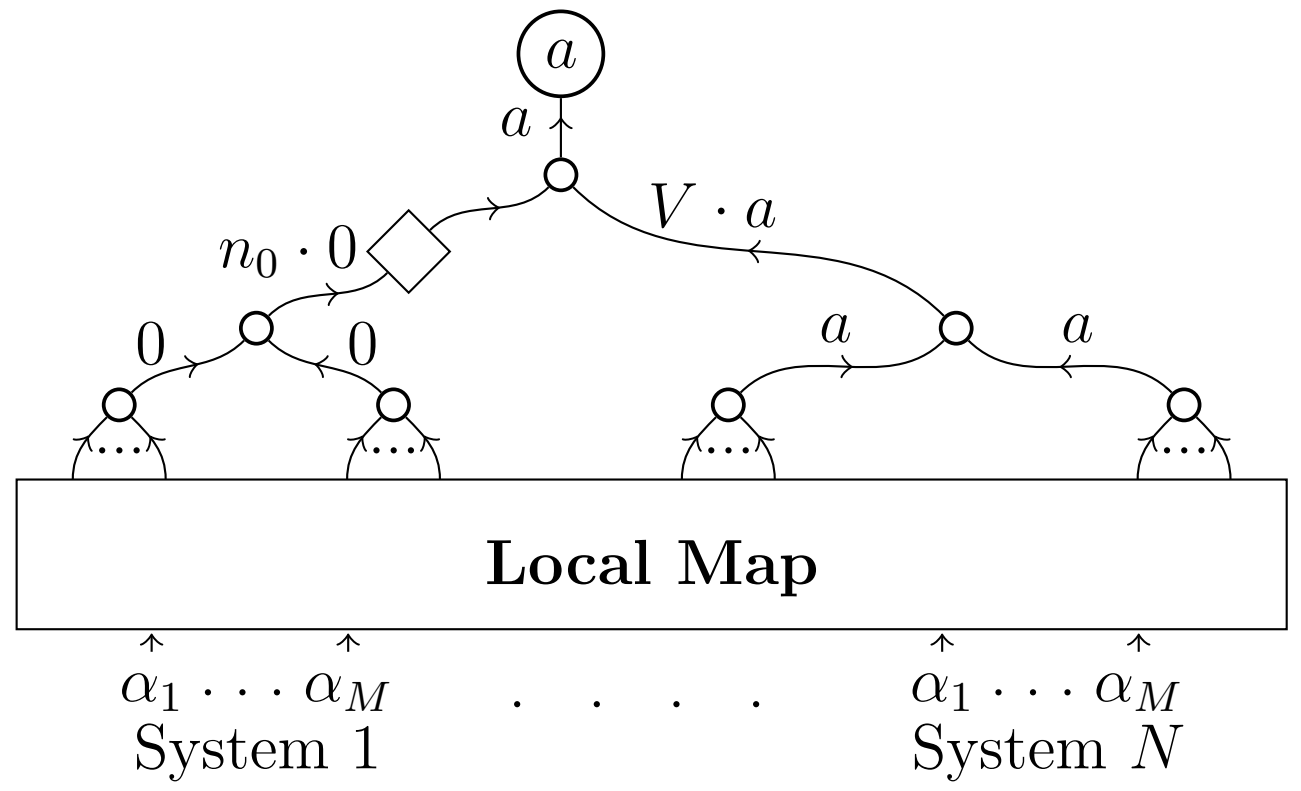

(5.12)

Since the ●-classical points form a basis, it is sufficient to show that the map from Diagram 5.12 always yields 0 when applied to ●-classical points. In the following diagram, the ○ nodes have been re-arranged using the spider theorem, so that the wiring of the local map can be written down explicitly in a clean way. The diagram also annotates the ●-classical values on the wires at each stage to aid in following the argument:

1. the values $b_0^1, ..., b_0^V$ for the 0 phases of systems 1 to $V$ are each duplicated $n_0 + n_0$ times and then added up to $b_0 := n_0 \cdot \sum i = 1^V b_0^i$ by the two ○ nodes

2. the values $b_1^i, ..., b_m^i$ for the $a_1, ..., a_m$ phases of each system $i = 1$ (for $i = 1, ..., V$) are each duplicated $n_k$ times (for $k = 1, ..., m$) and added up to $b^i := \sum_{r=1}^{m} n_r \cdot b_r^i$ by the respective ○ nodes

3. the values $b^1, ..., b^V$ are added up to $b := \sum_{i=1}^{V} b^i$

4. the value $b_0$ is added up to $b$
5. finally, the value $b_0$ is subtracted from $b$, and $b$ is tested against the $\bullet$-classical point $\langle a|$, obtaining the scalar $\langle a|b\rangle$ (which we want to be zero)

The steps are summarised by the following diagram:

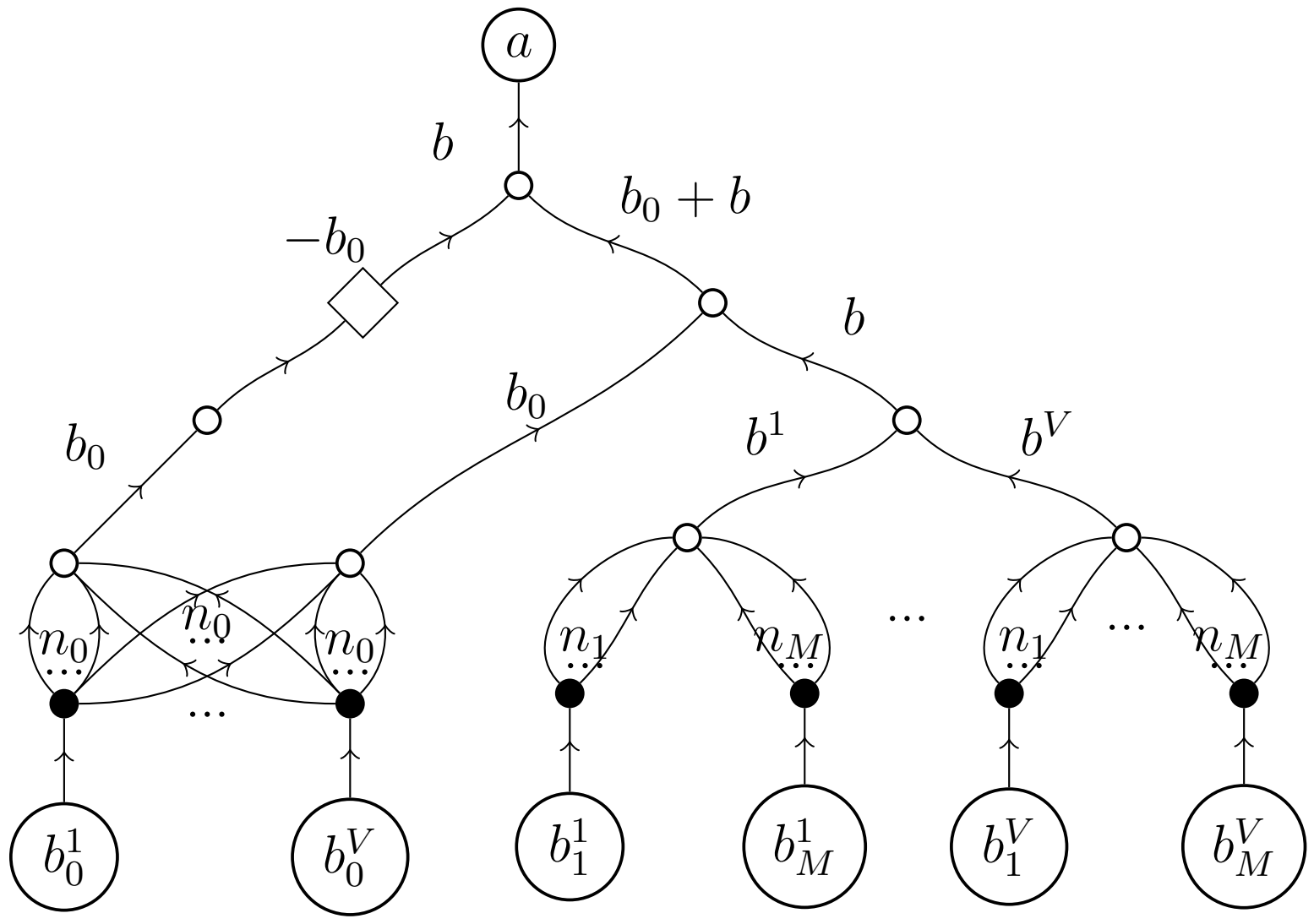

(5.13)

The $\bullet$-classical points $c$ that can be written as $c = \sum_{r=1}^{M} n_r \cdot c_r$ for some $\bullet$-classical points $c_1, ..., c_M$ form a subgroup $H$ of the group of $\bullet$-classical points. Indeed we have that $0 = \sum_{r=1}^{m} n_r \cdot 0$ and that $(\sum_{r=1}^{M} n_r \cdot c_r) + (\sum_{r=1}^{M} n_r \cdot d_r) = \sum_{r=1}^{M} n_r \cdot (c_r + d_r)$. Furthermore, by assumption we have that $H$ does not contain $a$, and as a consequence $\langle a|c\rangle = 0$ for all $c \in H$. Going back to Diagram 5.13, we see that $b^1, ..., b^V \in H$ (but $b_0$ need not be in $H$, hence the need to subtract it before testing against $a$). We thus conclude that $b \in H$ (since $H$ is closed under addition): hence the scalar $\langle a|b\rangle$ vanishes, concluding our proof that no local hidden variable can exist for our chosen measurement scenario. $\square$

**Theorem 5.3.8** (Mermin Locality)**.** *Let* $\mathbf{C}$ *be a* $\dagger$*-SMC. If for any strongly complementary pair* $(\circ, \bullet)$ *of* $\dagger$*-qSCFAs the group of* $\circ$*-phases is a trivial algebraic extension of the subgroup of* $\bullet$*-classical points (i.e. if there exist no algebraically non-trivial* $\circ$*-phases), then* $\mathbf{C}$ *is Mermin local.*

*Proof.* Consider an $N$-partite Mermin measurement scenario $\underline{\alpha}^s = (\alpha_1^s, ..., \alpha_N^s)_{s=1,...,S}$, and let $a_1, ..., a_M$ be the distinct $\circ$-phases appearing in it. Consider the system of equations $(\sum_{r=1}^{M} n_r^s \cdot x_r = c^s)_{s=1,...,S}$, where $n_r^s$ is the number of times phase $a_r$ appears in measurement $\underline{\alpha}^s$, and $c^s$ are the unique values making $x_r := a_r$ into a solution for the system. As the group of $\circ$-phases is a trivial algebraic extension of the subgroup of ●-classical points, there is a solution $x_r := b_r$ with $(b_r)_{r=1,...,M}$ ●-classical points. By using this, together with Lemma 5.2.4, we see that each measurement in the scenario is equal to the Mermin measurement obtained by replacing $a_r$ with $b_r$ for all $r = 1, ..., M$ (say $\beta_i^s := b_r$ if $\alpha_i^s = a_r$):

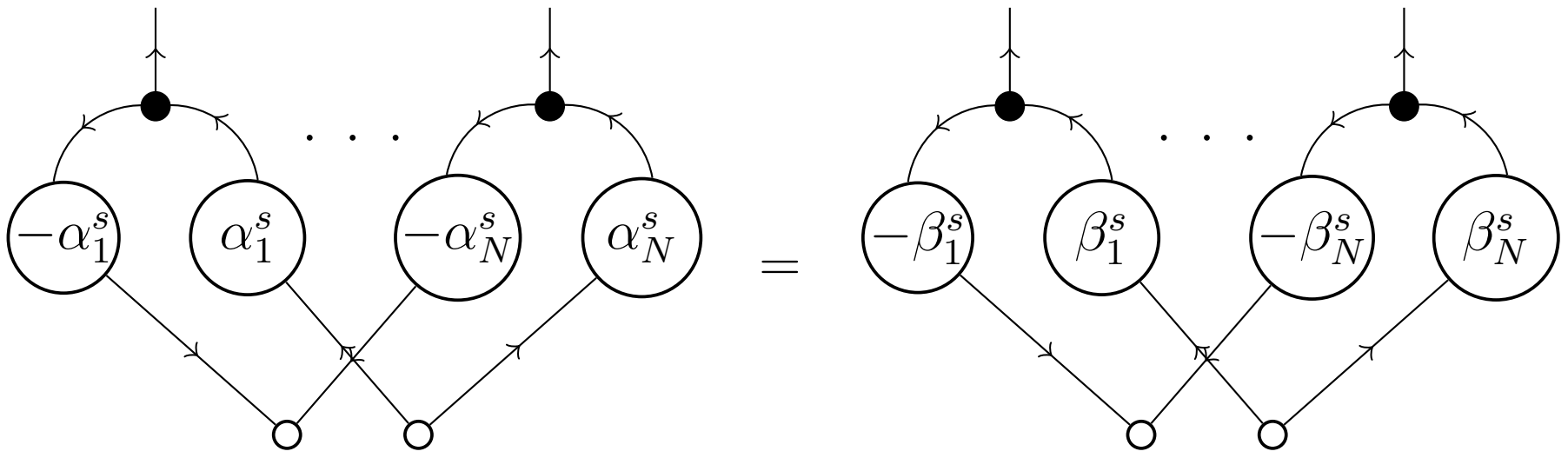

(5.14)

All phases are now induced by ●-classical points, and can thus be pushed up through the $\curlywedge$s:

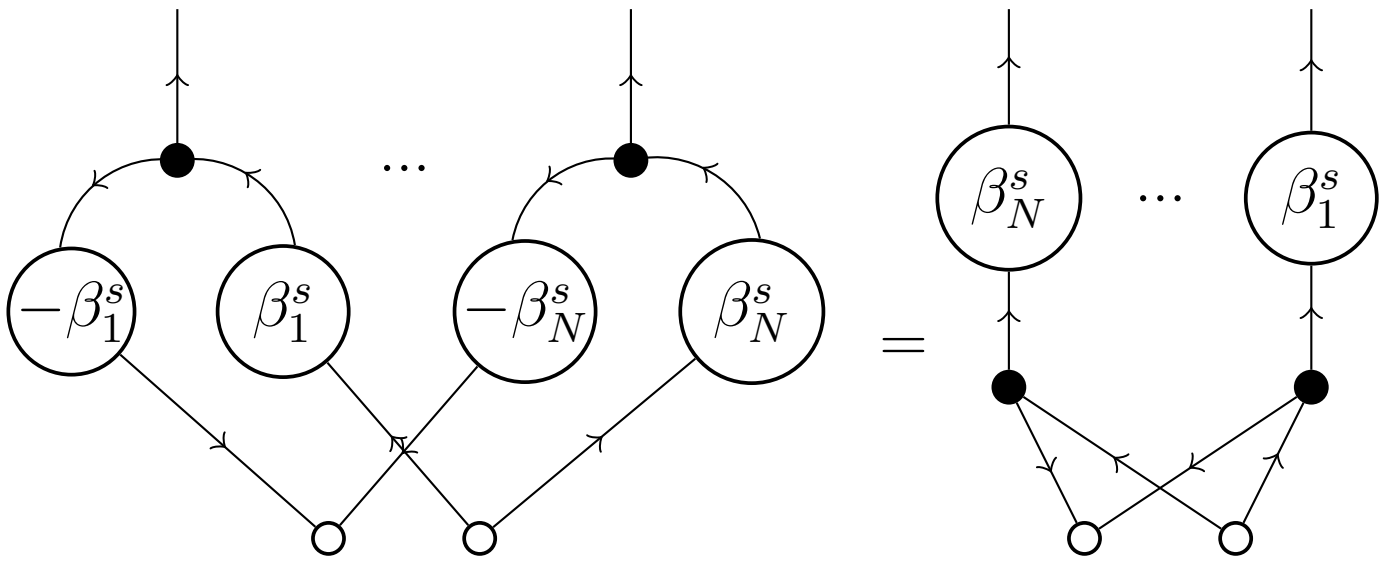

(5.15)

Now that each measurement of the scenario amounts to performing some set of ●-classical operations on the same state, it is no surprise that the following gives a local hidden variable model:

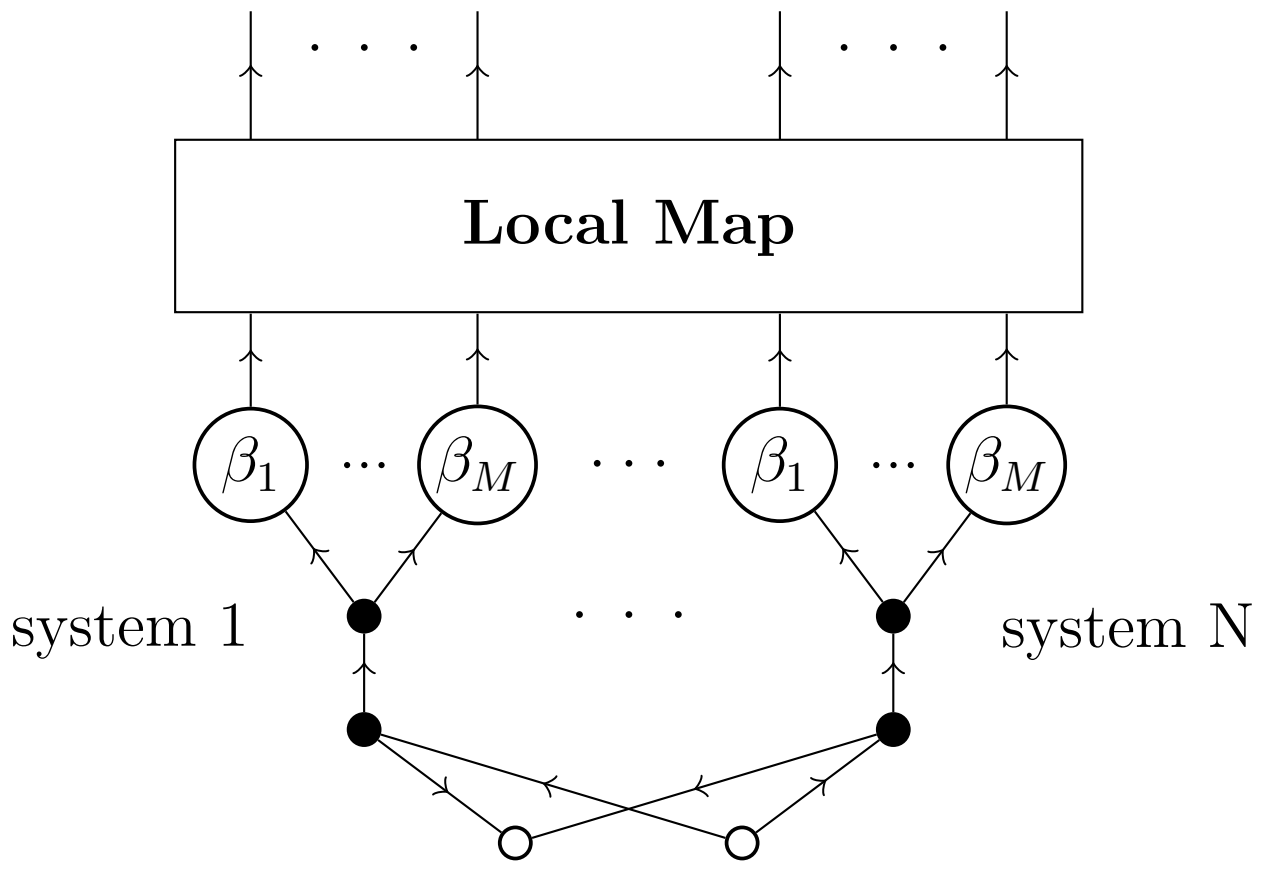

(5.16)

□

The abstract framework can now be applied to some particular examples of interest.

**Corollary 5.3.9.** *The restricted ZX calculus from [11, 31] for qubit stabilizer quantum mechanics (referred to as Stab in [32]) is Mermin non-local.*

*Proof.* Take ○ and ● to be the $Z$ and $X$ single-qubit observables in the ZX calculus. Then the group of ●-phases is $\mathbb{Z}_4$ and the subgroup of ●-classical points is $\mathbb{Z}_2$. Conclude using Theorem 5.3.7 and Example 5.3.4. □

**Corollary 5.3.10.** *The toy theory Spek from [32] is Mermin local.*

*Proof.* Same setup as in the previous corollary, but the phase group is now $\mathbb{Z}_2 \times \mathbb{Z}_2$. Conclude using Theorem 5.3.8 and Example 5.3.6 with $d = 2$. □

**Corollary 5.3.11.** *Qutrit stabilizer quantum mechanics from [90] is Mermin local.*

*Proof.* The phase group here is $\mathbb{Z}_3 \times \mathbb{Z}_3$. Conclude using Theorem 5.3.8 and Example 5.3.6 with $d = 3$. [1] □

**Corollary 5.3.12.** *The category* **FRel** *of finite sets and relations is Mermin local.*

*Proof.* See Section 4.3.2 for more details on strong complementarity in **FRel**. Any †-qSCFA on a set $\mathcal{H}$ in **FRel** is a groupoid: we write it in the form $\oplus_{h \in H} G_h$, where $H$ is a set, $G_h$ are disjoint groups and $\cup_{h \in H} G_h = \mathcal{H}$. Any strongly complementary pair ○, ● is in the form $(\oplus_{h \in H} G, \oplus_{g \in G} H)$, where both $G$ and $H$ are groups (seen as sets when indexing the groupoids), and we can w.l.o.g. write $\mathcal{H}$ as $G \times H$. Each ●-classical points is in the form $\{(g, h) \text{ s.t. } h \in H\}$ for some $g \in G$, while the ○-phases are in the form $\{(g_h, h) \text{ s.t. } h \in H\}$, for some family $(g_h)_{h \in H}$ of elements of $G$. Thus the group of ○-phases is the group $G^H$ of $H$-indexed vectors with values in $G$, and the subgroup of ●-classical points, isomorphic to $G$, is that of vectors with constant components. Conclude using Theorem 5.3.8 and Example 5.3.6. □

This latter result is particularly interesting for the following reasons:

1. Almost no †-qSCFAs in **FRel** have enough classical points (exactly one per space, out of a number that grows exponentially with system size).
2. The family of arguments from [32] fails in **FRel** (partially as a consequence of the previous point).

[1] This example was first constructed by Edwards in [41] without reference to the qutrit stabilizer formalism. This work also anticipated Example 5.3.6, using a specific construction.

3. There are plenty of strongly complementary pairs in **FRel**, and arbitrarily many more ○-phases than ● classical points, but the lack of *algebraically non-trivial* phases results in **FRel** being Mermin local.

4. As a consequence of point 3, quantum protocols relying only on Mermin non-locality will show no quantum advantage in **FRel**.

## 5.4 Quantum Secret Sharing: non-locality as a resource

The HBB CQ (N,N) family of Quantum Secret Sharing protocols originates in [72, 75], and has been abstractly formulated in a process theoretic context by Zamdzhiev [114]. We generalise this construction to arbitrary QPTs, leveraging its connection with Mermin non-locality.

This protocol requires a pair (○, ●) of strongly complementary observables, and an $(N+1)$-partite GHZ state shared by the dealer and the $N$ players. The dealer (and nobody else) knows the (classical) secret, in the form of a ●-classical point. The aim of the protocol is for the dealer to broadcast some information to all players on a public classical channel, and for the secret to be deterministically decodeable only if all $N$ players cooperate. The implementation, graphically summarised in (5.17), goes as follows:

1. the dealer and the players agree on a random set of ○-phases $\alpha_0, \alpha_1, ..., \alpha_N$ such that $\sum \alpha_j$ is some ●-classical point (call it $a$). This operation is done on a public channel.

2. the dealer measures his part of the system of the system with phase $\alpha_0$, and uses the resulting ●-classical data to encode the plaintext secret (classically adding the secret and the measurement data in the group $K_{\bullet}$; this generalises the original XOR operation, corresponding to $K_{\bullet} = \mathbb{Z}_2$ with addition mod 2) into a classical cyphertext. This operation is done locally and privately by the dealer.

3. the dealer broadcasts the cyphertext on a public classical channel to the players.

4. at some later stage, when they all agree to unveil the secret, the $N$ players measure their part of the system, each locally and privately.

5. all players broadcast the $\bullet$-classical results of their measurements on a public classical channel.

6. the broadcast results can be classically added in $K_{\bullet}$, then the result can be added to $a$ and finally to the cyphertext (again in the group $K_{\bullet}$) to recover the original $\bullet$-classical plaintext secret.

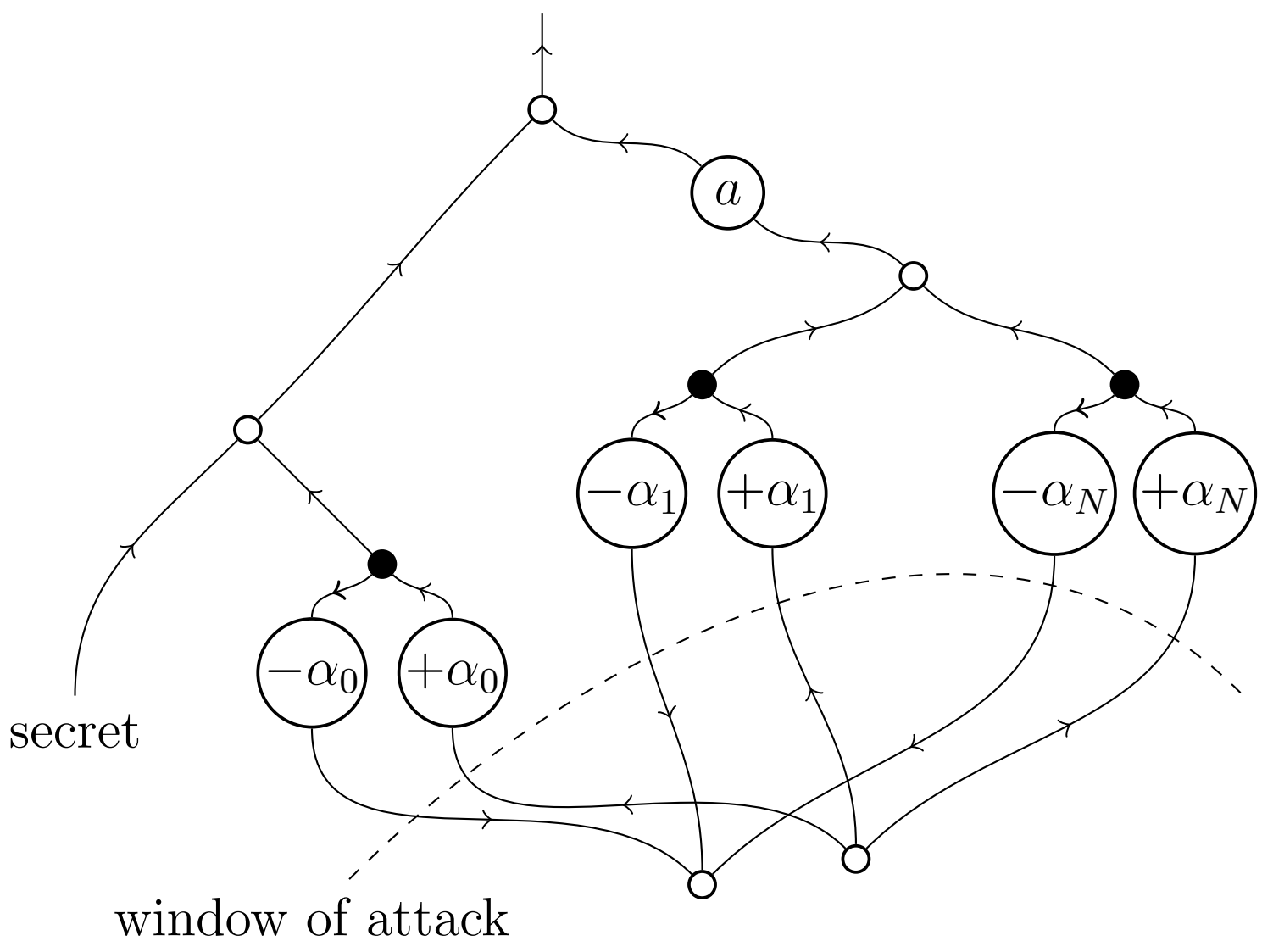

(5.17)

Most of the operations are either done locally and privately (all the measurements and the secret encoding), or broadcast by design on public classical channels, where one assumes that integrity of the message is guaranteed by appropriate classical protocols. There are many additional layers of quantum guarantees coming with this protocol, depending on the level of tampering allowed and on the phases chosen:

1. Assume no tampering happens anywhere. Then the refusal of (at least) one player to broadcast his or her measurement result makes the secret totally random to anyone else.

2. Assume that an attacker is allowed to tamper *only* with the GHZ state, and *before* the phases are chosen. Then the maximum amount of information she can gain is limited by (a) the random distribution on phases and (b) the amount of bias between the possible phases for each system. If $p_{max}$ is the highest probability appearing in the distribution of the phase choices (traditionally uniform with probability $1/2^N$)[2], and we let $k := |K_{\bullet}|$ be the dimensionality of the space (traditionally $k = 2$ for qubits), then optimal tampering reveals an average of $p_{max}$ $k$-its of classical information (on a secret of 1 $k$-it), in the case where the alternative measurements on each system are mutually unbiased (e.g. the traditional $X, Y$ pair). A more complicated failure expression can be worked out for arbitrary bases. This gain in information, however, is compensated by the introduction of a probability of failure for the entire protocol of $(1 - p_{max}) \cdot (1 - 1/k)$ (again in the mutually unbiased case), which can be detected by the players/dealer via statistical analysis of the outcomes.

[2]Not $1/2^{N+1}$, because of the parity requirement.

3. The kind of tampering allowed in the previous point does not give significant advantage to the attacker (at least for large number of players), and can be mitigated by appropriate statistical analysis of the measurement outputs; however, there is a stronger form of tampering that we can consider. Assume that the attacker is allowed to tamper with the GHZ state after the phases have been chosen, or even with the measurement devices of the dealer/player themselves, in a way that will ensure he knows the measurement outcomes with certainty beforehand; this is the model of attack assumed by device-independent security, pioneered by Barrett et al. [16]. Under this stronger model of attack, we can show that the protocol is secure if and only if the phases chosen by the players are algebraically non-trivial. Indeed, from the point of view of the dealer/players, the attack results in the measurement outcomes having a classical probability distribution:
   (a) if the phases are algebraically non-trivial, the probability distribution in the tampered case will never match, because of contextuality, that generated by the un-tampered protocol, and the attack can be detected by statistical analysis of the outcomes.
   (b) if the phases are algebraically trivial, on the other hand, they admit a probabilistic local hidden variable, and the attacker can generate her deterministic outcomes in a way to mimic the probability distribution of the un-tampered protocol.

To summarise, there are three distinct quantum resources playing complementary roles in the security of this protocol: the entanglement structure of the GHZ state, the amount of mutual complementarity of the available phases, and their algebraic non-triviality. Firstly, the entanglement structure of the GHZ state is the resource ensuring that the refusal of one player to cooperate results, if no tampering is allowed, into the inability for everyone else to recover the secret. Secondly, the amount of mutual complementarity of the available phases, e.g. the complementarity of the $X, Y$ pair, limits the maximum amount of information an attacker can gain by tampering with the state before phases are chose, and the minimum amount of disturbance introduced by the attack. Finally, Mermin non-locality, or equivalently algebraic non-triviality of the chosen phases, is the key resource ensuring device-independent security of the protocol.

## 5.5 Mermin in FHilb: beyond the complementary $XY$ pair

We now illustrate some corollaries of our results with a focus on **FHilb** and quantum mechanics. While in general we can have many different choices of measurement on each subsystem (see Definition 5.2.3), we shall restrict to the case of only two distinct measurements, i.e. $(N, M = 2, D)$ scenarios. In the case of qubits and $(N, 2, 2)$ scenarios, a pair of complementary measurements happen to be the only choices that will lead to a non-locality argument. One might then conjecture that the complementarity of the measurement pair will be the case for any dimension. In this section we show that this assumption is not the case. For $(N, 2, D)$ scenarios it is not necessary to have the two measurements be complementary, and there are many possible **Mermin effective** pairs in general.

**Definition 5.5.1.** A **two-measurement Mermin scenario** for $N$ systems (each with $D$ dimensions) and strongly complementary GHZ observable with $\circ$-phase group $G$ is denoted $G(N, 2, D)$. Each system has two possible measurement settings:

(a) the first measurement observable is the D-dimensional $X$ observable,

(b) and the second measurement observable $B$ is defined by a Z-phase gate applied to $X$.

In general, the form of $B$ can be specified by the $D$-dimensional Z-phase applied to $X$. This Z-phase is of the form $(1, e^{1b_1}, ..., e^{ib_{D-1}})^T$ with $D - 1$ degrees of freedom. A two-measurement Mermin scenario thus consists of $V$ variations each with $\beta$ measurements of the $B$ observable.

**Example 5.5.2.** For qubits there is only a single possible phase group: $\mathbb{Z}_2$. A Mermin argument for three qubits (denoted $\mathbb{Z}_2(3,2,2)$) has measurements of the usual $X$ observable and of the $B$ observable that is a phase applied to $X$, i.e. $\text{diag}(1, e^{ib_1})X$. In the traditional Mermin scenario $\mathbb{Z}_2(3,2,2)$ from [80], we have $V = 3$ and $\beta = 2$.

The state presented in Diagram 5.13 will be zero when the **control point** on the left is distinct from the **variations point** on the right. We can characterize this as a condition on $B$ in our two measurement scenario with the following theorem.

**Lemma 5.5.3.** *Measurements $X$ and $B$ allow a $(N,2,D)$ Mermin non-locality argument iff*

$$\sum_{j=1}^{D-1} e^{ic_j} = -1, \qquad \text{where } c_j = b_j\left(\bigoplus_{l=1}^{V}\beta\right), \tag{5.18}$$

*where the sum in $c_j$ is the group sum for the* $\circ$*-phase group $G$.*

*Proof.* Diagram 5.13 implies that the Mermin argument will succeed when the control point and variations point are distinct classical points. In **FHilb** this precisely means that they are orthogonal vectors. The vector that represents the control point is given by the $D$-dimensional unit for the $X$ observable, i.e. $1/\sqrt{D}(1,1,...,1)^T$. The variations point is then given by the group sum of other classical points specified by their phase. The phase for each classical point is given by the sum of phase accumulated by each $B$ measurement. As there are $\beta$ such measurements in each variation, their sum is given by

$$\frac{1}{\sqrt{D}}\begin{pmatrix} 1 \\ e^{i\beta b_1} \\ \vdots \\ e^{i\beta b_{D-1}} \end{pmatrix}_1 \oplus \begin{pmatrix} 1 \\ e^{i\beta b_1} \\ \vdots \\ e^{i\beta b_{D-1}} \end{pmatrix}_2 \oplus ... \oplus \begin{pmatrix} 1 \\ e^{i\beta b_1} \\ \vdots \\ e^{i\beta b_{D-1}} \end{pmatrix}_V = \frac{1}{\sqrt{D}}\begin{pmatrix} 1 \\ e^{ic_1} \\ \vdots \\ e^{ic_{D-1}} \end{pmatrix},$$

where the constants $c_j$ are defined as in Equation 5.18. Orthogonality between the control and variations points then requires

$$\begin{pmatrix} 1 & 1 & ... & 1 \end{pmatrix}\begin{pmatrix} 1 \\ e^{ic_1} \\ \vdots \\ e^{ic_{D-1}} \end{pmatrix} = 0 \qquad \Rightarrow \qquad \sum_{j=1}^{D-1} e^{ic_j} = -1$$

This exactly recovers Equation 5.18 and completes the proof. □

Note that in Mermin's original scenario measurement observables were necessarily complementary, but that in general this is not the case.

**Theorem 5.5.4.** *In* $(3, 2, 2)$ *three qubit Mermin scenarios, the two measurements must be complementary.*

*Proof.* We have $V = 3$, $\beta = 2$, $G = \mathbb{Z}_2$ and $D = 2$. Thus

$$c_j = \beta b_j \left( \bigoplus_{l=1}^{3} 1 \right) = 2b_j(3 \mod 2) = 2b_j$$

so that our condition on $B$ becomes

$$\sum_{j=1}^{D-1} e^{ic_j} = e^{i2b_1} = -1 \Rightarrow b_1 = \frac{\pi}{2}$$

with only a single solution. This means that in this scenario there is only one measurement that could be used with $X$. This is the $Y$ observable and it is complementary to $X$. □

**Theorem 5.5.5.** *For* $(N, 2, D)$ *scenarios the measurements need not be complementary.*

*Proof of Theorem 5.5.5.* We prove this by counterexample. Consider the three dimensional ($D = 3$) five party Mermin scenario. The phase group of the non-local state is then given by $G = \mathbb{Z}_3$. The control measurement is given by five systems all measured by the $X$ observable, i.e. $XXXXX$. The variations are

$$\begin{array}{ccccc} BBBXX & BBXBX & BXBBX & XBBBX & XBXBB \\ BBXXB & BXBXB & XBBXB & BXXBB & XXBBB \end{array}$$

so that $V = 10$ and $\beta = 3$. We calculate the coefficients

$$c_j = \beta b_j \left( \bigoplus_{l=1}^{10} 1 \right) = 3b_j(10 \mod 3) = 3b_j$$

Observable $B$ must then satisfy $e^{i3b_1} + e^{i3b_2} = -1$. Any $B$ observable satisfies this condition if $b_2 = -\frac{i}{3} \log \left[ -1 - e^{3ib_1} \right]$. Consider $b_1 = \frac{2\pi}{9} \Rightarrow b_2 = -\frac{2\pi}{9}$ and calculate (for $\omega = e^{2\pi i/3}$):

$$B :: \begin{pmatrix} 1 & 0 & 0 \\ 0 & e^{i2\pi/9} & 0 \\ 0 & 0 & e^{-i2\pi/9} \end{pmatrix} \frac{1}{\sqrt{3}} \begin{pmatrix} 1 & 1 & 1 \\ 1 & \omega & \omega^2 \\ 1 & \omega^2 & \omega^4 \end{pmatrix} = \frac{1}{\sqrt{3}} \begin{pmatrix} 1 & 1 & 1 \\ e^{2i\pi/9} & e^{8i\pi/9} & e^{-4i\pi/9} \\ e^{-2i\pi/9} & e^{-8i\pi/9} & e^{4i\pi/4} \end{pmatrix}$$

Observable $B$ is clearly not complementary to $X$ by simply checking the dot products of their basis vectors. □

Further we can exhibit numerical results that calculate the number of Mermin effective measurement pairs available for a particular scenario. For a given number of parties $N$ we have calculated the number of effective pairs maximized over all viable variation choices. Typically these maximum values are found for variations where $\beta$ is maximized. Figure 1a shows pair counts for $\mathbb{Z}_2(N, 2, 2)$ scenarios. Here it appears that the number of effective measurement pairings grows approximately linearly with the number of parties. Figure 1b shows pair counts for the more complex $\mathbb{Z}_3(N, 2, 3)$ scenarios. It is clear that there are many (in some cases thousands) more available measurement configurations than just those given by complementary observables. This vastly expands the number of experimental setups that will generate, with certainty, a non-locality violation. Indeed, combining this result with those of Section 5.4 opens up a large class of quantum secret sharing protocols based on non-complementary measurements.

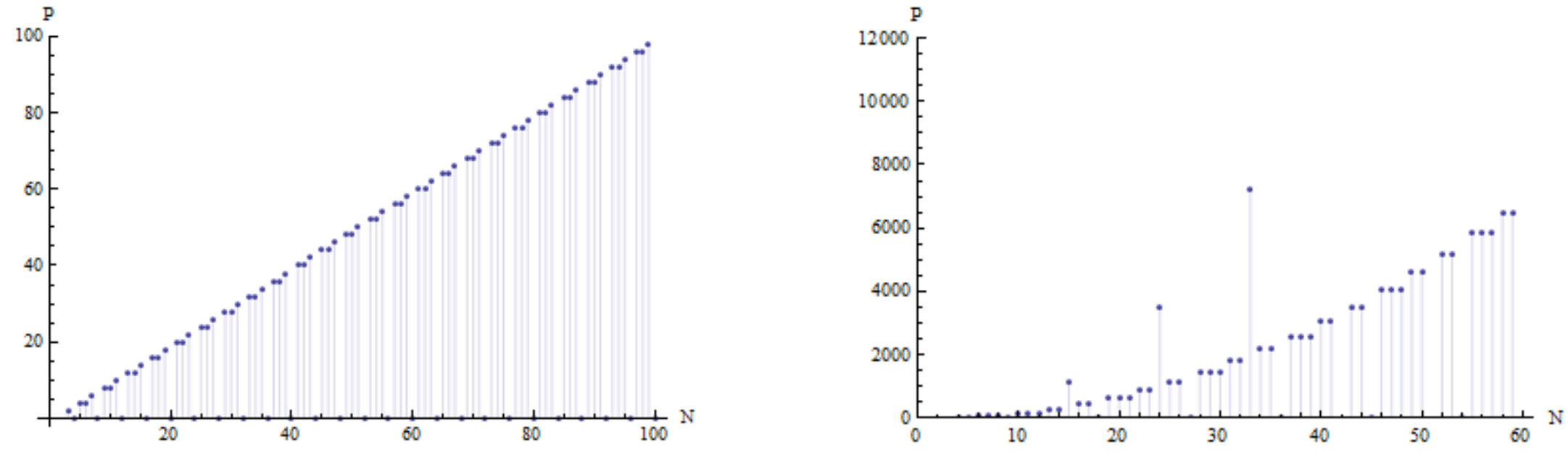

(a) (b)

Figure 5.1: (a) A plot of the number of Mermin effective measurement pairs $P$ vs. the number of parties in the Mermin scenario $N$ for $\mathbb{Z}_2(N, 2, 2)$ scenarios. (b) A plot of the number of effective pairs for $\mathbb{Z}_3(N, 2, 3)$ scenarios. These numbers were obtained by numerically counting solutions to (5.18).

## 5.6 Conclusions and future work

By using few, simple ingredients — †-SMCs, strongly complementary pairs, GHZ states, phases and classical points — we have generalised Mermin measurements to arbitrary GCTs. We have defined Mermin non-locality, and we have proven that a necessary and sufficient[3] condition for it is the existence of algebraically non-trivial phases, i.e. of phases which satisfy equations that classical points cannot. As a corollary, we have confirmed the well-known result that the stabilizer ZX calculus (and therefore **FHilb**) is Mermin non-local, and we have proven that **FRel**, a toy category of choice for quantum-like process theories, is Mermin local (despite its unboundedly large ratio of phases to classical points). This characterisation as the existence of certain phases opens the way to the treatment of Mermin non-locality as a resource in the abstract design of quantum protocols, as we have exemplified with the HBB CQ family of Quantum Secret Sharing protocols. Finally, the application of our general framework to Mermin-type experiment in quantum mechanics allows us to show that, even in the restricted case of two-measurement scenarios, complementary measurements are not necessary, leading to many more potential configurations than previously believed. We conclude with a few open questions for investigation:

(a) What are the minimal conditions under which algebraically non-trivial phases lead to non-locality?

(b) What is the exact connection between this framework and the one of Abramsky et al. [4] for generalised All-versus-Nothing arguments where measurement outcomes are elements of some general field?

(c) Is there a more informative group-theoretic formulation of the algebraic non-triviality used here?

(d) Our analysis focuses on non-locality paradoxes for a kind of GHZ state. It was recently shown by Tang et al. [102] that multipartite non-locality arguments can be constructed from any of a set of qudit graph states that they call GHZ graphs. What are the connections between these qudit graph states and the phase group formalism we present here?

(e) Which other quantum algorithms depend on Mermin non-locality as a resource to transcend classicality? Which process theories show these characteristics?

[3]Always necessary, sufficient under the assumption that classical points form a basis.

# Chapter 6

# Quantum Applications in Natural Language Processing

**Chapter Abstract**

We propose a new application of quantum computing to the field of natural language processing (NLP). Ongoing work in NLP attempts to incorporate grammatical structure into algorithms that compute meaning. In [28], Clark et al. introduce such a model (the CCS model) that is based on tensor product composition. While this algorithm has many advantages, its implementation is hampered by the large classical computational resources that are required. In this work we show how computational shortcomings of the CCS approach could be resolved using quantum computation. We address the value of a qRAM [45] for this model and extend an algorithm from Wiebe et al. [109] into a quantum algorithm to categorize similar sentences in CCS. Our new algorithm demonstrates a speedup over classical methods under certain conditions.

## 6.1 Introduction

As human computer interfaces become more advanced, natural language processing has grown to be a ubiquitous part of our world. Its techniques allow computers to understand natural language to perform tasks like automatic summarization, machine translation, information retrieval, and sentiment analysis. Most approaches to this problem, such as Google's search, understand strings of separate words in a 'bag of words' approach, ignoring any grammatical structure. This is certainly unsatisfactory, as we know that the meaning of a sentence is more than the meaning of its component words. Research in *distributional compositional semantics* (DisCo) seeks to address this by incorporating grammatical structure into bag-of-words models.

In [28], Clark et al. introduce a DisCo model (the CCS model) based on tensor product composition that gives a grammatically informed algorithm to compute the meaning of sentences and phrases. In the framework of this thesis, their model lives in the QPT **FVect**. While the algorithm has many advantages, its implementation is hampered by the large classical computational resources that it requires. This paper presents ways that quantum computers can solve some of these problems, making the CCS model of linguistics an attractive application for quantum computation.

We use the fact that quantum computation is naturally suited to managing high dimensional tensor product spaces. Recent literature has shown that quantum algorithms can thus provide advantages for machine learning [109, 91], inference [74], and regression [108, 105] tasks. These results are leveraged in two particular ways:

(a) We employ the scaling of quantum systems to address computational difficulties inherent in tensor-product based compositional semantics.

(b) Shared QPT structure makes algorithms in the CCS model especially amenable to quantum speedups. We specify a CCS sentence similarity algorithm that, under certain conditions, gives quadratic speedups for natural language tasks.

At an abstract level, we are taking a classical theory of NLP in the QPT **FVect** and computing outcomes in the QPT of quantum theory **FHilb**. The shared abstract structure makes the interface obvious.

In Section 6.2, we cover the basic framework of distributional compositional linguistics. Section 3 introduces the advantages of quantum representations for this framework. Sections 4 and 5 propose a quantum algorithm with quadratic speedup for calculating sentence similarity within CCS. Section 6 briefly discusses the noise tolerance of these methods.

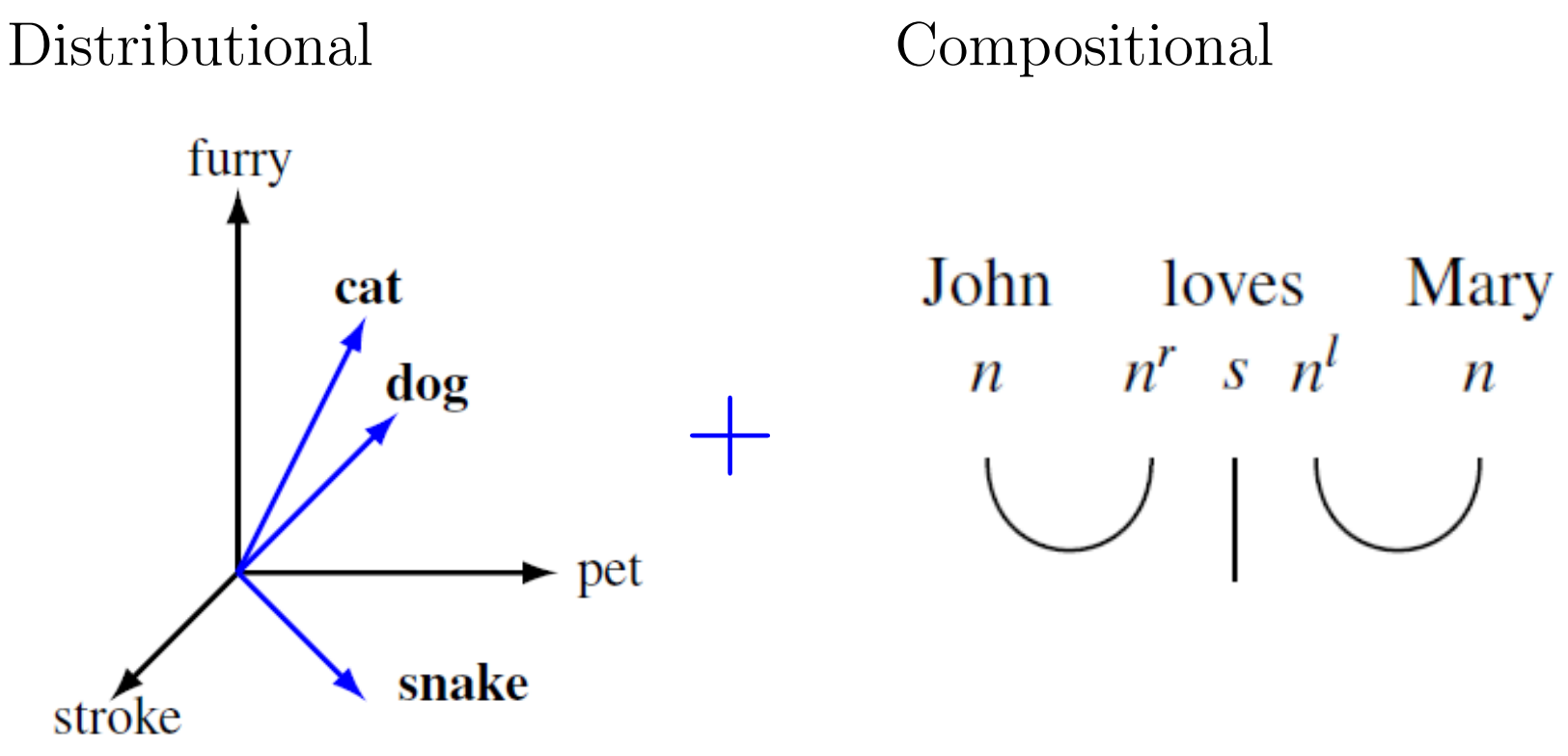

Figure 6.1: The DisCo approach combines word vectors with pregroup or combinatory categorical grammar. The diagram on the right shows which terms cancel in the derivation tree. It is drawn suggestively as explained in Section 6.2.

## 6.2 Distributional Compositional Semantics and the CSS model

In modern natural language processing, the **vector space model** is widely used to compute the meaning of individual words [97]. In this approach we first specify a set of context words, for example the 2000 most common words in a given corpus. These context words then form the basis of the vector space of word meanings in the following manner: for some given word, say "quantum", we look through a corpus and count the frequency with which each basis word appears 'near' to "quantum". It is likely that we would have a high frequency for words like "physics" and "information" for example. These frequencies then form the **word vector** for "quantum". Words are similar if the inner product of their word vectors is close to one. These 'bag of words' methods are typically referred to as **distributional**.

As the same sentence rarely occurs repeatedly, this distributional technique cannot be directly extended to calculate the meaning of longer phrases, sentences, paragraphs, etc. Instead, **compositional** semantics designs algorithms for deriving the meaning of a sentence or phrase from known meanings of component words, taking into account types and grammatical structure [73]. The **distributional compositional** semantic model (DisCo) combines both approaches to introduce grammatical form to the composition of word vectors [28]. See Figure 6.1.

In this model, each grammatical type is assigned a tensor product space based on Lambek's pregroup grammar [73] or combinatorial categorical grammar [55]. The meaning of nouns is, for example, calculated as in the distributional case, and we label their vector space $\mathcal{N}$. A transitive verb's meaning is then, following the grammar, a vector in the space $\mathcal{N} \otimes \mathcal{S} \otimes \mathcal{N}$, where $\mathcal{S}$ is the meaning space for sentences [28]. An intuition for this is that the transitive verb takes a subject noun as a left argument and an object noun as a right argument. An adjective lives in the space $\mathcal{N} \otimes \mathcal{N}$.

The QPT diagrammatic notation for vectors, tensors, and linear maps is commonly used in CCS. Here vertical composition (read bottom to top) represents composition of linear maps and horizontal composition represents the tensor product:

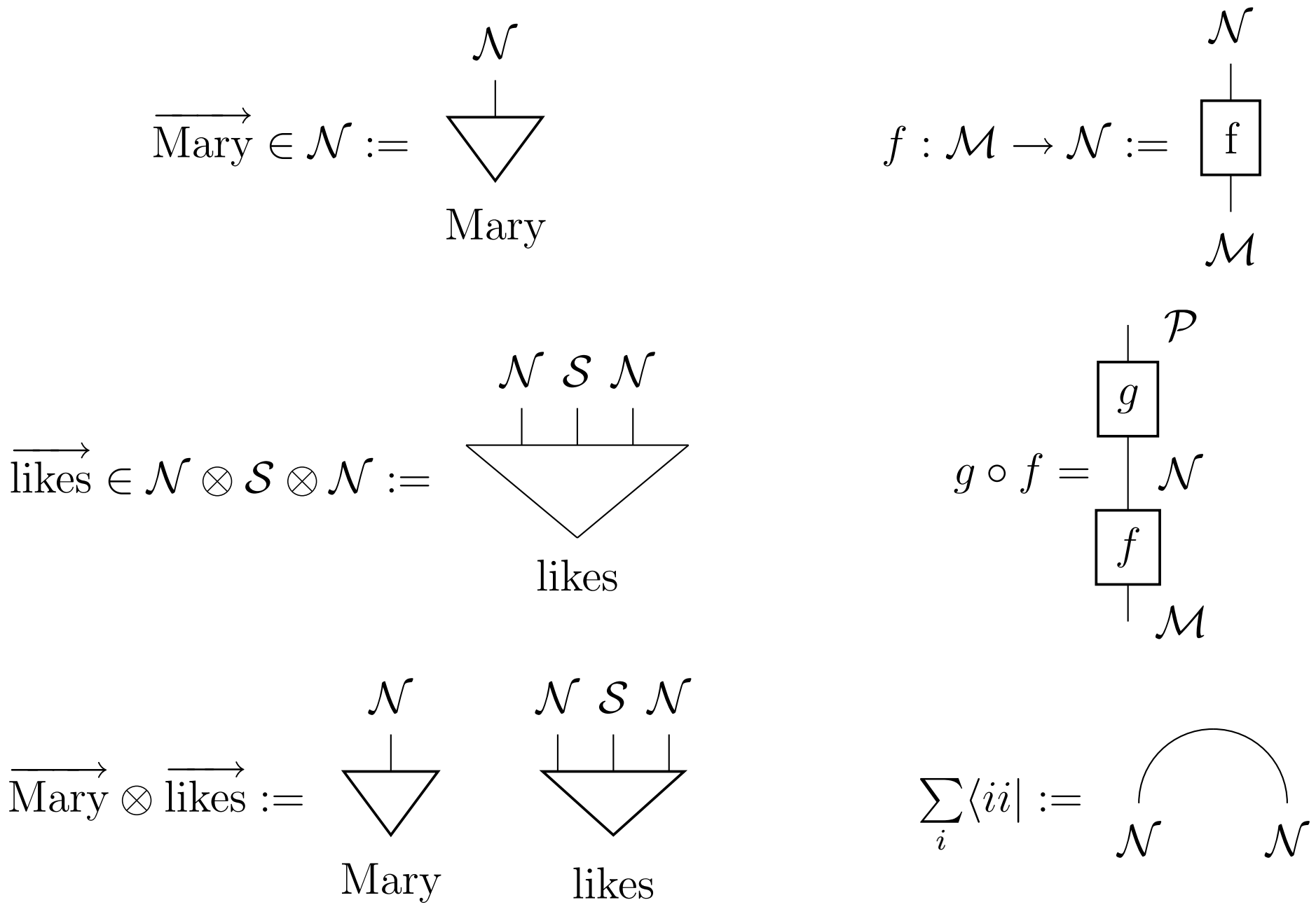

where $f : \mathcal{N} \to \mathcal{M}$ and $g : \mathcal{M} \to \mathcal{P}$ are linear maps and the linear map $\sum_i \langle ii|$ sums over all the basis vectors of $\mathcal{N}$ and is the usual QPT cap.

Given a well-typed sentence with meaning vectors $\vec{w_j}$ for each of $k$ words, the classical CCS algorithm for calculating a sentence's meaning is [27]:

(a) Compute the tensor product $\overrightarrow{\text{words}} = \vec{w_0} \otimes ... \otimes \vec{w_k}$ in the order that each word appears in the sentence.

(b) Construct a linear map that represents the type reduction by "wiring up" the vectors with the appropriate caps as in the following example:

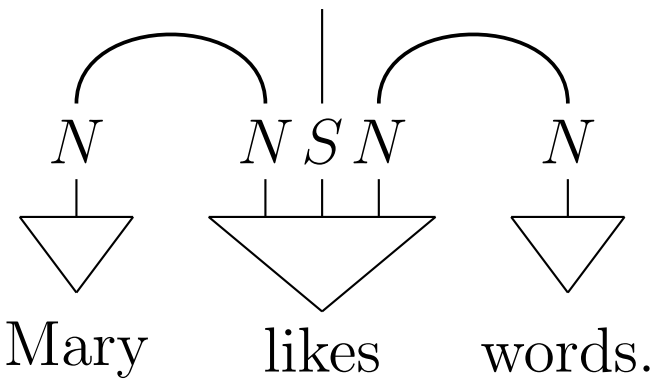

(6.1)

(c) Compute the meaning of the sentence by applying the linear map to the $\overrightarrow{\text{words}}$ vector. This results in a vector in $\mathcal{S}$ which corresponds to the meaning of the sentence.

We refer the reader to Coecke et al. [38] for a fuller description of the distributional compositional model and to Grefenstette and Sadrzadeh [50] and Kartsaklis [65] for experimental implementations.

These models suggest a promising approach to incorporate grammatical structure with vector space models of meaning, yet, as described by Grefenstette [49] they come with the computational challenges of large tensor product spaces. While there do exist some classical approaches to avoid the calculation of the full tensor product, such as the holographic reduced representations from Plate [88] or the use of dimensionality reduction by Polajnar et al. [89], these approaches introduce further assumptions and inaccuracies. For this reason, ameliorating the large computational costs introduced these large spaces through quantum computation is of particular interest.

## 6.3 Quantum computation for the CCS model

The most immediate advantage for quantum implementations of the CCS model is gained by storing meaning vectors in quantum systems. For $\alpha, \beta \in \mathbb{C}$ a two-level quantum system, a qubit, is defined by:

Qubit

$$\begin{aligned} |\psi\rangle &= \alpha|0\rangle + \beta|1\rangle \\ &= \alpha\begin{pmatrix} 1 \\ 0 \end{pmatrix} + \beta\begin{pmatrix} 0 \\ 1 \end{pmatrix} \end{aligned}$$

Qubits

$$|\psi_1\rangle \otimes |\psi_2\rangle = \begin{pmatrix} \alpha_1\alpha_2 \\ \alpha_1\beta_2 \\ \beta_1\alpha_2 \\ \beta_2\beta_2 \end{pmatrix}$$

and composition of qubits is given by the tensor product. This leaves each $n$-qubit system with $2^n$ degrees of freedom, indicating that $N$-dimensional classical vectors can be stored in $\log_2 N$ qubits. Consider a corpus whose word-meaning space is given by a basis of the 2,000 most common words. Even if we make the simplifying assumption that the sentence-meaning space is no larger than the word-meaning space we obtain the dramatic improvements details in Table 1.

| | One Transitive Verb | 10k tr. verbs |
|---|---|---|
| Classical | $8 \times 10^9$ bits | $8 \times 10^{13}$ bits |
| Quantum | 33 qubits | 47 qubits |

Table 6.1: Rough comparisons of the storage necessary for verbs in quantum and classical frameworks.

Further, these word meanings can be imported into a "bucket bridgade" quantum RAM that allows them to retrieved with a complexity linear in the number of qubits [45]. The general point is that because quantum systems compose via the tensor product they are a natural choice to store complex types and sentences that have the same compositional structure. We can then employ quantum algorithms on for natural language classification as presented in Section 6.5.

## 6.4 A quantum algorithm for the closest vector problem

Many tasks in computational linguistics such as clustering, text classification, phrase/word similarity, and sentiment analysis rely on computations that determine the closest vector to $\vec{s}$ out of some set of $N$-dimensional vectors $\{\vec{v}_0, \vec{v}_1, ...\vec{v}_{M-1}\}$. In clustering algorithms, for example, the set of vectors could be either the centroids of different clusters or the full set of training vectors, as in nearest neighbor clustering algorithms. Either the inner-product distance or Euclidean distance can be used. We will assume that all vectors are $N$-dimensional.

**Definition 6.4.1.** Given vector $\vec{s}$ and a set of $M$ vectors $U = \{\vec{v}_0, \vec{v}_1, ...\vec{v}_{M-1}\}$ the **closest vector problem** asks one to determine which $v_j$ has the smallest inner product distance with $\vec{s}$.

Direct calculation of the smallest vector would have complexity $\mathcal{O}(MN)$. In [109], Wiebe et al. introduce a quantum algorithm for this problem that, under certain conditions, demonstrate quadratic speedups over direct calculation and polynomial speedups over Monte-Carlo methods. Some proof of principle experiments have then demonstrated clustering of eight-dimensional vectors, based on these techniques, on a small photonic quantum computer [23]. This algorithm requires the following assumptions, where we write $v_{ji}$ for the $i^{\text{th}}$ entry of the $j^{\text{th}}$ vector:

(a) Vectors $\vec{v_j}$ and $\vec{s}$ are $d$-sparse, with no more than $d$ non-zero entries.

(b) The relevant vectors are stored in some kind of quantum memory so that the quantum computer can access their entries with the two oracles of the form:

$$\begin{aligned} \mathcal{O}|j\rangle|i\rangle|0\rangle &:= |j\rangle|i\rangle|v_{ji}\rangle, \\ \mathcal{F}|j\rangle|l\rangle &:= |j\rangle|f(j,l)\rangle, \end{aligned} \tag{6.2}$$

where $f(j,l)$ is the location of the $l^{\text{th}}$ non-zero entry of $v_j$. It is against these memory access oracles that the performance of our algorithm will be measured.

(c) $\max(|v_{ji}|^2) \leq r_{\max}$ for some known constant $r_{\max}$.

(d) All the vectors are normalized.

Under these assumptions we are able to run a quantum nearest-neighbor algorithm with complexity characterized by the following theorem:

**Theorem 6.4.2** ([109])**.** *We can find* $\max_j |\langle s \mid v_j\rangle|^2$ *with success probability* $1-\delta$ *and error* $\epsilon$ *using an expected number of* $\mathcal{O}$ *and* $\mathcal{F}$ *queries that is bounded above by*

$$1080\sqrt{M}\left\lceil\frac{4\pi(\pi+1)d^2r_{max}^4}{\epsilon}\right\rceil\left\lceil\frac{\ln\left(81M(\ln(M)+\gamma)\right)/\delta_0}{2(8/\pi^2-1/2)^2}\right\rceil, \tag{6.3}$$

*where* $\gamma \approx 0.5772$ *is Euler's constant.*

It is clear that for this quantum algorithm there is a quadratic improvement in scaling with $M$, the number of training vectors. While the dimension of the vectors $N$ is not explicitly included, in general it is implicitly there through the dependence on $d$. It is also clear that if $r_{\max} \propto 1/\sqrt{d}$, then the algorithm's dependence on both $d$ and $N$ drops out. As the vectors are normalized, this can be expected to occur if the vectors have sparsity that grows linearly with their size [109]. The authors further assume that for "typical" cases the error $\epsilon$ scales as $\Theta(1/\sqrt{N})$ so that the runtime for the quantum inner-product algorithm becomes $\mathcal{O}\left(\sqrt{NM}\ln(M)d^2r_{\max}^4\right)$.[1] This result shows a quadratic improvement over direct calculations and also shows improvement over Monte Carlo methods, whose complexity is $\mathcal{O}\left(NMd^2r_{\max}^4\right)$. These comparisons are summarized in Table 2.

| Type | Typical cases | Atypical cases |
|---|---|---|
| Classical Direct | $\mathcal{O}(NM)$ | $\mathcal{O}(NM)$ |
| Classical Monte Carlo | $\mathcal{O}\left(NMd^2r_{\max}^4\right)$ | $\mathcal{O}\left(Md^2r_{\max}^4/\epsilon^2\right)$ |
| Quantum | $\mathcal{O}\left(\sqrt{NM}\log(M)d^2r_{\max}^4\right)$ | $\mathcal{O}\left(\sqrt{M}\log(M)d^2r_{\max}^4/\epsilon\right)$ |

Table 6.2: Complexity comparisons for different closest vector algorithms. Adapted from [109].

In the following section we adapt this algorithm to sentence similarity calculations in the distributional compositional framework.

## 6.5 A quantum algorithm for CCS sentence similarity

The quantum algorithm from Section 6.4 can be used to advantage in natural language processing tasks however, the computational overhead of the CCS approach would dwarf this algorithm's advantages if it were naively applied. Throughout this section we will assume both that $r_{\max} \propto 1/\sqrt{d}$ and that the accuracy necessary for our calculation means $\epsilon$ scales as $\Theta(1/\sqrt{N})$. Consider the example of probabilistically classifying the meaning of a simple sentence. We illustrate this example with a noun-verb-noun sentence. The meaning of the nouns are vectors in an $N$-dimensional vector space and the meaning of the verb is a vector in the space $N \otimes S \otimes N$. We represent a derivation of the meaning of the full sentence with the following map:

$|\phi\rangle =$ 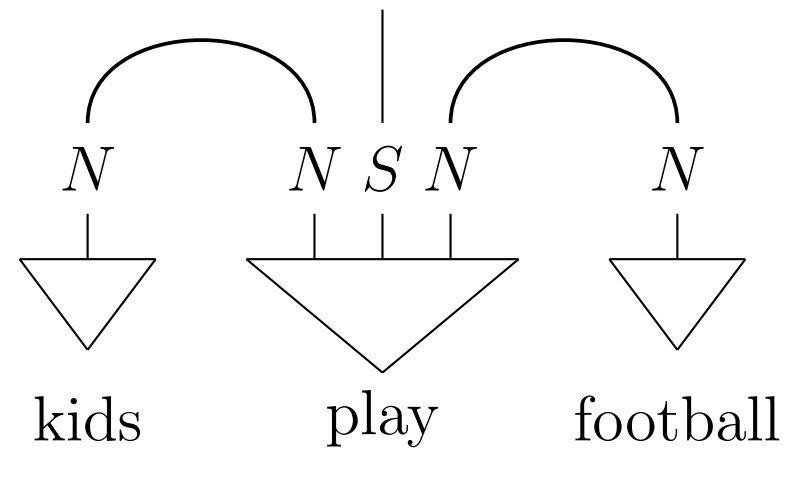

(6.4)

[1]This is argued for in Appendix D of [109] for a "typical" case where the vectors are uniformly distributed over the unit sphere.

From now on, we will take sentences to exist in the same meaning space as words, i.e. $S \simeq N$.

**Definition 6.5.1.** For meaning vector $\vec{s}$ and $M$ sets of meaning vectors, a **classification task** assigns $\vec{s}$ to the set containing the nearest-neighbor of $\vec{s}$.

An example task would be to determine if a sentence is about sports or politics or if a sentence expresses agreement or disagreement. If, to present a simplified example, we take each cluster to only contain a single vector ($\vec{v}$ and $\vec{w}$ respectively) then the sentence would be classified by computing

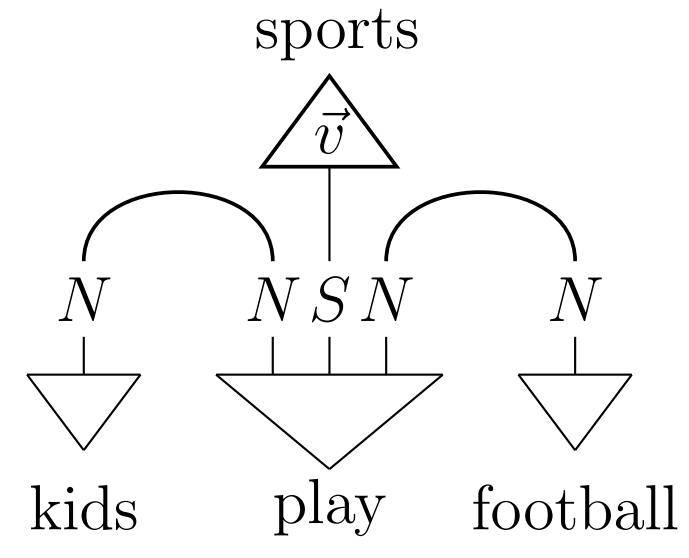

and

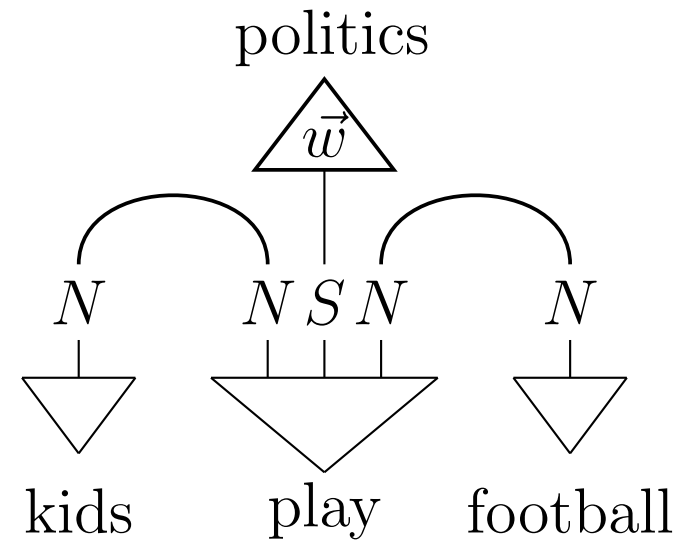

(6.5)

and assigning the sentence to the one of smaller value. We would proceed with two steps:

(a) Compute the derivation of $|\phi\rangle$, which, by classical direct calculation, takes $\mathcal{O}(3N)$ operations.

(b) See which of $\vec{v}$ and $\vec{w}$ is closest to $|\phi\rangle$. This is an instance of the closest vector problem where $\vec{s} = |\phi\rangle$, $M = 2$, and $U = \{\vec{v}, \vec{w}\}$. With direct calculation or Monte Carlo the second step requires[2] $\mathcal{O}(2N)$ to be compared with the quantum method at $\mathcal{O}(\sqrt{2N}\log 2)$. Even if we include the step to import the classical data from step one into quantum form, which can be done with $\mathcal{O}(\log_2 N)$ overhead [45], then we obtain a speedup for this step.

Still, despite the quantum speedup from step two, the full algorithm for general $M$ runs in $\mathcal{O}(3N\sqrt{M}\log M)$, remaining linear in $N$.

In order to recover a speedup we refine the application of the quantum algorithm by posing a version of the closest vector problem that avoids the

[2]If we assume the appropriate $d$-sparsity scaling.

initial calculation of $|\phi\rangle$ altogether. Note the equivalence of the calculations in Equation 6.5 with

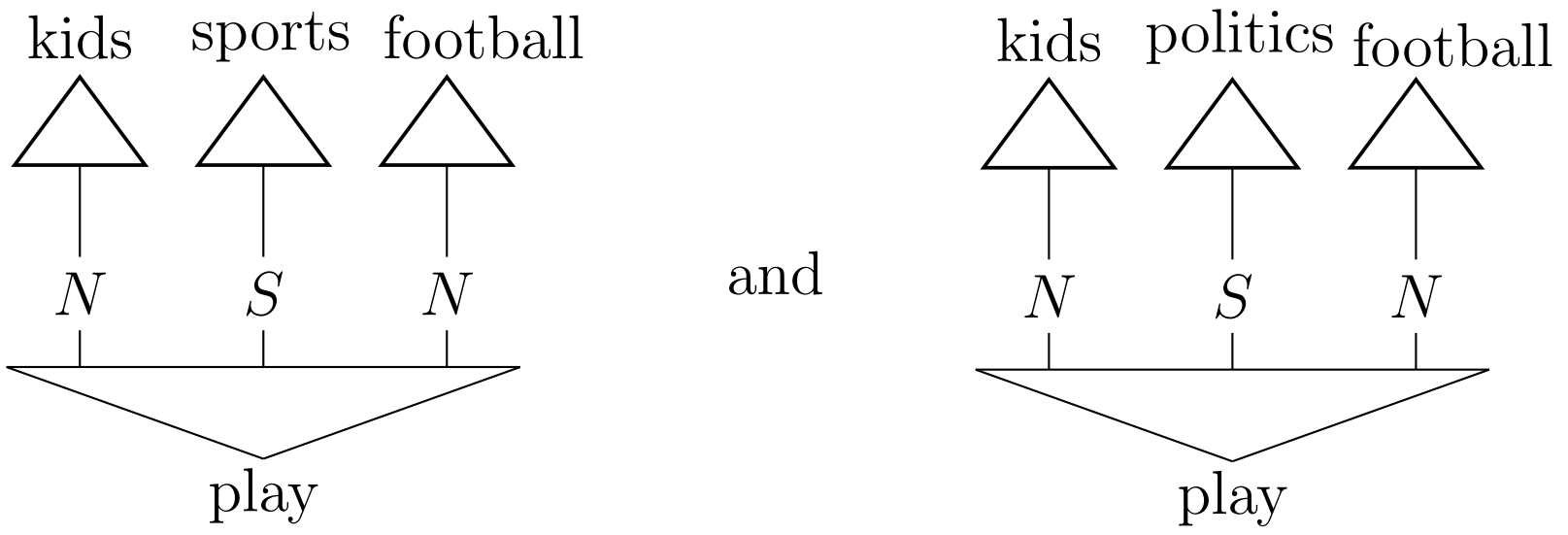

(6.6)

Rather than directly calculating $|\phi\rangle$, which is not relevant to the classification task, we can formulate a closest vector problem where $\vec{s} = |play\rangle$, $M = 2$ and $U = \{|kids\rangle \otimes |v\rangle \otimes |football\rangle, |kids\rangle \otimes |w\rangle \otimes |football\rangle\}$. The runtime of this **deferred quantum algorithm**, including import, will then be $\mathcal{O}(\sqrt{MN})$, showing our desired quadratic speedup in both variable.

We extend this to result to include general sentences in the CCS model with the following theorem.[3]

**Theorem 6.5.2.** *For an $N$-dimensional noun meaning space, there exists a quantum algorithm to classify any CCS model sentence composed of $n$ tensors $\vec{w_0}, \vec{w_1}, ..., \vec{w_{n-1}}$ whose maximum dimension is $N^k$ into $M$ classes with time $\mathcal{O}(\sqrt{MN^{k(n-1)/2}} \log M)$. This is not always an improvement on the classical algorithm which runs in $\mathcal{O}(MN)$, but will be if $k(n-1) < 4$ (very short sentence fragments) or if $M$ is much larger than $N$, thought this lasts case is unlikely in practice.*

*Proof.* The trick we play in Equation 6.6 amounts to splitting the sentence derivation into a bipartite graph. As the CCS connections are based on a pregroup derivation, the connections will always form a tree, taking words as nodes and connections as edges. Trees can always be partitioned into bipartite graphs, thus, up to the ordering of inputs on each tensor which can be kept track of, we can always give a deferred quantum algorithm with associated speedup for any such CCS sentence. The following procedure explicitly details how to construct this bipartite partitioning.

For every CCS sentence there is one word that acts as the **head** of the derivation. This is the word whose output $S$ wire contains the sentence meaning following its derivation's linear map. In Equation 6.4 this is the word "play". Connect the dangling wire of the head word $\vec{w_h}$ with the vector $\vec{v_i}$ against which similarity is being measured. Starting with this head word we then separate the sentence into a top layer and a bottom layer with the following steps. Assign the head word to the top layer. Place every word it is connected to on the bottom layer. Next for every word just assigned to the bottom, take all their connected words, which are not yet assigned, and assign them to the top. Continue this procedure while alternating top and bottom until all words are assigned. This gives a simple two-coloring of the derivation graph. □

[3]The authors would like to thank an anonymous reviewer for correcting an error in the original statement of the algorithm.

**Example 6.5.3.** Consider the following example sentence from [64]:

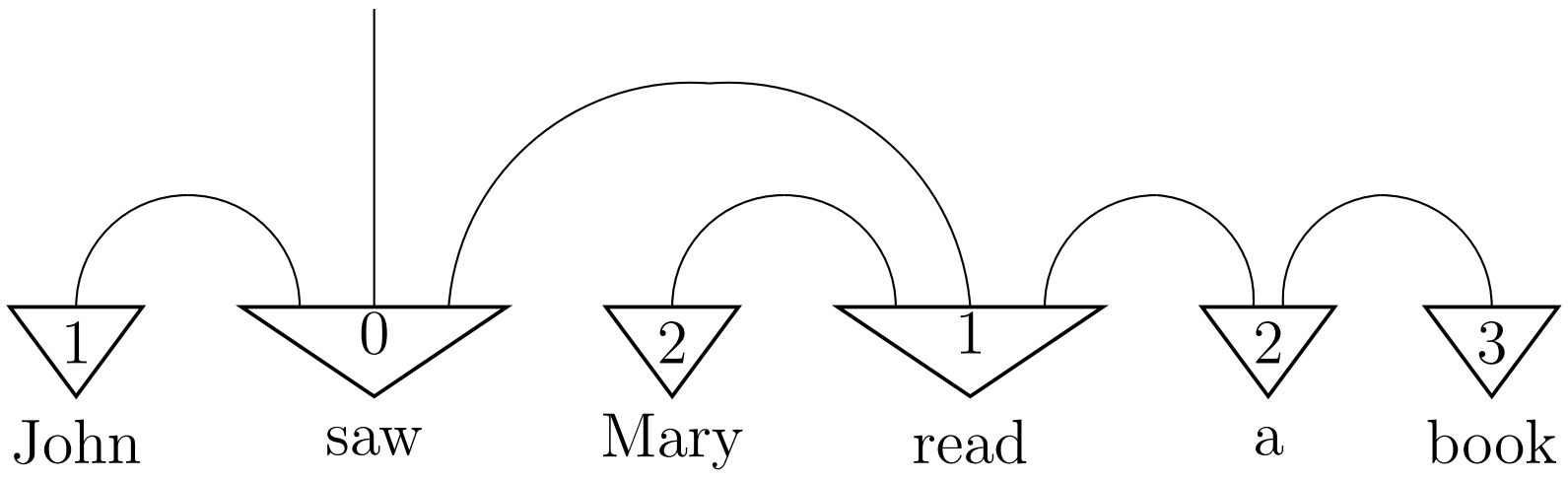

where we have labeled the vectors based on their depth in the derivation tree. The two-layer form assigns even vectors to the top layer and odd vectors to the bottom:

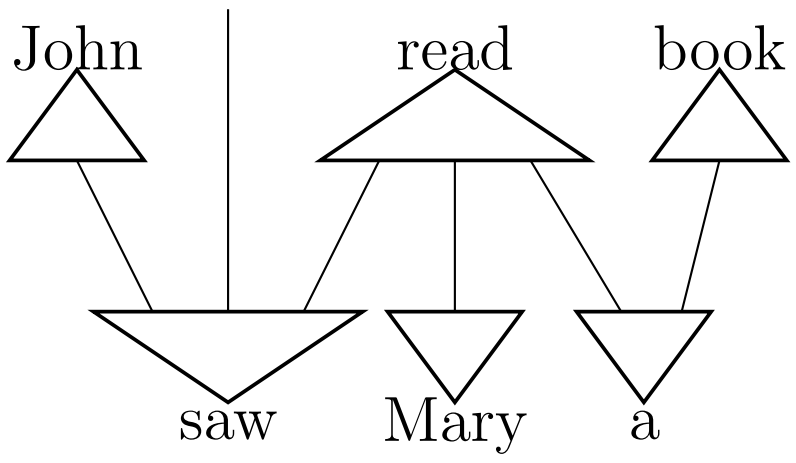

Hooking the dangling wire up to a classifying vector reduces the similarity computation to the calculation of a single inner product. Note that this procedure works for any derivation tree,[4] so sentence fragments, such as noun phrases, can be easily analyzed in exactly the same manner.

It is also very likely that other algorithms can be found, which decompose the sentence in other ways, and that give other time complexities.

[4]Even non-pregroup and non-CCG models will work as long as there is some tree derivation.

## 6.6 Noise tolerance and Conclusion

To reap the benefits of quantum algorithm in the domain of natural language processing, we apply a new technique to defer the calculation of a sentence's compositional meaning, eliminating the overhead costs. By this method we are able to introduce a quantum algorithm for calculating sentence similarity that offers quadratic speedup over classical direct calculation and Monte-Carlo methods. These kinds of algorithms are particularly attractive for practical applications of quantum computing as noisy results can be tolerated: in our case when the desired errors is lower bounded by $1/\sqrt{N}$. Vector space models are already inherently noisy and typical tasks allow for errors in results, so this restriction does not affect their efficacy.

An additional point is that the density matrix formalism of Piedeleu et al. [87] can also be naturally modeled by mixed states of quantum systems. In fact, this analogy was the genesis for the theory of disambiguation presented there, as another example of the shared structure that led to the results presented here. At a basic level, our work exploits the abstract connection between natural language processing and quantum information. More formally, we can see both quantum computation in the category of finite dimensional Hilbert spaces and linear maps [5, 57] and CCS in the product category of pregroup grammar and finite dimensional vectors spaces [38]. The connection between these two (as process theories) makes the application of one to the other apparent.

# Chapter 7

# Outlook

The goal of this thesis is to apply the abstract process theoretic framework for dagger symmetric monoidal categories to quantum algorithms and protocols. In the preceding chapters, we have presented results that leverage the structure of process theories to make the following original contributions:

(a) In Section 4.1, we used strong complementarity to construct the abelian Fourier transform over finite groups in arbitrary †-SMCs. This indicates that it is the presence of strongly complementary observables in quantum theory that makes the Fourier transform algorithm structurally native to quantum computation.

(b) In Section 4.2, we connected unitary oracles to complementary observables in arbitrary process theories. We then used these unitary oracles to construct a quantum algorithm for a new blackbox problem GROUPHOMID, with a speedup over classical algorithms for the problem, and investigated its quantum optimality.

(c) In Section 4.3, a quantum-like process theory in **Rel** is used to build models of the Deutsch-Jozsa, single-shot Grover's, and GROUPHOMID quantum blackbox algorithms in the sets and relations of non-deterministic classical computation. In the case of Grover's algorithm, this leads to a non-trivial classical algorithm.

(d) In Chapter 5, we characterized necessary and sufficient conditions for the Mermin locality (and Mermin non-locality) of an arbitrary quantum-like process theory. These results answered open questions regarding the connection between phase groups and non-locality. Further, we extended this framework to present new experimental tests of Mermin non-locality for any number of parties with access to an arbitrary number of measurements on systems of any size. We indicated some applications of these setups in quantum secret sharing.

(e) Lastly, in Chapter 6, we used the shared process theoretic structure of natural language processing and quantum information to adapt a general quantum machine learning algorithm into a domain specific sentence classification algorithm while maintaining a quantum speedup.

We conclude with a brief discussion of this work's future outlook.

## 7.1 Blackbox algorithms

Besides those already considered in this text, there are a slew of other algebraic blackbox problems for which there are known super-polynomial speedups. These are all candidates for study using our process theoretic framework. Such work would further clarify the relationships amongst these algorithms as well as open new avenues for generalization and combination.

In particular the hidden shift problem and hidden non-linear structure problems, as described by Childs and van Dam [25], look like promising places to start. In the hidden shift problem we are given a (not necessarily abelian) group $G$ along with two injective functions $f_0 : G \to S$ and $f_1 : G \to S$ for some set $S$ with the promise that

$$f_0(g) = f_1(sg) \qquad \text{for some } s \in G. \tag{7.1}$$

We are then tasked to find the hidden $s$. This problem is already known to be related to the hidden subgroup problem (HSP) in several ways: (i) for abelian $G$, then hidden shift is equivalent to the HSP[1] (ii) for non-abelian $G$ hidden shift is related to a HSP over a wreath product group and, when $G$ is a symmetric group, hidden shift is equivalent to testing the isomorphism of rigid graphs [25]. A polynomial time algorithm for the abelian hidden shift problem is not known, but there is a quantum speedup from $2^{\mathcal{O}(\log|G|)}$ to $2^{\mathcal{O}(\sqrt{\log|G|})}$ that can be achieved using the Kuperberg sieve [25, 71].

Within our QPT framework, we can easily represent the promise on the oracle. Let $\mathbb{G}$ be a system with an internal group of strongly complementary classical

[1] Take the hidden subgroup problem for $G \rtimes_\phi \mathbb{Z}/2\mathbb{Z}$ where the homomorphism $\phi : \mathbb{Z}/2\mathbb{Z} \to \mathrm{Aut}(G)$ is given by $\phi(0)(x) = x$ and $\phi(1)(x) = x^{-1}$.

structures ( $\circ$, $\bullet$ ) where $|s\rangle$ is a classical state of $\bullet$. Let $S$ be another system with a classical structure $\circ$. The functions $f_0, f_1$ are then self-conjugate comonoid homomorphisms $\mathbb{G} \to S$ promised to obey:

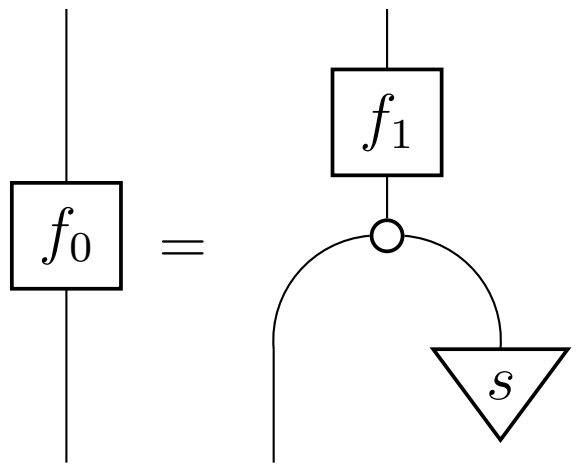

(7.2)

where $\circ$ is the group multiplication in $G$. This promise specification could be used to further analyze the known quantum hidden shift algorithms as well as propose further generalizations beyond the $M$-generalized hidden shift problems of [25]. In that paper Childs and van Dam emphasize that little is known about the non-abelian hidden shift problem in general and that may serve as a slightly easier version of the graph isomorphism problem in the case of symmetric groups. All of these facts make the hidden shift problem an attractive next step for process theoretic analysis, though there are competing candidates.

## 7.2 Complexity and Categories

Here we will propose directions for connecting the categorical reasoning used for verification of quantum algorithms to analyses of their computational complexity. In this thesis our analyses were limited in several ways:

(i) We were limited to query complexity analyses where we could manually count the number of query calls in a diagram.

(ii) We were limited to an analysis of deterministic protocols and lack a categorified notion of bounded error protocols. This limitation occurred both in our analysis of algorithms and in our analysis of non-locality protocols. Ideally we would have a notion of bounded error success internal to process theories.

(iii) In many ways because of the first issue, we are unable to directly connect the structure of process theories to the complexity class of problems that their processes can solve. This sort of result would be analogous to the connection between strongly complementary observables and Mermin non-locality tests from Chapter 5. At present we can only show what process theoretic structures enable particular *implementations* of algorithms as is done in Chapter 4. We would prefer that complexity classes which are efficiently accessible from different process theories should be directly related to their categorical structure.

While these are fundamental challenges, the perspective given in this thesis gives a foothold from which to begin addressing these important problems. One interesting avenue of approach would be to introduce some notion of repeatability and scaling into the diagrams of a process theory. Kissinger and Quick [68, 69] introduce a notion of !-boxes (pronounced bang boxes) to formalize the intuitive notion of the dot-dot-dot (...) that often appears in our morphisms definitions. These !-boxes appear as (sometime colored) boxes around elements of a diagram that can be arbitrarily repeated zero or more times, as in the following example:

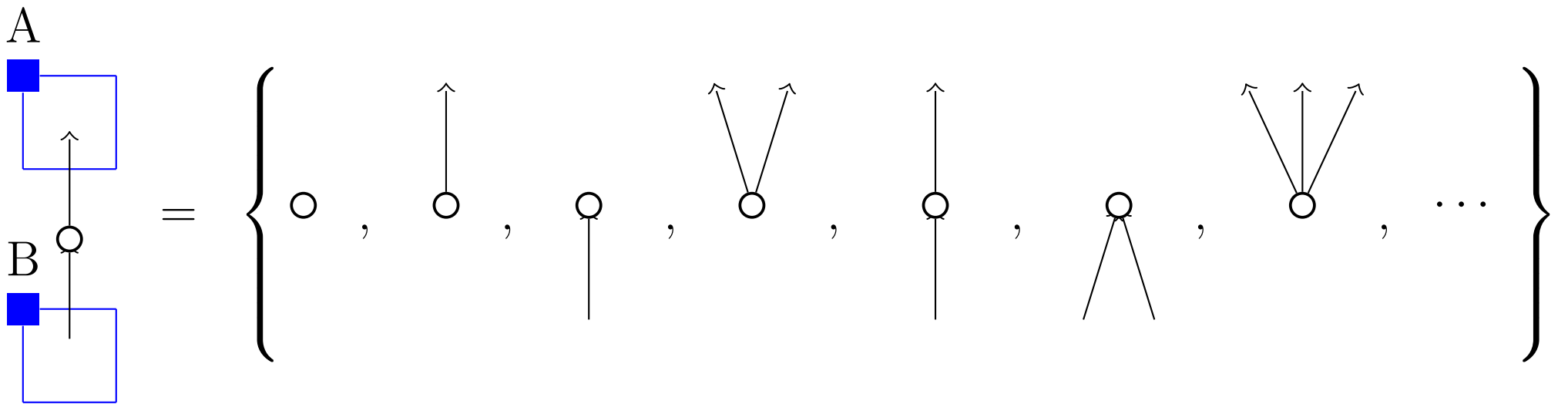

(7.3)

We could imagine labelled !-boxes with different asymptotic scalings and then defining their own set of rewrite rules, e.g.:

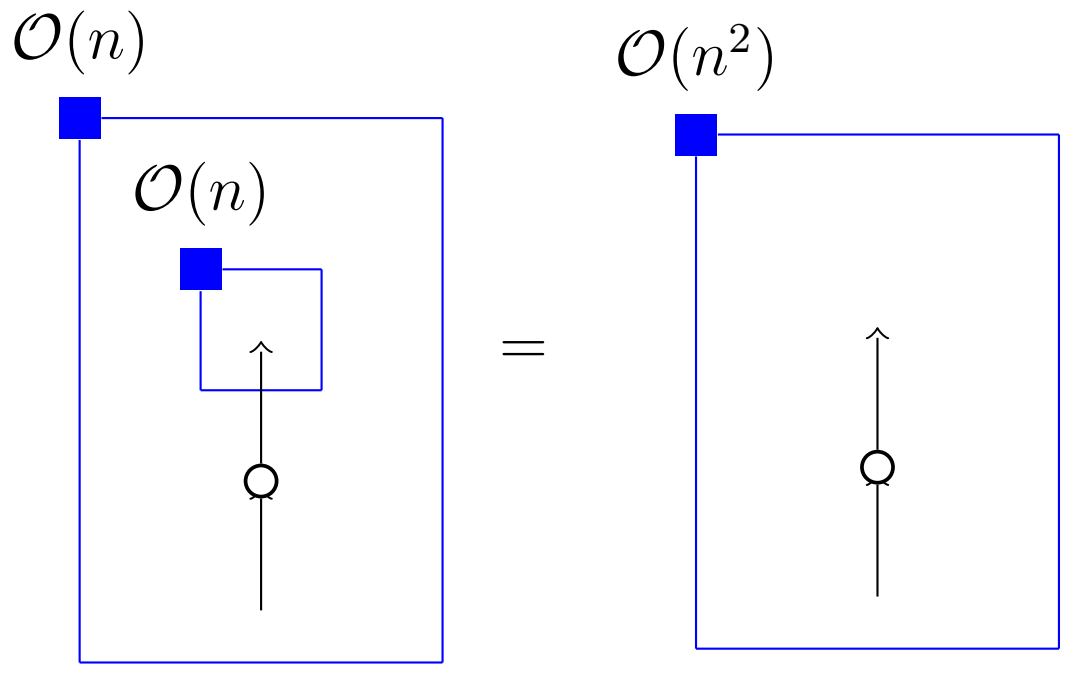

(7.4)

Another way of thinking of these scaling labels is as the enrichment of a process theory in some suitably defined *category of asymptotic scalings*. A process theory enriched in this way could then be related to categories of programs and algorithms as presented, for example, by Yanofsky [110]. Such an approach would begin to address the first and third limitations described. Further, another perspective on enrichment would allow the labelling of processes with their success probability, addressing the second limitation. While these are still speculative ideas, we hope this indicates that there are many different and concrete directions to expand process theoretic analysis of algorithms in general. This structural perspective adds to the investigation of the many open questions regarding the complexity class separations of different theories.

## 7.3 Quantum Programming Languages

The development of more powerful quantum programming languages will allow us to better control and apply rapidly developing quantum hardware. Prominent examples of current quantum programming languages such as Quipper [48] and LIQU$|\rangle$ [106] are based on the quantum circuit model. While this structural simplicity is useful initially, it quickly becomes unwieldy for complex quantum algorithms for both the users and in the hardware. For users this situation is analogous to classical programming using only boolean circuits and so clearly inefficient. Secondly, limitations in the first generation of classical control hardware for quantum processors make it advantageous to avoid having to load an entire fault-tolerant compiled circuit all at once. The ScaffCC framework is one recent approach to dealing with this problem of large *flattened* quantum circuit programs with sometimes $10^{12}$ gates for current applications [60]. If we can dynamically compile to quantum circuits from higher level concepts, then we can have access to larger algorithms. A first example of this comes from reversible computation, where methods can be "uncomputed" [112]. This would allow the loading of some circuit to implement the unitary $U$, along with a command to then run $U^\dagger$ that doesn't require an explicit circuit representation of $U^\dagger$ in advance. The control flow of these kinds of languages can be completely represented in structured reversible flow circuits [113], e.g. Figure 7.1.

These circuits live in not only a process theory, but a quantum-like process theory, where the reversibility corresponds to the existence of a dagger functor. The development of reversible quantum programming languages can be viewed as a first step beyond the process theory of quantum circuits (Section 2.2.1) and into quantum-like process theories (Chapter 3). This perspective gives a roadmap for including more of the structures from quantum-like QPTs into the design of higher-level quantum programming languages, e.g. classical structures, internal groups and representations, and CPM measurements. As we seek more applications for quantum systems, these improved languages we be transformative tools.

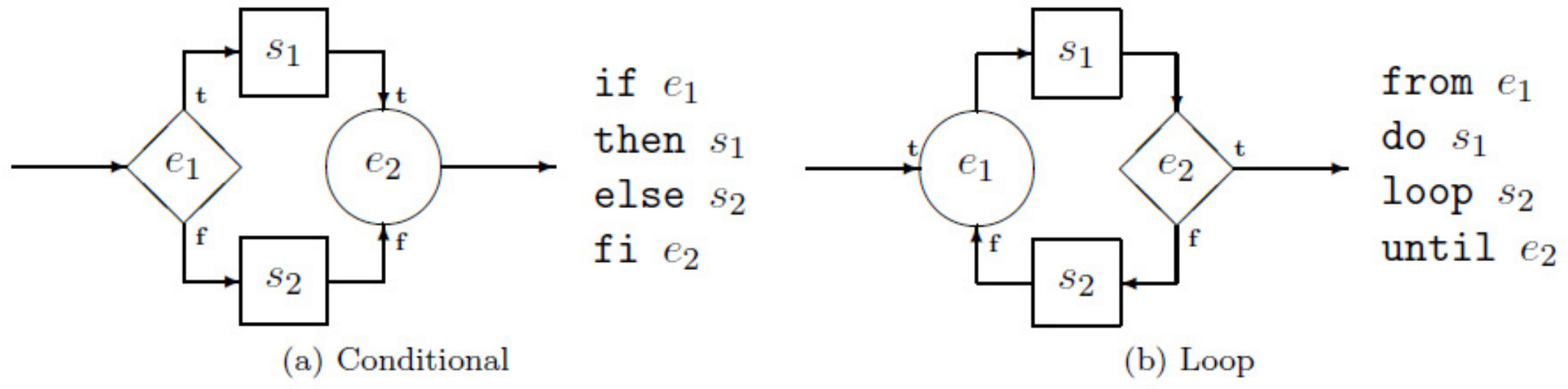

Figure 7.1: Examples of reversible control circuits. These circuits are read from left to right. Figure from [111].

# Appendix A

# Classical Relations

In this appendix we list examples of classical relations as calculated by a Mathematica package available at: https://github.com/willzeng/GroupoidHomRelations

Classical relations $\mathbb{Z}_3 \to \mathbb{Z}_3$:

{(0,0),(0,1),(0,2)}
{(0,0),(1,1),(2,2)}
{(0,0),(1,2),(2,1)}

Classical relations $\mathbb{Z}_4 \to \mathbb{Z}_4$:

{(0,0),(0,1),(0,2),(0,3)}
{(0,0),(1,1),(2,2),(3,3)}
{(0,0),(2,1),(0,2),(2,3)}
{(0,0),(3,1),(2,2),(1,3)}

The classical relations from $\mathbb{Z}_2 \oplus \mathbb{Z}_2 \to \mathbb{Z}_2 \oplus \mathbb{Z}_2$ are:

| | |
|---|---|
| {(0,2),(2,2),(1,3),(3,3)} | {(0,0),(1,1),(2,2),(3,3)} |
| {(0,2),(2,2),(1,3),(2,3)} | {(0,0),(1,1),(2,2),(2,3)} |
| {(0,2),(2,2),(0,3),(3,3)} | {(0,0),(0,1),(2,2),(3,3)} |
| {(0,2),(2,2),(0,3),(2,3)} | {(0,0),(0,1),(2,2),(2,3)} |
| {(2,0),(3,1),(0,2),(1,3)} | {(0,0),(2,0),(1,1),(3,1)} |
| {(2,0),(3,1),(0,2),(0,3)} | {(0,0),(2,0),(1,1),(2,1)} |
| {(2,0),(2,1),(0,2),(1,3)} | {(0,0),(2,0),(0,1),(3,1)} |
| {(2,0),(2,1),(0,2),(0,3)} | {(0,0),(2,0),(0,1),(2,1)} |

## A.1 Mathematica Code

```
(* This code generates groupoid homomorphisms in the category
 of relations *)
(* by William Zeng 2015 *)
(* william.zeng@cs.ox.ac.uk *)
(* Available at https://github.com/willzeng/GroupoidHomRelations *)

Groupoid Definitions

These multiplication tables define whatever groupoid
homomorphisms you are considering. It is prepopulated
with some usual small ones

(* Define the cyclic group multiplication *)
(* INPUT: Integers 1,1 *)
(* OUTPUT: Integer 2*)
z3[x_, y_] := ({
      {0, 1, 2},
      {1, 2, 0},
      {2, 0, 1}
     }) [[x + 1]][[y + 1]];
(* This defines multiplication on a set of inputs *)
(* INPUT: Integer sets {1,2},{2}*)
(* OUTPUT: Integer set {0,2}*)
z3Set[X_, Y_] := Union[Flatten[
    Table[
     Table[
      z3[X[[x]], Y[[y]]]
      , {x, 1, Length[X]}]
     , {y, 1, Length[Y]}]
    ]];
z4[x_, y_] := ({
      {0, 1, 2, 3},
      {1, 2, 3, 0},
      {2, 3, 0, 1},
      {3, 0, 1, 2}
     }) [[x + 1]][[y + 1]];
z4Set[X_, Y_] := Union[Flatten[
    Table[
     Table[
      z4[X[[x]], Y[[y]]]
      , {x, 1, Length[X]}]
     , {y, 1, Length[Y]}]
    ]];

(* a is undefined *)
z2z2[x_, y_] := ({
      {0, 1, {}, {}},
      {1, 0, {}, {}},
      {{}, {}, 2, 3},
      {{}, {}, 3, 2}
     }) [[x + 1]][[y + 1]];
```

```
z2z2Set[X_, Y_] := Union[Flatten[
    Table[
     Table[
      z2z2[X[[x]], Y[[y]]]
      , {x, 1, Length[X]}]
     , {y, 1, Length[Y]}]
    ]];

z2z2z2[x_, y_] := ({
       {0, 1, {}, {}, {}, {}},
       {1, 0, {}, {}, {}, {}},
       {{}, {}, 2, 3, {}, {}},
       {{}, {}, 3, 2, {}, {}},
       {{}, {}, {}, {}, 4, 5},
       {{}, {}, {}, {}, 5, 4}
      }) [[x + 1]][[y + 1]];
z2z2z2Set[X_, Y_] := Union[Flatten[
    Table[
     Table[
      z2z2z2[X[[x]], Y[[y]]]
      , {x, 1, Length[X]}]
     , {y, 1, Length[Y]}]
    ]];

(* Relational evaluation method *)
(* INPUT: Relation R = {{0,0},{1,1},{2,2},{2,3}};
and argument x=2*)
(* OUTPUT: List of Integers {2,3}*)
eval[R_, x_] := Map[#[[2]] &, Select[R, #[[1]] == x &]];
(* All possible relations *)
(*allRelations = Subsets[Tuples [{0,1,2,3},2]]; *)

Choose Groupoids

(*Set the size of the domain and coDomain*)
nD = 4; (*Size of the domain *)
ncoD = 4; (*Size of the coDomain *)

(* YOU NEED TO SET THE GROUPOIDS HERE*)
(* Test to see if monoid multiplication is preserved. *)
(* Need this and unitality condition to define a groupoid
homomorphism*)
(* INPUT: Relation R = {{0,0},{1,1},{2,2}};*)
(* OUTPUT: True or False *)
isMonHom[R_] := If[Length[Position[Flatten[
      Table[
       Table[
        (*Specify which groupoids are domain and
        coDomain here*)
        (*Use the multiplication version that is
        defined on sets of elements for the coDomain*)
        eval[R, z2z2[x, y]] == z2z2Set[eval[R, x], eval[R, y]]
        , {y, 0, nD - 1}]
       , {x, 0, nD - 1}]]
```

```
    , False]] == 0, True, False]

Print["Be careful. A full search requires checking ",
  2^(nD*ncoD), " relations."];

(* R = {{0,0},{1,1},{2,2}};*)
(* check which relations are comonoid homs*)
(*passed =Select[allRelations [[1;;1000]],
isCoHom[#]&
];
Grid[%]*)

Be careful . A full search requires checking 65536 relations .

Search for Monoid Homomorphisms

Print["Note that this outputs a Monoid Homomorphism relation. If you are interested in a classical relations (comonoid homomorphism relations), you will need to take the relational converse of this output.  See arXiv:1503.05857"];

(* All possible pairs of elements one in domain and one in
coDomain *)
pairings = Tuples[{Range[0, nD − 1], Range[0, ncoD − 1]}];

(* Specify the range of relations to search over *)
(* They are defined as binary strings from 0 to 2^(nD*ncoD)*)
(* and searched sequentially*)
(* BE SURE TO HARD CODE IN THE UNITALITY CONDITION *)
start = 0;
end = 2^(nD*ncoD); (* MAX is 2^(nD*ncoD)*)

results = {};
For[i = start, i < end, i++,
 (* We use binary numbers to index the different possible*)
 (* relations where a 0 or a 1 indicates if
 (*one of the nD*ncoD pairings is present *)
 binaryIndex = PadLeft[IntegerDigits[i, 2], nD*ncoD];
 relation = Map[pairings[[#]] &, Flatten[Position[binaryIndex, 1]]];

 (* Check the unitality condition *)
 (* BE SURE TO HARD CODE IN THE UNITALITY CONDITION HERE*)
 (* EXAMPLES *)
 (* Z3 −> 3 should be unitalCondition=Union[eval[relation,0]]=={0};*)
 (* Z4 −> Z4 should be unitalCondition=Union[eval[relation,0]]=={0};*)
 (* Z2Z2 −> Z2Z2 should be unitalCondition=Union[eval[relation,0],eval[relation
     ,2]]=={0,2}; *)
 unitalCondition = Union[eval[relation, 0], eval[relation, 2]] == {0, 2};

 (* Return True if both the unitality condition and preservation of monoid
mult are met*)
 If[unitalCondition && isMonHom[relation],
  Print[relation];
  results = Append[results, relation];
  ];
```

```
  ]

Note that this outputs a Monoid Homomorphism relation.
        If you are interested in a classical relations (comonoid homomorphism
            relations), you will need to take the relational
        converse of this output. See arXiv:1503.05857

Tests − If you want to dive deeper into an example

(* Prints the multiplication table for relation R *)
(* BE SURE TO SPECIFY THE SAME GROUPOIDS*)
(* INPUT: Relation R = {{0,0},{1,0}}; *)
(* OUTPUT: {{0,0},True} {{0,1},True} {{0,2},True}    {{0,3},True}
{{1,0},True}    {{1,1},True}    {{1,2},True}    {{1,3},True}
{{2,0},True}    {{2,1},True}    {{2,2},True}    {{2,3},True}
{{3,0},True}    {{3,1},True}    {{3,2},True}    {{3,3},True}

 *)
mTable[R_] := Grid[Table[
   Table[
    (* Specify groupoids as in isMonHom *)
    {{x, y}, eval[R, z2z2[x, y]] == z2z2z2Set[eval[R, x], eval[R, y]]}
    , {y, 0, nD − 1}]
   , {x, 0, nD − 1}]]

(* Pick a relation to test and see its multiplication table*)
R = {{0, 0}, {0, 4}, {1, 0}, {1, 4}, {2, 2}, {3, 2}};
mTable[R]
Union[eval[R, 0], eval[R, 2]] == {0, 2, 4}
```